\documentclass[twocolumn,aps,floatfix]{revtex4}
\usepackage{graphicx}
\usepackage{amsmath}
\usepackage{color}
\newcommand{\be}{\begin{equation}}
\newcommand{\ee}{\end{equation}}
\newcommand{\bea}{\begin{eqnarray}}
\newcommand{\eea}{\end{eqnarray}}
\newcommand{\bs}{\begin{split}}
\newcommand{\bse}{\begin{subequations}}
\newcommand{\ese}{\end{subequations}}

\begin{document}

\title{Influence of classical anisotropy fields on the properties of Heisenberg antiferromagnets within unified molecular field theory}
\author {David C. Johnston} 
\affiliation {Ames Laboratory and Department of Physics and Astronomy, Iowa State University, Ames, Iowa 50011, USA}

\date{\today}

\begin{abstract}

A comprehensive study of the influence of classical anisotropy fields on the magnetic properties of Heisenberg antiferromagnets within unified molecular field theory versus temperature~$T$, magnetic field~$H$, and anisotropy field parameter~$h_{\rm A1}$ is presented for systems comprised of identical crystallographically-equivalent local moments. The anisotropy field for collinear $z$-axis antiferromagnetic (AFM) ordering is constructed so that it is aligned in the direction of each ordered and/or field-induced thermal-average moment with a magnitude proportional to the moment, whereas that for XY~anisotropy is defined to be in the direction of the projection of the moment onto the $xy$ plane, again with a magnitude proportional to the moment.  Properties studied include the zero-field N\'eel temperature~$T_{\rm N}$, ordered moment, heat capacity and anisotropic magnetic susceptibility of the AFM phase versus~$T$ with moments aligned either along the $z$~axis or in the $xy$~plane.  Also determined are the high-field magnetization perpendicular to the axis or plane of collinear or planar noncollinear AFM ordering, the high-field magnetization along the $z$~axis of a collinear $z$-axis AFM, spin-flop (SF), and paramagnetic (PM) phases, and the free energies of these phases versus $T$, $H$, and $h_{\rm A1}$. Phase diagrams at $T=0$ in the $H_z$--$h_{\rm A1}$ plane and at $T>0$ in the $H_z$--$T$ plane are constructed for spins~$S=1/2$.  For $h_{\rm A1}=0$ the SF phase is stable at low field and the PM phase at high field with no AFM phase present.  As $h_{\rm A1}$ increases, the phase diagram contains the AFM, SF and PM phases.  Further increases in $h_{\rm A1}$ lead to the disappearance of the SF phase and the appearance of a tricritical point on the AFM--PM transition curve.  Applications of the theory to extract $h_{\rm A1}$ from experimental low-field magnetic susceptibility data and high-field magnetization versus field isotherms for single crystals of AFMs are discussed.

\end{abstract}

\maketitle

\section{Introduction}

Collinear and planar noncollinear Heisenberg antiferromagnets (AFMs) always have at least a small amount of some type of magnetocrystalline anisotropy present that establishes the axis or plane, respectively, along which the ordered magnetic moments are aligned with respect to the crystal axes.  These include single-ion anisotropy, spin exchange anistropy in spin space and anisotropy due to classical magnetic dipole interactions.  These anisotropies are known to change the AFM ordering (N\'eel) temperature $T_{\rm N}$ as well as the magnetic and thermal properties of the spin system \cite{Yosida1951,Kanamori1963}.  Recently we carried out comprehensive studies of the influence of dipolar and uniaxial quantum $DS_z^2$ magnetocrystalline anisotropies on the thermal and magnetic properties of Heisenberg AFMs  containing identical crystallographically-equivalent spins \cite{Johnston2016,Johnston2017}, where the Heisenberg interactions are treated within  unified molecular-field theory (MFT) \cite{Johnston2012,Johnston2015, Sangeetha2016}.  In this MFT the properties of collinear and planar noncollinear AFMs are calculated on the same footing and the theory is expressed in terms of directly measurable quantities instead of exchange interactions or molecular-field coupling constants \cite{Johnston2012,Johnston2015}.  The theory for $DS_z^2$ anisotropy applies  only to spins $S\geq 1$, a serious limitation, since the magnetic properties of $S=1/2$ systems are of great interest.

A generic classical uniaxial anisotropy field has been used sporadically in the past \cite{Keffer1966} to study the effects of anisotropy, but a comprehensive formulation of it and study of its influence on the thermal and static magnetic properties of Heisenberg AFMs are lacking. Here we report results from such investigations.  An important  advantage of this type of anisotropy is that such uniaxial and planar (XY) anisotropies apply to systems with $S=1/2$ in addition to $S\geq1$.  Another is that the anisotropy parameter in a system is much more easily derived from experimental magnetic data on single crystals compared to that for single-ion anisotropy.  The Heisenberg exchange interactions are treated within the unified MFT, again assuming identical crystallographically-equivalent spins. 

Results from the unified MFT of Heisenberg AFMs that are needed to develop the theory incorporating classical anisotropy fields are summarized in Appendix~\ref{Sec:UnifiedMFT}.  A summary of notation and thermodynamics expressions used in the paper are given in Sec.~\ref{Sec:Prelims}.  We use two forms of anisotropy field depending on whether the anisotropy field induces collinear AFM ordering along the $z$~axis or collinear or planar noncollinear AFM ordering in the $xy$~plane.  A detailed discussion of these is presented in Sec.~\ref{Sec:ClassAnisField}.  

Calculations of the AFM ordering (N\'eel) temperature~$T_{\rm N}$ and ordered moment versus temperature~$T$ in the presence of both the exchange and anisotropy fields in zero applied field~$H$ are given in Sec.~\ref{TNOrdMomCmagSmagFmag} for arbitrary antiferromagnets containing identical crystallographically-equivalent spins.  Laws of corresponding states for these properties and others are the same for all AFMs and ferromagnets (FMs) when expressed in terms of the universal reduced parameters of the unified MFT\@.  Expressions for the magnetic internal energy, heat capacity, entropy, and free energy of the AFM phase in zero field for both uniaxial and planar anisotropy are also derived and plotted in Sec.~\ref{TNOrdMomCmagSmagFmag}.  The anisotropic magnetic susceptibilities~$\chi$ arising from the classical anisotropy field are derived for the paramagnetic (PM) phase in Sec.~\ref{Sec:ChiPM} and for the AFM phase in Sec.~\ref{Sec:ChiAFM}, and the perpendicular high-field magnetizations for the PM and AFM states are calculated in Sec.~\ref{Sec:HiHPerpMAnis}.

The high-field magnetization parallel to the easy axis of a collinear AFM is of special interest. This is derived for the PM phase together with its free energy $F_{\rm mag}$ versus $H$ in Sec.~\ref{Sec:PMMomentFmag}.  The spin-flop (SF) phase is treated in Sec.~\ref{Sec:SpinFlop}, in which are presented the ordered moment versus $T$ in $H=0$, the thermal-average moment $\mu_{iz}$ versus $H$ using two different approaches, the spin-flop critical field $h_{\rm cSF}$ at which the SF phase exhibits a second-order transition to the PM phase with increasing $H$, the zero-field internal energy $U_{\rm mag}$ versus $T$, and the (Helmholtz) free energy $F_{\rm mag}$ versus $T$ and $H$.  The more involved calculations of the magnetic properties of the AFM phase in high longitudinal fields are given separately in Sec.~\ref{Sec:HiHParMAnisII}, including the $z$-axis sublattice, average and staggered moments, and $F_{\rm mag}$ versus~$T$, ~$H$, and anisotropy parameter~$h_{\rm A1}$.

Phase diagrams are constructed in Sec.~\ref{Sec:PhaseDiags}.  We start with the determination of the low-temperature properties of the AFM, SF, and PM phases and their dependences on the parameters of the MFT in Sec.~\ref{Sec:T0PhaseDiags}.  The $H_z$ versus $h_{\rm A1}$ phase diagrams at $T=0$ in the $H_z$--$h_{\rm A1}$ plane are then constructed.  In addition, $\mu_z$ versus $H_z$ plots are provided for various values of~$h_{\rm A1}$ to compare with experimental data at $T\ll T_{\rm N}$.  In this section, phase diagrams in the $H_\perp$--$h_{\rm A1}$ plane for fields $H_\perp$ perpendicular to the easy $z$~axis of a collinear AFM or easy plane of a planar noncollinear AFM are presented. 

We then move on to construct phase diagrams in the $H_z$--$T$ plane in Sec.~\ref{Sec:Hz-T phase diagrams} from free energy minimization with respect to the SF and AFM phases (the PM phases are high-field extensions of these phases beyond their respective critical fields).  Representative phase diagrams are presented for spins $S=1/2$ for six values of~$h_{\rm A1}$.  For $h_{\rm A1}=0$, the only stable phases with increasing~$H_z$ are the SF and higher-field PM phases, as expected.  With increasing $h_{\rm A1}$, the AFM phase appears at low fields for $T\leq T_{\rm N}$ followed by the SF and PM phases with increasing field.  Further increasing $h_{\rm A1}$ results in the gradual disappearance of the SF phase and appearance of  a tricritical point on the AFM--PM phase boundary.  When $h_{\rm A1}$ is sufficiently large, the SF phase disappears, leaving only the AFM and PM phases in the phase diagram with both first- and second-order transitions between them along the transition curve with a tricritical point separating the two regions.  At $T=0$ the AFM to PM transition is a 180$^\circ$ spin-flip transition of the moment initially opposite in direction to the field to being parallel to the field, whereas at finite~$T$ the transition is a ``gradual'' spin-flip where the magnitude of the initially oppositely-directed moment smoothly decreases to zero and then that moment increases with field in the direction of the field, eventually becoming the same in a second-order transition to the PM phase as that of the moment that was initially in the direction of the field.

A summary is given in Sec.~\ref{Sec:Summary}.  We discuss in depth how $h_{\rm A1}$ and another parameter $f_J$ can be derived from experimental data using our formulas for different magnetic properties. Also discussed are the relationships between the formulas for $T_{\rm N}$ and the Weiss temperature $\theta_{\rm p}$ in the Curie-Weiss law for the present classical anisotropy field treatment with those with $DS_z^2$ anisotropy \cite{Johnston2017} and arrive at a proportional relationship between $h_{\rm A1}$ and $D$ for small values of~$D$.  In general, magnetic anisotropy data are much easier to analyze in terms of the present classical anisotropy field than in terms of $DS_z^2$ anisotropy.

\section{\label{Sec:Prelims} Notation and Thermodynamics}

\subsection{Notation Summary}

Henceforth we designate two parameters changed by the presence of the anisotropy field by removing the subscript~$J$ to indicate that these values contain the contribution of the anisotropy field in zero applied field:
\bse
\label{Eqs:RedPars}
\be
T_{{\rm N}J} \to T_{\rm N},\qquad  \theta_{{\rm p}J} \to  \theta_{\rm p}.
\ee
The $T_{{\rm N}J}$, $\theta_{{\rm p}J}$ and $f_J$ parameters retain their meanings in terms of the Heisenberg exchange constants and magnetic structure as given in Eqs.~(\ref{Eq:TmGeneral}), (\ref{Eq:WeissTemp}) and~(\ref{Eq:fRatioDef}), respectively.  We normalize energies, fields and temperatures by $T_{{\rm N}J}$ in this paper, as given in the following summary and definitions of parameters. 

\bea
\bar{\mu}_\alpha &=& \frac{\mu_\alpha}{\mu_{\rm sat}} = \frac{\mu_\alpha}{gS\mu_{\rm B}},\label{Eq:barmualphaDef2}\\*
h_\alpha &\equiv& \frac{g\mu_{\rm B}H_\alpha}{k_{\rm B}T_{{\rm N}J}},\label{Eq:halphaDef}\\*
T_{\rm A1} &\equiv& \frac{g\mu_{\rm B}H_{\rm A1}}{k_{\rm B}},\label{Eq:TA1Dev}\\*
h_{\rm A1} &\equiv& \frac{T_{\rm A1}}{T_{{\rm N}J}} = \frac{g\mu_{\rm B}H_{\rm A1}}{k_{\rm B}T_{{\rm N}J}} \geq 0,\label{Eq:hA1Def}\\*
f_J &\equiv& \frac{\theta_{{\rm p}J}}{T_{{\rm N}J}},\label{Eq:fJDef}\\*
t &\equiv& \frac{T}{T_{{\rm N}J}},\\*
T_{\rm N} &=& T_{{\rm N}J}+T_{\rm A1}\label{Eq:TNDef}\\*
\frac{T_{\rm N}}{T_{{\rm N}J}} &=& 1 + \frac{T_{\rm A1}}{T_{{\rm N}J}} = 1 + h_{\rm A1},\label{Eq:TNATNRatio}\\*
t_{\rm A} &\equiv& \frac{T}{T_{\rm N}} = \frac{t}{1+h_{\rm A1}},\label{Eq:tADef}
\eea
\ese

The magnetic susceptibility per spin $\chi_\alpha$ in the $\alpha$ principal-axis direction is rigorously defined in the absence of a ferromagnetic component to the magnetization as
\be
\chi_\alpha =\lim_{H_\alpha\to0}\mu_\alpha(H_\alpha)/H_\alpha.
\ee
We define two reduced magnetic susceptibilities in the $\alpha$ principal-axis direction.  The first is
\bse
\label{Eqs:ReducedChi}
\be
\chi_\alpha^* \equiv \frac{\bar{\mu}_\alpha}{h_\alpha}\Big|_{h_\alpha\to0}.
\label{Eq:chiAlphaStarDef}
\ee
The second is
\be
\bar{\chi}_\alpha \equiv \frac{\chi_\alpha T_{{\rm N}J}}{C_1} = \left(\frac{3}{S+1}\right)\chi_\alpha^*,
\label{Eq:chiRedDef}
\ee
\ese
where the single-spin Curie constant~$C_1$ is given in Eq.~(\ref{Eq:CurieConst2}).

\subsection{Thermodynamics}

In this section we give thermodynamics expressions needed in this paper assuming that the ordered and/or induced moment of a representative spin $\vec{\mu}_i$ versus field and temperature has already been determined within the unified MFT as outlined in Appendix~\ref{Sec:UnifiedMFT} in the case of zero applied and anisotropy fields.

The magnetic internal energy $U_{\rm mag}$ of spin~$i$ for a local magnetic induction ${\bf B}_i$ in the $\alpha$ principal-axis direction is
\be
U_{{\rm mag}i} =-\mu_{i\alpha} B_{i\alpha},
\label{Eq:Umagi}
\ee
where here $B_{i\alpha}$ is written in general as
\be
B_{i\alpha} = \frac{1}{2}(H_{{\rm exch}i\alpha} + H_{{\rm A}i\alpha}) + H_\alpha,
\label{Eq:Bialpha}
\ee
and $H_{{\rm A}i\alpha}$ is the local anisotropy field seen by spin~$i$ discussed later. We have seen that the exchange field seen by a spin is proportional to $\mu_{i\alpha}$.  This is also true for the anisotropy field by assumption in Sec.~\ref{Sec:ClassAnisField} below.   Thus the parts of $U_{{\rm mag}i}$ associated with these fields are both proportional to $\mu_{i\alpha}^2$, indicating that they both ultimately arise from interactions between pairs of spins, hence the prefactor of 1/2 in the first term of Eq.~(\ref{Eq:Bialpha}) as discussed in regard to Eq.~(\ref{Eq:Umag0hA10}) where only the exchange field was present.  We write the sum of the exchange and anisotropy fields as
\be
H_{{\rm exch}i\alpha} + H_{{\rm A}i\alpha} = a\mu_{i\alpha},
\ee
where the constant~$a$ contains the parameters associated with these fields.  Then Eq.~(\ref{Eq:Bialpha}) becomes
\be
B_{i\alpha} = a\mu_{i\alpha} + H_\alpha.
\label{Eq:Bialpha2}
\ee

\subsubsection{Properties in Zero Applied Field}

When $H_\alpha=0$, Eqs.~(\ref{Eq:Umagi}) and~(\ref{Eq:Bialpha}) yield the internal energy per spin as 
\be
U_{\rm mag}(H_\alpha=0,T) =-\frac{a}{2}\mu_{\alpha}^2(T).
\ee
We always assume that the spins are identical and crystallographically equivalent, so the subscript~$i$ is suppressed when $H_\alpha=0$.  Then the magnetic heat capacity per spin $C_{\rm mag}$ is
\be
C_{\rm mag}(H_\alpha=0,T) = \frac{dU_{\rm mag}(H_\alpha=0,T)}{dT} = -a\mu_{\alpha}\frac{d\mu_{\alpha}}{dT}.
\ee
The magnetic entropy $S_{\rm mag}(H_\alpha=0,T)$ per spin is then obtained as
\bea
S_{\rm mag}(H_\alpha=0,T) &=& S_{\rm mag}(H_\alpha=0,T=0)\label{Eq:SmagH0T}\\*
&& +\ \int_0^T\frac{C_{\rm mag}(H_\alpha=0,T)}{T}dT,\nonumber
\eea
and the (Helmholtz) free energy $F_{\rm mag}(H_\alpha=0,T)$ as
\bea
F_{\rm mag}(H_\alpha=0,T) &=& U_{\rm mag}(H_\alpha=0,T) \label{Eq:FmagH0T}\\*
&& -\ T S_{\rm mag}(H_\alpha=0,T).\nonumber
\eea

\subsubsection{Properties at Nonzero Temperature and Nonzero Applied Field}

It is most convenient in this paper to calculate the thermodynamic properties in the $H_\alpha$-$T$ plane by choosing the path from $(H_\alpha=0,T=0)$ to $(H_\alpha=0,T)$ as in the previous section and then at constant~$T$ from $(H_\alpha=0,T)$ to $(H_\alpha,T)$.  The differential of the free energy for the second part of the path at constant~$T$, $dF_{\rm mag} = -S_{\rm mag}dT - \mu_\alpha dH_\alpha$ with $dT=0$, yields
\be
dF_{\rm mag}(H_\alpha,T) = - \mu_\alpha dH_\alpha.
\ee
Then using Eq.~(\ref{Eq:FmagH0T}) one obtains
\be
F_{\rm mag}(H_\alpha,T) = F_{\rm mag}(H_\alpha=0,T) - \int_0^{H_\alpha}\mu_\alpha(H_\alpha,T)dH_\alpha,
\label{Eq:FmagUnreduced}
\ee
where $F_{\rm mag}(H_\alpha=0,T)$ is found as described above.

The variation of the magnetic entropy with field at constant temperature is found from the Maxwell relation
\be
(dS_{\rm mag})_T = \left(\frac{\partial \mu_\alpha(H_\alpha,T)}{\partial T}\right)_{H_\alpha}dH_\alpha.
\label{Eq:dSmagFixT}
\ee
Then using Eq.~(\ref{Eq:SmagH0T}) one obtains
\bea
S_{\rm mag}(H_\alpha,T) &=& S_{\rm mag}(H_\alpha=0,T) \\*
&& +\ \int_0^{H_\alpha}\left(\frac{\partial \mu_\alpha(H_\alpha,T)}{\partial T}\right)_{H_\alpha}dH_\alpha.\nonumber
\eea

An increment of internal energy is
\be
dU_{\rm mag} = T dS_{\rm mag}-\mu_\alpha dH_\alpha.
\ee
Using Eq~(\ref{Eq:dSmagFixT}) for~$dS_{\rm mag}$ at fixed~$T$ gives 
\be
(dU_{\rm mag})_T = \left[T \left(\frac{\partial \mu_\alpha(H_\alpha,T)}{\partial T}\right)_{H_\alpha}-\mu_\alpha\right] dH_\alpha,
\ee
and hence
\bea
U_{\rm mag}(H_\alpha,T) &=& U_{\rm mag}(H_\alpha=0,T) \\*
&& \hspace{-0.3in} +\ \int_0^{H_\alpha}\left[T \left(\frac{\partial \mu_\alpha(H_\alpha,T)}{\partial T}\right)_{H_\alpha}-\mu_\alpha\right] dH_\alpha.\nonumber
\eea

In the free-energy expression~(\ref{Eq:FmagUnreduced}), the integral of $(\partial\bar{\mu}_\alpha(h_\alpha,t)/\partial t)_{h_\alpha}$ over $h_\alpha$ in $S_{\rm mag}$ and~$U_{\rm mag}$ is not present because it cancelled out in the definition $F_{\rm mag} = U_{\rm mag}-TS_{\rm mag}$.

\subsubsection{Expressions in Reduced Variables}

In order to formulate laws of corresponding states for the thermodynamic properties, we normalize all energies by $k_{\rm B}T_{{\rm N}J}$, where $T_{{\rm N}J}$ is the N\'eel temperature in zero field arising from exchange interactions alone as discussed in Appendix~\ref{Sec:UnifiedMFT}.  We also define the following dimensionless reduced variables
\bse
\bea
b_\alpha &=& \frac{g\mu_{\rm B}B_\alpha}{k_{\rm B}T_{{\rm N}J}},\\*
A &=& \frac{a}{k_{\rm B}T_{{\rm N}J}},\\*
b_\alpha &=& A\bar{\mu}_\alpha + h_\alpha.
\eea
\ese
Then also using Eqs.~(\ref{Eqs:RedPars}), the expressions in the above two subsections become
\bse
\label{Eqs:Thermodynamics}
\bea
\frac{U_{\rm mag}(h_\alpha=0,t)}{k_{\rm B}T_{{\rm N}J}} &=&-\frac{AS}{2}\bar{\mu}_{\alpha}^2(h_\alpha=0,t),\label{Eq:UmagGen}\\*
\frac{C_{\rm mag}(h_\alpha=0,t)}{k_{\rm B}} &=&-AS\bar{\mu}_{\alpha}(t)\frac{d\bar{\mu}_{\alpha}(h_\alpha=0,t)}{dt},\\*
\frac{S_{\rm mag}(h_\alpha=0,t)}{k_{\rm B}} &=&\frac{S_{\rm mag}(h_\alpha=0,t=0)}{k_{\rm B}}\\*
&&  +\ \int_0^t\frac{C_{\rm mag}(h_\alpha=0,t)/k_{\rm B}}{t}dt,\nonumber\\
\frac{F_{\rm mag}(h_\alpha=0,t)}{k_{\rm B}T_{{\rm N}J}} &=& \frac{F_{\rm mag}(h_\alpha=0,t)}{k_{\rm B}T_{{\rm N}J}} - t\frac{S_{\rm mag}(h_\alpha=0,t)}{k_{\rm B}}.\nonumber\\
\label{Eq:FmagHz0}\\
\frac{F_{\rm mag}(h_\alpha,t)}{k_{\rm B}T_{{\rm N}J}} &=& \frac{F_{\rm mag}(h_\alpha=0,t)}{k_{\rm B}T_{{\rm N}J}} \label{Eq:FmagGen}\\*
&& -\ S\int_0^{h_\alpha}\bar{\mu}_\alpha(h_\alpha,t)dh_\alpha,\nonumber\\*
\frac{S_{\rm mag}(h_\alpha,t)}{k_{\rm B}} &=&\frac{S_{\rm mag}(h_\alpha=0,t)}{k_{\rm B}} \\*
&& +\ S\int_0^{h_\alpha}\left(\frac{\partial\bar{\mu}_\alpha(h_\alpha,t)}{\partial t}\right)_{h_\alpha}dh_\alpha,\nonumber\\*
\frac{U_{\rm mag}(h_\alpha,t)}{k_{\rm B}T_{{\rm N}J}} &=& \frac{U_{\rm mag}(h_\alpha=0,t)}{k_{\rm B}T_{{\rm N}J}} \\*
&&\hspace{-0.9in} +\ S\int_0^{h_\alpha}\left[t\left(\frac{\partial\bar{\mu}_\alpha(h_\alpha,t)}{\partial t}\right)_{h_\alpha}-\bar{\mu}_\alpha(h_\alpha,t)\right]dh_\alpha.\nonumber
\eea
\ese

\section{\label{Sec:ClassAnisField} AFM Ordering in a Classical Anisotropy Field}

\begin{figure}
\includegraphics [width=1.5in]{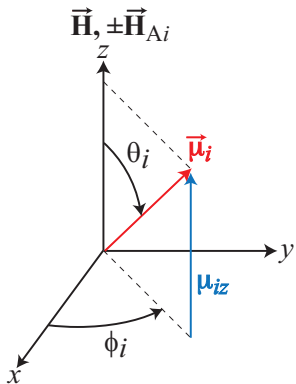}
\caption{(Color online) The orientation of a representative magnetic moment $\vec{\mu}_i$ described by spherical coordinates $\theta_i$ and $\phi_i$ in an applied magnetic field ${\bf H} = H_z\,\hat{\bf k}$ and a generic classical anisotropy field ${\bf H}_{{\rm A}i}$ directed along the $\pm z$-axis.  For such an anisotropy field collinear AFM ordering along the $z$~axis is favored if $H_z = 0$.   }
\label{Fig:chiParallel}
\end{figure}

\begin{figure}
\includegraphics [width=1.5in]{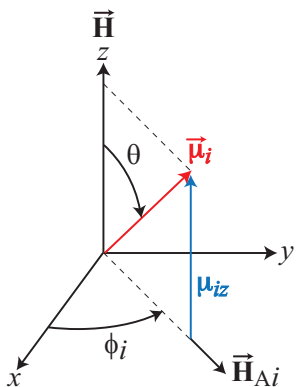}
\caption{(Color online) The orientation of a representative magnetic moment $\vec{\mu}_i$ in an applied magnetic field ${\bf H} = H\_z,\hat{\bf k}$ and an anisotropy field ${\bf H}_{{\rm A}}$ in the $xy$~plane that is directed along the projection of $\vec{\mu}_i$ onto the $xy$-plane as shown.  For such an anisotropy field collinear or planar noncollinear AFM ordering within the $xy$ plane is favored if $H_z = 0$. The azimuthal angle $\phi_i$ is in general different for different moments but the value for each moment is not affected by {\bf H}\@. }
\label{Fig:chiPerp2}
\end{figure}

The lowest-order uniaxial anisotropy free energy $F_{{\rm A}i}$ per spin associated with a uniaxial or planar anisotropy symmetry as in Figs.~\ref{Fig:chiParallel} and~\ref{Fig:chiPerp2}, respectively, for an ordered and/or magnetic field-induced thermal-average magnetic moment $\vec{\mu}_i$ is written as \cite{Kanamori1963}
\be
F_{{\rm A}i}=K_{1i}\sin^2\theta_i, 
\label{Eq:K1Def}
\ee
where $\theta_i$ is the polar angle between $\vec{\mu}_i$ and the uniaxial $z$-axis.  Here we assume that this relation is valid for the entire angular region $0\leq\theta\leq\pi/2$.  The $z$~axis for $F_{{\rm A}i}$ from which $\theta_i$ is defined is assumed to be a uniaxial axis of the lattice, and hence the anisotropy is fundamentally magnetocrystalline in origin.  This generic model is assumed to apply to spin systems with any spin angular momentum quantum number~$S$ (in units of $\hbar$ which is Planck's constant divided by $2\pi$) and can therefore treat systems with $S=1/2$ for which a magnetocrystalline $DS_z^2$ term in the Hamiltonian gives no anisotropy.  The anisotropy constant $K_1$ is in general different for different moments $\vec{\mu}_i$ because of their different magnitudes as discussed below, hence the subscripts~$i$ in Eq.~(\ref{Eq:K1Def}).  If $K_{1i}$ is positive and {\bf H} = 0, then the lowest free energy of a system occurs with $\sin\theta_i=0$ for all $\vec{\mu}_i$, for which the ordered moments are collinear and aligned parallel or antiparallel to the uniaxial $z$~axis, whereas if $K_{1i}$ is negative the lowest free energy occurs when $\sin\theta_i=90^\circ$ for all $\vec{\mu}_i$, resulting in collinear or coplanar ordering in the $xy$~plane.  Using Eq.~(\ref{Eq:K1Def}), the magnitude $\tau_{{\rm A}i}$ of the torque on each $\vec{\mu}_i$ by its anisotropy field ${\bf H}_{{\rm A}i}$ (see below) has the same form for all moments and is given by
\be
\tau_{{\rm A}i} = \left|\frac{\partial F_{{\rm A}i}}{\partial \theta}\right| =  2\left|K_{1i}\sin\theta_i\cos\theta_i\right|.
\label{Eq:tauA}
\ee

\subsection{\label{Sec:cAxisAnis} Collinear Ordering along the z~Axis: Uniaxial Anisotropy}

\begin{figure*}
\includegraphics [width=3.in]{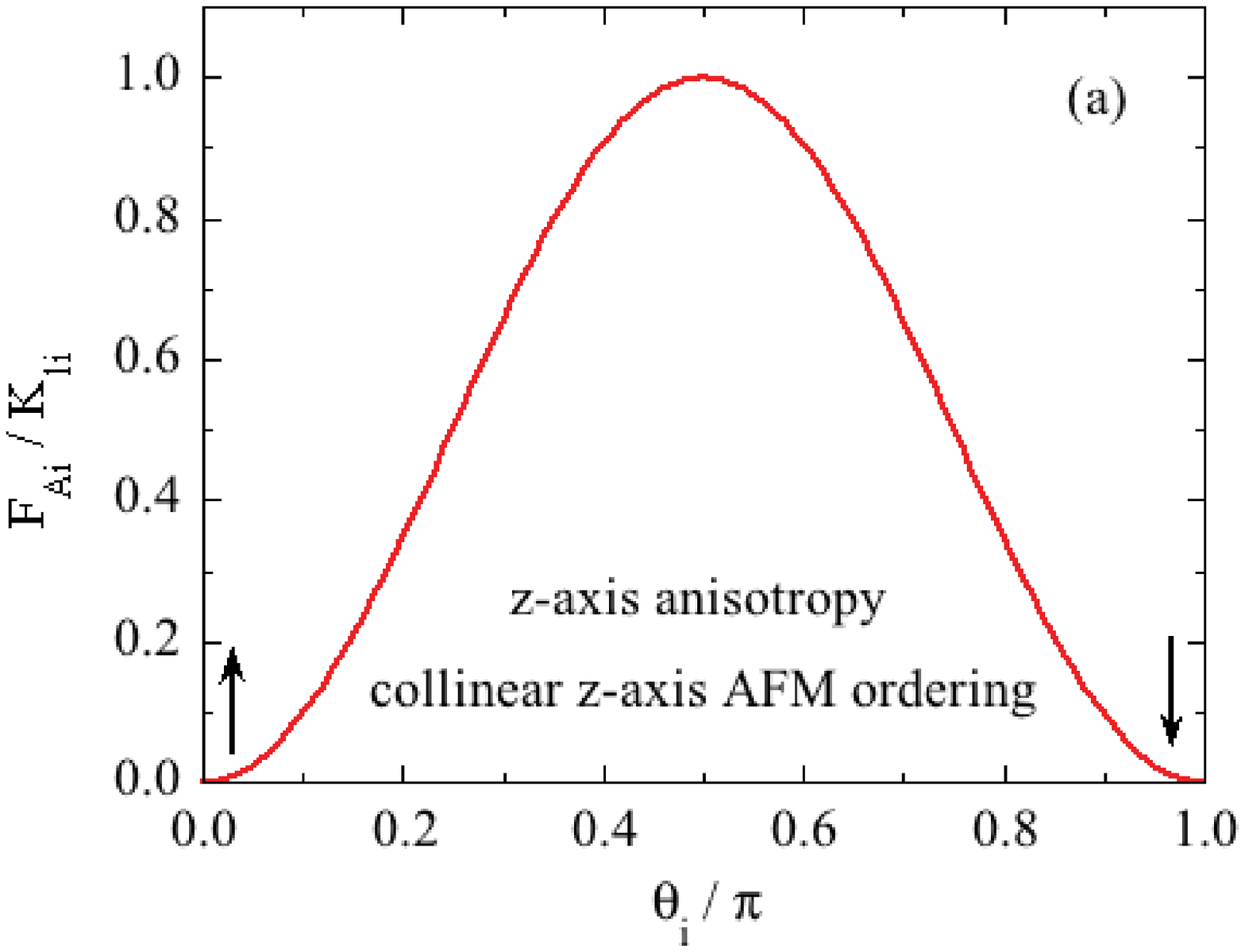}\includegraphics [width=3.in]{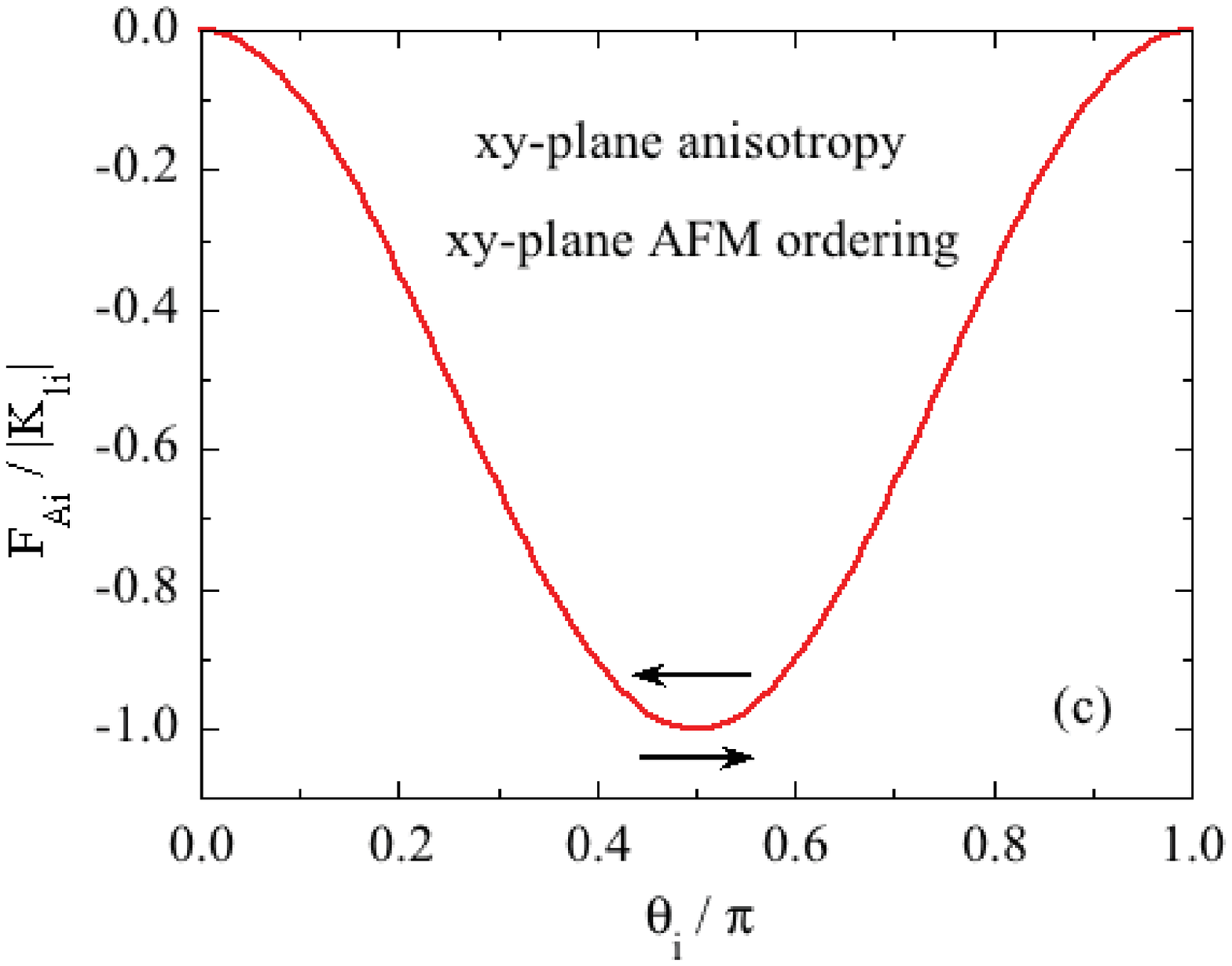}
\includegraphics [width=3.in]{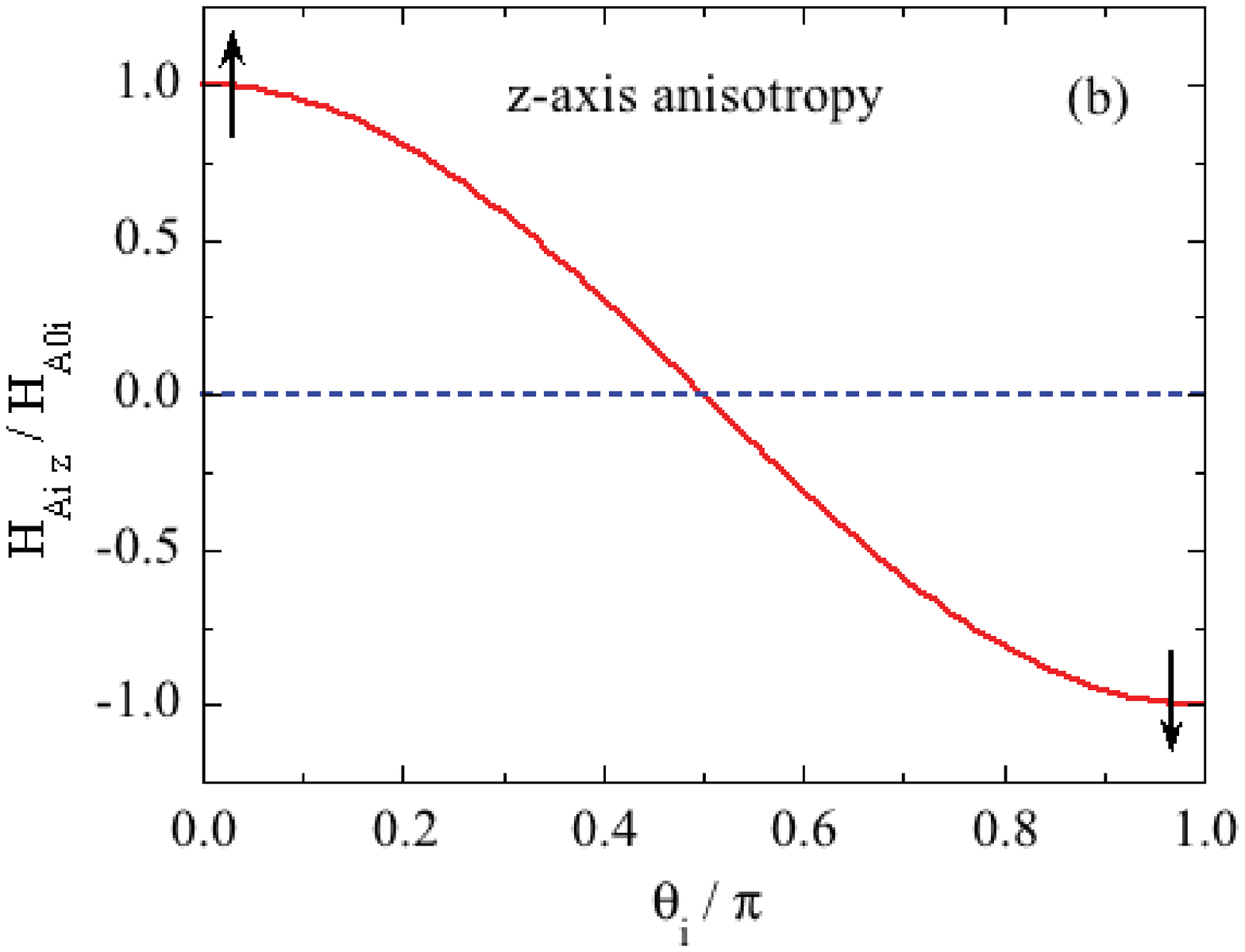}\includegraphics [width=3.in]{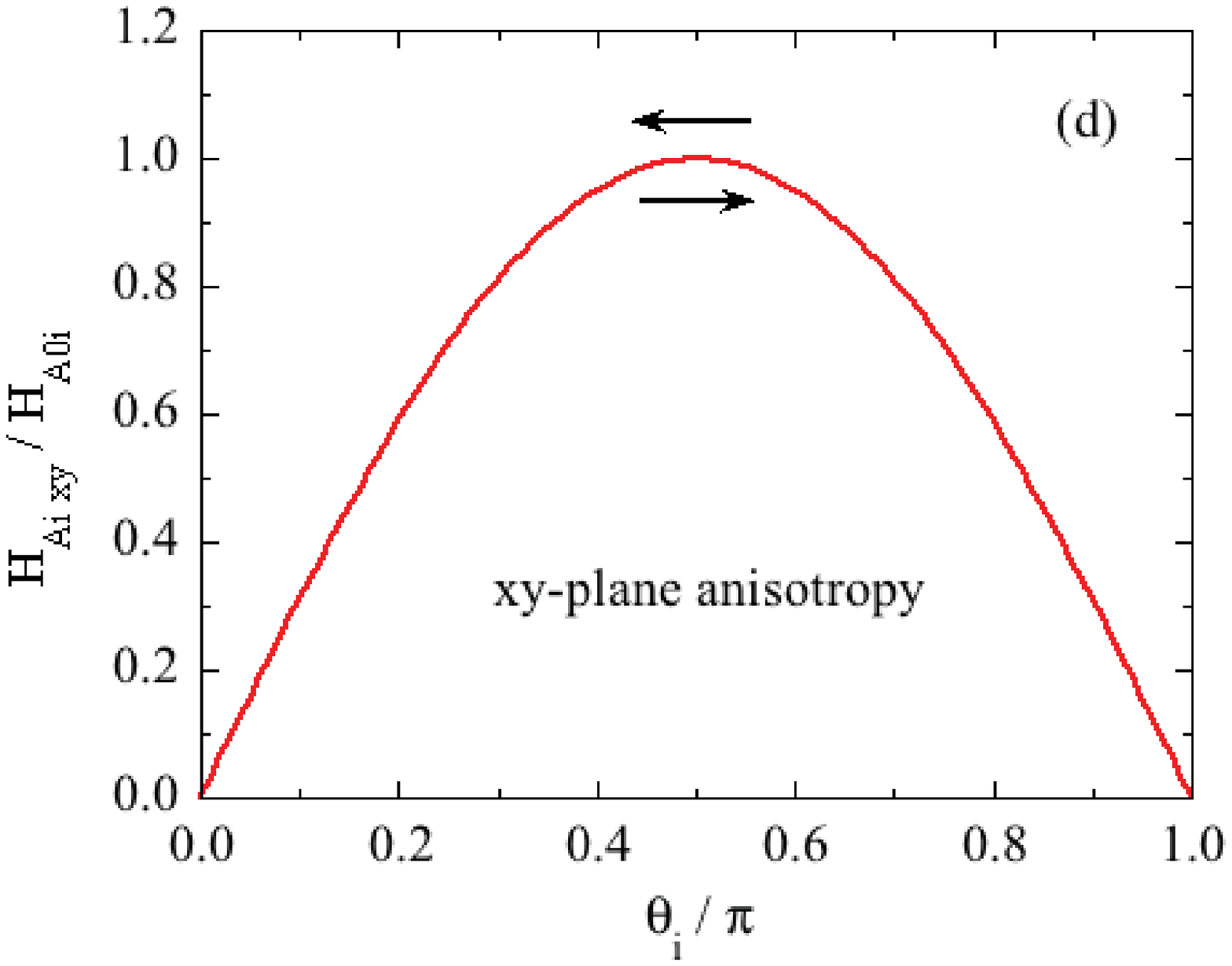}
\caption{(Color online) Comparisons of the free energy and anisotropy fields, respectively, for (a,b)~axial $z$-axis anisotropy and ordering and (c,d)~$xy$-plane anisotropy and ordering.  The anisotropy free energy per spin $F_{{\rm A}i}$ normalized by $|K_{1i}|$ is given in Eq.~(\ref{Eq:K1Def}), where $K_{1i}>0$ for axial anisotropy and $K_{1i}<0$ for planar anisotropy.  The anisotropy fields $H_{{\rm A}i}$ for axial and planar anisotropies are given in Eqs.~(\ref{Eqs:HAiAxial}) and~(\ref{Eqs:HAiGen}), respectively. }
\label{Fig:Anis_Free_energy}
\end{figure*}

For collinear AFM ordering along the $z$~axis in $H=0$ with uniaxial anisotropy, one has $\theta_i=0\ {\rm or}\ 180^\circ$ in Fig.~\ref{Fig:chiParallel}.  The anisotropy field ${\bf H}_{{\rm A}i}$ along the $z$~axis in such a collinear AFM is defined to be in the same direction $\pm \hat{\bf k}$ as that of the ordered moment $\vec{\mu}_i$, which can be written as
\bse
\label{Eqs:HAiAxial}
\bea
{\bf H}_{{\rm A}i} &=& H_{{\rm A0}i}\cos\theta_i\,\hat{\bf k},\label{Eq:HAiAxial}\\*
H_{{\rm A}iz} & = & H_{{\rm A0}i}\cos\theta_i,\label{Eq:HAiAxialb}
\eea
\ese
where $H_{{\rm A0}i}\geq 0$ is the amplitude of the anisotropy field for axial anisotropy.  For uniaxial ordering $K_{1i}>0$ in Eq.~(\ref{Eq:K1Def}), so that the minimum free energy $F_{{\rm A}i} = 0$ occurs for collinear AFM ordering with the moments oriented along the $z$~axis as shown in Fig.~\ref{Fig:Anis_Free_energy}(a).  If the moments all rotate with increasing field into a ``spin flop'' phase to give $\theta_i \lesssim 90^\circ$ for each spin, then from Eq.~(\ref{Eq:K1Def}) and Fig.~\ref{Fig:Anis_Free_energy}(a) the anisotropy free energy of each moment increases to $\approx K_{1i}$.   

Using Eq.~(\ref{Eq:HiHMuTEqs}) for a representative moment $\vec{\mu}_i$, the torque due to the anisotropy field on the moment tilted by an angle~$\theta$ with respect to the $z$~axis is
\bse
\label{Eqs:TauAi}
\be
\vec{\tau}_{{\rm A}i} = \vec{\mu}_i\times {\bf H}_{{\rm A}i} = \mu_i H_{{\rm A}0i}\sin\theta_i\cos\theta_i[\sin{\phi_i}\,\hat{\bf i} - \cos{\phi_i}\,\hat{\bf j}],
\ee
with magnitude
\be
\tau_{{\rm A}i} = \left|\mu_i H_{{\rm A0}i}\sin\theta_i\cos\theta_i\right|,
\label{Eq:tauAAxial}
\ee
\ese
where $\mu_i$ is the magnitude of the (thermal-average) $\vec{\mu}_i$ and $\theta_i$ is the polar angle in Fig.~\ref{Fig:chiParallel}.  Comparing Eqs.~(\ref{Eq:tauAAxial}) and~(\ref{Eq:tauA}) gives the anisotropy constant for moment~$i$ as
\be
K_{1i} = \frac{\mu_i H_{{\rm A0}i}}{2} > 0,
\label{Eq:K1Par}
\ee
where $K_{1i}$ is positive for uniaxial collinear ordering in zero field as discussed above.  As noted above, $K_1$ can depend on the specific moment~$i$ if the magnitude $\mu_i$ is not the same for all moments.

The maximum magnitude of ${\bf H}_{{\rm A}i}$ from Eqs.~(\ref{Eqs:HAiAxial}) occurs at $\theta_i=0$ or $180^\circ$, at which the anisotropy free energy in Eq.~(\ref{Eq:K1Def}) is minimum (zero) as shown in Fig.~\ref{Fig:Anis_Free_energy}(a).  A plot of $H_{{\rm A}iz}/H_{{\rm A0}i}$ versus~$\theta_i$ from Eq.~(\ref{Eq:HAiAxialb}) is shown in Fig.~\ref{Fig:Anis_Free_energy}(b), which by comparison with Fig.~\ref{Fig:Anis_Free_energy}(a) demonstrates that the maximum magnitude of the anisotropy field occurs at the ordering angles for collinear AFM ordering, for which the free energy is minimum.   

\subsection{Collinear or Planar Noncollinear Ordering in the xy~Plane: Planar Anisotropy}

When planar (XY) anisotropy is present, the ordered AFM  structure in $H=0$ can be either a collinear structure or a planar noncollinear structure with the ordered moments aligned in the $xy$~plane for both structures.  In either case the polar angle for the orientations of all ordered moments for $H=0$ is $\theta_i = 90^\circ$ in Fig.~\ref{Fig:chiPerp2}.  In order that these magnetic structures have a lower magnetic free energy than for collinear AFM ordering along the $z$~axis requires that 
\be
K_{1i} < 0
\label{Eq:K1neg}
\ee
in Eq.~(\ref{Eq:K1Def}), as shown in Fig.~\ref{Fig:Anis_Free_energy}(c).

{\it From Fig.~\ref{Fig:chiPerp2}, ${\bf H}_{{\rm A}i}$ is directed along the projection of $\vec{\mu}_i$ onto the $xy$~plane instead of along the $z$~axis as described in Eq.~(\ref{Eq:HAiAxial}) for uniaxial anisotropy.}  Therefore, instead of Eq.~(\ref{Eq:HAiAxial}), we now write ${\bf H}_{{\rm A}i}$ in spherical coordinates as
\bse
\label{Eqs:HAiGen}
\bea
{\bf H}_{{\rm A}i} &=& H_{{\rm A0}i}\sin\theta_i(\cos\phi_i\,\hat{\bf i}+\sin\phi_i\,\hat{\bf j}),\label{Eq:HAiGen}\\*
H_{{\rm A}i\,xy} &\equiv& H_{{\rm A0}i}\sin\theta_i, \label{Eq:HAiGenb}
\eea
\ese
where $H_{{\rm A0}i}$ is the magnitude of ${\bf H}_{{\rm A}i}$ when $\theta_i = 90^\circ$.  
The torque exerted by ${\bf H}_{{\rm A}i}$ on $\vec{\mu}_i$ is obtained from Eqs.~(\ref{Eq:HAiGen}) and~(\ref{Eqs:muimujExpand}) as
\be
\vec{\tau}_{{\rm A}i} = \vec{\mu}_i\times {\bf H}_{{\rm A}i} = -\mu_i H_{{\rm A0}i}\sin\theta_i\cos\theta_i(\sin\phi_i\,\hat{\bf i}- \cos\phi_i\,\hat{\bf j}),
\label{Eq:HAtorque}
\ee
with magnitude
\be
\tau_{{\rm A}i} = \left|\mu_i H_{{\rm A0}i}\sin\theta_i\cos\theta_i\right|.
\label{Eq:tauA2}
\ee
This is the same expression as in Eq.~(\ref{Eq:tauAAxial}) for collinear AFM ordering along the $z$~axis, but here the zero-torque condition applies to $\theta_i=\pi/2$ instead of~0 or~$\pi$ as appropriate for $z$-axis collinear ordering.

Comparing Eqs.~(\ref{Eq:tauA2}) and~(\ref{Eq:tauA}) and using~(\ref{Eq:K1neg}) gives
\be
K_{1i} = -\frac{\mu_i H_{{\rm A0}i}}{2} < 0,
\label{Eq:K1}
\ee
which is the same as in Eq.~(\ref{Eq:K1Par}) for axial anisotropy except for the sign.  A plot of $H_{{\rm A}i\,xy}/H_{{\rm A0}i}$ versus~$\theta_i$ from Eq.~(\ref{Eq:HAiGen}) is shown in Fig.~\ref{Fig:Anis_Free_energy}(d), which by comparison with Fig.~\ref{Fig:Anis_Free_energy}(c) demonstrates that the anisotropy field is maximum at the ordering angle $\theta_i = \pi/2$ for planar AFM ordering for which the free energy is minimum.

\subsection{Fundamental Anisotropy Field $H_{\rm A1}$}

In the present treatment of either uniaxial or planar anisotropy, we write the anisotropy field amplitude $H_{{\rm A0}i}\geq 0$ in Eqs.~(\ref{Eqs:HAiAxial}) and~(\ref{Eqs:HAiGen}) as
\bse
\be
H_{{\rm A0}i}(T) = \frac{3H_{\rm A1}}{S+1}\, \bar{\mu}_i(T) =  \frac{3H_{\rm A1}}{g\mu_{\rm B}S(S+1)}\, {\mu}_i(T),
\label{Eq:HA0}
\ee
where the subsidiary anisotropy field
\be
H_{\rm A1}\geq0
\ee
\ese
does not depend on the moment~$\vec{\mu}_i$ or on~$T$ and is therefore a more fundamental anisotropy field than $H_{{\rm A0}i}$.  The reason for including the factor $3/(S+1)$ in Eq.~(\ref{Eq:HA0}) is explained in Sec.~\ref{TNOrdMomCmagSmagFmag} below.  The reduced ordered moment $\bar{\mu}_i \equiv \mu_i/\mu_{\rm sat}$ can be numerically calculated for all moments in $H=0$ using Eq.~(\ref{Eq:barmuA}) below but the value can be different for different moments if $H\neq 0$.  Inserting Eq.~(\ref{Eq:HA0}) into~(\ref{Eq:K1Par}) or~(\ref{Eq:K1})  gives
\be
|K_{1i}| = \frac{3g\mu_{\rm B}SH_{\rm A1}}{2(S+1)} \bar{\mu}_i^2(T)= \frac{3H_{\rm A1}}{2g\mu_{\rm B}S(S+1)} \mu_i^2(T),\label{Eq:K12}
\ee
where we used Eq.~(\ref{Eq:barmualphaDef2}).  Since $\bar{\mu}_i(T = T_{\rm N})=0$ if $H=0$ where $T_{\rm N}$ is the N\'eel temperature in the presence of both exchange and anisotropy fields (see below), one has $K_{1i}(T\to T_{\rm N}^-) = 0$ if $H=0$ \cite{Oguchi1958}.  However, for $H > 0$ a field-induced thermal-averaged moment $\mu_i$ arises in the paramagnetic state at $T\geq T_{\rm N}$, and this anisotropy therefore influences both the AFM and PM (FM-aligned) states.

\section{\label{TNOrdMomCmagSmagFmag} N\'eel Temperature, Ordered Moment, Internal Energy, Heat Capacity, Entropy, and Free Energy of the Antiferromagnetic Phase in Zero Applied Field}

The definition of the anisotropy field ${\bf H}_{{\rm A}i}$ in Eq.~(\ref{Eq:HAiAxial}) for collinear AFM ordering along the $z$~axis ($\theta_i = 0$ or~$180^\circ$) and in Eq.~(\ref{Eq:HAiGen}) for ordering in the $xy$-plane shows that for $H=0$, ${\bf H}_{{\rm A}i}$ is parallel to each ordered magnetic moment $\vec{\mu}_i$ in the ordered state below $T_{\rm N}$, just as the exchange field ${\bf H}_{{\rm exch}\,i}$ is.  Since the local exchange and anisotropy fields are both in the same direction as that of the respective ordered moment in the AFM state in $H=0$, they reinforce each other, and also have the same values for each moment because all moments are identical and crystallographically equivalent by assumption.

For $H=0$ the parameters $\mu_0,\ \bar{\mu}_0,\ K_1$ and $H_{\rm A0}$ do not depend on the spin~$i$ and hence we drop the subscript~$i$ when discussing these quantities for $H=0$.  Here the parameters $\mu_0$ and $\bar{\mu}_0$ respectively refer to the ordered moment and reduced ordered moment in $H=0$ but in the presence of both the exchange and anisotropy fields as appropriate. 

From Eqs.~(\ref{Eqs:Hexchi050}) for the exchange field in $H=0$ together with Eq.~(\ref{Eq:CurieConst2}), one obtains 
\be
\frac{g\mu_{\rm B}H_{\rm exch0}}{k_{\rm B}T} = \frac{3T_{{\rm N}J}}{(S+1)T}\ \bar{\mu}_0.
\label{Eq:gmuHexchkT}
\ee
Using Eq.~(\ref{Eq:HA0}), a similar expression for the anisotropy field is
\be
\frac{g\mu_{\rm B}H_{\rm A0}}{k_{\rm B}T} = \frac{3g\mu_{\rm B}H_{\rm A1}}{(S+1)k_{\rm B}T}\ \bar{\mu}_0 = \frac{3T_{A1}}{(S+1)T}\ \bar{\mu}_0,
\label{Eq:gmuHA0kT}
\ee
where the anisotropy temperature $T_{\rm A1}$ (not a real temperature) is defined in terms of $H_{\rm A1}$ in Eq.~(\ref{Eq:TA1Dev}).  For $H=0$, the magnetic induction obtained by MFT that is seen by each moment is $B = H_{{\rm exch0}}+H_{\rm A0}$.  Using Eqs.~(\ref{Eq:gmuHexchkT}) and~(\ref{Eq:gmuHA0kT}), $\bar{\mu}_0$ is governed by the Brillouin function $B_S(y)$ according to Eqs.~(\ref{Eq:BS(y)}) as
\bea
\bar{\mu}_0 &=& B_S(y_0),\label{Eq:BrillH0HA}\\*
y_0 &=& \frac{3}{S+1}\left(T_{{\rm N}J} + T_{\rm A1}\right)\frac{\bar{\mu}_0}{T}.\nonumber
\eea

The ordering temperature occurs as $\bar{\mu}_0\to0$.  Using the first-order Taylor series expansion term of the Brillouin function in Eq.~(\ref{Eq:BSyTaylor}), Eq.~(\ref{Eq:BrillH0HA}) gives the N\'eel temperature $T = T_{\rm N}$ in the presence of both the exchange and anisotropy fields as
\label{Eqs:TA1Defs}
\be
T_{\rm N} = T_{{\rm N}J} + T_{\rm A1} = T_{{\rm N}J}(1+h_{\rm A1}),
\label{Eq:TNADef}
\ee
where $h_{\rm A1}$ is defined in terms of $T_{\rm A1}$ and $H_{\rm A1}$ in Eqs.~(\ref{Eq:hA1Def}) and~(\ref{Eq:TNATNRatio}).  Thus the presence of the reinforcing anisotropy field $h_{\rm A1}>0$ increases the N\'eel temperature, as expected.  From Eq.~(\ref{Eq:TNADef}), the fractional increase in the N\'eel temperature due to the anisotropy field, $\frac{T_{\rm N}}{T_{{\rm N}J}}-1$, is equal to $h_{\rm A1}$, an appealing physical interpretation of $h_{\rm A1}$.  This behavior is comparable to the influence of a $DS_z^2$ anisotropy on~$T_{\rm N}$ at small~$D$  where $T_{\rm N}$ is proportional to $D$, but is very different from the behavior of $T_{\rm N}$ versus~$D$ at larger~$D$ where $T_{\rm N}$ varies nonlinearly with $D$ \cite{Johnston2017}.  However, for the classical anisotropy treated in this paper both the ordering temperature~$T_{\rm N}$ and the Weiss temperature $\theta_{\rm p}$ (see below) vary linearly with $h_{\rm A1}$ in the same way for arbitrary values of $h_{\rm A1}$.

\begin{figure}
\includegraphics [width=3.in]{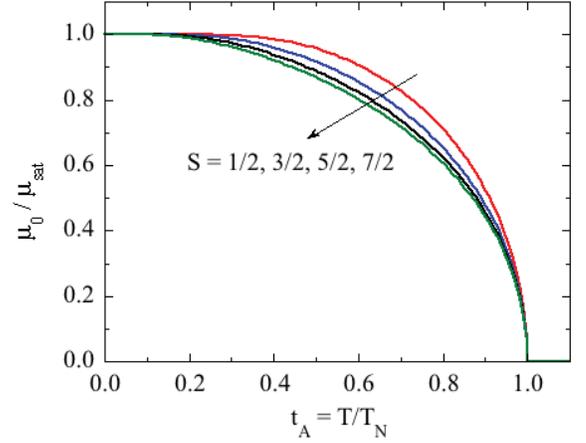}
\caption{(Color online) Reduced ordered moment $\bar{\mu}_0$ versus reduced temperature $t_{\rm A}$ in zero applied field but in the presence of an anisotopy field. These behaviors are valid within MFT for both uniaxial and planar anisotropies for any type of magnetic ordering of identical crystallographically-equivalent spins.}
\label{Fig:mubarVStA}
\end{figure}

To determine the zero-field ordered moment versus temperature for $T \leq T_{\rm N}$, we use Eqs.~(\ref{Eq:tADef}) and~(\ref{Eq:TNADef}) and Eq.~(\ref{Eq:BrillH0HA}) becomes
\bse
\label{Eqs:barmuA}
\bea
\bar{\mu}_0 &=& B_S(y_0),\label{Eq:barmuA}\\*
y_0 &=& \frac{3\bar{\mu}_0}{(S+1)t_{\rm A}}\label{Eq:y0A}.
\eea
\ese
This equation, which is used to numerically calculate $\bar{\mu}_0(t_{\rm A})$, has the same form as Eq.~(\ref{Eq:mubar0}) for $H=H_{\rm A1}=0$, except with $t_{\rm A}\equiv T/T_{\rm N}$ in Eq.~(\ref{Eq:tADef}) replacing $t\equiv T/T_{{\rm N}J}$ as shown in Fig.~\ref{Fig:mubarVStA} \cite{Johnston2011}.  {\it Hence the reason we introduced the factor of $3/(S+1)$ in the definition of the anisotropy field $H_{{\rm A0}i}$ in Eq.~(\ref{Eq:HA0}) was to require Eqs.~(\ref{Eqs:barmuA}) to have the same form as Eqs.~(\ref{Eq:mubar0})}.

To determine $\bar{\mu}_0$ in terms of $t = T/T_{{\rm N}J}$ instead of $t_{\rm A}=T/T_{\rm N}$, one can use Eqs.~(\ref{Eq:tADef}) and~(\ref{Eq:barmuA}) to obtain
\be
\bar{\mu}_0 = B_S\left[\frac{3\bar{\mu}_0(1+h_{\rm A1})}{(S+1)t}\right].
\label{Eq:barmu0TNJ}
\ee
Setting $h_{\rm A1}=0$, one recovers Eqs.~(\ref{Eq:mubar0}) for the case of zero anisotropy.

\begin{figure}
\includegraphics [width=3.in]{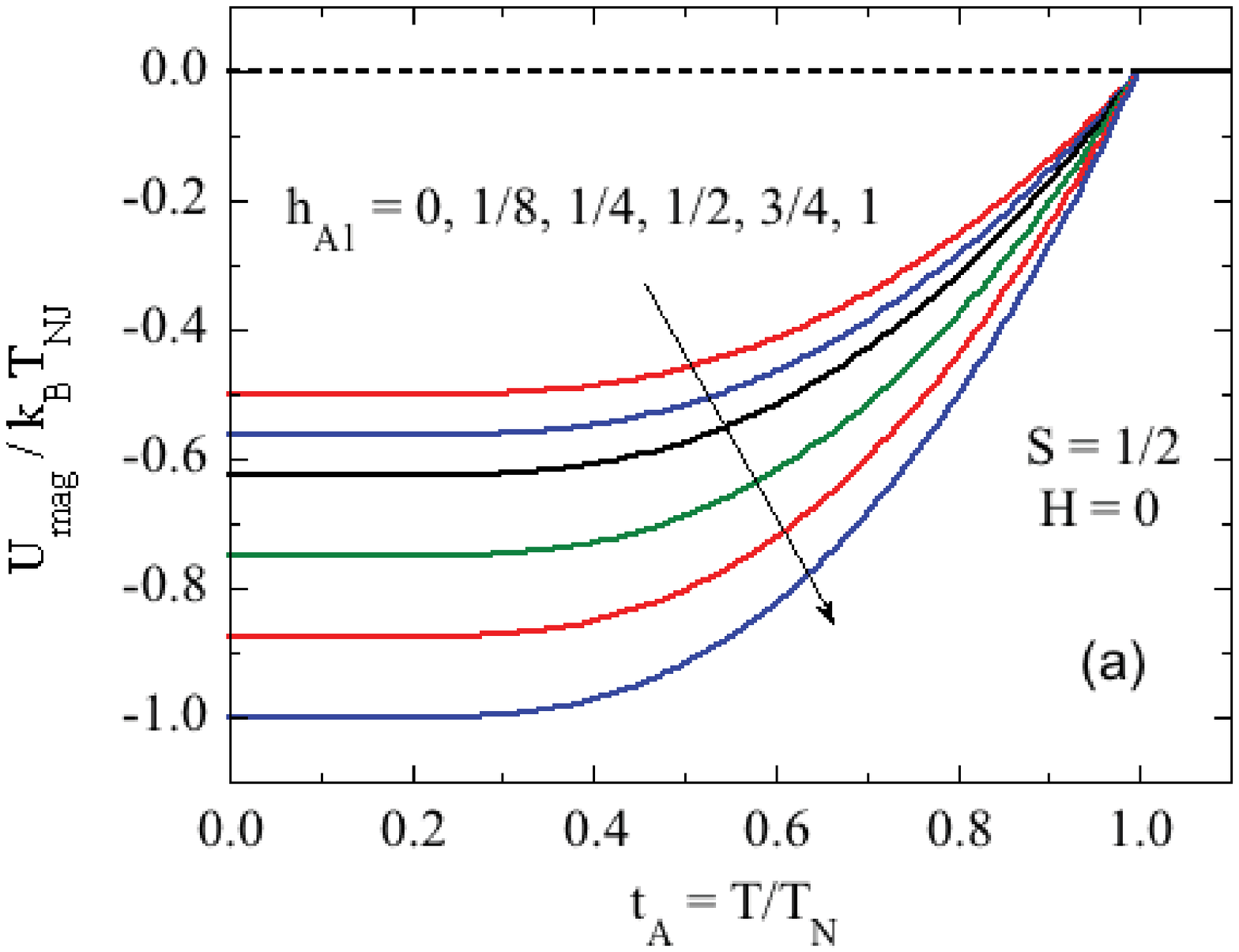}
\includegraphics [width=3.in]{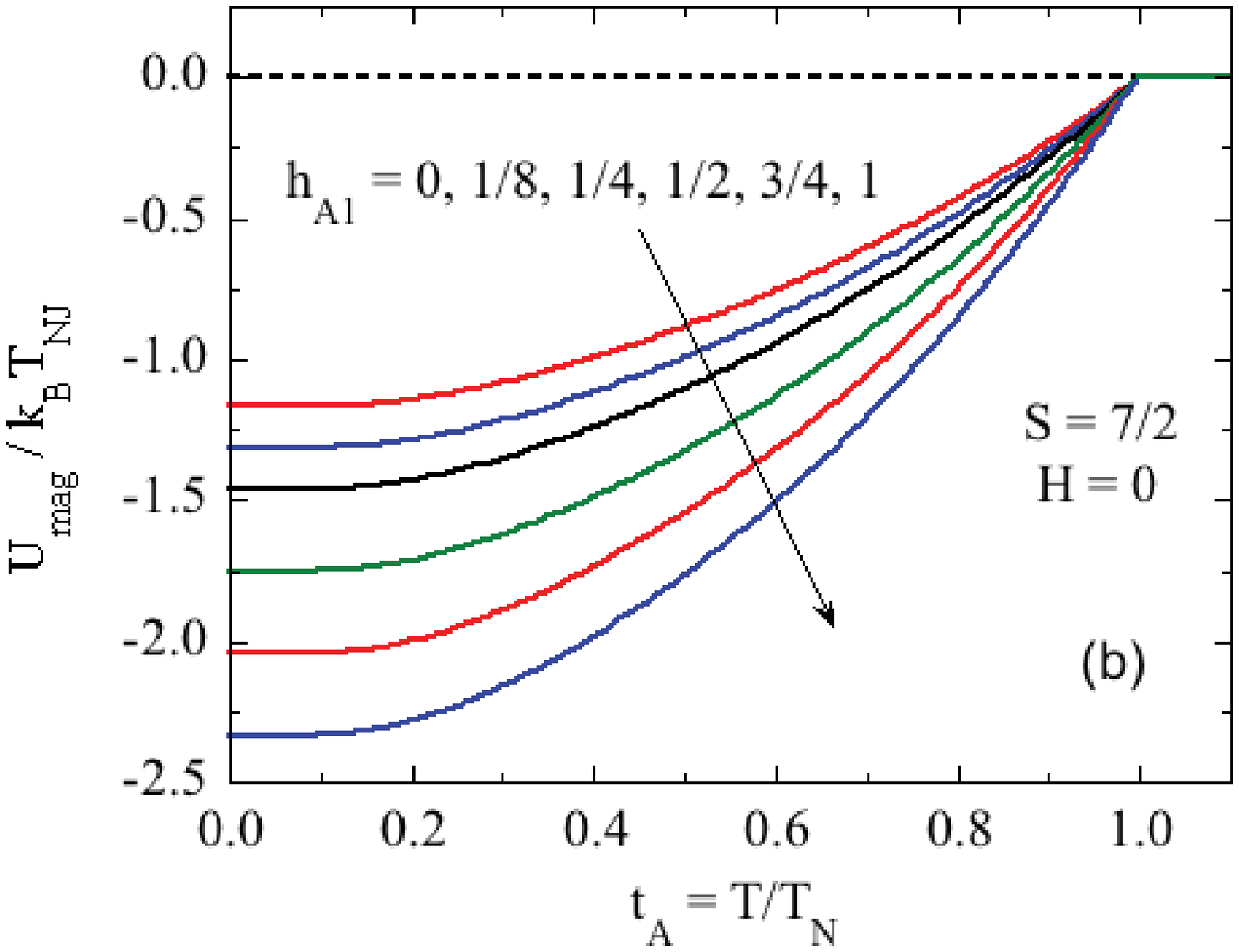}
\caption{(Color online) Magnetic internal energy per spin $U_{\rm mag}$ normalized by $k_{\rm B}T_{{\rm N}J}$ of the AFM phase versus reduced temperature $t_{\rm A}$ in zero applied field in the presence of a reduced anisotopy fields~$h_{\rm A1}=0$ to1 for spins (a)~$S=1/2$ and (b)~$S=7/2$ obtained using Eqs.~(\ref{Eqs:barmuA}) and~(\ref{Eq:Ui}). }
\label{Fig:UmagVstAS12hA1xx}
\end{figure}

In zero field all spins have the same internal energy per spin~$U_i$ according to Eq.~(\ref{Eq:Bialpha}), which has two contributions for either $z$-axis or $xy$-plane ordering given by
\bse
\label{Eqs:IntEnH0}
\bea
U_i &=& U_{{\rm exch}0} + U_{{\rm A}i} \label{Eq:EiSum}\\*
U_{{\rm exch}i} &=& -\frac{1}{2}\mu H_{{\rm exch0}},\label{Eexch0}\\*
U_{{\rm A}i} &=&  -\frac{1}{2}\mu H_{{\rm A0}i}.\label{Eq:E_A}
\eea
\ese
Normalizing the energies by $k_{\rm B}T_{{\rm N}J}$, Eqs.~(\ref{Eq:UmagAFMH0}), (\ref{Eqs:RedPars}), (\ref{Eq:HAiAxialb}) or (\ref{Eq:HAiGenb}), and~(\ref{Eq:HA0}) yield
\bse
\bea
\frac{U_{\rm exch0}}{k_{\rm B}T_{{\rm N}J}} &=& -\frac{3S}{2(S+1)}\bar{\mu}_0^2,\label{Eq:Eexch0a}\\*
\frac{U_{{\rm A}i}}{k_{\rm B}T_{{\rm N}J}} &=& -\frac{3S}{2(S+1)}h_{\rm A1}\bar{\mu}_0^2,\label{Eq:EAb}\\
\frac{U_i}{k_{\rm B}T_{{\rm N}J}} &=& -\frac{3S}{2(S+1)}(1+h_{\rm A1})\bar{\mu}_0^2.\label{Eq:Ui}
\eea
\ese

Shown in Fig.~\ref{Fig:UmagVstAS12hA1xx} are plots of $U_i/k_{\rm B}T_{{\rm N}J}$ versus reduced temperature~$t_{\rm A}$ for a range of reduced anisotropy parameters $h_{\rm A1}=0$ to~1 and for spins~$S=1/2$ and~$S=7/2$ obtained using Eqs.~(\ref{Eqs:barmuA}) and~(\ref{Eq:Ui}).  One sees that the zero-temperature internal energy decreases (becomes more stable)  with increasing~$h_{\rm A1}$ as expected.  Also, the internal energy goes to zero when the ordered moment goes to zero with increasing temperature.

\begin{figure}
\includegraphics [width=3.in]{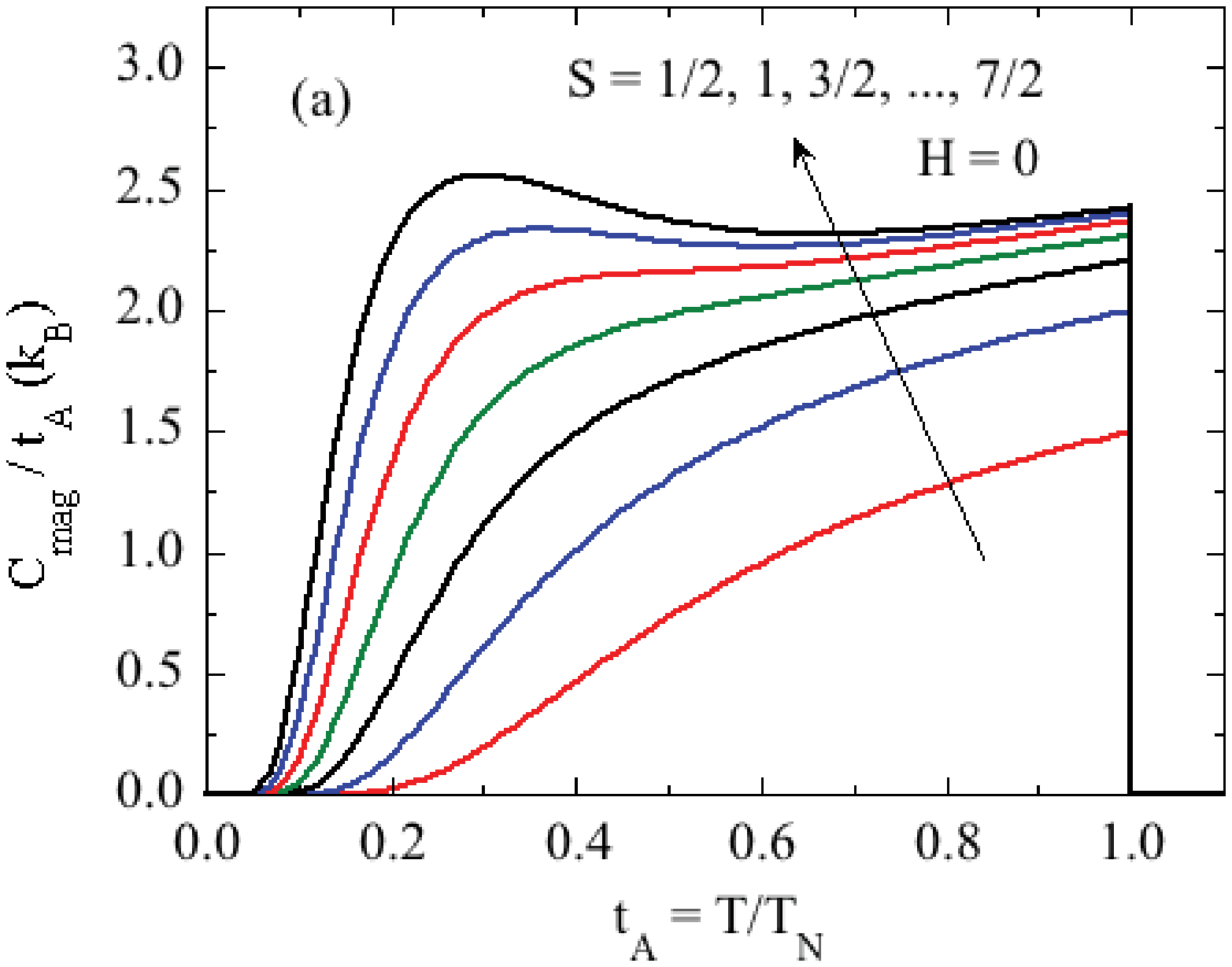}
\includegraphics [width=3.in]{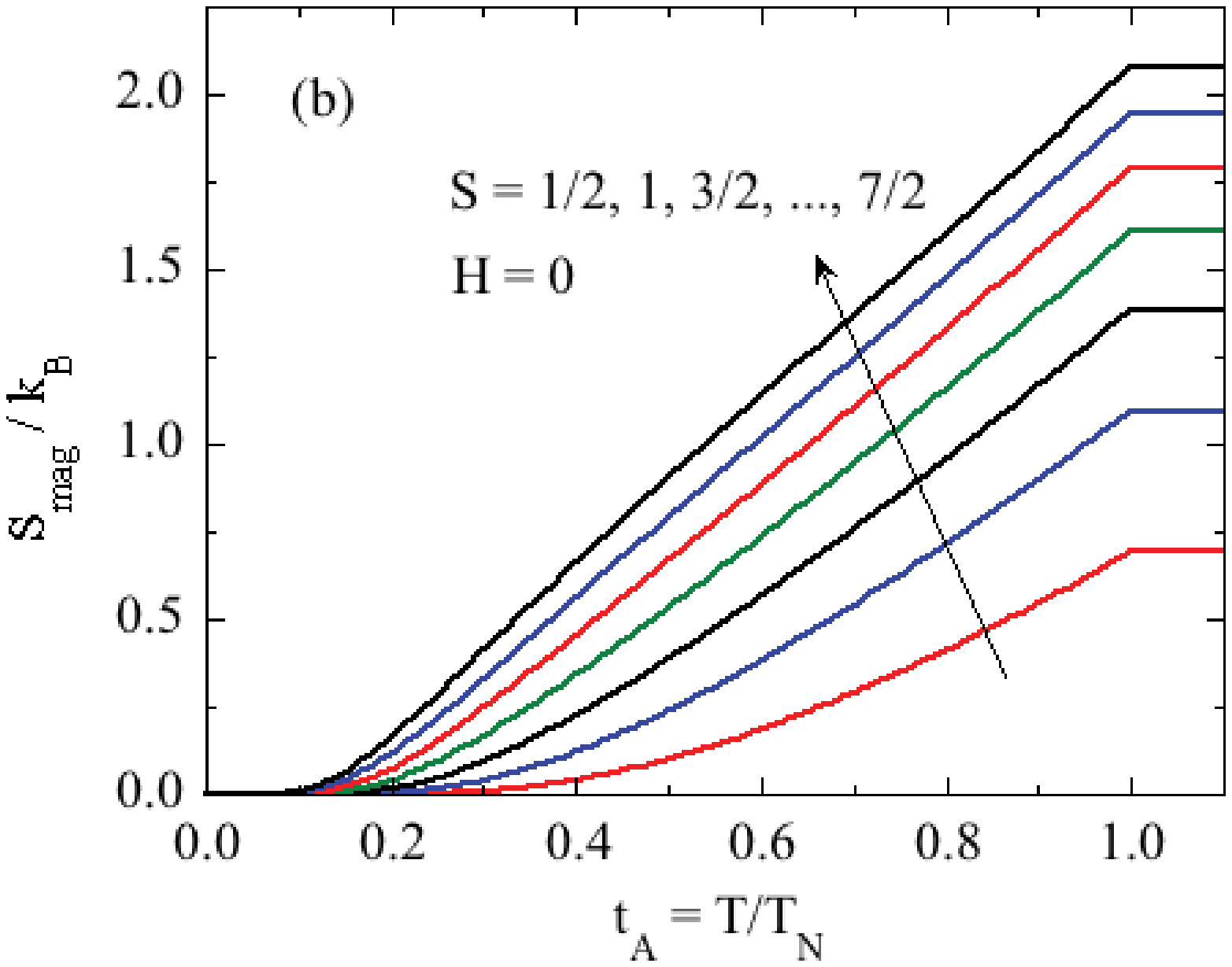}
\caption{(Color online) Magnetic heat capacity per spin $C_{\rm mag}$ of the AFM phase versus reduced temperature $t_{\rm A}$ in zero applied field for any reduced anisotopy field~$h_{\rm A1}\geq0$ and spins $S=1/2$ to~7/2 in half-integer increments. The hump that develops with increasing~$S$ at a temperature $\sim t_{\rm A}/4$ is intrinsic to molecular-field theory. (b)~Magnetic entropy per spin $S_{\rm mag}/k_{\rm B}$ versus~$t_{\rm A}$ for the same parameters as in~(a). }
\label{Fig:CmagSmagOntAFM_vs_tAhA10Sxx}
\end{figure}

The magnetic heat capacity per spin is
\bea
\frac{C_{\rm mag}}{k_{\rm B}} &=& \frac{d(U_i/k_{\rm B}T_{{\rm N}J})}{dt}\nonumber\\*
&=& -\left(\frac{3S}{S+1}\right)(1+h_{\rm A1})\bar{\mu}_0(t_{\rm A})\frac{d\bar{\mu}_0(t_{\rm A})}{dt},\nonumber\\*
&=& -\left(\frac{3S}{S+1}\right)\bar{\mu}_0(t_{\rm A})\frac{d\bar{\mu}_0(t_{\rm A})}{dt_{\rm A}},\label{Eq:CmagH0}
\eea
where we used Eq.~(\ref{Eq:tADef}) to obtain the third equality, $\bar{\mu}_0(t_{\rm A})$ is obtained by solving Eqs.~(\ref{Eqs:barmuA}) and $d\bar{\mu}_0(t_{\rm A})/dt_{\rm A}$ is obtained from  Eq.~(\ref{Eq:dBSy0}) where $y=y_0$ is given in Eq.~(\ref{Eq:y0A}).  Equation~(\ref{Eq:CmagH0}) for $C_{\rm mag}$ is identical in form to the equation for $C_{\rm mag}$ with $h_{\rm A1}=0$ and with $t$ replacing $t_{\rm A}$ \cite{Johnston2011}. The presence of $h_{\rm A1}$ in Eq.~(\ref{Eq:CmagH0}) is therefore equivalent to the replacements $T_{{\rm N}J}\to T_{\rm N}$ and~$t\to t_{\rm A}$ in the equation for $h_{\rm A1}=0$. Plots of $C_{\rm mag}/t_{\rm A}$ versus~$t_{\rm A}$ are shown for $S=1/2$ to $S=7/2$ in Fig.~\ref{Fig:CmagSmagOntAFM_vs_tAhA10Sxx}(a).  One sees that with increasing~$S$, on approaching $T_{\rm N}$ from below $C_{\rm mag}/t_{\rm A}$ approaches a constant value for increasing~$S$ given by
\be
\frac{C_{\rm mag}(t_{\rm A}\to1,S\to\infty)}{k_{\rm B}} = 5/2, 
\ee
consistent with the exact expression for finite~$S$ \cite{Johnston2015}
\be
\frac{C_{\rm mag}(t_{\rm A}\to1)}{k_{\rm B}} = \frac{5S(S+1)}{1+2S(S+1)}.
\ee 
The broad hump that develops in $C_{\rm mag}/k_{\rm B}t_{\rm A}$ at $t_{\rm A} \sim 1/4$ for large~$S$ is intrinsic to the MFT\@. It arises from a practical point of view in order that the statistical mechanics value for the magnetic entropy per spin at $T_{\rm N}$, given by
\be
S_{\rm mag}(t_{\rm A}=1)/k_{\rm B} = \ln(2S+1),
\label{Eq:SmagAtTN}
\ee
continues to increase with increasing~$S$, since as just stated the $C_{\rm mag}(t_{\rm A}\sim 1)$ is bounded with increasing~$S$ and hence the increasing entropy must arise by increasing $C_{\rm mag}$ at lower and lower temperatures with increasing~$S$\@.

The $S_{\rm mag}/k_{\rm B}$ versus~$t_{\rm A}$ for $h_{\rm A1}>0$ is obtained using
\be
\frac{S_{\rm mag}(t_{\rm A})}{k_{\rm B}} = \int_0^{t_{\rm A}}\frac{C_{\rm mag}(t_{\rm A})/k_{\rm B}}{t_{\rm A}}dt_{\rm A},
\ee
where $S_{\rm mag}(t_{\rm A}=0)=0$ because the energy levels are nondegenerate at $t_{\rm A}=0$ due to the presence of nonzero $H_{\rm exch}$ and~$H_{\rm A}$, and $C_{\rm mag}(t_{\rm A})/k_{\rm B}$ is obtained as described above.  The $S_{\rm mag}$ is plotted versus~$t_{\rm A}$ for $S=1/2$ to~$S=7/2$ in Fig.~\ref{Fig:CmagSmagOntAFM_vs_tAhA10Sxx}(b), where the high-$T$ limit in Eq.~(\ref{Eq:SmagAtTN}) is indeed obtained for each value of $S$ for $T\geq T_{\rm N}$\@.  

\begin{figure}
\includegraphics [width=3.in]{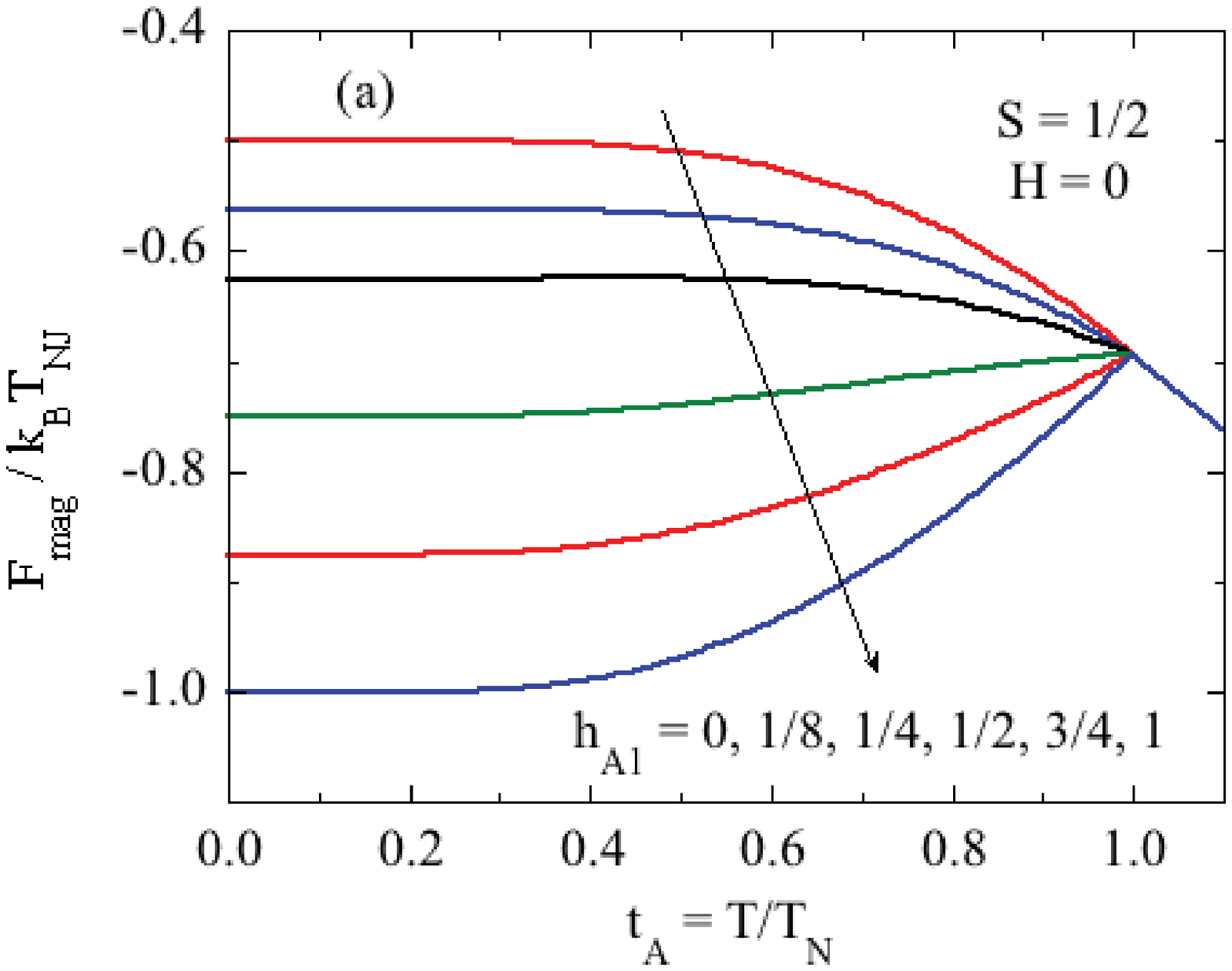}
\includegraphics [width=3.in]{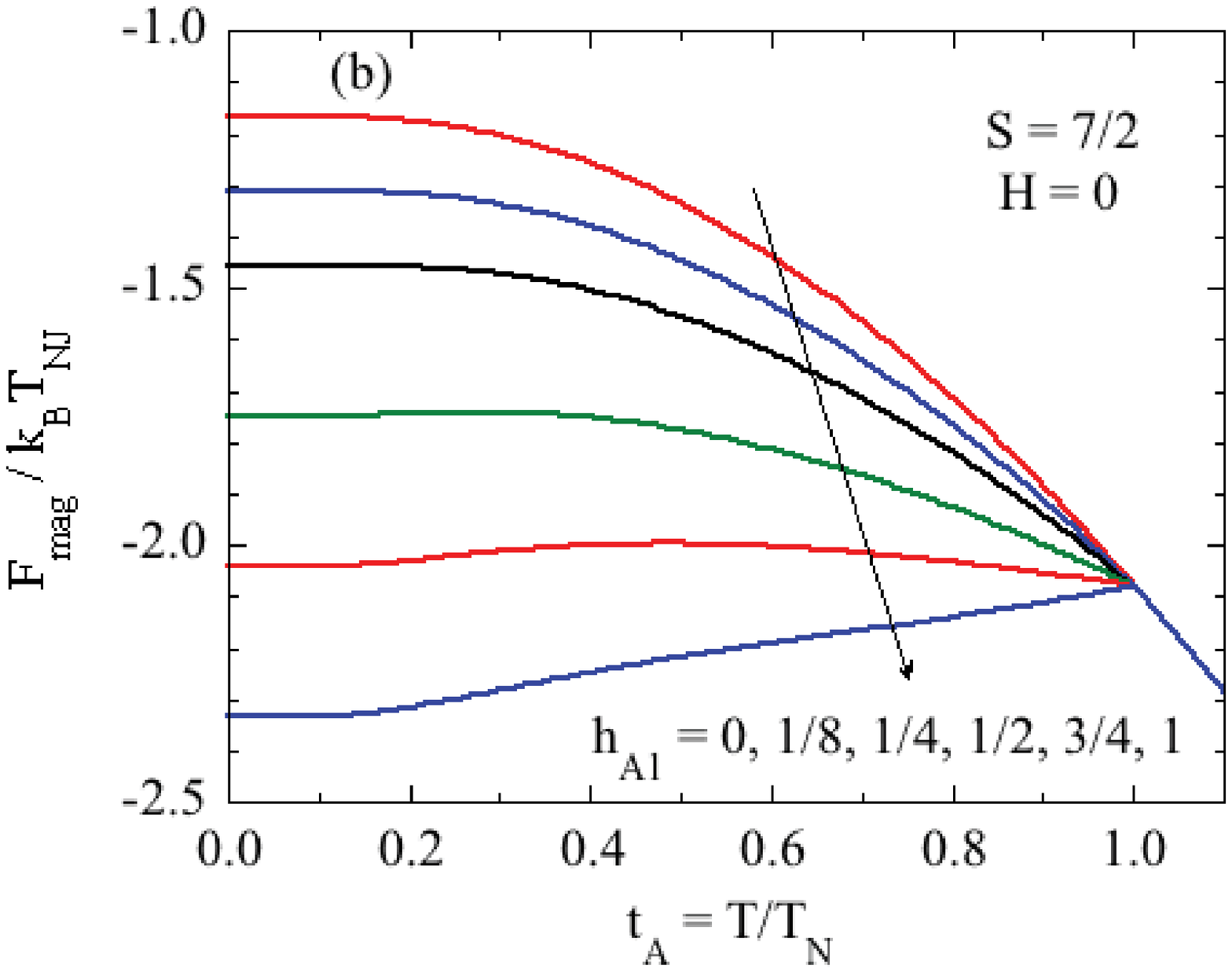}
\caption{(Color online) Reduced magnetic free energy per spin $F_{\rm mag}/k_{\rm B}T_{{\rm N}J}$ of the AFM phase versus reduced temperature $t_{\rm A}$ in zero applied field for anisotopy fields~$h_{\rm A1}$ as listed and spins (a)~$S=1/2$ and (b)~$S=7/2$, obtained using Eq.~(\ref{Eq:fmagH0}) and the data in Figs.~\ref{Fig:UmagVstAS12hA1xx} and~\ref{Fig:CmagSmagOntAFM_vs_tAhA10Sxx}.}
\label{Fig:FmagVstAS72hA1xx}
\end{figure}

The reduced Helmholtz free energy per spin versus reduced temperature~$t_{\rm A}$ is given in general by
\be
\frac{F_{\rm mag}}{k_{\rm B}T_{{\rm N}J}} = \frac{U_{\rm mag}}{k_{\rm B}T_{{\rm N}J}} - t_{\rm A}\frac{S_{\rm mag}}{k_{\rm B}}.
\label{Eq:fmagH0}
\ee
Shown in Fig.~\ref{Fig:FmagVstAS72hA1xx} are plots of $F_{\rm mag}/k_{\rm B}T_{{\rm N}J}$ for $H=0$ versus $t_{\rm A}$ with $h_{\rm A1}$ values from~0 to~1 for spins $S=1/2$ and $S=7/2$ obtained from the data in Figs.~\ref{Fig:UmagVstAS12hA1xx} and~\ref{Fig:CmagSmagOntAFM_vs_tAhA10Sxx}.  One sees that $F_{\rm mag}$ varies monotonically with $t_{\rm A}$, but that the sign of the slope depends on the value of $h_{\rm A1}$.  Another important feature is that $F_{\rm mag}$ is independent of $h_{\rm A1}$ for $t_{\rm A}\geq 1$ because $U_{\rm mag}=0$ in that temperature range and $S_{\rm mag}$ versus~$t_{\rm A}$ is independent of $h_{\rm A1}$ for a given value of the spin~$S$ because the influence of $h_{\rm A1}$ is already included via its effect on~$T_{\rm N}$ in the definition $t_{\rm A}\equiv T/T_{\rm N}$\@.

\section{\label{Sec:ChiPM} Magnetic Susceptibility of the Paramagnetic Phase}

In the paramagnetic (PM) phase at $T\geq T_{\rm N}$, there is no ordered or induced moment in the absence of a field {\bf H} applied along a principal-axis direction.  When $H_\alpha>0$, the field-induced thermal-average moment of each spin points in the direction of {\bf H}.  From Eq.~(\ref{Eq:Hexch:z0}), the magnitude of the exchange field seen by each moment is
\be
H_{{\rm exch}\alpha} = \frac{3k_{\rm B}\theta_{{\rm p}J}}{g\mu_{\rm B}(S+1)}\,\bar{\mu}_\alpha,
\ee
where $\theta_{{\rm p}J}$ is the Weiss temperature due to the exchange interactions alone, which is defined in terms of the exchange constants in the spin system in Eq.~(\ref{Eq:WeissTemp}), and $\bar{\mu}_\alpha = \mu_\alpha/\mu_{\rm sat} = \mu_\alpha/gS\mu_{\rm B}$ is the normalized thermal-average moment induced by $H_\alpha$ in the $\alpha$ direction.

\subsection{Anisotropic Paramagnetic Susceptibility with a Uniaxial Anisotropy Field Along the $z$~Axis}

\subsubsection{${\bf H} \perp z$.}

Here we consider a uniaxial anisotropy field ${\bf H}_{{\rm A}i}$ along the $z$~axis as in Eq.~(\ref{Eq:HAiAxial}) and Fig.~\ref{Fig:chiParallel} with the induced moments in the PM state with $T\geq T_{\rm N}$ aligned perpendicular to the $z$~axis due to an infinitesimal {\bf H} applied in the $xy$~plane. According to Eqs.~(\ref{Eqs:TauAi}) with $\theta_i=90^\circ$, the  torque of ${\bf H}_{{\rm A}i}$ on $\vec{\mu}_i$ is zero.  Hence the anisotropy field has no influence on $\mu_\perp$, where the $\hat{\perp}$ direction is perpendicular to the easy axis or plane for AFM ordering.  Therefore the low-field susceptibility $\chi_\perp$ follows the Curie-Weiss law given by Eq.~(\ref{Eq:ChialphaPM}) for exchange interactions alone as
\bse
\label{Eqs:ChiabAxAnis}
\be
\chi_{\rm \perp PM}(T\geq T_{\rm N}) \equiv \chi_{xy}(T\geq T_{\rm N}) = \frac{C_1}{T-\theta_{{\rm p}J}}.
\label{Eq:chiABAxial}
\ee
The $xy$-plane susceptibility at $T_{\rm N}$ is thus
\be
\chi_{\rm \perp PM}(T_{\rm N}) = \frac{C_1}{T_{\rm N}-\theta_{{\rm p}J}} = \frac{C_1}{T_{{\rm N}J} + T_{\rm A1}-\theta_{{\rm p}J}},
\label{Eq:chiABAxialTNA}
\ee
\ese
where we used Eq.~(\ref{Eq:TNADef}) for~$T_{\rm N}$ to obtain the second equality. The presence of the infinitesimal $H_\perp$ does not measurably affect $T_{\rm N}$.  The reduced susceptibilities defined in Eqs.~(\ref{Eqs:ReducedChi}) are
\bse
\bea
\bar{\chi}_{\rm \perp PM}(T\geq T_{\rm N})  &\equiv& \frac{\chi_{\rm \perp PM} T_{{\rm N}J}}{C_1} \\*
&=& \frac{1}{t-f_J} = \frac{1}{t_{\rm A}(1+h_{\rm A1})-f_J},\nonumber\\*
\bar{\chi}_{\rm \perp PM}(T= T_{\rm N}) &=&\frac{1}{1+h_{\rm A1}-f_J},\label{Eq:barchiPerpPMTTN}\\*
\chi_{\rm \perp PM}^*(T\geq T_{\rm N}) &\equiv& \frac{\bar{\mu}_\perp}{h_\perp} = \left(\frac{S+1}{3}\right)\bar{\chi}_{\rm \perp PM}(T\geq T_{\rm N})  \nonumber\\*
\\*
&=& \frac{S+1}{3(t-f_J)} = \frac{S+1}{3[t_{\rm A}(1+h_{\rm A1})-f_J]},\nonumber\\*
\chi_{\rm \perp PM}^*(T= T_{\rm N}) &=& \frac{S+1}{3(1+h_{\rm A1}-f_J)}.
\eea
\ese

\subsubsection{${\bf H} \parallel z$.}

If {\bf H} is along the $z$-axis, then an anisotropy field in the direction of {\bf H} and of the induced moment is present with magnitude $H_{\rm A0}$ given by Eq.~(\ref{Eq:HA0}).  The normalized induced moment in the $z$-direction ($\bar{\mu}_\parallel$) is given by Eqs.~(\ref{Eq:BS(y)}), (\ref{Eq:Hexch:z}), (\ref{Eq:HA0}) and~(\ref{Eq:TA1Dev}) as
\bea
\bar{\mu}_{\rm \parallel PM}&=& B_S\left[\frac{g\mu_{\rm B}}{k_{\rm B}T}(H_{\rm exch} + H_{\rm A0} + H_z)\right]\label{Eq:mucPar}\\*
&=& B_S\left[\frac{3}{S+1}\left(\theta_{{\rm p}J} + T_{\rm A1}\right)\frac{\bar{\mu}_\parallel}{T} + \frac{g\mu_{\rm B}H_z}{k_{\rm B}T}\right].\nonumber
\eea
Using the first-order term in the Taylor series expansion of the Brillouin function in Eq.~(\ref{Eq:BSyTaylor}) one obtains the Curie-Weiss law
\bse
\label{Eqs:chicA}
\bea
\chi_{\rm \parallel PM}(T\geq T_{\rm N}) &=& \frac{\mu_\parallel}{H_z} = \frac{C_1}{T-\theta_{\rm p}},\label{Eq:Chicaxial}\\*
\chi_{\rm \parallel PM}(T_{\rm N}) &=& \frac{C_1}{T_{\rm N}-\theta_{\rm p}}\label{Eq:ChicaxialTNA}\\*
&=& \frac{C_1}{T_{{\rm N}J}(1+h_{\rm A1})-\theta_{\rm p}},
\eea
where the Weiss temperature in the presence of the anisotropy is
\bea
\theta_{\rm p} &=& \theta_{{\rm p}J} + T_{\rm A1} = \theta_{{\rm p}J} + \theta_{\rm pA}\label{Eq:thetapJ+thetapA} \\*
&=&\theta_{{\rm p}J} + h_{\rm A1}T_{{\rm N}J},\nonumber\\*
\theta_{\rm pA} &=& T_{\rm A1} = h_{\rm A1}T_{{\rm N}J}.\label{Eq:thetpAhA1}
\eea
\ese
Equations~(\ref{Eqs:chicA}) yield the reduced forms~(\ref{Eqs:ReducedChi}) as
\bse
\bea
\bar{\chi}_{\rm \parallel PM}(T\geq T_{\rm N}) &=& \frac{1}{(1+h_{\rm A1})t_{\rm A}-f_J-h_{\rm A1}},\\*
\bar{\chi}_{\rm \parallel PM}(T = T_{\rm N}) &=& \frac{1}{1-f_J},\label{Eq:barchiparPMTN}\\*
\chi_{\rm \parallel PM}^*(T\geq T_{\rm N}) &=& \frac{S+1}{3[(1+h_{\rm A1})t_{\rm A}-f_J-h_{\rm A1}]},\\*
\chi_{\rm \parallel PM}^*(T = T_{\rm N}) &=& \frac{S+1}{3(1-f_J)}.
\eea
\ese

Thus the Weiss temperatures from the exchange interactions and from the anisotropy are additive.  This additivity also occurs for anisotropy arising from the magnetic dipole interaction \cite{Johnston2016} and from the uniaxial $DS_z^2$ single-ion anisotropy at small $D$ \cite{Johnston2017}.  From Eqs.~(\ref{Eq:Chicaxial}) and~(\ref{Eq:thetapJ+thetapA}), one sees that the $z$-axis anisotropy field in the direction of ${\bf H}$ increases $\chi_{\rm \parallel PM}$ at fixed $T$, as expected since the anisotropy field increases the magnitude of the local magnetic induction seen by each induced moment.

In addition, one finds that $T_{\rm N}$ in Eq.~(\ref{Eq:TNADef}) and $\theta_{\rm p}$ in Eq.~(\ref{Eq:thetapJ+thetapA}) for {\bf H} directed along the $z$~axis are both shifted towards positive values by the same amount due to the anisotropy field, and therefore
\be
T_{\rm N}-\theta_{\rm p} = T_{{\rm N}J}-\theta_{{\rm p}J} \qquad ({\bf H}\parallel {\rm easy\ axis}).
\label{EqTNATN}
\ee
By comparing Eqs.~(\ref{Eq:chiABAxial}) and~(\ref{Eq:Chicaxial}), the Weiss temperatures are seen to be different for $\chi_{\rm \perp PM}$ and $\chi_{\rm\parallel PM}$ and hence Eq.~(\ref{EqTNATN}) applies for ${\bf H}\parallel z$ but not for ${\bf H}\perp z$.  From the definition for~$f_J$ in Eq.~(\ref{Eq:fRatioDef}) together with Eq.~(\ref{EqTNATN}), Eq.~(\ref{Eq:ChicaxialTNA}) can alternatively be written as
\be
\chi_\parallel(T_{\rm N})  = \frac{C_1}{T_{{\rm N}J}-\theta_{{\rm p}J}} = \frac{C_1}{T_{{\rm N}J}(1-f_J)},
\label{Eq:chiCAxialTN}
\ee
as is also apparent from Eq.~(\ref{Eq:barchiparPMTN}).

Since $T_{\rm N}>T_{{\rm N}J}$, one sees by comparison of Eqs.~(\ref{Eq:barchiPerpPMTTN}) and~(\ref{Eq:barchiparPMTN}) that $\chi_\parallel(T_{\rm N}) > \chi_\perp(T_{\rm N})$ if the $g$~values for fields in the two directions are the same.

\subsection{Anisotropic Paramagnetic Susceptibility with XY Planar Anisotropy}

If the anisotropy field is in the $xy$~plane as in Fig.~\ref{Fig:chiPerp2}, one cannot identify a unique easy-axis direction.  Hence we specify the anisotropic susceptibilities as $\chi_z$ and $\chi_{xy}$ instead of $\chi_\perp$ and $\chi_\parallel$, respectively.  In the presence of an applied field in some direction in the $xy$~plane, the induced moments in the PM state are aligned in the same direction.

Following the same steps as in the previous section, we find that $\chi_z(T\geq T_{\rm N})$ is the same as $\chi_\perp(T\geq T_{\rm N})$ in Eqs.~(\ref{Eqs:ChiabAxAnis}), i.e.,
\begin{subequations}
\bea
\chi_z(T\geq T_{\rm N}) &=& \frac{C_1}{T-\theta_{{\rm p}J}},\label{Eq:chiCPlanar}\\*
\chi_z(T_{\rm N}) &=& \frac{C_1}{T_{\rm N}-\theta_{{\rm p}J}} = \frac{C_1}{T_{{\rm N}J} + T_{\rm A1}-\theta_{{\rm p}J}},\hspace{0.5in}\label{Eq:chiCPlanarTNA}
\eea
where $T_{\rm A1}$ is defined in  Eq.~(\ref{Eq:TA1Dev}).

Similarly $\chi_{xy}(T\geq T_{\rm N})$ is the same as $\chi_\parallel(T\geq T_{\rm N})$ in Eq.~(\ref{Eq:Chicaxial}):
\be
\chi_{xy}(T\geq T_{\rm N}) = \frac{C_1}{T-\theta_{\rm p}} = \frac{C_1}{T -T_{\rm A1}- \theta_{{\rm p}J}},
\label{Eq:ChiAPlanar}
\ee
\end{subequations}
Therefore at the N\'eel temperature, using Eq.~(\ref{EqTNATN}) one obtains
\be
\chi_{xy}(T_{\rm N}) = \frac{C_1}{T_{\rm N}-\theta_{\rm p}} = \frac{C_1}{T_{{\rm N}J }-\theta_{{\rm p}J}}.
\label{Eq:ChiAPlanarTN}
\ee

{\it Thus in the paramagnetic state with $T\geq T_{\rm N}$, if one has $z$-axis uniaxial anisotropy then $\chi_z > \chi_{xy}$, whereas for $xy$ planar anisotropy one has $\chi_{xy} > \chi_z$.}  These relationships are expected, since a uniaxial anisotropy field helps to align the moments along the $z$~axis, whereas an $xy$ planar anisotropy field helps to align the moments in the $xy$~plane.

\section{\label{Sec:ChiAFM} Anisotropic Magnetic Susceptibility of the Antiferromagnetic Phase}

\subsection{\label{Eq:ChiPerpAFM} Perpendicular Susceptibility}

\begin{figure}
\includegraphics [width=1.5in]{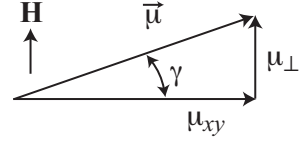}
\caption{Figure showing that the influence of an infinitesimal magnetic field {\bf H} along the $\perp$-axis on each spin in the $xy$~plane.  The {\bf H} induces a tilting of each ordered magnetic moment $\vec{\mu}$ towards the magnetic field direction by an infinitesimal angle $\gamma$, which results in an induced $\perp$-axis component $\mu_\perp$ of $\vec{\mu}$.  The angle $\gamma$ in the figure is greatly exaggerated for clarity.  To first order in $\gamma$ and~$H$ the magnitude of the ordered moment is unaffected by the presence of {\bf H}.}
\label{Fig:chiPerp}
\end{figure}

To calculate $\chi_{\perp{\rm AFM}}(T\leq T_{\rm N})$ in the presence of ${\bf H}_{\rm A}$ we assume here the presence of a planar XY anisotropy as in Fig.~\ref{Fig:chiPerp2} with the ordered moments aligned in the $xy$~plane for $H=0$.  The expression for $\chi_{\perp{\rm AFM}}$ in Eq.~(\ref{Eq:ChiPerpAnis2}) below is valid for both collinear and planar noncollinear AFM structures.  We calculate the infinitesimal angle $\gamma$ in Fig.~\ref{Fig:chiPerp}  for which the total torque on a representative moment $\vec{\mu}_i$ is zero, and from that $\chi_{\perp{\rm AFM}}(T\leq T_{\rm N})$ is obtained.

From Fig.~\ref{Fig:chiPerp}, one finds that the ordered moment magnitude $\mu_0$ in $H_\perp=0$ does not change to first order in~$H_\perp$ and the radian angle~$\gamma$.  Thus using spherical coordinates, the magnetic moment $\vec{\mu}_i$ to first order in~$\gamma$ is
\be
\vec{\mu}_i = \mu_0\left(\cos\phi_i\,\hat{\bf i} + \sin\phi_i\, \hat{\bf j} + \gamma\,\hat{\perp}\right),
\ee
where $\phi_i$ is the angle between $\vec{\mu}_i$ and the positive $x$~axis in $H=0$.  The torque contribution due to the exchange field is obtained writing $\theta=\frac{\pi}{2}-\gamma$ and thus $\sin\theta\cos\theta= \gamma$ in Eq.~(\ref{Eq:Term1}) and then using Eqs.~(\ref{Eq:CurieConst2}) and~(\ref{Eq:barmualphaDef2}), yielding 
\bea
\vec{\mu}_i\times {\bf H}_{{\rm exch}\,i} &=& -\frac{\gamma\mu_0^2}{C_1}(\sin\phi_i\,\hat{\bf i}-\cos\phi_i\,\hat{\bf j})(T_{{\rm N}J} - \theta_{{\rm p}J})\nonumber\\*
&=& -\frac{\gamma\mu_0^2}{\chi_{\perp J}}(\sin\phi_i\,\hat{\bf i}-\cos\phi_i\,\hat{\bf j}),
\label{Eq:mucrossHexchi2}
\eea
where Eq.~(\ref{Eq:ChiPerpJ}) was used to obtain the second equality.  The contribution of the applied magnetic field to the torque to first order in $H_\perp$ is
\be
\vec{\mu}_i\times {\bf H} = \mu_0 H_\perp (\sin\phi_i\,\hat{\bf i}-\cos\phi_i\,\hat{\bf j}).
\label{muCrossHperp}
\ee
The torque on $\vec{\mu}_i$ exerted by ${\bf H}_{{\rm A}i}$ to first order in $\gamma = 90^\circ - \theta$ is given by Eq.~(\ref{Eq:HAtorque}) as
\be
\vec{\mu}_i\times {\bf H}_{{\rm A}i} = -\gamma\mu_0 H_{\rm A0}(\sin\phi_i\,\hat{\bf i}- \cos\phi_i\,\hat{\bf j}).
\label{muCrossHA}
\ee
Then setting the sum of the three torques to zero, solving for $\gamma\mu_0 = \mu_\perp$ and using Eqs.~(\ref{Eq:CurieConst2}), (\ref{Eq:chiTNJHeis}), (\ref{Eq:HA0}) and~(\ref{Eq:TNADef}), one obtains the perpendicular susceptibility $\chi_{\perp {\rm AFM}} = \mu_\perp/H_\perp $ in the AFM state as
\be
\chi_{\perp {\rm AFM}}(T\leq T_{\rm N}) = \frac{C_1}{T_{\rm N} - \theta_{{\rm p}J}} = \frac{C_1}{T_{{\rm N}J} + T_{\rm A1}-\theta_{{\rm p}J}},
\label{Eq:ChiPerpAnis2}
\ee
which agrees with Eq.~(\ref{Eq:chiABAxialTNA}) for the PM state at~$T_{\rm N}$.  Thus $\chi_{\perp {\rm AFM}}$ is independent of $T$ below $T_{\rm N}$ with the value $\chi_{\perp{\rm PM}}(T_{\rm N})$.  From Eq.~(\ref{Eq:ChiPerpAnis2}), one sees that $\chi_{\perp {\rm AFM}}(T\leq T_{\rm N})$ is reduced compared to the pure Heisenberg case in which $T_{\rm A1}$ would be zero, since that anisotropy field resists the tilting of the moments out of the $xy$~plane by $H_\perp$.  The same $T$~independence of $\chi_\perp$ for $T\leq T_{\rm N}$ was found for AFM ordering in the presence of magnetic dipole interactions with or without the presence of exchange interactions~\cite{Johnston2016}.  In contrast, when quantum uniaxial $DS_z^2$ anisotropy is present in a Heisenberg spin system, $\chi_\perp$ decreases with decreasing~$T$ below~$T_{\rm N}$~\cite{Johnston2017}.

\subsection{Parallel Susceptibility of Collinear z-Axis Antiferromagnets below $T_{\rm N}$}

In this section we calculate $\chi_\parallel(T\leq T_{\rm N})$ in the presence of a uniaxial anisotropy field along the easy $z$-axis as in Fig.~\ref{Fig:chiParallel}.  Here we follow the approach of Ref.~\cite{Johnston2017} in which the influence of quantum $DS_z^2$ anisotropy was studied instead of the present generic classical anisotropy.  In the collinear ordered state, we consider two sublattices.  Sublattice $\vec{\mu}_i = \mu_i\,\hat{\bf k}$ is taken to point in the direction of the field~$H_z$ and sublattice $\vec{\mu}_j = -\mu_j\,\hat{\bf k}$ to point in the opposite direction in zero field.  

The exchange field seen by a spin on sublattice~$i$ is \cite{Johnston2017}
\bse
\be
{\bf H}_{{\rm exch}i} = \frac{3k_{\rm B}T_{{\rm N}J}}{2g^2\mu_{\rm B}^2S(S+1)}\big[\vec{\mu}_i(1+f_J) - \vec{\mu}_j(1-f_J)\big].
\label{Eq:Hexchi2sub}
\ee
If $H_z=0$, one has $\vec{\mu}_j = -\vec{\mu}_i$ and $\mu_i=\mu_0$ for all spins, yielding
\be
{\bf H}_{{\rm exch}i0} = \frac{3k_{\rm B}T_{{\rm N}J}\bar{\mu}_0}{g\mu_{\rm B}(S+1)}
\label{Eq:Hexchi2subH0}
\ee
and
\be
y_{\rm exch0} \equiv \frac{g\mu_{\rm B}H_{{\rm exch}i}}{k_{\rm B}T} = \frac{3}{(S+1)t}\bar{\mu}_0.
\ee
\ese  

The anisotropy field seen by $\vec{\mu}_i$ in the $z$~direction is
\bse
\be
H_{{\rm A0}iz} = \frac{3H_{\rm A1}}{g\mu_{\rm B}S(S+1)}\mu_{iz} = \frac{3H_{\rm A1}}{S+1}\bar{\mu}_0,
\ee
yielding
\be
y_{\rm A0} \equiv \frac{g\mu_{\rm B}H_{{\rm A0}i}}{k_{\rm B}T} = \frac{3h_{\rm A1}}{(S+1)t}\bar{\mu}_0.
\label{Eq:yA0}
\ee
\ese
Thus the parameter~$y_0$ is
\be
y_0 = y_{\rm exch0} + y_{\rm A0} = \frac{3}{(S+1)t}(1+h_{\rm A1})\bar{\mu}_0
\ee
But $t_{\rm A} = t/(1+h_{\rm A1})$, so one can also write
\be
y_0 = \frac{3}{(S+1)t_{\rm A}}\bar{\mu}_0
\ee
Then the reduced ordered moment in zero field $\bar{\mu}_0$ is obtained at each~$t$ or $t_{\rm A}$ by solving
\be
\bar{\mu}_0 = B_S(y_0).
\ee
When a field $H_z$ is present, one has
\be
y_H \equiv \frac{g\mu_{\rm B}H_z}{k_{\rm B}T} = \frac{h_z}{t}.
\label{Eq:y_H}
\ee

If~$H_z$ is infinitesimal as needed to calculate $\chi_\parallel$, one must go back to Eq.~(\ref{Eq:Hexchi2sub}) to obtain the infinitesimal change in the exchange field.  In this case one has $d\vec{\mu}_j = -d\vec{\mu}_i$ and Eq.~(\ref{Eq:Hexchi2sub}) gives
\be
dH_{{\rm exch}iz} = \frac{3k_{\rm B}T_{{\rm N}J}f_J}{g\mu_{\rm B}(S+1)}d\bar{\mu}_{iz}.
\label{Eq:dHexchi2sub}
\ee
Then one obtains
\be
dy_{{\rm exch}i} = \frac{3f_J}{(S+1)t}d\bar{\mu}_{iz}.
\ee
From Eqs.~(\ref{Eq:yA0}) and~(\ref{Eq:y_H}) one also has
\be
dy_{{\rm A}i} = \frac{3h_{\rm A1}}{(S+1)t}d\bar{\mu}_{iz},\qquad dy_H = \frac{dh_z}{t}.
\ee
The sum of the three changes in~$dy_i$ is
\be
dy_i = \frac{3}{(S+1)t}(f_J+h_{\rm A1})d\bar{\mu}_{iz} + \frac{dh_z}{t}.
\label{Eq:dyi}
\ee

The change $d\bar{\mu}_{iz}$ in the reduced moment on sublattice~$i$ is governed by the Brillouin function, i.e., 
\be
d\bar{\mu}_{iz} = B_S^\prime(y_0)dy_i.
\label{Eq:dbarmuiz}
\ee
Substituting $dy_i$ from Eq.~(\ref{Eq:dyi}) into~(\ref{Eq:dbarmuiz}) and solving for $d\bar{\mu}_{iz}$ gives the reduced $z$-axis susceptibility per spin according to Eq.~(\ref{Eq:chiRedDef}) as
\be
\bar{\chi}_{\rm \parallel AFM}(t) = \frac{1}{\tau^* - (f_J+h_{\rm A1})},
\label{Eq:barchiPar}
\ee
where
\be
\tau^*(t) = \frac{(S+1)t}{3B_S^\prime(y_0)}.
\ee
If $h_{\rm A1}=0$, one recovers the $\bar{\chi}_\parallel$ expression for the pure Heisenberg case given in Refs.~\cite{Johnston2012, Johnston2015}.

Using Eq.~(\ref{Eq:tADef}), one can also calculate $\bar{\chi}_{\rm \parallel AFM}$ in Eq.~(\ref{Eq:barchiPar}) versus $t_{\rm A} = T/T_{\rm N}$ instead of versus $t=T/T_{{\rm N}J}$ from
\be
\bar{\chi}_{\rm\parallel AFM}(t_{\rm A}) = \frac{1}{\tau_{\rm A}^* - (f_J + h_{\rm A1})},
\label{Eq:barchiPar*}
\ee
where
\be
\tau_{\rm A}^*(t_{\rm A}) = \frac{(S+1)t_{\rm A}(1+h_{\rm A1})}{3B_S^\prime(y_0)}.
\ee
We find
\be
\bar{\chi}_\parallel(t_{\rm A}=1) = \frac{1}{1-f_J},
\ee
so from Eq.~(\ref{Eq:barchiPar*}) one obtains
\be
\frac{\bar{\chi}_{\rm\parallel AFM}(t_{\rm A})}{\bar{\chi}_\parallel(t_{\rm A}=1)} = \frac{1-f_J}{\tau_{\rm A}^*(t_{\rm A}) - (f_J + h_{\rm A1})},
\label{Eq:ChiParRatio}
\ee
where $\tau_{\rm A}^*(t_{\rm A}=1) = 1+h_{\rm A1}$ and hence the ratio in Eq.~(\ref{Eq:ChiParRatio}) at $t_{\rm A}=1$ is equal to unity as required.

\subsection{\label{Sec:ChiCollinAFMs} Summary: Anisotropic Susceptibility of Collinear \lowercase{z}-Axis Antiferromagnets in Reduced Parameters}

Using the definition of the reduced susceptibility in Eq.~(\ref{Eq:chiRedDef}), together with Eqs.~(\ref{Eqs:RedPars}), (\ref{Eq:chiABAxial}), (\ref{Eqs:chicA}), and~(\ref{Eq:barchiPar*}), the anisotropic reduced susceptibilities versus $t_{\rm A}\equiv T/T_{\rm N}$ for the PM and AFM phases are summarized as
\bse
\label{Eqs:ChiParPerpAnis}
\bea
\bar{\chi}_\perp  &=&  \Bigg\{
\begin{array}{ll}
\frac{1}{1+h_{\rm A1}-f_J} & \qquad ({\rm AFM},\ t_{\rm A}\leq 1)\vspace{0.1in}\\*
\frac{1}{(1+h_{\rm A1})t_{\rm A} -f_J} & \qquad ({\rm PM},\ t_{\rm A}\geq 1),
\end{array}
\label{Eq:ChiPerpAnis3}
\eea 
\bea
\bar{\chi}_\parallel  &=&  \Bigg\{
\begin{array}{ll}
\frac{1}{\tau^\ast_{\rm A} - (f_J + h_{\rm A1})} &  ({\rm AFM},\ t_{\rm A}\leq 1)\vspace{0.1in}\\*
\frac{1}{(1+h_{\rm A1})t_{\rm A} - (f_J + h_{\rm A1})} &  ({\rm PM},\ t_{\rm A}\geq 1),
\end{array}
\label{Eq:ChiParAnis}
\eea 
\be
\frac{\chi_\parallel(T_{\rm N})}{\chi_\perp(T_{\rm N})} = 1 + \frac{h_{\rm A1}}{1-f_J},
\label{Eq:ChiParPerpTNARatio}
\ee
where
\bea
\tau_{\rm A}^\ast &=& \frac{(S+1)(1+h_{\rm A1})t_{\rm A}}{3B_S^\prime(y_0)},\\*
y_0 &=& \frac{3\bar{\mu}_0}{(S+1)t_{\rm A}}, \\*
\bar{\mu}_0 &=& B_S(y_0). 
\eea
\ese

\begin{figure}
\includegraphics [width=3.3in]{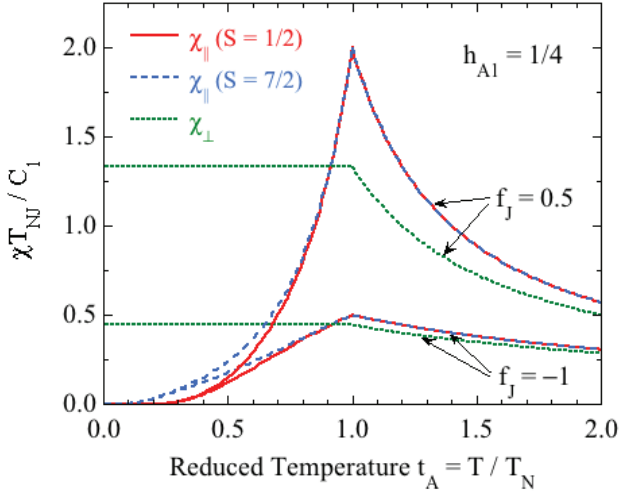}
\caption{(Color online) Anisotropic reduced magnetic susceptibilities $\bar{\chi}_\parallel$ and~$\bar{\chi}_\perp$ versus reduced temperature $t_{\rm A}$ for two different $S$ values, two different $f_J$ values and for a fixed anisotropy field $h_{\rm A1} = 1/4$, according to Eqs.~(\ref{Eqs:ChiParPerpAnis}).}
\label{Fig:MFT_Chi_Coll_AF_w_Anis}
\end{figure}

In these reduced susceptibility units, $\bar{\chi}_\perp(t_{\rm A})$ is independent of $S$ for all $t_{\rm A}$, and $\bar{\chi}_{\rm\parallel AFM}(t_{\rm A}<1)$ is dependent of $S$ since $\tau^\ast_{\rm A}$ depends on $S$\@.  These features are illustrated in plots of $\bar{\chi}_\perp(t_{\rm A})$ and $\bar{\chi}_\parallel(t_{\rm A})$ in Fig.~\ref{Fig:MFT_Chi_Coll_AF_w_Anis} for $S=1/2$ and~7/2 and for $f_J=-1$ and $f_J=0.5$, all with a fixed value of the reduced anisotropy parameter $h_{\rm A1}=1/4$.  An important feature of the temperature dependences is that $\chi_{\rm \parallel PM} > \chi_{\perp{\rm PM}}$ at $t_{\rm A}\geq 1$, but a crossover occurs where $\chi_{\rm \parallel AFM} < \chi_{\perp{\rm AFM}}$ at lower $t_{\rm A}$.  

From Eq.~(\ref{Eq:ChiParPerpTNARatio}), as $f_J$ increases algebraically towards its upper limit of unity at a fixed value of $h_{\rm A1}$, the ratio $\chi_\parallel(T_{\rm N})/\chi_\perp(T_{\rm N})$ increases, as observed in Fig.~\ref{Fig:MFT_Chi_Coll_AF_w_Anis}.

\section{\label{Sec:HiHPerpMAnis} High-Field Perpendicular Magnetization of the Antiferromagnetic and Paramagnetic Phases}

In this section the ``perpendicular'' direction $\hat{\perp}$ of an applied field {\bf H} refers to a direction perpendicular to the easy axis (for a collinear AFM) or plane (for a planar noncollinear AFM) of the anisotropy field ${\bf H}_{\rm A}$.

\subsection{\label{Sec:muPerpAFM} Antiferromagnetic Phase}

The $\chi_{\rm \perp AFM}(T\leq T_{\rm N})$ for fields $H_\perp\to0$ was calculated in Sec.~\ref{Eq:ChiPerpAFM}. Here we determine the magnetization in high perpendicular magnetic fields for both collinear and planar noncollinear AFMs at fields below the perpendicular critical field $H_{\rm c\perp AFM}\equiv \mu_0(H=0,T\leq T_{\rm N})/\chi_{\rm \perp AFM}$.  We find that $\mu_{\rm \perp AFM}$ is proportional to $H_\perp$ up to $H_{\rm c\perp AFM}$ with the same $T$-independent slope $\chi_{\rm \perp AFM}$ as for $H_\perp\to0$ in Eq.~(\ref{Eq:chiABAxialTNA}), and that the ordered moment $\mu_0(T)$ is independent of $H_\perp$ in the AFM phase.

\begin{figure}
\includegraphics [width=2.in]{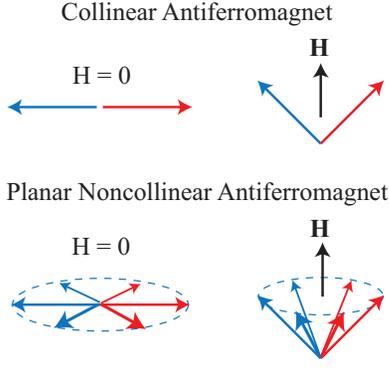}
\caption{(Color online) Influence on the generic magnetic structure due to a high magnetic field applied perpendicular to the easy axis of a collinear antiferromagnet (AFM) (top panel) and to the easy plane of a planar noncollinear AFM  (bottom panel).  Hodographs of the zero-field magnetic moment vectors are shown on the left.  In high fields as shown on the right, the AFM structures become canted towards the field.  The ordered moments of the collinear AFM are now coplanar, whereas those of the noncollinear AFM now lie on the surface of a cone with the axis of the cone along the magnetic field axis as shown.  At a sufficiently high field $H=H_{\rm c\perp AFM}$ given by Eq.~(\ref{Eq:HcperpDef}), the moments in either case become parallel to each other and a second-order transition from the canted AFM to the PM state occurs.}
\label{Fig:High_Perp_Field_Structs}
\end{figure}

For collinear AFMs, at high fields the canted moments lie in a plane defined by the initial parallel axis and the applied field as shown in the top panel of Fig.~\ref{Fig:High_Perp_Field_Structs}.  In contrast, for a planar noncollinear structure at $H=0$, in large fields the moments in a hodograph lie on the surface of a cone with the tails of the moment vectors at the apex and the axis of the cone along the applied field axis as shown in the bottom panel of Fig.~\ref{Fig:High_Perp_Field_Structs}.  We can therefore treat both the collinear and planar noncollinear cases simultaneously, where the anisotropy field is in the plane perpendicular to the applied field as shown in Fig.~\ref{Fig:chiPerp2}.

From Fig.~\ref{Fig:chiPerp2}, the torque on $\vec{\mu}_i$ due to a perpendicular field {\bf H} in Eq.~(\ref{muCrossHperp}) is the same as that due to ${\bf H}_{{\rm A}i}$ in Eq.~(\ref{muCrossHA}) except for the scalar prefactor and the opposite direction.  Therefore comparing Eqs.~(\ref{muCrossHperp}) and~(\ref{muCrossHA}) one can include the influence of ${\bf H}_{{\rm A}i}$ on the value of the induced moment $\mu_\perp$ by setting $H = H_\perp - H_{\rm A0}\cos\theta$ in the expression setting the net torque equal to zero in the absence of $H_{\rm A0}$ \cite{Johnston2017}.  Then using the definitions $\mu_{\perp} = \mu\cos\theta$, $\bar{\mu}=\mu/(gS\mu_{\rm B})$ and  $H_{\rm A0}$ in terms of $H_{\rm A1}$ in Eq.~(\ref{Eq:HA0}) gives
\be
\mu_\perp = \frac{1}{T_{{\rm N}J} - \theta_{{\rm p}J}}\left[C_1H_\perp - C_1\frac{3H_{\rm A1}\mu_\perp}{gS(S+1)\mu_{\rm B}}\right],
\ee
where the single-spin Curie constant $C_1$ is given in Eq.~(\ref{Eq:CurieConst2}).  Solving for $\mu_\perp$ gives
\be
\mu_\perp = \frac{C_1H_\perp}{T_{\rm N} - \theta_{{\rm p}J}},
\ee
where to obtain this equation we used the expression for $T_{\rm N}$ in Eq.~(\ref{Eq:TNADef}) and the definition of $T_{\rm A1}$ in Eq.~(\ref{Eq:TA1Dev}).  Hence 
\bea
\mu_\perp(T\leq T_{\rm N}) &=&\chi_{\rm \perp AFM} H_\perp \qquad (\mu_\perp \leq \mu_0),\nonumber\\*
\chi_{\rm \perp AFM} &=& \frac{C_1}{T_{\rm N} - \theta_{{\rm p}J}}, \label{Eq:ChiPerpWAnis}
\eea
where $\chi_{\rm \perp AFM}$ is seen to be the same as the zero-field perpendicular susceptibility already obtained in Eq.~(\ref{Eq:ChiPerpAnis2}), which in turn is the same as $\chi_{\rm \perp PM}(T_{\rm N})$ in Eq.~(\ref{Eq:chiCPlanarTNA}).

This independence of $\mu_\perp/H_\perp$ with respect to~$H_\perp$ in the AFM phase indicates that the magnitude $\mu$ of the moments is independent~$H_\perp$ and in particular is equal to the zero-field value, i.e., $\mu(T)=\mu_0(T)$.  Thus the $T$-dependent critical field $H_{\rm c\perp AFM}$ is given by the field at which $\mu_\perp = \mu_0(T)$, i.e.,
\be
H_{\rm c\perp AFM}(T\leq T_{\rm N}) = \frac{\mu_0(T)}{\chi_{\rm \perp AFM}}.
\label{Eq:HcperpDef}
\ee

Using Eq.~(\ref{Eq:chiRedDef}) together with the variable definitions in Eqs.~(\ref{Eqs:RedPars}), Eq.~(\ref{Eq:ChiPerpWAnis}) gives
\be
\bar{\chi}_{\rm \perp AFM} = \frac{1}{1+h_{\rm A1} - f_J},
\label{Eq:barchiperp}
\ee
which reproduces the first entry in Eqs.~(\ref{Eq:ChiPerpAnis3}).  Using Eq.~(\ref{Eq:chiRedDef}) one obtains
\be
\chi_{\rm \perp AFM}^* = \frac{S+1}{3}\bar{\chi}_{\rm \perp AFM} = \frac{S+1}{3(1+h_{\rm A1} - f_j)}.
\label{Eq:barmuperpvshperp}
\ee
Then using the definition $\bar{\mu}_\perp = \chi_{\rm \perp AFM}^* h_\perp$ from Eq.~(\ref{Eq:chiAlphaStarDef}) and setting $\bar{\mu}_\perp = \bar{\mu}_0$ yields the reduced critical field
\be
h_{\rm c\perp AFM}(t_{\rm A}) = \left[\frac{3\bar{\mu}_0(t_{\rm A})}{S+1}\right](1+h_{\rm A1}-f_J),
\label{Eq:hcPerpRed}
\ee
where $\bar{\mu}_0(t_{\rm A})$ is found by solving Eqs.~(\ref{Eqs:barmuA}) and $\bar{\mu}_0(t_{\rm A}\geq 1) = 0$.  The dependence of $h_{\rm c\perp AFM}$ on~$t_{\rm A}$ is thus the same as that of $\bar{\mu}_0$ on~$t_{\rm A}$ shown above in Fig.~\ref{Fig:mubarVStA}.  For given values of $t_{\rm A}$, $h_{\rm A1}$, and $f_J$,  $h_{\rm c\perp AFM}(t_{\rm A}=0)$ decreases with increasing spin~$S$\@.  At $t_{\rm A} = 0$ one has $\bar{\mu}_0=1$.  Then Eq.~(\ref{Eq:hcPerpRed}) gives
\be
h_{\rm c\perp AFM}(t_{\rm A}=0)  = \frac{3(1+h_{\rm A1}-f_J)}{S+1}.
\label{Eq:hcperT0}
\ee

\subsection{Paramagnetic Phase}

The paramagnetic (PM) phase can be reached from the AFM phase by increasing the field to $H_\perp>H_{\rm c\perp AFM}$ at $T<T_{\rm N}$ or by increasing the temperature to $T>T_{\rm N}$ at $H_\perp=0$.  In either case, the thermal-average moment induced by the applied magnetic field {\bf H} is in the direction of {\bf H} if {\bf H} is in a principal axis direction as considered in this paper.  In this section both {\bf H} and the field-induced PM moment $\mu_\perp$ are in the same $\hat{\perp}$ direction that is perpendicular to the easy axis of a collinear AFM or to the easy plane of a planar noncollinear AFM\@.  Then  according to Eq.~(\ref{Eq:HAiAxial}) and Fig.~\ref{Fig:chiParallel} or Eq.~(\ref{Eq:HAiGen}) and Fig.~\ref{Fig:chiPerp2}, respectively, the anisotropy field $H_{\rm A}$ is zero in either case.  Therefore Eq.~(\ref{Eq:muvsBrill2}) and the definitions of the reduced variables in Eq.~(\ref{Eqs:RedPars}) immediately give
\bea
\bar{\mu}_{\rm \perp PM} &=& B_S\left[\frac{3f_J T_{{\rm N}J}\bar{\mu}_{\rm \perp PM}}{(S+1)T} + \frac{g\mu_{\rm B}H_\perp}{k_{\rm B}T}\right]\label{Eq:muvsBrillPerpPM}\\*
 &=& B_S\left[\frac{3f_J \bar{\mu}_{\rm \perp PM}}{(S+1)t} + \frac{h_\perp}{t}\right].
\eea
Even though $H_{\rm A}=0$ for the perpendicular moment orientation, one still has $T_{\rm N} > T_{{\rm N}J}$ if $h_{\rm A1}>0$.  Therefore to compare with experimental data we reexpress the reduced temperature as $t \to (1+h_{\rm A1})t_{\rm A}$ using Eq.~(\ref{Eq:tADef}), yielding
\be
\bar{\mu}_{\rm \perp PM} = B_S\left\{\frac{1}{1 + h_{\rm A1}}\left[\frac{3f_J\bar{\mu}_{\rm \perp PM}}{(S+1)t_{\rm A}} + \frac{h_\perp}{t_{\rm A}}\right]\right\}.
\label{Eq:muvsBrillPerpPM3}
\ee
The $\bar{\mu}_{\rm \perp PM}$ for given values of $h_{\rm A1},\ f_J$ and~$t_{\rm A}$ is determined by numerically solving Eq.~(\ref{Eq:muvsBrillPerpPM3}).

\begin{figure}
\includegraphics [width=3.in]{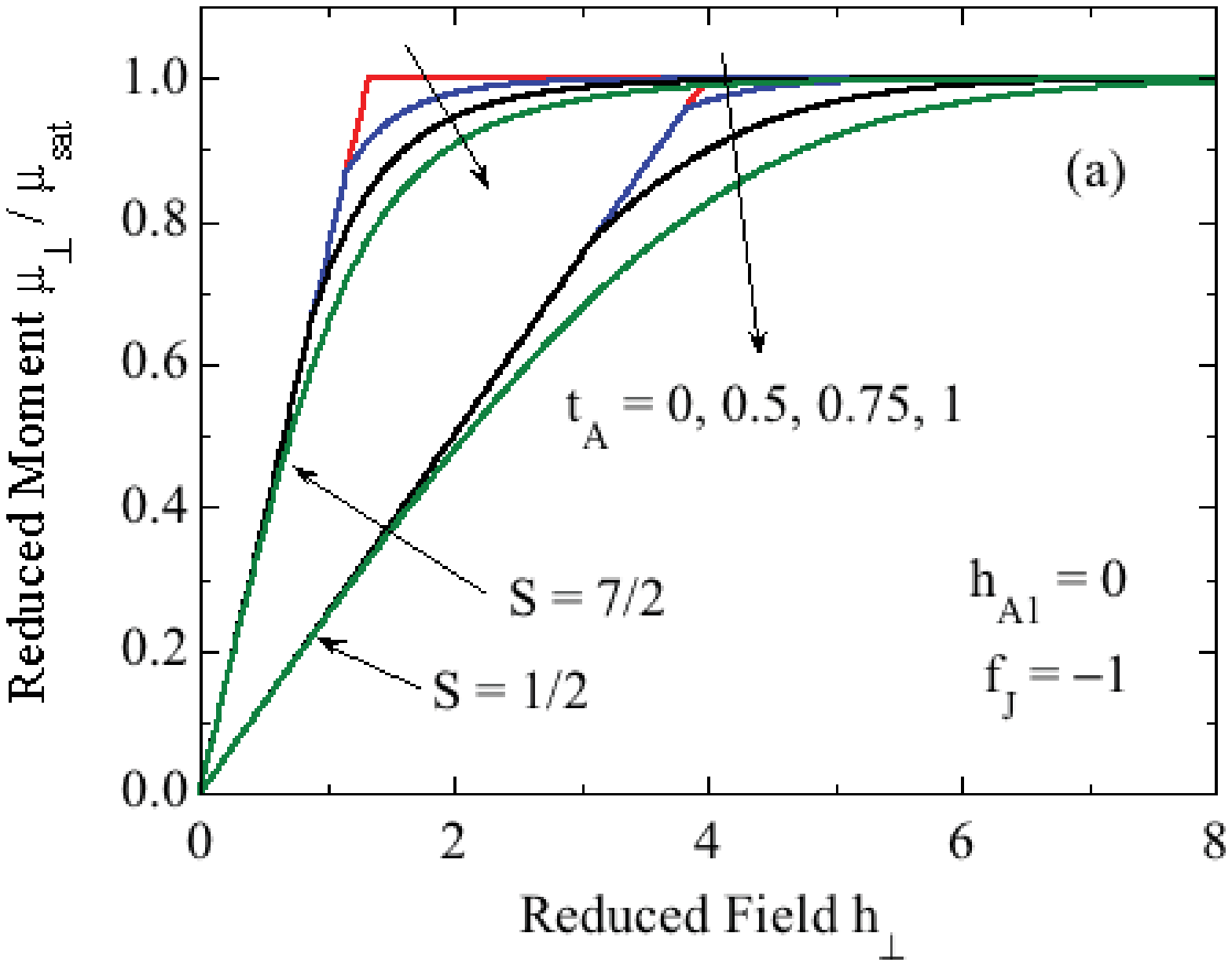}
\includegraphics [width=3.in]{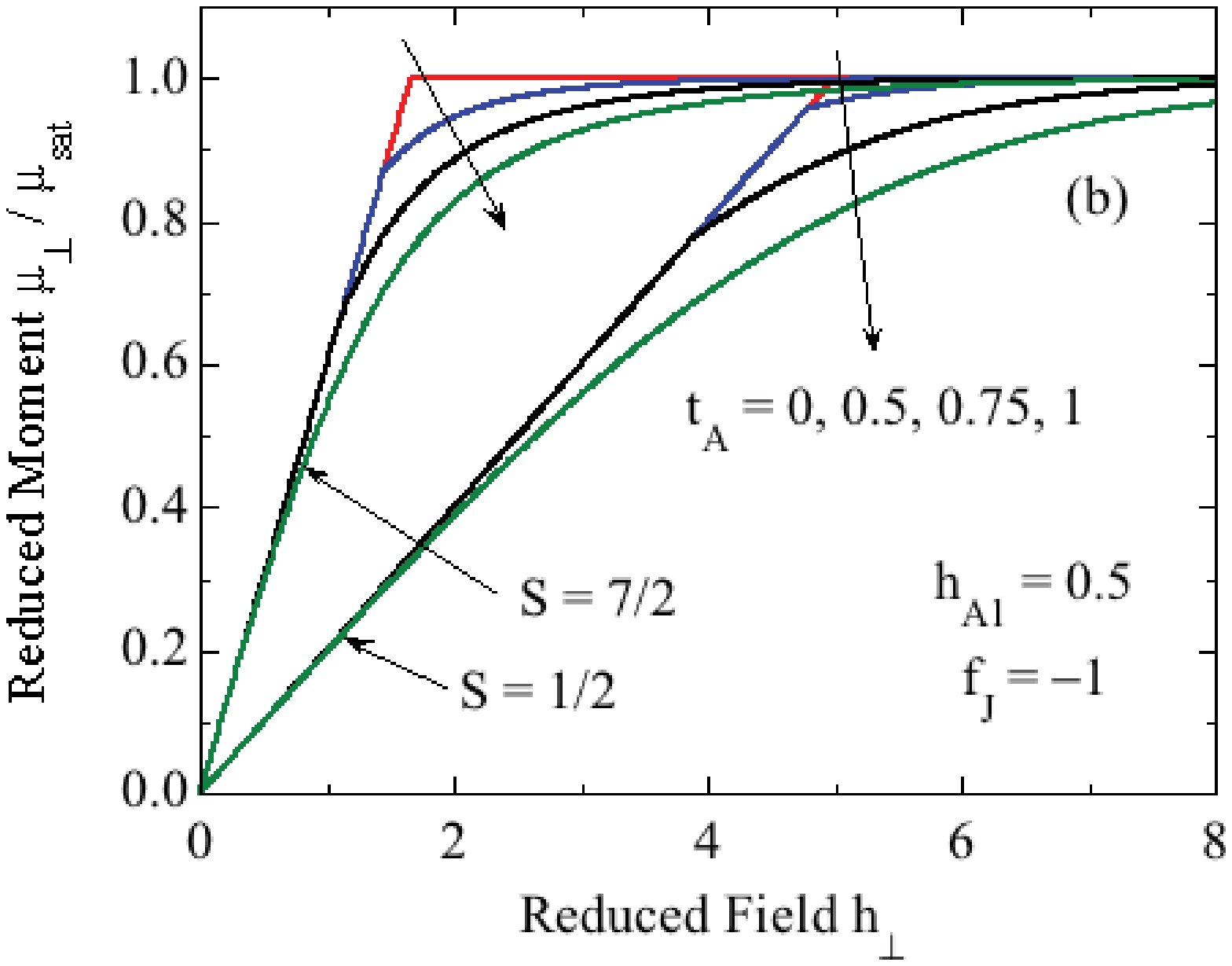}
\caption{(Color online) Reduced perpendicular moment $\bar{\mu}_\perp \equiv \mu_\perp/\mu_{\rm sat}$ versus reduced perpendicular field~$h_\perp$ for spins $S=1/2$ and $S=7/2$ at the reduced temperatures~$t_{\rm A}=T/T_{\rm N}$ indicated for parameters $f_J = -1$ and reduced anisotropy fields (a) $h_{\rm A1} = 0$ and (b) $h_{\rm A1} = 1/2$, according to Eqs.~(\ref{Eq:muPerpArray}).  Discontinuities in slope at fields $h_{\rm c\perp}(T)$ are seen as the system undergoes second-order transitions from the canted AFM state to the PM state with increasing field. The reduced critical fields at $t_{\rm A}=0$ for $h_{\rm A1}=0$ are $h_{\rm c\perp AFM} = 4/3$ and~4 for $S=7/2$ and $S=1/2$, respectively, and for $h_{\rm A1}=1/2$ are $h_{\rm c\perp AFM} = 5/3$ and~5 for $S=7/2$ and $S=1/2$, respectively.  Both are in agreement with Eq.~(\ref{Eq:hcPerpRed}).}
\label{Fig:muPerpS1272fJm1hA10}
\end{figure}

The results for the two cases $h_\perp \leq h_{\rm c\perp AFM}(t_{\rm A})$ and $h_\perp \geq h_{\rm c\perp AFM}(t_{\rm A})$ are summarized respectively as
\be
\bar{\mu}_\perp(h_\perp) = 
\begin{cases}
\frac{(S+1)h_\perp}{3(1+h_{\rm A1}-f_J)} & \hspace{-0.6in} ({\rm AFM,}\ h_\perp \leq h_{\rm c\perp AFM})\vspace{0.05in}\\*
B_S\left\{\frac{1}{1 + h_{\rm A1}}\left[\frac{3f_J\bar{\mu}_\perp}{(S+1)t_{\rm A}} + \frac{h_\perp}{t_{\rm A}}\right]\right\}\\*
& \hspace{-0.5in} ({\rm PM,}\ h_\perp \geq h_{\rm c\perp AFM})
\end{cases}
\label{Eq:muPerpArray}
\ee
where $h_{\rm c\perp AFM}$ is given in Eq.~(\ref{Eq:hcPerpRed}).  Using Eqs.~(\ref{Eq:muPerpArray}), the $\bar{\mu}_\perp$ versus $h_\perp$ curves for spin $S=1/2$ and~7/2 with $f_J=-1$ at four reduced temperatures and $h_{\rm A1} = 0$ and~1/2 are plotted in Fig.~\ref{Fig:muPerpS1272fJm1hA10}.  A discontinuity in the slope of $\bar{\mu}_\perp$ versus $h_\perp$ is seen at $h_\perp = h_{\rm c\perp AFM}$ for each reduced temperature~$t_{\rm A}$, reflecting a second-order transition from the AFM to the PM phase.

\section{\label{Sec:HiHParMAnisI} High-Field Parallel Magnetization of \lowercase{z}-Axis Collinear Antiferromagnets: \hspace{0.5in} Paramagnetic and Spin-Flop Phases}

\begin{figure}
\includegraphics [width=2.in]{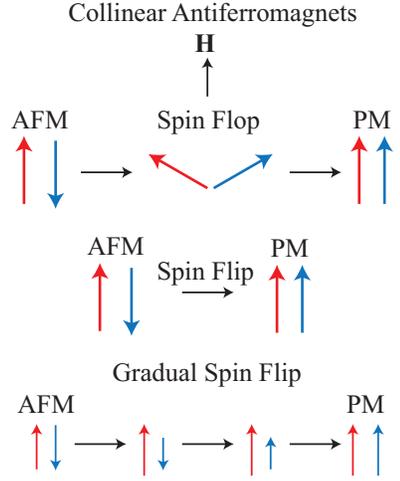}
\caption{(Color online) Phase transitions that can potentially occur in collinear antiferromagnets (AFM) when a magnetic field {\bf H} is applied along the easy axis.  The magnitude $H$ of the field increases from left to right.  The top panel shows a first-order spin-flop (SF) transition that occurs from a collinear AFM structure to a SF phase at a SF field $H_{\rm SF}$, which is a canted AFM structure.  At higher fields, the angle between the two sublattice magnetic moments goes continuously to zero, corresponding to a second-order transition from the SF phase to a paramagnetic (PM) phase at a critical field $H_{\rm cSF}$.  Alternative scenarios with increasing $H$ include either a first-order spin-flip transition directly from the AFM to the PM phase as shown in the middle panel, or a continuous evolution (``gradual spin flip'') of the AFM phase into the PM phase via a second-order phase transition as illustrated in the bottom panel.}
\label{Fig:MparMperp}
\end{figure}

\subsection{Introduction}

When a collinear AFM is placed in a magnetic field parallel to the easy axis (defined to be the $z$-axis here), different $T$-dependent behaviors can occur.  A first-order spin-flop (SF) transition may occur from the AFM phase to a SF phase as shown in the top panel of Fig.~\ref{Fig:MparMperp}, where the orientations of the ordered moments aligned along the $z$~axis flop with increasing field to an approximately perpendicular canted perpendicular orientation \cite{Stryjewski1977}.  It is common to use the term ``spin flop'' to denote both the magnetic phase and the magnetic phase transition.  Upon further increasing the field a second-order spin-flop to paramagnetic (PM) phase transition occurs in which all moments then point in the direction of the field.  

The PM phase is sometimes called a ``ferromagnetic phase'' in the literature because the magnetic structure of the field-induced PM phase has ferromagnetic (FM) alignment of the field-induced moments.  However, we reserve the term ``ferromagnetic phase'' for a ferromagnetic structure that is caused by the interactions between the moments in zero applied magnetic field, not by the field.  Indeed, a thermodynamic transition from a PM phase to a FM phase cannot occur versus $T$ in finite $H$ because the FM order parameter (the net magnetization) is never nonzero in a finite $H$ at a finite $T$\@.

A first-order spin-flip transition may occur with increasing field directly from the AFM phase to the PM phase if the anisotropy field along the $z$~axis is sufficiently strong, as shown in the middle panel of Fig.~\ref{Fig:MparMperp}.  Within MFT the magnitude and direction of the initially antiparallel moment can also vary smoothly with field, resulting eventually in a second-order AFM to PM transition as shown in the bottom panel of Fig.~\ref{Fig:MparMperp}.

\subsection{\label{Sec:PMMomentFmag} $z$-Axis Induced Moment and Free Energy of the Paramagnetic (PM) Phase}

In this section, we change notation for the PM phase from $\mu_{\parallel}$ to~$\mu_{z{\rm PM}}$.  The general high-field expression for the PM phase was already obtained in Eq.~(\ref{Eq:mucPar}).  Utilizing Eqs.~(\ref{Eqs:RedPars}), Eq.~(\ref{Eq:mucPar}) can be written in reduced variables as
\bea
\bar{\mu}_{z{\rm PM}} &=& B_S(y_{\rm PM}), \label{Eq:muParPM}\\*
y_{\rm PM} &=& \frac{3(f_J + h_{\rm A1})}{(S+1)t}\bar{\mu}_{z{\rm PM}} + \frac{h_z}{t}\nonumber\\*
&=& \frac{1}{(1+h_{\rm A1})t_{\rm A}}\left[\frac{3(f_J+h_{\rm A1})}{S+1}\bar{\mu}_{z{\rm PM}} + h_z\right].\nonumber
\eea
When the reduced temperature is taken to be~$t$, one can write
\be
y_{\rm PM} = \frac{b_z}{t},
\ee
where the reduced magnetic induction $b_z$ seen by a representative spin is
\be
b_z = \frac{3(f_J + h_{\rm A1})}{S+1}\bar{\mu}_{z{\rm PM}} + h_z.
\label{Eq:bzPM}
\ee

\begin{figure}
\includegraphics [width=3.in]{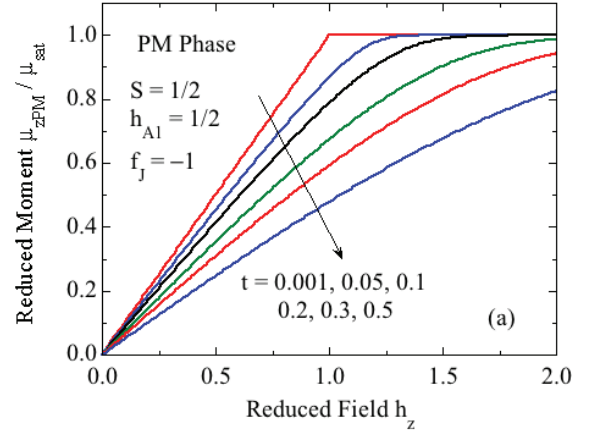}
\includegraphics [width=3.in]{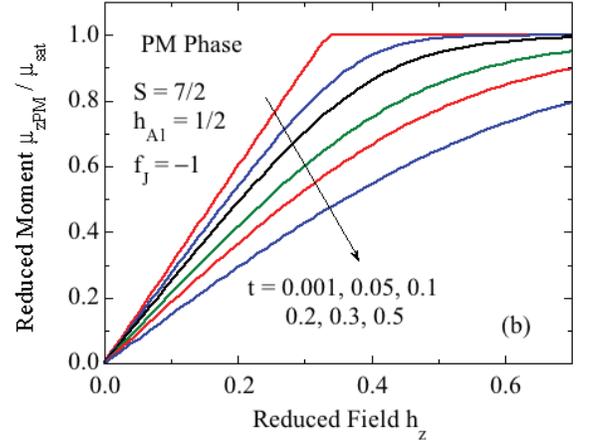}
\caption{(Color online) Reduced $z$-axis moment $\mu_{z{\rm PM}}/\mu_{\rm sat}$ of the paramagnetic (PM) phase versus reduced field~$h_z=g\mu_{\rm B}H_z/k_{\rm B}T_{{\rm N}J}$ for spins (a)~$S=1/2$ and (b)~$S=7/2$ at the indicated reduced temperatures~$t=T/T_{{\rm N}J}$ and for $f_J = -1$ and $h_{\rm A1} = 1/2$, according to Eqs.~(\ref{Eq:muParPM}).}
\label{Fig:muzPMvsHzS12fJm1hA150}
\end{figure}

\begin{figure}
\includegraphics [width=3.in]{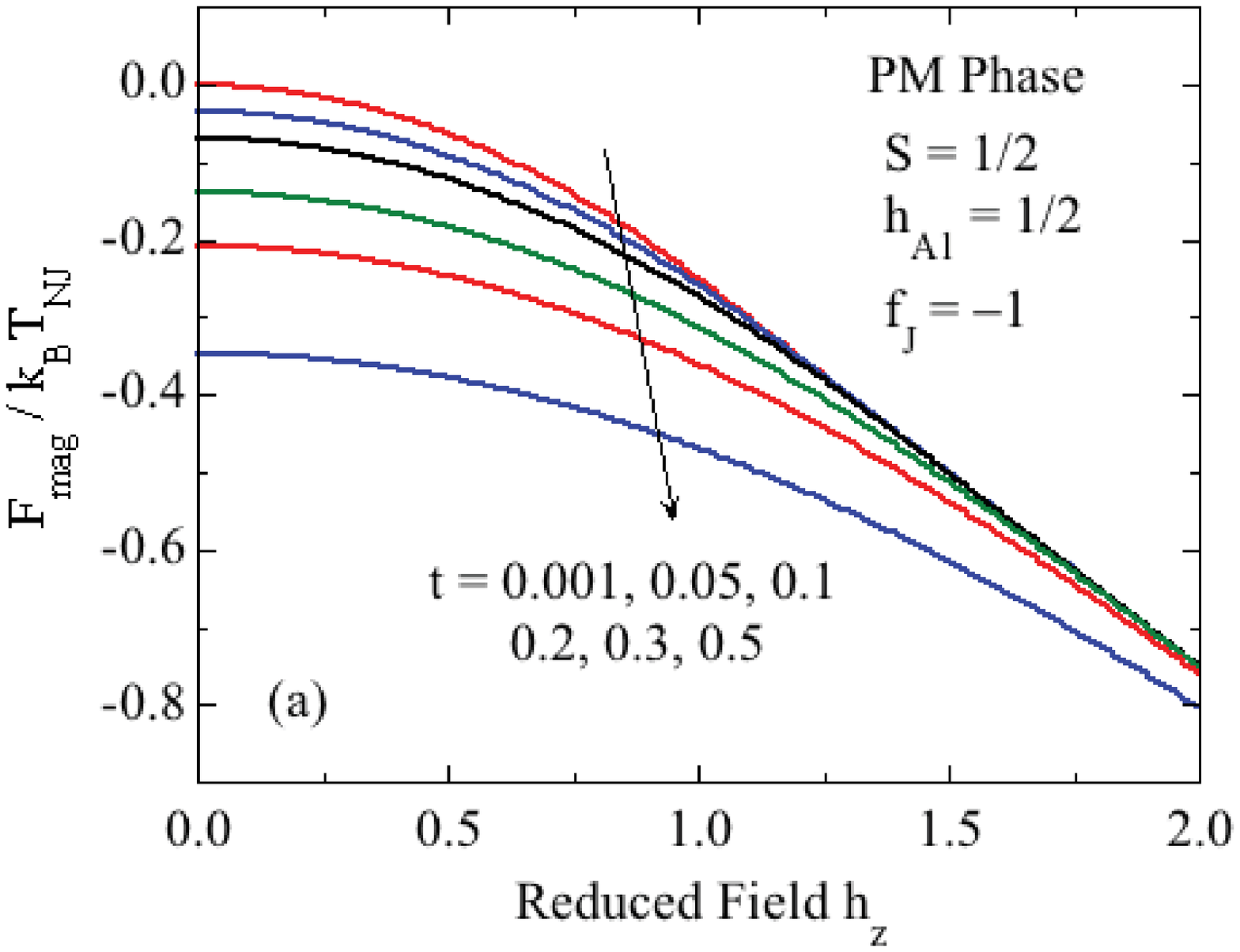}
\includegraphics [width=3.in]{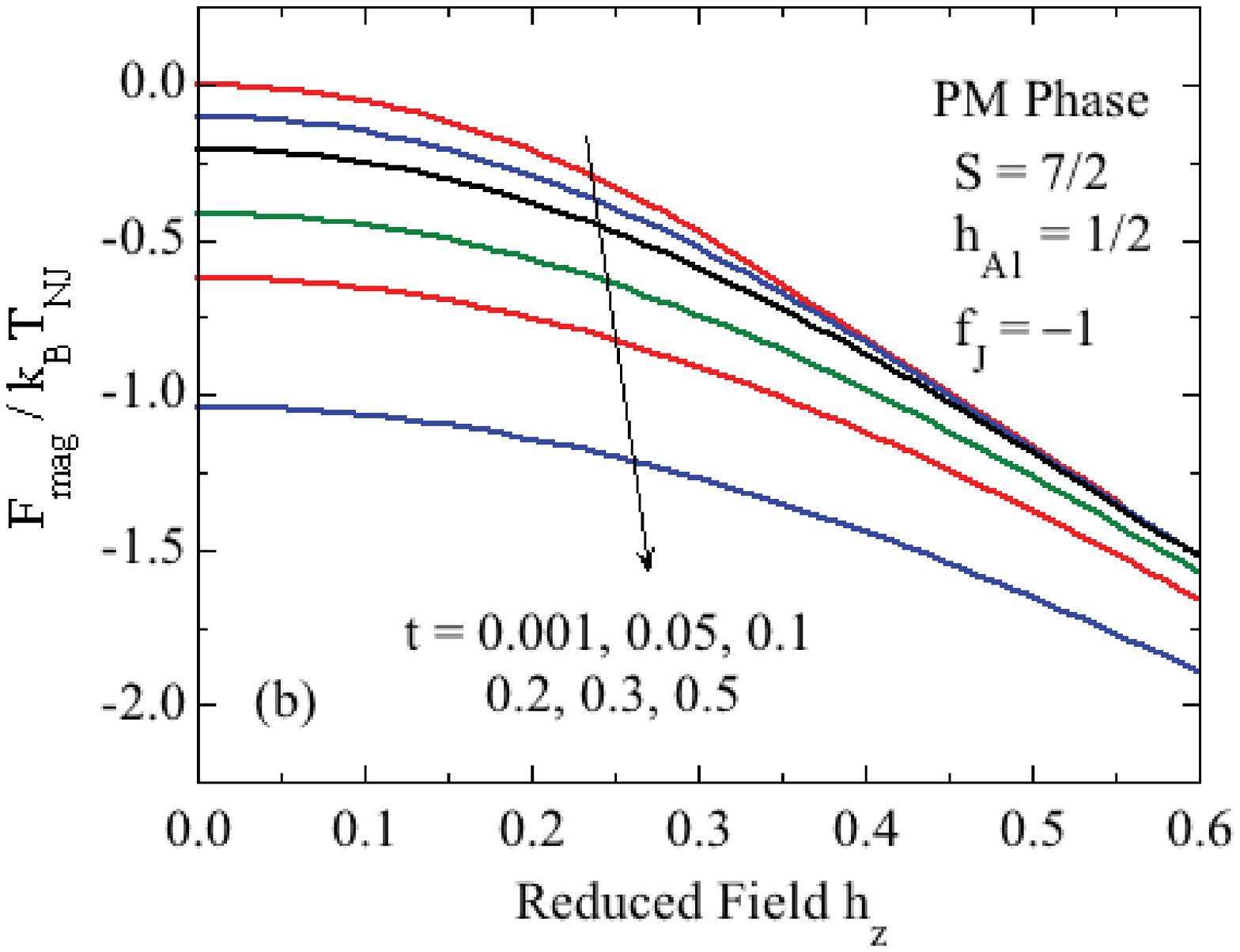}
\caption{(Color online) Reduced free energy $F_{\rm mag}/k_{\rm B}T_{{\rm N}J}$ of the paramagnetic (PM) phase versus reduced field~$h_z$ for spins (a)~$S=1/2$ and (b)~$S=7/2$ at the indicated reduced temperatures~$t$ and for $f_J = -1$ and $h_{\rm A1} = 1/2$, obtained using Eq.~(\ref{Eq:FmagPM}) and the data in Fig.~\ref{Fig:muzPMvsHzS12fJm1hA150}.}
\label{Fig:FmagPMvsHzS12fJm1hA150}
\end{figure}

Shown in Fig.~\ref{Fig:muzPMvsHzS12fJm1hA150}(a) are plots of $\bar{\mu}_{z{\rm PM}}$ versus reduced field~$h_z$ obtained from Eqs.~(\ref{Eq:muParPM}) for parameters~$f_J=-1$ and $h_{\rm A1}=1/2$, each for spins~$S=1/2$ and~$S=7/2$, at reduced temperatures $t=T/T_{{\rm N}J}$ as indicated.  Perhaps unexpectedly, $\bar{\mu}_{z{\rm PM}}$ for $t\to0$ is seen to be proportional to~$h_z$ from $h_z=0$ to a critical field $h_{\rm cPM}$ at which $\bar{\mu}_{z{\rm PM}}$ saturates to the value of unity and continues to have that value at higher fields.  The scale of the abscissa is reduced by about a factor of 3 for $S=7/2$ compared to that for $S=1/2$.  However, the shapes of the plots for the two spin values are very similar for the same reduced temperature.

In $h_z=0$, one sees from Fig.~\ref{Fig:muzPMvsHzS12fJm1hA150} that $\bar{\mu}_{z{\rm PM}}=0$, so Eq.~(\ref{Eq:UmagGen}) gives the internal energy per spin as
\be
\frac{U_{\rm mag}(h_z=0,t)}{k_{\rm B}T_{{\rm N}J}} = 0.
\ee
Also, the PM phase in $h_z=0$ is completely disordered at all temperatures, so the entropy per spin is
\be
\frac{S_{\rm mag}(h_z=0,t)}{k_{\rm B}} = \ln(2S+1).
\ee
Thus the free energy in $h_z=0$ is given by Eq.~(\ref{Eq:FmagHz0}) as
\be
\frac{F_{\rm mag}(h_z=0,t)}{k_{\rm B}T_{{\rm N}J}} = -t\ln(2S+1).
\ee
Now including the field dependence using Eq.~(\ref{Eq:FmagGen}) gives
\be
\frac{F_{\rm mag}(h_z,t)}{k_{\rm B}T_{{\rm N}J}} = -t\ln(2S+1) -S\int_0^{h_z}\bar{\mu}_z(h_z,t)dh_z.
\label{Eq:FmagPM}
\ee

The reduced free energy is plotted versus~$h_z$ for spins~$S=1/2$ and~$S=7/2$ in Fig.~\ref{Fig:FmagPMvsHzS12fJm1hA150} with the same parameters as in Fig.~\ref{Fig:muzPMvsHzS12fJm1hA150}, obtained using Eq.~(\ref{Eq:FmagPM}).

\subsection{\label{Sec:SpinFlop} Spin-Flop Phase of Collinear Antiferromagnets}

\subsubsection{\label{OrdMomH0SF} Ordered Moment in Zero Field}

\begin{figure}
\includegraphics [width=3.in]{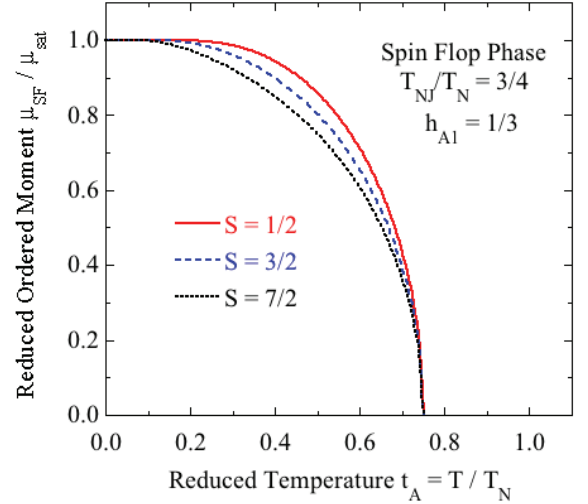}
\caption{(Color online) Reduced ordered moment $\bar{\mu}_{\rm SF} = \mu_{\rm SF}/\mu_{\rm sat}$ of the spin flop phase versus reduced temperature $t_{\rm A}$ for spins $S = 1/2$, 3/2 and 7/2 with $h_{\rm A1}= 1/3$, calculated from Eq.~(\ref{Eq:OrdMomFlop}). The ordered moment of the SF phase does not depend on $f_J$ or on applied field for $h_z\leq h_{\rm cSF}$. }
\label{Fig:OrderedMomentMFT_HA}
\end{figure}

The magnetic structure and magnetic field orientation in the spin flop~(SF) phase in the top panel of Fig.~\ref{Fig:MparMperp} with nonzero anisotropy field ${\bf H}_{\rm A}$ along the easy axis are the same as those used for calculation of the high-field perpendicular magnetization in Appendix~\ref{Sec:UnifiedMFT} for the case of zero anisotropy field $H_{\rm A}=0$.  In that case we obtained Eq.~(\ref{Eq:ordMoment}) in which the reduced ordered moment $\bar{\mu}\equiv \mu/\mu_{\rm sat}$ depends only on $t\equiv T/T_{{\rm N}J}$ and not on the applied field $H_\perp$ if $H_\perp\leq H_{\rm c\perp}$.  Equation~(\ref{Eq:ordMoment}) is identical to Eq.~(\ref{Eq:mubar0}) for determining $\bar{\mu}_0(t)$ for $H = H_{\rm A} =0$.  Similarly, in the spin flop phase, {\bf H} and ${\bf H}_{\rm A}$ are in the same direction perpendicular to the $H=0$ AFM ordering plane and hence the ordered moment again cannot depend on $H_z$ or $H_{\rm A}$ and is therefore given by the same Eqs.~(\ref{Eq:ordMoment}) and~(\ref{Eq:mubar0}).  We have confirmed this conclusion from detailed calculations that will not be presented here.  Thus Eq.~(\ref{Eq:ordMoment}) in the case of the SF phase reads
\bse
\label{Eqs:mubarFlop}
\bea
\bar{\mu}_{\rm SF} &=& B_S\left[\frac{3\bar{\mu}_{\rm SF}}{(S+1)t}\right]\label{Eq:barmuSFvst}\\*
&=& B_S\left[\frac{3\bar{\mu}_{\rm SF}}{(S+1)(1+h_{\rm A1})t_{\rm A}}\right],\label{Eq:OrdMomFlop}
\eea
\ese
where to obtain the second equality we used Eq.~(\ref{Eq:tADef}).  The ordered moment in the SF phase goes to zero at a temperature $T_{{\rm N}J}$ below the N\'eel temperature $T_{\rm N}$, as shown in Fig.~\ref{Fig:OrderedMomentMFT_HA} for spins $S = 1/2$, 3/2 and~7/2 with $h_{\rm A1}=1/3$ for which $T_{{\rm N}J}/T_{\rm N} = 3/4$ according to Eq.~(\ref{Eq:TNATNRatio}).  This feature is critically important to the construction of the phase diagrams in the $h_z$--$t_{\rm A}$ plane that are presented in   Fig.~\ref{Fig:PhaseDiagram_S12fJm1hA10} below.

The total derivative of $\bar{\mu}_{\rm SF}$ with respect to reduced temperature $t_{\rm A}$ is obtained by substituting \mbox{$t \to (1+h_{\rm A1})t_{\rm A}$} from Eq.~(\ref{Eq:tADef}) into Eq.~(\ref{Eq:dbarmu0dt}), yielding 
\bse
\label{Eqs:dbarmuFlop}
\be
\frac{d\bar{\mu}_{\rm SF}}{dt_{\rm A}} = -\frac{\bar{\mu}_{\rm SF}}{t_{\rm A}\Big[\frac{(S+1)(1+h_{\rm A1})t_{\rm A}}{3B_S^\prime(y_{\rm A})} - 1\Big]},
\label{Eq:dbarMuFlop}
\ee
where
\be
y_{\rm A} = \frac{3\bar{\mu}_{\rm SF}}{(S+1)(1+h_{\rm A1})t_{\rm A}},
\label{Eq:DefineyA}
\ee
\ese
$\bar{\mu}_{\rm SF}(t_{\rm A})$ is obtained by numerically solving Eq.~(\ref{Eq:OrdMomFlop}) and the $B_S(y_{\rm A})$ and $B_S^\prime(y_{\rm A})$ functions are given in Eqs.~(\ref{Eqs:BS}).  For $h_{\rm A1}=0$, Eq.~(\ref{Eq:dbarMuFlop}) reduces to Eq.~(\ref{Eq:dbarmu0dt}) (with $t_{\rm A}=t$), as required.

\subsubsection{\label{Sec:muzVShzSF} Magnetization versus $z$-Axis Field}

The magnetic susceptibility $\chi_{z{\rm SF}}$ along the easy $z$~axis of the SF phase shown in the top panel of Fig.~\ref{Fig:MparMperp} is not the same as $\chi_\perp$ of the AFM phase in Eq.~(\ref{Eq:ChiPerpAnis2}) obtained when the applied field is perpendicular to the easy axis or plane as in Fig.~\ref{Fig:High_Perp_Field_Structs}.  The reason for this difference is that when the applied field is along the $z$~axis in the SF phase, this field and the anisotropy field are in the same direction for all magnetic moments, whereas in the AFM case the anisotropy field lies within the $xy$ plane and hence these two fields are perpendicular to each other.  Thus the reduced critical field for the spin flop phase $h_{\rm cSF}$, at which the ordered moments become parallel to the field with increasing field, is smaller than $h_{\rm c\perp AFM}$ of the AFM phase in a perpendicular field in the presence of an anisotropy field.

\subsubsection*{Torque Calculation}

To calculate the $z$-axis susceptibility of the SF phase we use a similar calculation as in Sec.~\ref{Sec:muPerpAFM}, but with the replacement
\be
H_\perp\to H_z+H_{\rm A} = H_z + \frac{3H_{\rm A1}\bar{\mu}_{\rm SF}\cos\theta}{S+1},
\ee 
where we have used Eqs.~(\ref{Eq:HAiAxial}) and~(\ref{Eq:HA0}) to express $H_{\rm A}$ in terms of $H_{\rm A1}$ and have set $\theta_i\to\theta$ and $\bar{\mu}_i,\ \bar{\mu}\to\bar{\mu}_{\rm SF}$.  Inserting this expression into Eq.~(\ref{Eq:tauequalszero}) gives
\be
\frac{3k_{\rm B}}{S+1}(T_{{\rm N}J}-\theta_{{\rm p}J})\bar{\mu}_{\rm SF}\cos\theta = g\mu_{\rm B}H_z + \frac{3g\mu_{\rm B}H_{\rm A1}\bar{\mu}_{\rm SF}\cos\theta}{S+1}.
\label{Eq:tauequalszero2}
\ee
Then solving for $\bar{\mu}_{z{\rm SF}} = \bar{\mu}_{\rm SF}\cos\theta$ gives
\bse
\bea
\bar{\mu}_{z{\rm SF}} &=& \frac{(S+1)h_z}{3(1-f_J-h_{\rm A1})},\label{Eq:barmuzSF}\\*
{\rm or}\ h_z &=& \frac{3(1-f_J-h_{\rm A1})}{S+1}\bar{\mu}_{z{\rm SF}},\label{Eq:hzVsbarmuz}
\eea
\ese
where we used $T_{{\rm N}J}-\theta_{{\rm p}J}=T_{{\rm N}J}(1-f_J)$, the reduced anisotropy field $h_{\rm A1}$ was defined in Eq.~(\ref{Eq:TNATNRatio}), and similarly for the reduced applied field $h_z$.  Thus $\bar{\mu}_z\propto h_z$ in the SF phase.  Since $\bar{\mu}_z\geq0$,  the maximum physical range of $h_{\rm A1}$ is
\be
0 \leq h_{\rm A1} < 1-f_J.
\label{Eq:hA1Range}
\ee
The reduced susceptibilities defined in Eqs.~(\ref{Eq:chiAlphaStarDef}) and~(\ref{Eq:chiRedDef}) are then
\bea
\chi_{z{\rm SF}}* &=& \frac{S+1}{3(1-f_J-h_{\rm A1})}, \\*
\bar{\chi}_{z{\rm SF}} &=& \frac{1}{1-f_J-h_{\rm A1}}.\label{Eq:chibarzSF}
\eea
One sees by comparison with Eq.~(\ref{Eq:barchiperp}) that \mbox{$\bar{\chi}_{z{\rm SF}}>\bar{\chi}_{\rm\perp AFM}$}. This inequality was qualitatively explained previously by Buschow and de~Boer \cite{Buschow2004}.  

\subsubsection*{Alternate Hamiltonian Diagonalization Calculation}

In this section we give an alternative derivation of the field-induced moment of the SF phase.  The energy~$E_i$ of a representative spin~$i$ in a magnetic induction ${\bf B}_i$ is
\be
E_i = -\vec{\mu}_i\cdot {\bf B}_i = g\mu_{\rm B}{\bf S}\cdot{\bf B}_i,
\label{Eq:EiQ}
\ee
where in the second equality we used the expression for the magnetic moment operator
\be
\vec{\mu} = -g\mu_{\rm B}{\bf S},
\ee
the negative sign comes from the negative charge on the electron, and {\bf S} is the spin operator.  As usual, we normalize all energies by $k_{\rm B}T_{{\rm N}J}$, so Eq.~(\ref{Eq:EiQ}) becomes
\be
\epsilon \equiv \frac{E_i}{k_{\rm B}T_{{\rm N}J}} = {\bf S}\cdot{\bf b}_i,
\label{Eq:EiQ2}
\ee
where the reduced induction~${\bf b}_i$ is defined as in Eq.~(\ref{Eq:halphaDef}), and ${\bf b}_i$ is the sum of the reduced applied, anisotropy and exchange fields.

Using Eqs.~(\ref{Eq:HexchiDef}), (\ref{Eqs:TNqpJ}), (\ref{Eqs:muimujExpand}), and~(\ref{Eq:EiQ}), the exchange part of the reduced Hamiltonian for ${\bf S}_i$, assumed without loss of generality to lie in the $xz$~plane, is 
\bse
\bea
\frac{{\cal H}_{{\rm exch}i}}{k_{\rm B}T_{{\rm N}J}} &=& \frac{3\bar{\mu}_{\rm SF}}{S+1}(S_x\sin\theta+f_JS_z\cos\theta)\\*
&=&  \frac{3}{S+1}(S_x\bar{\mu}_{x{\rm SF}}+f_JS_z\bar{\mu}_{z{\rm SF}}),\nonumber
\eea
where we used the relations $\bar{\mu}_{\rm SF}=\mu_{\rm SF}/g\mu_{\rm B}S$, $\bar{\mu}_{x{\rm SF}} = \bar{\mu}_{\rm SF}\sin\theta$ and $\bar{\mu}_{z{\rm SF}} = \bar{\mu}_{\rm SF}\cos\theta$, and $\mu_{\rm SF}$ is the magnitude of the ordered moment of each spin.  Here $S_x$ is the usual combination of raising and lowering operators $S_x=(S_++S_-)/2$ and $S_z$ is diagonal in the $|S,S_z\rangle$ Hilbert space.  Similarly, the parts of the Hamiltonian for the anisotropy and applied fields are
\bea
\frac{{\cal H}_{{\rm A}i}}{k_{\rm B}T_{{\rm N}J}} &=& \frac{3h_{\rm A1}}{S+1}\bar{\mu}_zS_z\\*
\frac{{\cal H}_{Hi}}{k_{\rm B}T_{{\rm N}J}} &=& S_zh_z.
\eea
\ese
We thereby obtain the total reduced Hamiltonian
\bse
\label{Eqs:SFHam}
\bea
\frac{\cal H}{k_{\rm B}T_{{\rm N}J}} &=& \left(\frac{3\bar{\mu}_{x{\rm SF}}}{S+1}\right)S_x\nonumber\\*
&& +\ \left[\left(\frac{3\bar{\mu}_{z{\rm SF}}}{S+1}\right)(f_J+h_{\rm A1})+h_z\right]S_z\nonumber\\*
&\equiv& b_xS_x+b_zS_z,
\eea
where
\bea
b_x &=& \frac{3\bar{\mu}_{x{\rm SF}}}{S+1},\label{Eqs:bxby}\\*
b_z &=& \left(\frac{3\bar{\mu}_{z{\rm SF}}}{S+1}\right)(f_J+h_{\rm A1})+h_z.\nonumber
\eea
\ese
The reduced magnetic moment operators for eigenenergies $n = 1$ to $2S+1$ are \cite{Johnston2017}
\bse
\bea
\bar{\mu}_x^{\rm op} &=& -\frac{1}{S}\frac{\partial\epsilon_n}{\partial b_x}\Big|_{b_x = 3\bar{\mu}_x/(S+1)},\\*
\bar{\mu}_z^{\rm op} &=& -\frac{1}{S}\frac{\partial\epsilon_n}{\partial b_z}\Big|_{b_z = [3\bar{\mu}_z/(S+1)](f_J+h_{\rm A1}) + h_z}.
\eea
\ese
Then the thermal-average reduced moments $\bar{\mu}_{x{\rm SF}}$ and $\bar{\mu}_{z{\rm SF}}$ for the SF phase are calculated by solving the simultaneous equations
\bse
\label{Eqs:SolveMuxMuzSF}
\bea
\bar{\mu}_{x{\rm SF}} &=& \frac{1}{Z_S}\sum_{n=1}^{2S+1}\bar{\mu}_x^{\rm op}e^{-\epsilon_n/t},\label{Eqs:barmuxbarmuz}\\*
\bar{\mu}_{z{\rm SF}} &=& \frac{1}{Z_S}\sum_{n=1}^{2S+1}\bar{\mu}_z^{\rm op}e^{-\epsilon_n/t},\nonumber
\eea
where
the partition function is
\be
Z_S = \sum_{n=1}^{2S+1}e^{-\epsilon_n/t},
\label{Eq:ZSFinal}
\ee
the reduced magnitude of the ordered moment is
\be
\bar{\mu}_{\rm SF} = \sqrt{\bar{\mu}_{x{\rm SF}}^2+\bar{\mu}_{z{\rm SF}}^2},
\ee
\ese
and in this section we use the reduced temperature $t\equiv T/T_{{\rm N}J}$.  The two Eqs.~(\ref{Eqs:barmuxbarmuz}) are solved iteratively for $\bar{\mu}_{x{\rm SF}}$ and~$\bar{\mu}_{z{\rm SF}}$ for each desired combination of $f_J$, $h_{\rm A1}$, $h_z$ and $t$ for a fixed spin~$S$~\cite{Johnston2017}.

\begin{figure}
\includegraphics [width=3.in]{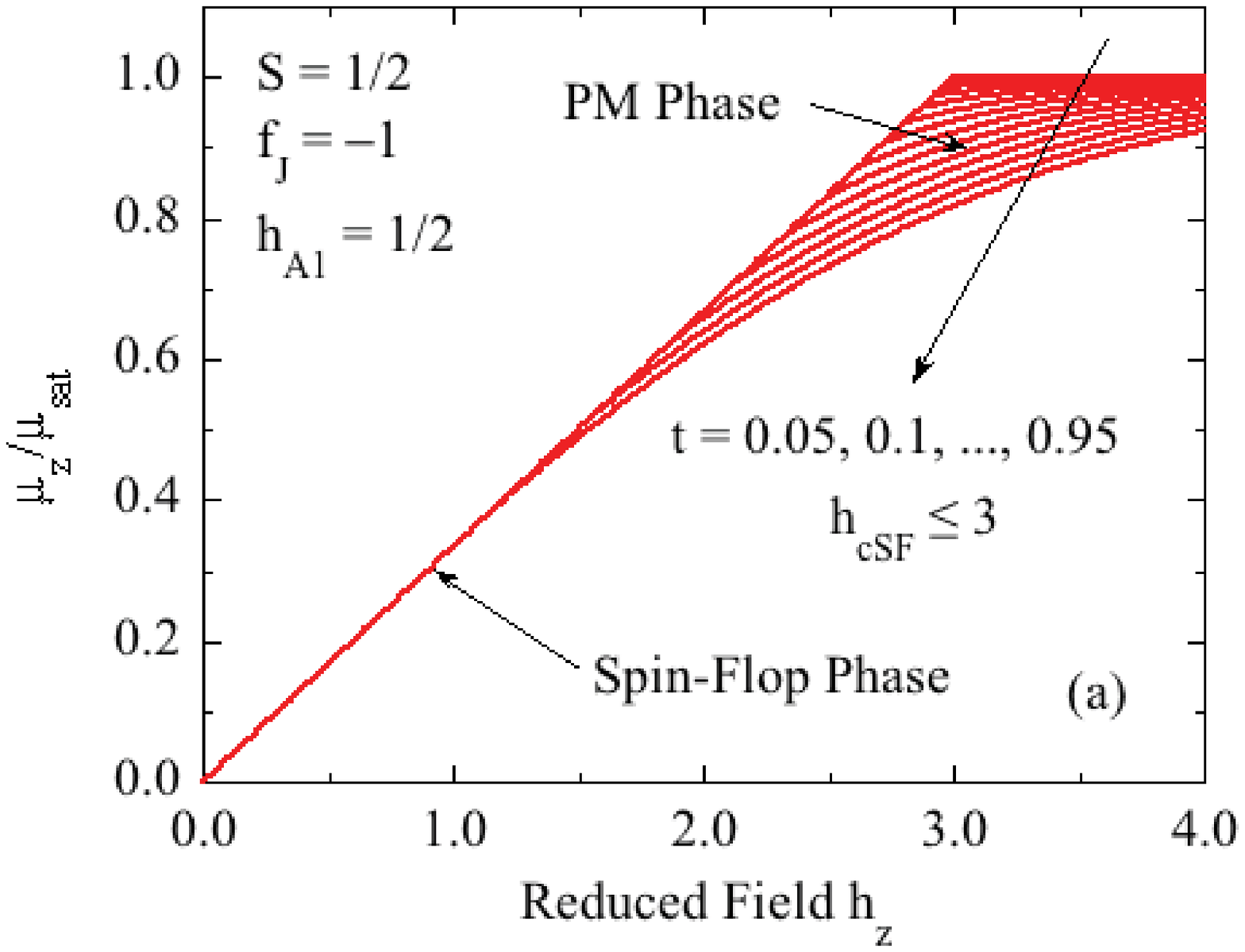}
\includegraphics [width=3.in]{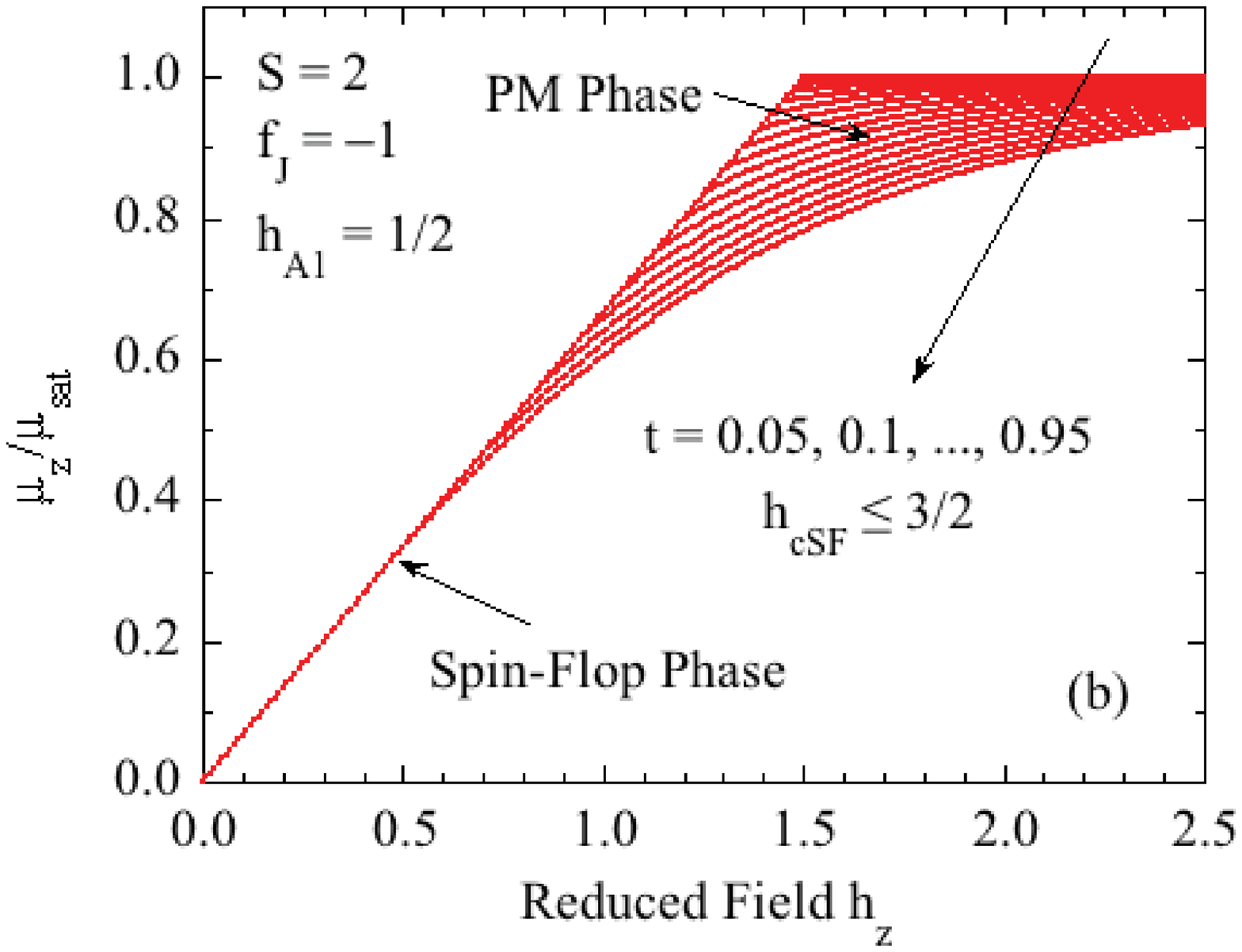}
\includegraphics [width=3.in]{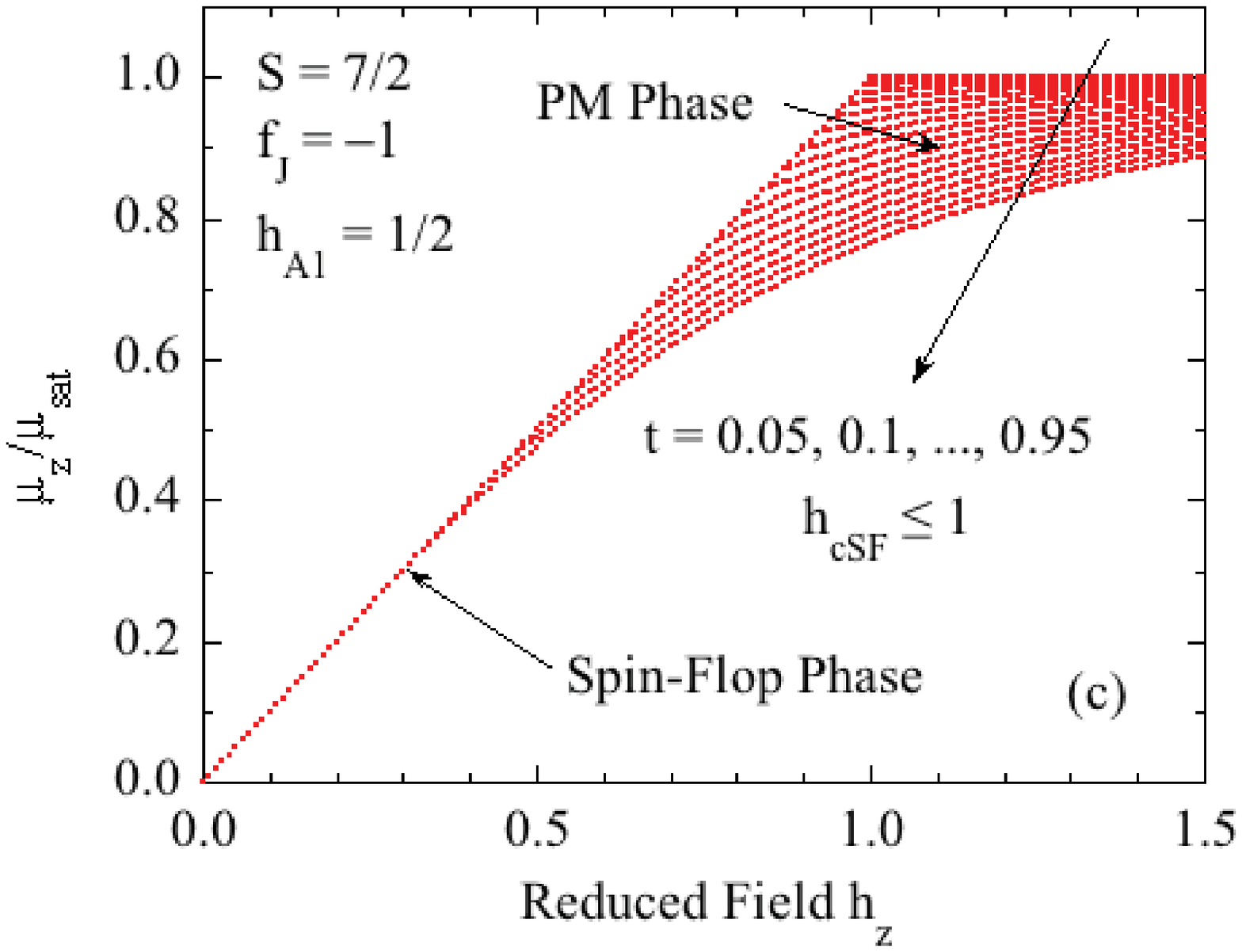}
\caption{(Color online) Reduced induced moment per spin $\bar{\mu}_z \equiv \mu_{z}/\mu_{\rm sat}$ for the low-field spin-flop (SF)  and high-field paramagnetic (PM) phases of a collinear or planar noncollinear antiferromagnet versus reduced field~$h_z$ for reduced anisotropy field $h_{\rm A1} = 1/2$ and $f_J=-1$ at reduced temperatures $t = T/T_{{\rm N}J}$ from 0.05 to 0.95 for spins (a) $S = 1/2$, (b)~$S=2$, and (c)~$S=7/2$ calculated using Eqs.~(\ref{Eqs:SolveMuxMuzSF}). The SF and PM field ranges are separated by a break in slope in $\bar{\mu}_z$ versus~$h_z$ at the reduced critical field $h_z = h_{\rm cSF}(t)$ in Eq.~(\ref{Eq:HcFlopDef}).  Note the different abscissa scales in panels (a)--(c).}
\label{Fig:muVsHtSFhA150fJm1S12}
\end{figure}

Calculations of $\bar{\mu}_{z{\rm SF}}$ versus~$h_z$ isotherms at many $t$ values obtained using Eqs.~(\ref{Eqs:SolveMuxMuzSF}) are shown in Fig.~\ref{Fig:muVsHtSFhA150fJm1S12} for spins $S = 1/2$, 2 and~7/2 with $f_J=-1$ and $h_{\rm A1} = 1/2$, where the data for the PM phase at $h_z\geq h_{\rm cSF}$ (below) are obtained automatically.  These results agree with what would have been obtained from the results in the previous section based on torque calculations.

We also find that the magnitude of the reduced ordered moment $\bar{\mu}_{\rm SF}$ is independent of $h_z$ for the SF phase (over the proportional part of the $\bar{\mu}_z$ versus~$h_z$ isotherm) at each temperature.

\subsubsection{Critical Field}

\begin{figure}
\includegraphics [width=3.in]{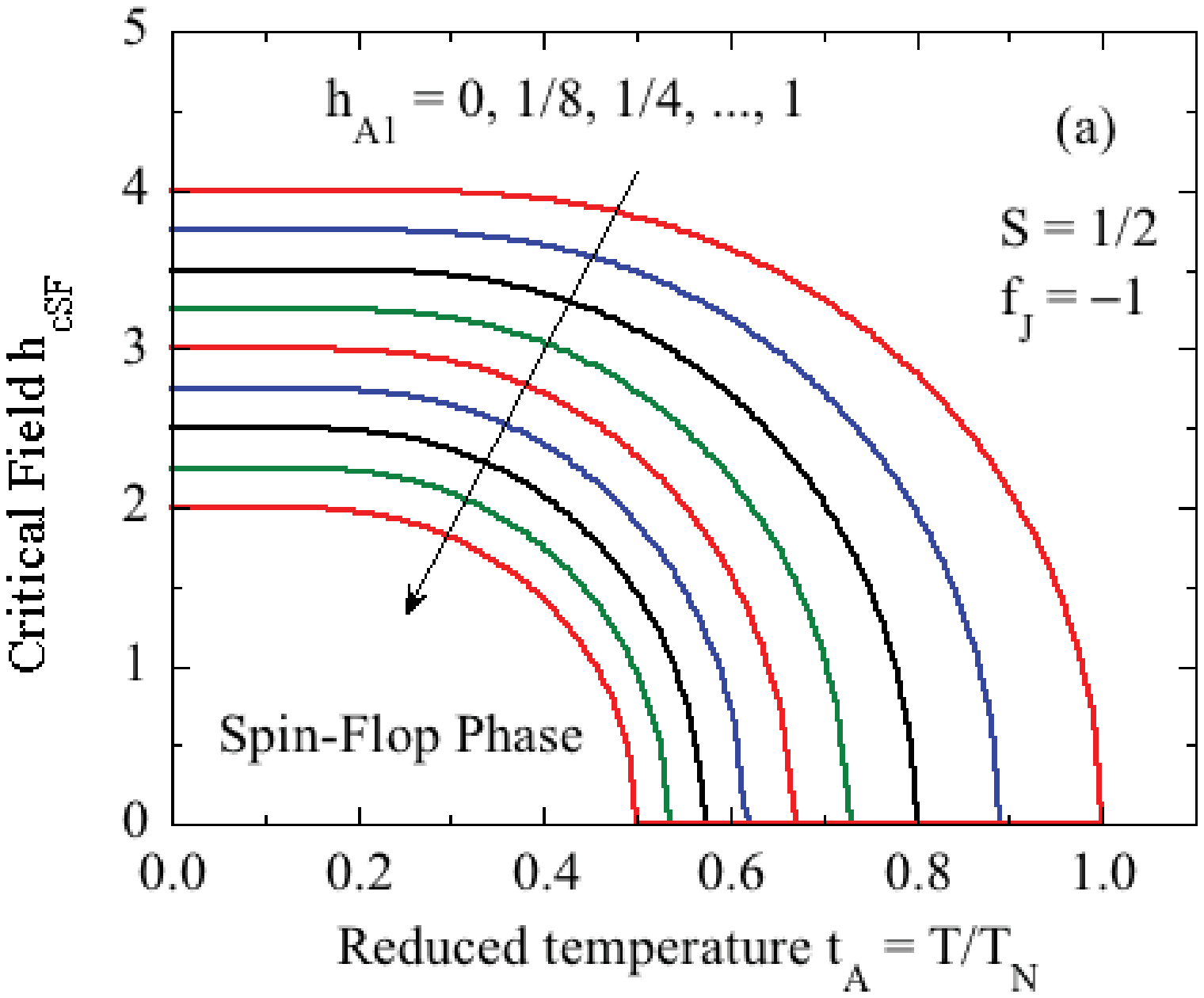}
\includegraphics [width=3.in]{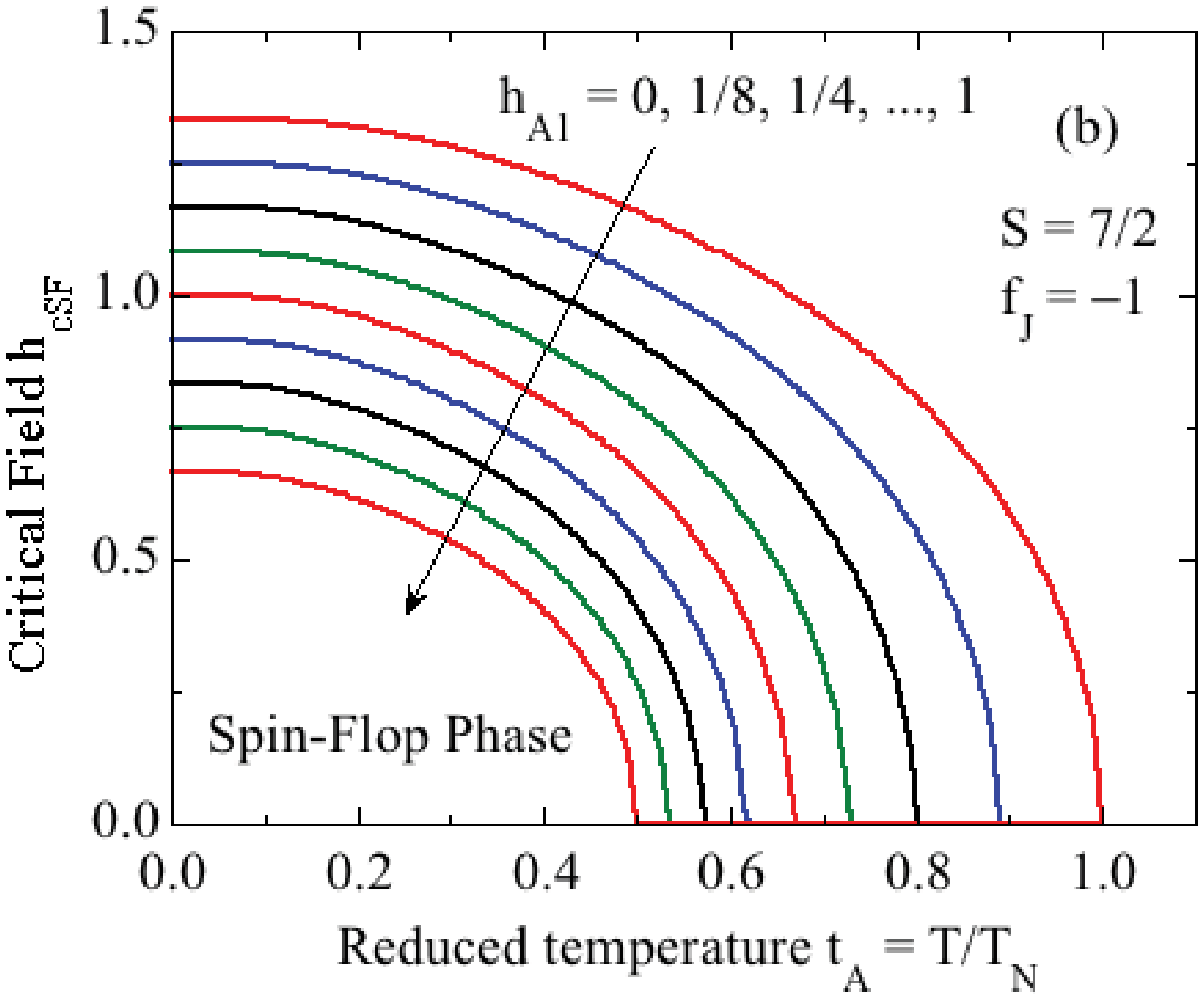}
\caption{(Color online) Reduced critical field $h_{\rm cSF}$ for the spin-flop phase of a collinear antiferromagnet versus reduced temperature $t_{\rm A}$ with $f_J=-1$ and $h_{\rm A1}= 0$ to~1 for spins (a)~$S = 1/2$ and (b)~$S=7/2$, calculated using Eq.~(\ref{Eq:HcFlopDef}). }
\label{Fig:MFT_Flop_hcFlopS12fm1}
\end{figure}

The critical field $H_{\rm cSF}$ of the spin flop phase is defined as the value of the applied field $H_z$ at which all the magnetic moments become aligned with the field, as in the right-hand side of the top panel of Fig.~\ref{Fig:MparMperp}.  Since $\mu_{z{\rm SF}}/H_z$ is independent of $H_z$ within the SF phase, this criterion and~Eq.~(\ref{Eq:barmuzSF}) gives the reduced critical field
\be
h_{\rm cSF} = \frac{3(1-f_J-h_{\rm A1})}{S+1}\bar{\mu}_{\rm SF},
\label{Eq:HcFlopDef}
\ee
where $\bar{\mu}_{\rm SF}$ versus~$t$ or~$t_{\rm A}$ is obtained by solving the first or second of Eqs.~(\ref{Eqs:mubarFlop}), respectively.  The $h_{\rm cSF}$ is dependent on temperature because $\bar{\mu}_{\rm SF}$ is.  Since $0 \leq \bar{\mu}_{\rm SF}\leq 1$, the physically relevant range for positive $h_{\rm cSF}$ is
\be
0 \leq h_{\rm cSF} \leq \frac{3(1-f_J-h_{\rm A1})}{S+1}.
\label{Eq:HcNorm2}
\ee
For $h_z \geq h_{\rm cSF}$, the system is in the PM phase with all induced moments having the same magnitude $\bar{\mu}_{z{\rm PM}}$ and pointing in the direction of {\bf H}\@.  

Shown in Fig.~\ref{Fig:MFT_Flop_hcFlopS12fm1} are plots of $h_{\rm cSF}$ versus~$t_{\rm A}$ for $f_J=-1$ and spins $S=1/2$ and $S=7/2$, each with anisotropy parameters $h_{\rm A1} = 0$ to~1.  The shapes of the curves are significantly different for the two spin values.  One also sees that the critical fields are much smaller for $S = 7/2$ than for $S=1/2$, consistent with Eq.~(\ref{Eq:HcFlopDef}).

\subsubsection{Spin-Flop and Paramagnetic Phase Magnetization Summary}

To summarize, the field dependences of the magnetization for the low-field SF and high-field PM phases are given by Eqs.~(\ref{Eq:barmuzSF}) and~(\ref{Eq:muParPM}), respectively, as  
\bse
\label{Eqs:muVsHAFPMFlop}
\bea
\bar{\mu}_{z{\rm SF}} &=& \bar{\mu}_z = \frac{(S+1)h_z}{3(1-f_J-h_{\rm A1})} \qquad (h_z \leq h_{\rm cSF}),\hspace{0.5in}\label{Eq:muVstAFlop}\\*
\bar{\mu}_{z{\rm PM}} &=& B_S(y_{\rm PM}) \\*
y_{\rm PM} &=& \frac{1}{(1+h_{\rm A1})t_{\rm A}}\left[\frac{3(f_J+h_{\rm A1})}{S+1}\bar{\mu}_{z{\rm PM}} + h_z\right]\nonumber\\*
&& \hspace{1.6in} (h \geq h_{\rm cSF}),\nonumber
\eea
\ese
where $h_{\rm cSF}$ is given in Eq.~(\ref{Eq:HcFlopDef}) and $\bar{\mu}_{\rm SF}$ is obtained by solving Eq.~(\ref{Eq:OrdMomFlop}).  Note that the slope of $\bar{\mu}_{z{\rm SF}}$ versus $h_z$ for the SF phase in Eq.~(\ref{Eq:muVstAFlop}) depends on $S$, $f_J$, and $h_{\rm A1}$, and not on the temperature.  The temperature only determines the maximum field at which the proportionality occurs.

\begin{figure}
\includegraphics [width=3.in]{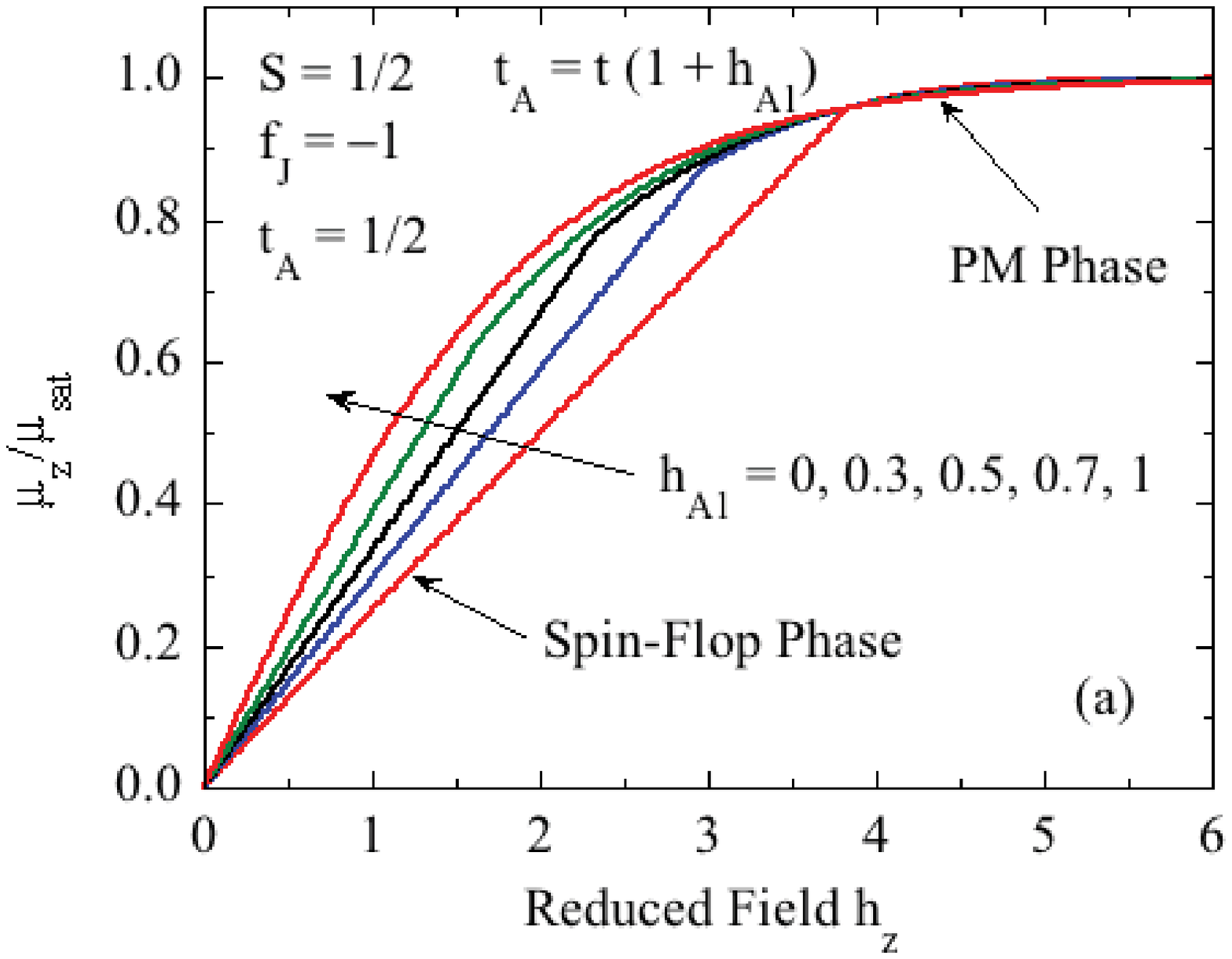}
\includegraphics [width=3.in]{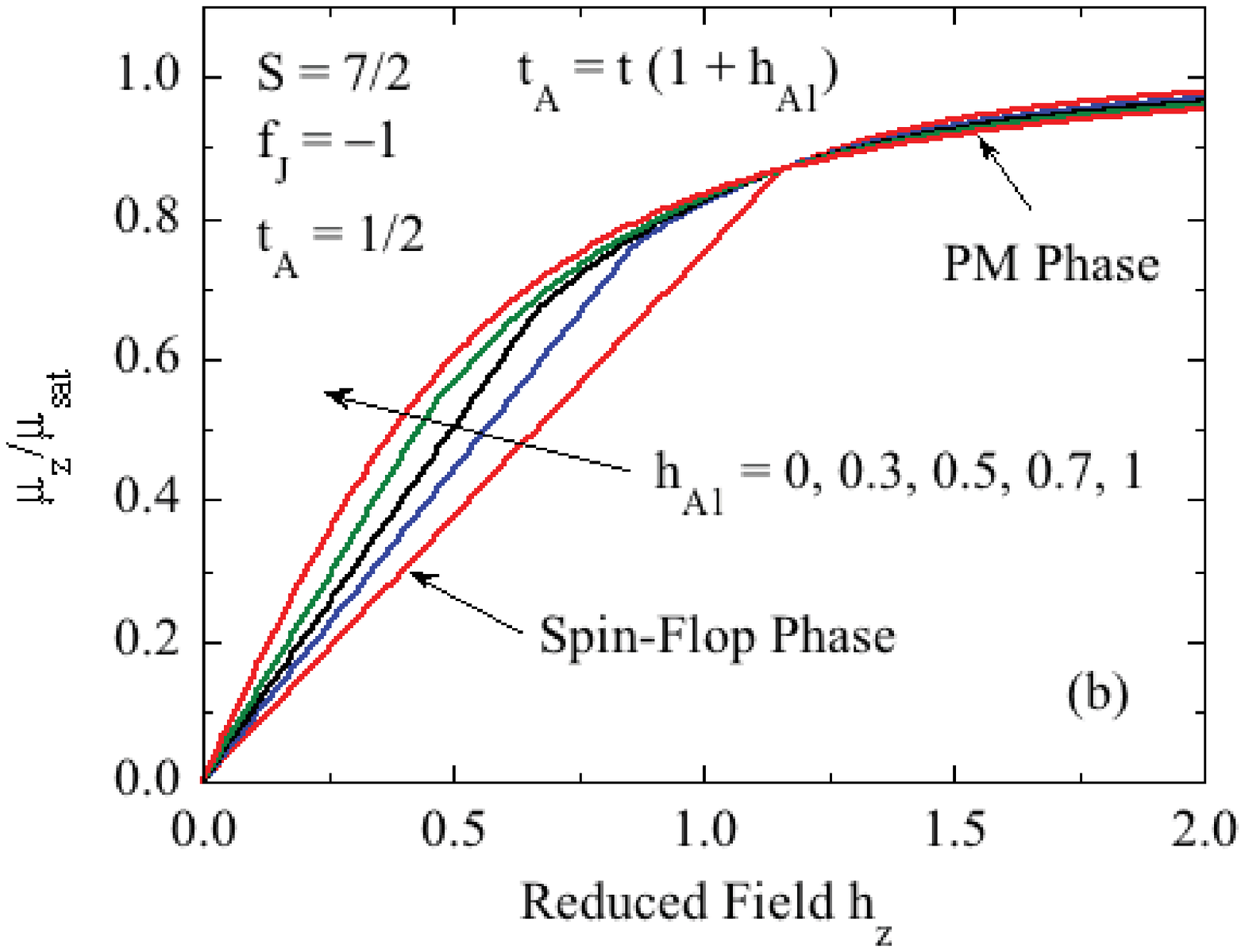}
\caption{(Color online) Reduced ordered moment $\bar{\mu}_z \equiv \mu_z/\mu_{\rm sat}$ versus reduced field~$h_z$ for the spin-flop (SF) and subsequent paramagnetic (PM) phases of a collinear or planar noncollinear antiferromagnet at reduced temperature $t_{\rm A} = T/T_{\rm N} = t(1+h_{\rm A1}) = 1/2$ with $f_J=-1$ and $h_{\rm A1}= 0$ to~1 for spins (a) $S = 1/2$ and (b)~7/2, calculated using Eqs.~(\ref{Eq:HcFlopDef}) and~(\ref{Eqs:muVsHAFPMFlop}). The SF and PM ranges are separated by a break in slope in $\mu_z/\mu_{\rm sat}$ versus~$h_z$ at $h_z = h_{\rm cSF}$.  However, the curve in each of (a) and~(b) with $h_{\rm A1}=1$ is paramagnetic over the full field ranges shown.}
\label{Fig:MFT_Flop_muzVShS12fm1}
\end{figure} 

\begin{figure}
\includegraphics [width=3.in]{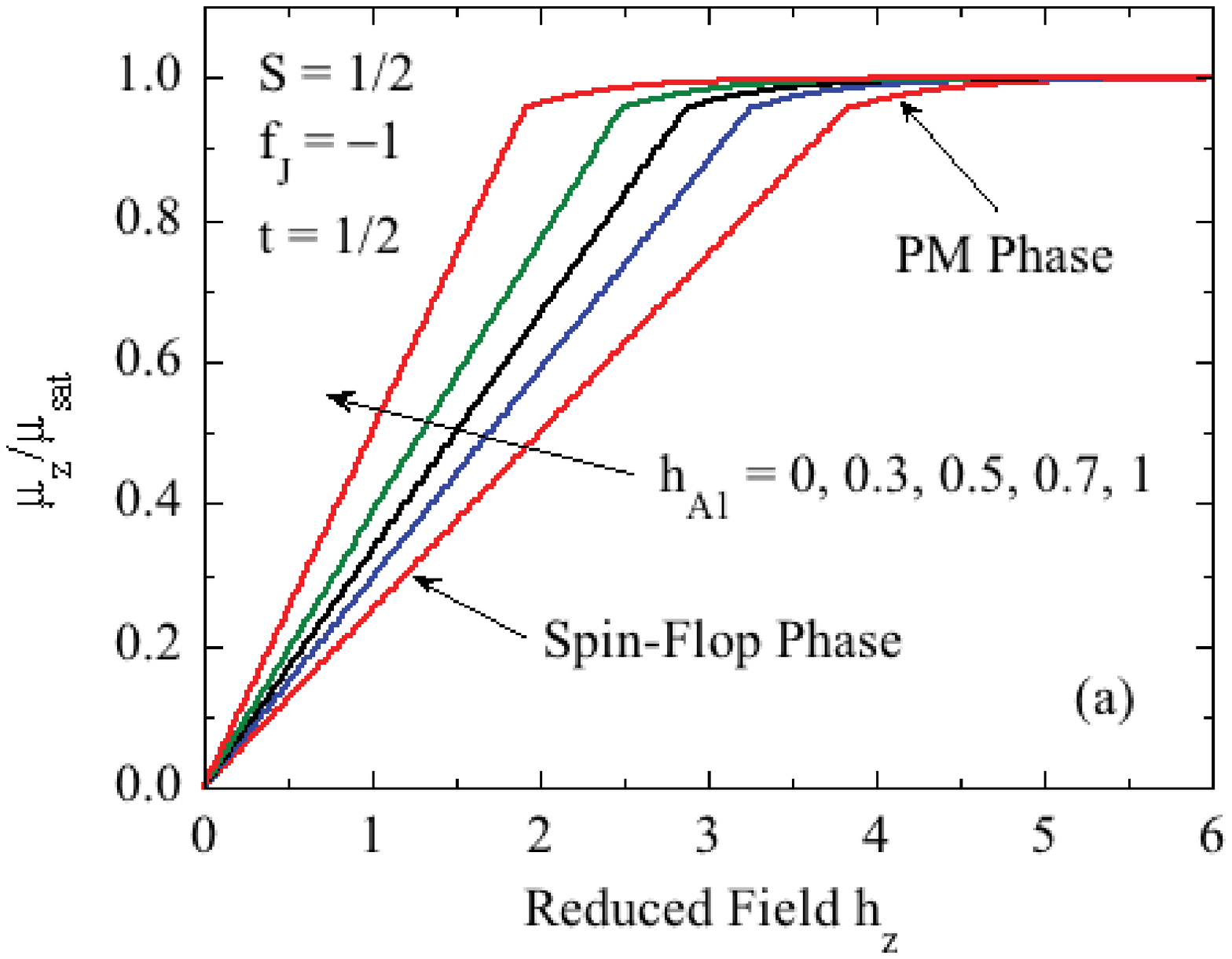}
\includegraphics [width=3.in]{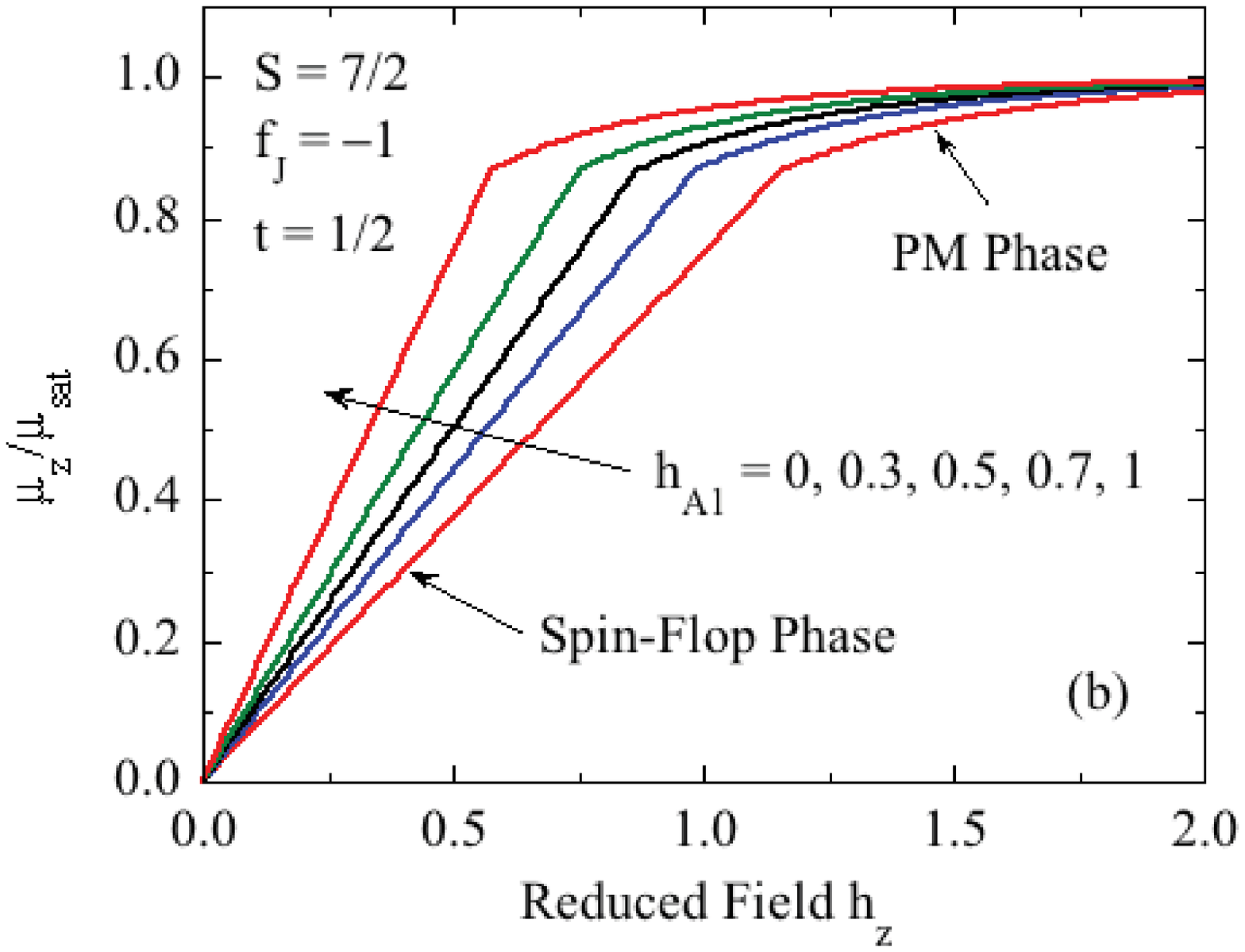}
\caption{(Color online) Same as Fig.~\ref{Fig:MFT_Flop_muzVShS12fm1}, except that the reduced temperature $t=T/T_{{\rm N}J}$ is fixed at the value of~1/2 instead of $t_{\rm A} = T/T_{\rm N} = 1/2$.  The plots are different than in Fig.~\ref{Fig:MFT_Flop_muzVShS12fm1} because $T_{\rm N}$ depends on $h_{\rm A1}$.}
\label{Fig:MFT_SFmuzVShzfJm1t50S12}
\end{figure}

The reduced $z$-axis moment of the SF phase $\bar{\mu}_{z{\rm SF}}$ is plotted versus the reduced fiield~$h_z$ in Fig.~\ref{Fig:MFT_Flop_muzVShS12fm1} for $t_{\rm A}=1/2$ and for $S=1/2$ and $S=7/2$ with $h_{\rm A1} = 0$ to~1.  The low-field SF portion is proportional to~$h_z$ but then undergoes a second-order phase transition via a slope reduction to the PM state for which $\bar{\mu}_{z{\rm SF}}$ exhibits negative curvature.  For $h_{\rm A1}=1$ only the PM phase occurs for both spin values, as seen in Fig.~\ref{Fig:MFT_Flop_muzVShS12fm1}, because one can show that $h_{\rm cSF}=0$ for any $S$ if $h_{\rm A1} = 0.5$, $f_J=-1$ and~$t_{\rm A}=0.5$ as illustrated in Fig.~\ref{Fig:MFT_Flop_hcFlopS12fm1} for $S=1/2$ and $S=7/2$.  It is important to note here that $t_{\rm A}$ is not proportional to the absolute temperature, since it depends on $h_{\rm A1}$ according to the formula in the figures.  Therefore in Fig.~\ref{Fig:MFT_SFmuzVShzfJm1t50S12} the same quantities are plotted as in Fig.~\ref{Fig:MFT_Flop_muzVShS12fm1}, but where the reduced temperature $t=T/T_{{\rm N}J}$, proportional to the absolute temperature~$T$, is fixed to the same value of 1/2.  Qualitative differences are seen between the two figures.

\subsubsection{\label{Sec:mu0SF} Internal Energy versus Temperature}

\begin{figure}
\includegraphics [width=3.in]{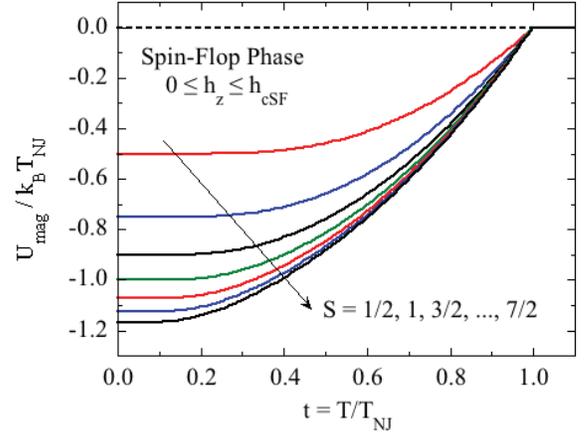}
\caption{(Color online) Internal energy per spin $U_{\rm mag}$ normalized by $k_{\rm B}T_{{\rm N}J}$ versus reduced temperature $t$ for the spin-flop phase with spins $S=1/2$ to~7/2, obtained from Eq.~(\ref{UmagAFMHz0}).  $U_{\rm mag}$ is independent of field in the field range of stability of the SF phase with respect to the PM phase, given by $0\leq h_z\leq h_{\rm cSF}$, where $h_{\rm cSF}$ is given in Eq.~(\ref{Eq:HcFlopDef}).}
\label{Fig:UmagVStSFSxx}
\end{figure}

We established in Sec.~\ref{Sec:muzVShzSF} that the ordered moment $\bar{\mu}_{\rm SF}$ is independent of field within the SF phase, i.e., for $0\leq h_z\leq h_{\rm cSF}(t)$.  For $h_z=0$, the ordered moments are oriented in the $xy$~plane for which the anisotropy field is zero as inferred from Eq.~(\ref{Eqs:HAiAxial}) and Fig.~\ref{Fig:Anis_Free_energy}(b). Hence the magnetic induction seen by a spin is identical to that of a spin in an AFM in zero applied and anisotropy fields, and therefore the internal energy per spin is given by Eq.~(\ref{Eq:UmagAFMH0}) or by  Eq.~(\ref{Eq:Ui}) with~$h_{\rm A1}=0$, i.e.,
\be
\frac{U_{\rm mag}}{k_{\rm B}T_{{\rm N}J}} = -\frac{3S}{2(S+1)}\bar{\mu}_0^2,
\label{UmagAFMHz0}
\ee
where $\bar{\mu}_0(t)$ is obtained by solving Eq.~(\ref{Eq:mubar0}).  At $t=0$, one has $\bar{\mu}_0=1$, yielding
\be
\frac{U_{\rm mag}(h_z\leq h_{\rm cSF},t=0)}{k_{\rm B}T_{{\rm N}J}} = -\frac{3S}{2(S+1)}.
\label{UmagAFMHz0t0}
\ee

Shown in Fig.~\ref{Fig:UmagVStSFSxx} are plots of $U_{\rm mag}/k_{\rm B}T_{{\rm N}J}$ versus~$t$ for spins~$S=1/2$ to~$S=7/2$ in half-integer increments.  The internal energy for all spin values goes to zero at the same temperature $T=T_{{\rm N}J}$ because $\bar{\mu}_0$ does. One also sees that Eq.~(\ref{UmagAFMHz0t0}) is satisfied for all spin values.

\subsubsection{\label{Sec:FmagSF}  Free Energy}

\begin{figure}
\includegraphics [width=3.in]{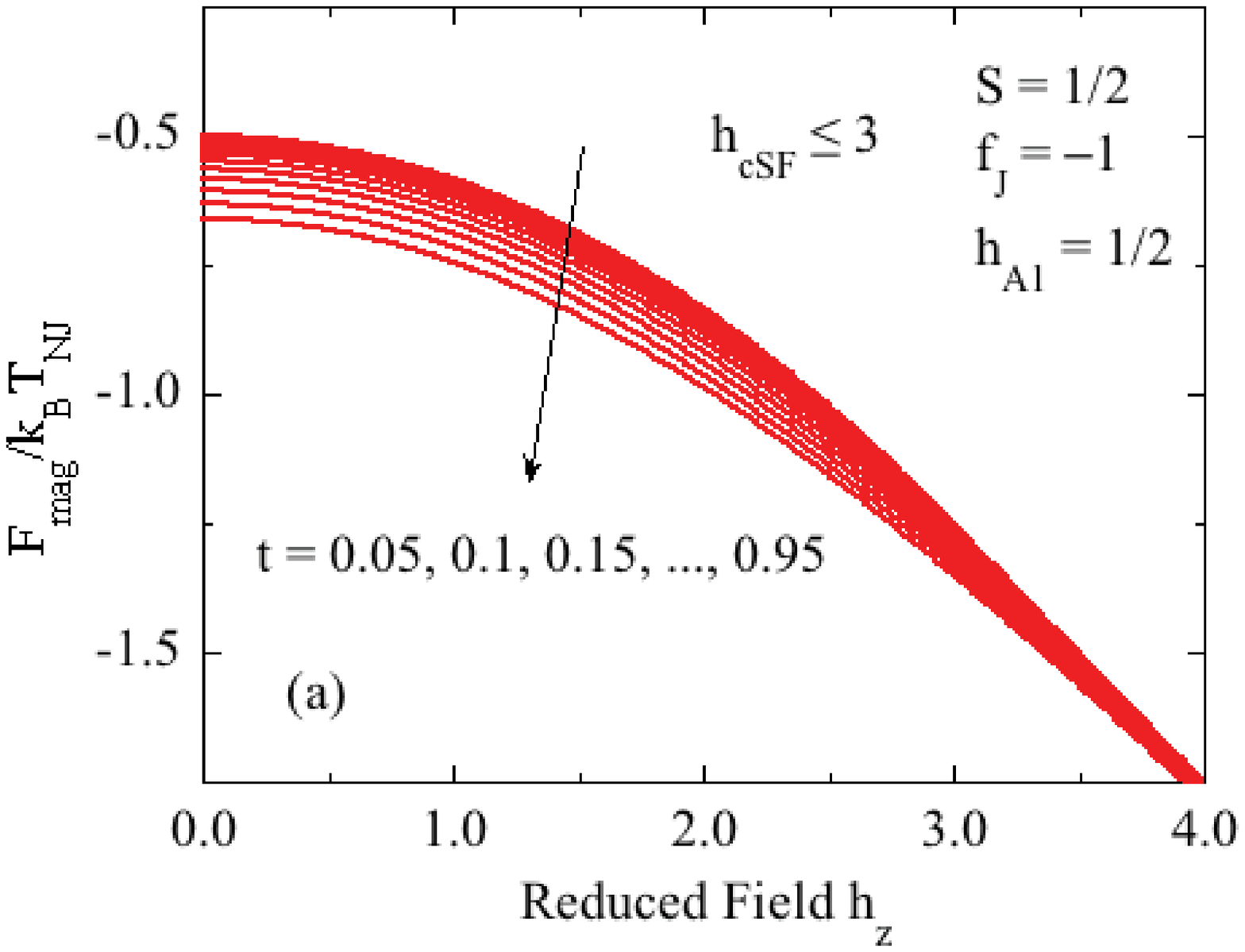}
\includegraphics [width=3.in]{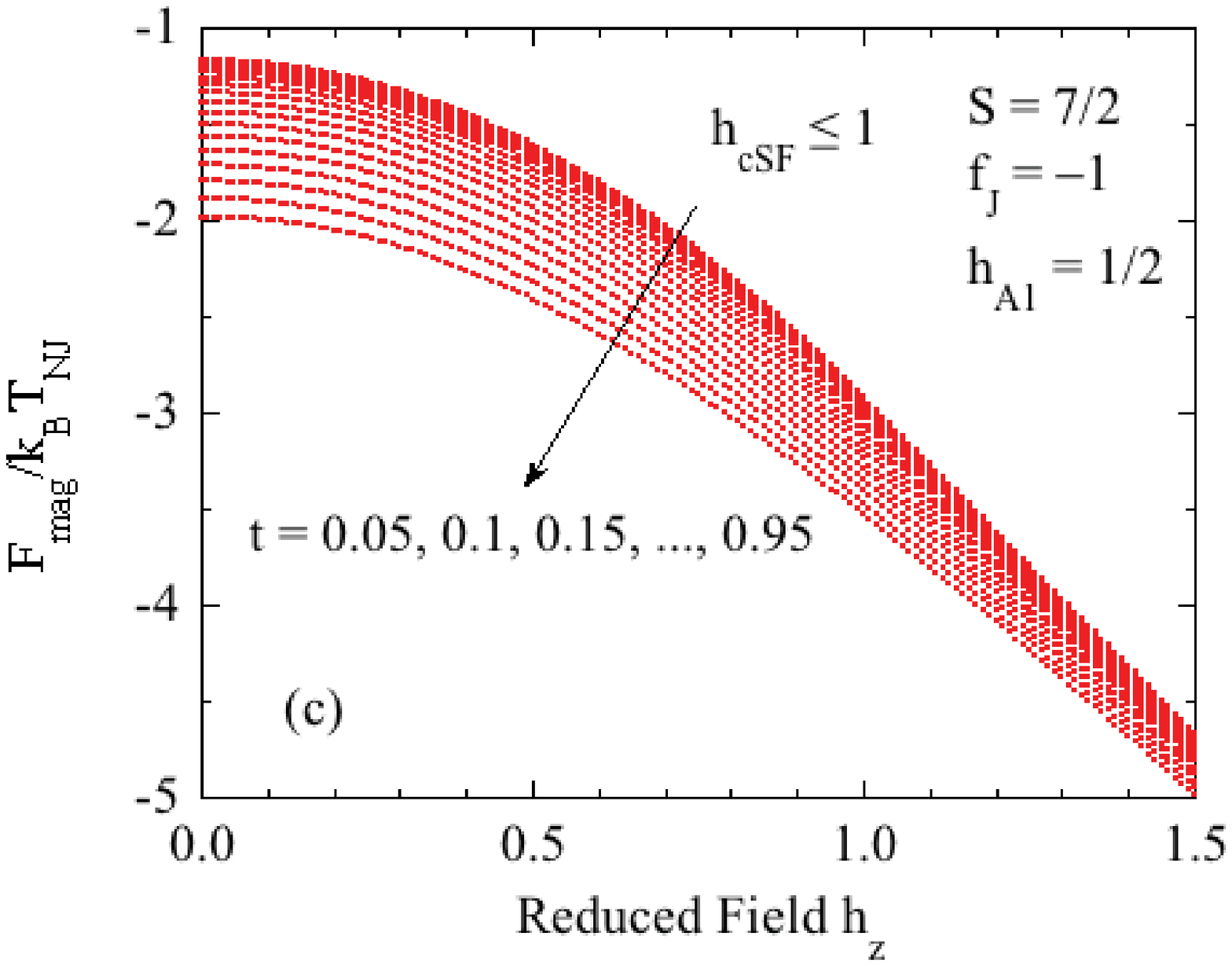}
\caption{(Color online) Reduced magnetic free energy per spin $F_{\rm mag}/k_{\rm B}T_{{\rm N}J}$ for the low-field spin-flop (SF) and high-field paramagnetic (PM) phases of a collinear or planar noncollinear antiferromagnet versus reduced field~$h_z$ for reduced anisotropy parameter $h_{\rm A1} = 1/2$ and $f_J=-1$ at reduced temperatures $t = T/T_{{\rm N}J}$ from 0.05 to 0.95 for spins (a) $S = 1/2$ and (b)~$S=7/2$, calculated using Eqs.~(\ref{Eq:FmagGen}), (\ref{Eqs:muVsHAFPMFlop}), and~(\ref{UmagAFMHz0}) togetheer with $\bar{\mu}_z(h_z,t)$ data such as iin Figs.~\ref{Fig:muVsHtSFhA150fJm1S12} and~\ref{Fig:MFT_SFmuzVShzfJm1t50S12}. Note the different axis scales for the two panels.  The second-order phase transitions from the SF to the PM phase occur at $h_z = h_{\rm cSF}(t)$ in Eq.~(\ref{Eq:HcFlopDef}) and Fig.~\ref{Fig:MFT_Flop_hcFlopS12fm1} and are not obvious in either panel.  The respective value of $h_{\rm cSF}(t=0)$ [the upper limit of $h_{\rm cSF}(t)$] is shown in each panel.}
\label{Fig:FmagVsHSFhA150fJm1S12}
\end{figure}

The free energy~$F_{\rm mag}$ is calculated from Eqs.~(\ref{Eqs:Thermodynamics}) using Eq.~(\ref{UmagAFMHz0}) and $U_{\rm mag}$ data such as in Fig.~\ref{Fig:UmagVStSFSxx} and $\bar{\mu}_z(h_z,t)$ data such as illustrated in Figs.~\ref{Fig:muVsHtSFhA150fJm1S12} and~\ref{Fig:MFT_SFmuzVShzfJm1t50S12}.  Plots of $F_{\rm mag}/k_{\rm B}T_{{\rm N}J}$ versus $h_z$ at fixed values of $t=T/T_{{\rm N}J}$ from~0.05 to~1 for spins $S=1/2$ and~$S=7/2$ are shown in Fig.~\ref{Fig:FmagVsHSFhA150fJm1S12}.  Because the free energy in Eq.~(\ref{Eq:FmagGen}) is derived from an integral of $\bar{\mu}_z(h_z,t)$ over $h_z$, the second-order transitions between the SF and PM states at $h_z = h_{\rm cSF}$ are not obvious from the figure.  The value of $h_{\rm cSF}(t=0)$ for each spin value is given in the respective panel. 

\section{\label{Sec:HiHParMAnisII} High-Field Parallel Magnetization of \lowercase{z}-Axis Collinear Antiferromagnets: Antiferromagnetic Phase}

Here we consider the general behavior of a collinear AFM where the field is applied along the easy $z$-axis of the AFM structure at finite temperatures. By definition, in the collinear AFM phase the ordered moments are always aligned along the $z$~axis.

\subsection{Preliminaries}

When the magnetization along the easy axis of a collinear AFM becomes nonlinear in finite fields, one must define two different sublattices~1 and~2 because in general the magnitudes of the ordered moments parallel and antiparallel to the applied field {\bf H} are different by amounts greater than infinitesimal.  Sublattice~1 is defined to consist of all moments that are parallel to {\bf H} and sublattice~2 consists of the  moments that are antiparallel to {\bf H} when $H_z=0$.  When $H_z$ increases, the magnitudes of the $z$-components $\mu_{1z}$ and $\mu_{2z}$ are in general not the same, which gives a net uniform magnetization in the direction of the field.  However, within the unified MFT we do not require the two sublattices to be bipartite, where the exchange interactions only connect spins of one sublattice with those on the other.  The exchange interactions can connect further neighbors and can be nonfrustrating and/or frustrating for AFM order.  An anisotropy field along the uniaxial $z$~axis is present, as shown in Fig.~\ref{Fig:chiParallel}.

For moments $\vec{\mu}_i$ and~$\vec{\mu}_j$ on the same (``s'') sublattice of a collinear AFM structure, as defined above, the angle between the moments is $\phi_{ji}=0$ in Eq.~(\ref{Eq:HexchiDef}) and for a pair of moments on different (``d'') sublattices, the angle between them in $H_z=0$ is $\phi_{ji}=180^\circ$.  We then write the expressions~(\ref{Eq:TmGeneral}) and~(\ref{Eq:WeissTemp}) for $T_{{\rm N}J}$ and $\theta_{{\rm p}J}$ at $H_z = 0$ for the two-sublattice collinear AFM, respectively, as
\begin{subequations}
\label{Eqs:TNthetaXX}
\bea
T_{{\rm N}J} &=& -\frac{S(S+1)}{3k_{\rm B}} \Big({\sum_j}^{\rm s}J_{ij} - {\sum_j}^{\rm d}J_{ij}\Big),\\*
\theta_{{\rm p}J} &=& -\frac{S(S+1)}{3k_{\rm B}} \Big({\sum_j}^{\rm s}J_{ij} + {\sum_j}^{\rm d}J_{ij}\Big).
\eea
\end{subequations}
Solving these simultaneous equations for the two sums gives
\be
\bs
{\sum_j}^{\rm s}J_{ij} &= -\frac{3k_{\rm B}(T_{{\rm N}J}+\theta_{{\rm p}J})}{2S(S+1)} =  -\frac{3k_{\rm B}T_{{\rm N}J}(1+f_J)}{2S(S+1)},\\*
{\sum_j}^{\rm d}J_{ij} &= \frac{3k_{\rm B}(T_{{\rm N}J}-\theta_{{\rm p}J})}{2S(S+1)} = \frac{3k_{\rm B}T_{{\rm N}J}(1-f_J)}{2S(S+1)},
\end{split}
\label{Eqs:TNthetaSums}
\ee
where $f_J\equiv\theta_{{\rm p}J}/T_{{\rm N}J}$ is defined in Eq.~(\ref{Eq:fRatioDef}).  We  emphasize that $T_{{\rm N}J}$, $\theta_{{\rm p}J}$ and $f_J$ are defined, even in the presence of the anisotropy field, only in terms of the exchange constants and magnetic structure by the above equations, whereas $T_{\rm N}$ and $\theta_{\rm p}$ are the actual N\'eel and Weiss temperatures in the presence of a uniaxial anisotropy field and zero or infinitesimal magnetic field that are both aligned along the easy $z$~axis.

In the following, we parameterize the high-field magnetization using the variables $f_J$, which only depends on the exchange constants and AFM structure, and the reduced anisotropy field $h_{\rm A1}$ defined in Eq.~(\ref{Eq:hA1Def}).  This choice of variables allows one to separate the effects on the magnetization due to the anisotropy field from those due to the exchange interactions and AFM structure.

\subsection{Exchange, Anisotropy and Applied Fields}

For a collinear AFM in a parallel applied field $H_z$ along the easy $z$-axis, only the $z$-components of the moments and the exchange fields are relevant.  Using the definition $\bar{\mu}_{iz}\equiv \mu_{iz}/\mu_{\rm sat} = \mu_{iz}/(gS\mu_{\rm B})$ for the two sublattices $i=1,\,2$, and Eqs.~(\ref{Eq:HexchiDef}) and~(\ref{Eqs:TNthetaSums}), the $z$-component of the exchange field seen by each moment on sublattice~1 is
\begin{subequations}
\label{Eqs:ExchangeFields1}
\be
\bs
H_{{\rm exch}\,1z} &= -\frac{1}{g^2\mu_{\rm B}^2}\Big(\mu_{1z}{\sum_j}^{\rm s}J_{ij}+\mu_{2z}{\sum_j}^{\rm d}J_{ij}\Big)\\*
&= \frac{3k_{\rm B}T_{{\rm N}J}}{2g\mu_{\rm B}(S+1)}\big[\bar{\mu}_{1z}(1+f_J) - \bar{\mu}_{2z}(1-f_J)\big].
\label{Eq:Hexch1zRed}
\end{split}
\ee
We express the magnetic fields in reduced for using Eq.~(\ref{Eq:halphaDef}). For the local exchange field seen by a spin in  sublattice~1 in Eq.~(\ref{Eq:Hexch1zRed}), the reduced field is
\be
h_{{\rm exch}\,1z} \equiv \frac{g\mu_{\rm B}H_{{\rm exch}\,1z}}{k_{\rm B}T_{{\rm N}J}} = \frac{3\big[\bar{\mu}_{1z}(1+f_J) - \bar{\mu}_{2z}(1-f_J)\big]}{2(S+1)}.
\label{Eq:Hexch1onT}
\ee
\ese
Similarly, the exchange field for a spin in sublattice~2 is
\bse
\label{Eqs:ExchangeFields2}
\be
\bs
H_{{\rm exch}\,2z} &= -\frac{1}{g^2\mu_{\rm B}^2}\Big(\mu_{1z}{\sum_j}^{\rm d}J_{ij}+\mu_{2z}{\sum_j}^{\rm s}J_{ij}\Big)\\*
&= \frac{3k_{\rm B}T_{{\rm N}J}}{2g\mu_{\rm B}(S+1)}\big[-\bar{\mu}_{1z}(1-f_J) + \bar{\mu}_{2z}(1+f_J)\big],
\end{split}
\ee
yielding the reduced exchange field
\be
h_{{\rm exch}\,2z} \equiv \frac{g\mu_{\rm B}H_{{\rm exch}\,2z}}{k_{\rm B}T_{{\rm N}J}} = \frac{3\big[-\bar{\mu}_{1z}(1-f_J) + \bar{\mu}_{2z}(1+f_J)\big]}{2(S+1)}.
\label{Eq:Hexch2onT}
\ee
\ese
Using Eqs.~(\ref{Eq:HAiAxialb}), (\ref{Eq:HA0}), (\ref{Eq:barmualphaDef}), and the expression $\bar{\mu}_i\cos\theta=\bar{\mu}_{iz}$, one obtains the anisotropy field
\bse
\label{Eqs:gmuBB/kT}
\be
H_{{\rm A}iz} = \frac{3H_{\rm A1}}{S+1}\bar{\mu}_{iz},
\ee
yielding the reduced anisotropy field
\be
h_{{\rm A}iz} \equiv \frac{g\mu_{\rm B}H_{{\rm A}iz}}{k_{\rm B}T_{{\rm N}J}} = \frac{3h_{\rm A1}}{S+1}\bar{\mu}_{iz}.
\ee
One also has the reduced applied field
\be
h_z \equiv \frac{g\mu_{\rm B}H_z}{k_{\rm B}T_{{\rm N}J}} .
\ee
\ese

The total reduced local magnetic inductions seen by spins in sublattices~$i=1,\ 2$ are then
\be
b_{iz} \equiv \frac{g\mu_{\rm B}B_{iz}}{k_{\rm B}T_{{\rm N}J}} = h_{{\rm exch}iz} + h_{{\rm A}iz} + h_z.
\ee
Inserting the above expressions for the components on the right-hand side gives
\bse
\label{Eqs:b12zAFM}
\bea
b_{1z} &=& \frac{3[\bar{\mu}_{1z}(1+f_J+2h_{\rm A1}) - \bar{\mu}_{2z}(1-f_J)]}{2(S+1)} + h_z,\nonumber\\*
\\*
b_{2z} &=& \frac{3[-\bar{\mu}_{1z}(1-f_J) + \bar{\mu}_{2z}(1+f_J+2h_{\rm A1})]}{2(S+1)} + h_z.\nonumber\\*
\eea
\ese

\subsection{Coupled Equations for the Two Sublattice Magnetizations}

The values of $\bar{\mu}_{iz}\ (i=1,\ 2)$ versus~$H$ and~$T$ are governed by separate Brillouin functions for the two sublattices as in Eqs.~(\ref{Eq:BS(y)}).  One thus has two simultaneous consistency relations
\be
\bar{\mu}_{iz} = B_S\left(\frac{b_{iz}}{t}\right)\quad(i=1,\ 2).
\label{Eq:barmuiz}
\ee
Substituting Eqs.~(\ref{Eqs:b12zAFM}) into~(\ref{Eq:barmuiz}) gives
\bse
\label{Eqs:mu12zvsH}
\be
\bar{\mu}_{1z} = B_S\Bigg\{\frac{3[\bar{\mu}_{1z}(1+f_J+2h_{\rm A1}) - \bar{\mu}_{2z}(1-f_J)]}{2(S+1)t} + \frac{h_z}{t}\Bigg\} ,
\label{Eq:barmu1z}
\ee 
\be
\bar{\mu}_{2z} = B_S\Bigg\{\frac{3[-\bar{\mu}_{1z}(1-f_J) + \bar{\mu}_{2z}(1+f_J+2h_{\rm A1})]}{2(S+1)t} + \frac{h_z}{t}\Bigg\} .
\label{Eq:barmu2z}
\ee 
\ese
When $H_z=0$ and $T\leq T_{\rm N}$, one has $\bar{\mu}_{2z} = -\bar{\mu}_{1z}$ and Eqs.~(\ref{Eq:barmu1z}) and~(\ref{Eq:barmu2z}) each reduce to the same general expression~(\ref{Eq:barmuA}) for the ordered moment versus temperature, as required.  For the PM regime $T\geq T_{\rm N}$, $\bar{\mu}_{1z} = \bar{\mu}_{2z}$ and Eqs.~(\ref{Eq:barmu1z}) and~(\ref{Eq:barmu2z}) each reduce to the $z$-axis magnetic moment of the PM state of the AFM given by Eqs.~(\ref{Eq:muParPM}), as also required.

\subsection{\label{Sec:AFMzAxisHiFields} Sublattice, Average and Staggered Moments and Free Energy versus Magnetic Field, Temperature, and Anisotropy Parameter}

Two important quantities can be obtained from Eqs.~(\ref{Eqs:mu12zvsH}) from which the thermal-average sublattice magnetic moments $\bar{\mu}_{1z}$ and $\bar{\mu}_{2z}$ versus temperature, magnetic field and anisotropy parameter are calculated.  The first is the net average magnetic moment, normalized by the saturation moment, which is
\begin{subequations}
\label{Eqs:muzaveMuzdag2}
\be
\bar{\mu}_{z\,{\rm ave}} = \frac{\bar{\mu}_{1z}+\bar{\mu}_{2z}}{2}.
\label{Eq:muzave}
\ee
This is the uniform magnetization along the easy axis measured in a conventional magnetometer.  The second important quantity is the AFM order parameter $\bar{\mu}_z^\dagger$,  which is the average $z$-axis staggered moment in the $z$-direction normalized by the saturation moment, given by
\be
\bar{\mu}_z^\dagger= \frac{\bar{\mu}_{1z}-\bar{\mu}_{2z}}{2}.
\label{Eq:muzdagger}
\ee
\end{subequations}
By assumption $\bar{\mu}_{1z}\geq \bar{\mu}_{2z}$, so $\bar{\mu}_z^\dagger \geq 0$.  The spin system is in the AFM phase when $\bar{\mu}_z^\dagger>0$ and is in the associated high-field PM phase when $\bar{\mu}_z^\dagger=0$.

The potential phase transitions between collinear AFM and PM states discussed below will be preempted if the free energy of the AFM phase for some combination of $t,\ h_z,$ and~$h_{\rm A1}$ is higher than that of the SF phase, and conversely.  Therefore in this section we eventually determine the free energy of the AFM phase versus temperature from the values of the thermal-average moments $\bar{\mu}_{1z}$ and $\bar{\mu}_{2z}$ in the presence of the anisotropy and applied fields for comparison with the free energy of the SF phase found previously in Sec.~\ref{Sec:FmagSF}.

\begin{figure*}
\includegraphics [width=3.in]{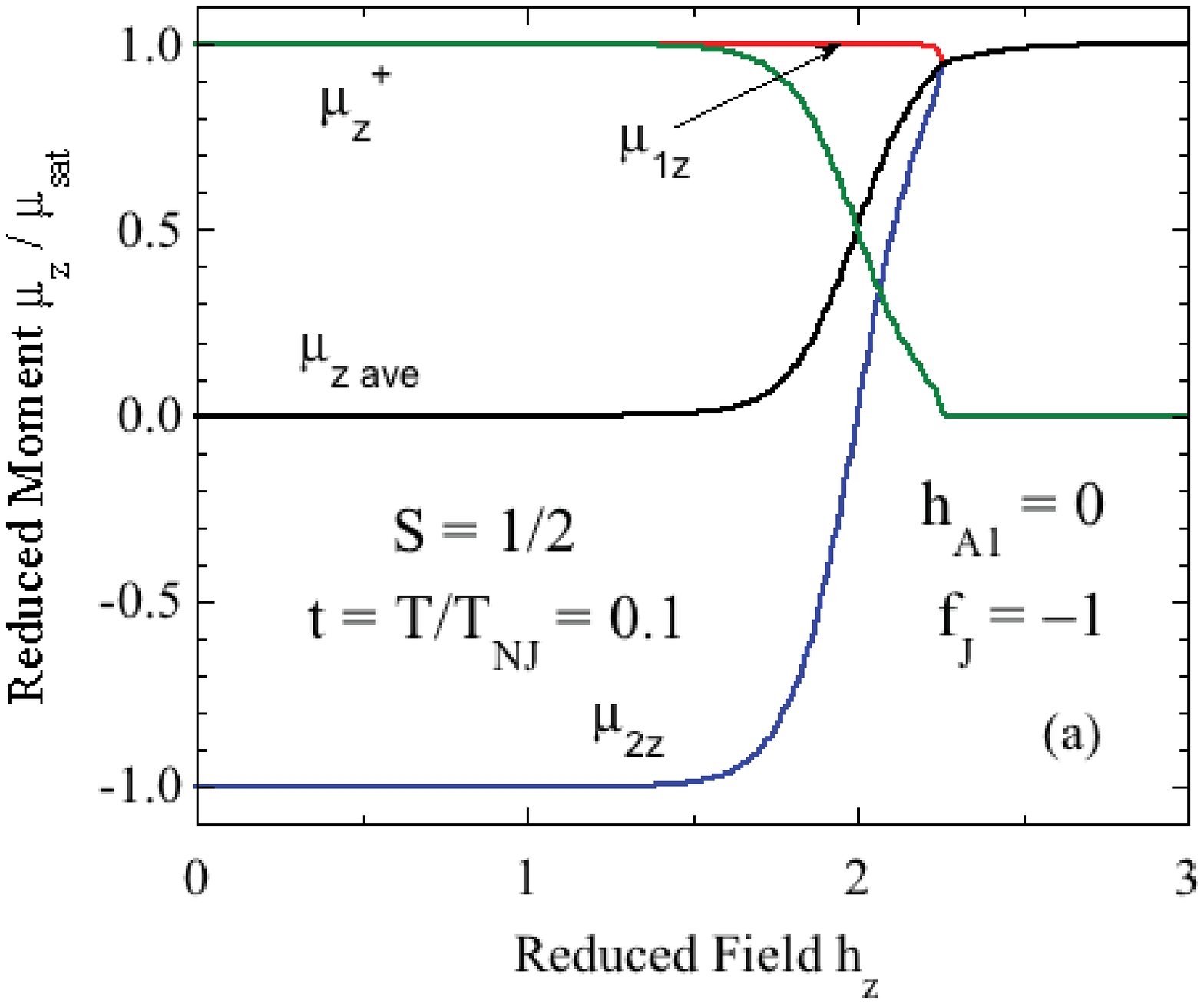}\includegraphics [width=3.in]{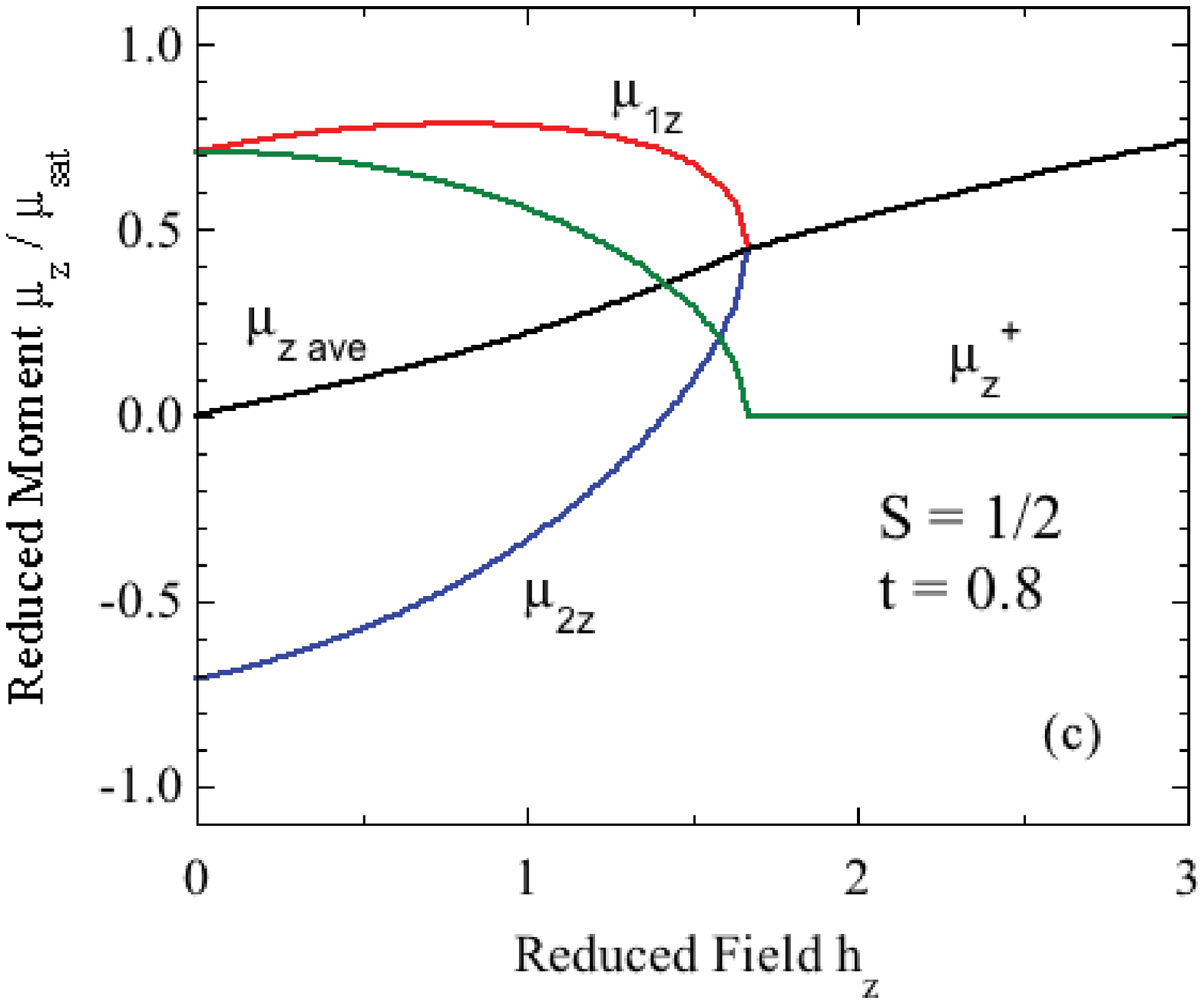}
\includegraphics [width=3.in]{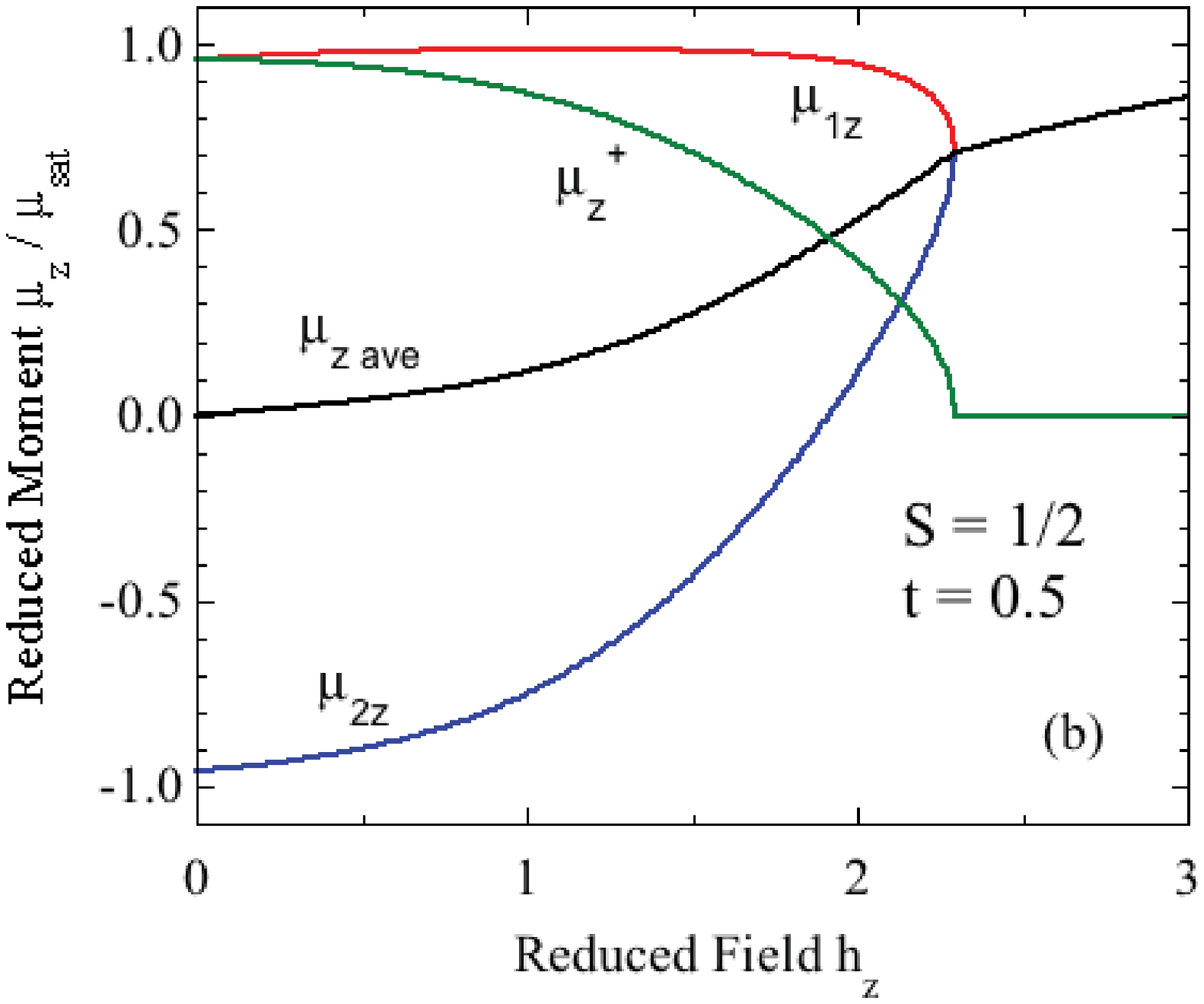}\includegraphics [width=3.in]{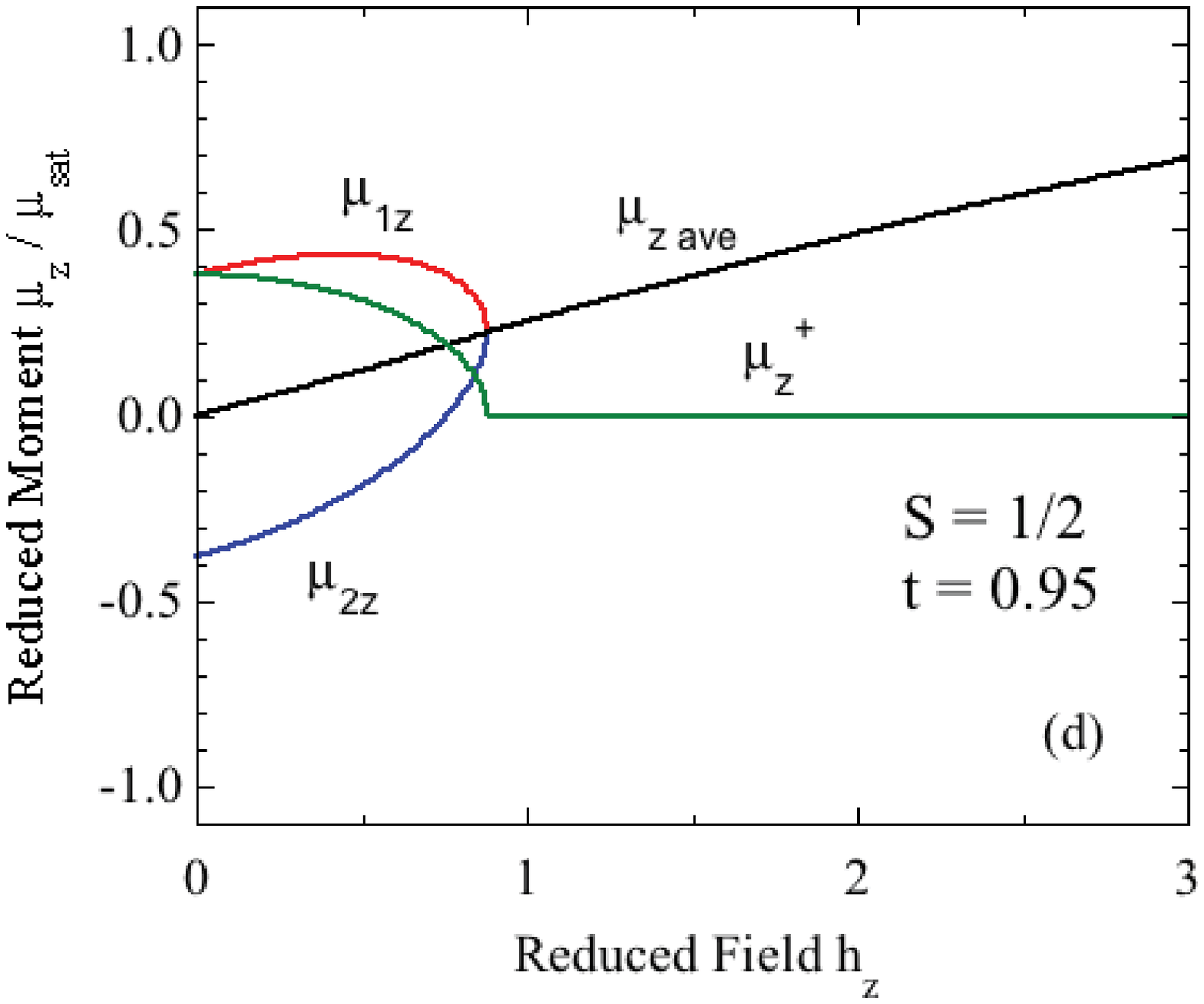}
\caption{(Color online) Ordered moments $\bar{\mu}_{iz} \equiv \mu_{iz}/\mu_{\rm sat}\ (i=1$, 2) of the two magnetic sublattices along with the AFM order parameter $\bar{\mu}_z^\dagger \equiv (\bar{\mu}_{1z}-\bar{\mu}_{2z})/2$ and the average ordered moment $\bar{\mu}_{z\,{\rm ave}}\equiv (\bar{\mu}_{1z}+\bar{\mu}_{2z})/2$ for spin $S=1/2$, $f_J=-1$ and $h_{\rm A1}=0$, all versus the reduced applied magnetic field $h_z$ along the easy $z$~axis for reduced temperatures $t \equiv T/T_{{\rm N}J}$ of (a)~0.1, (b)~0.5, (c)~0.8 and (d)~0.95.  The AFM regime is defined by the region where $\bar{\mu}^\dagger_z > 0$, and the PM regime is defined by $\bar{\mu}^\dagger_z = 0$. The transition field between these two regimes is defined as the criticial field $h_{\rm cAFM}$.  Only second-order transitions are observed for $0 < t_{\rm A}<1$ with $f_J=-1$ and $h_{\rm A1}=0$.}
\label{Fig:muzAFMhA10fJm1S12}
\end{figure*}

\begin{figure*}
\includegraphics [width=3.in]{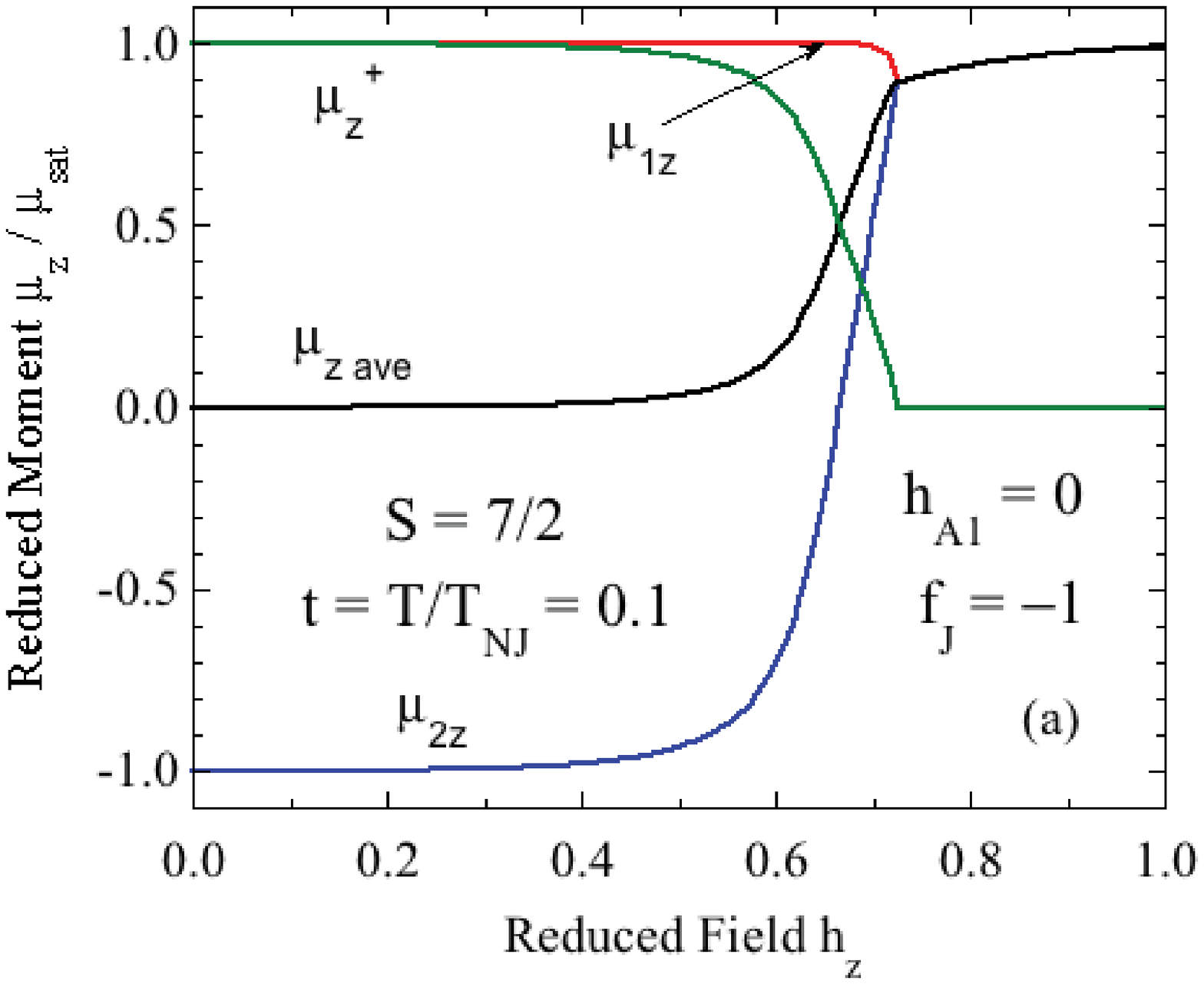}\includegraphics [width=3.in]{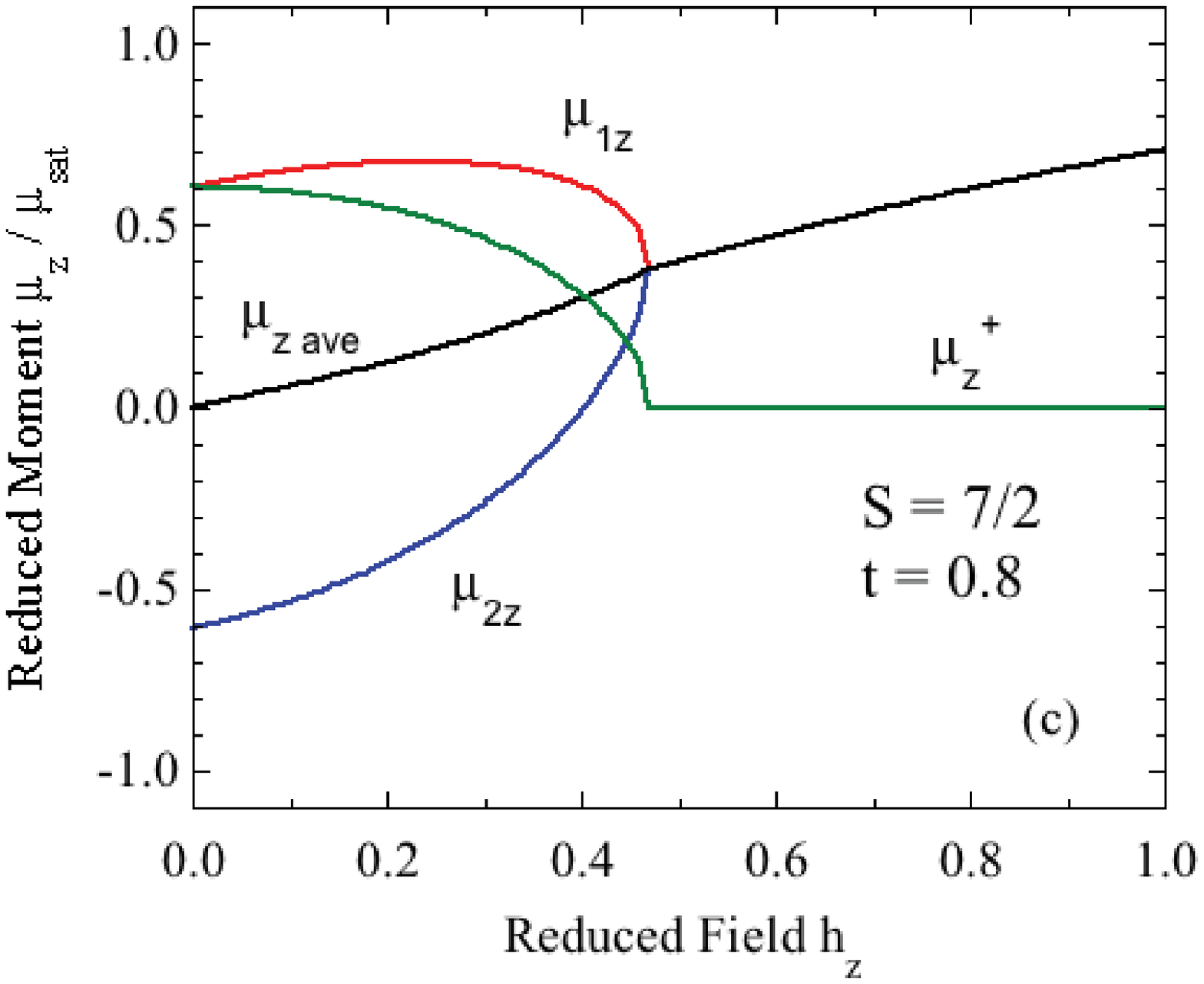}
\includegraphics [width=3.in]{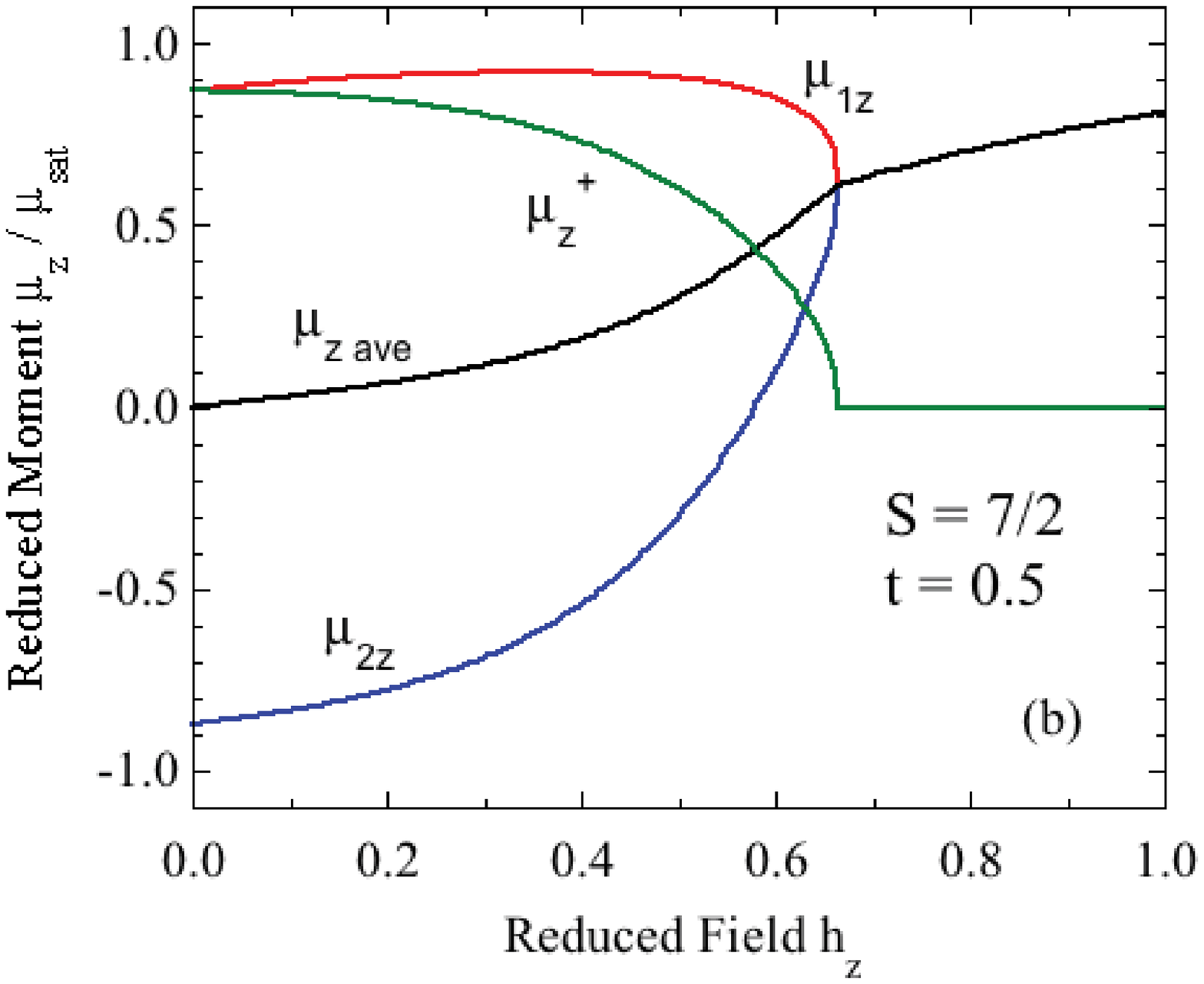}\includegraphics [width=3.in]{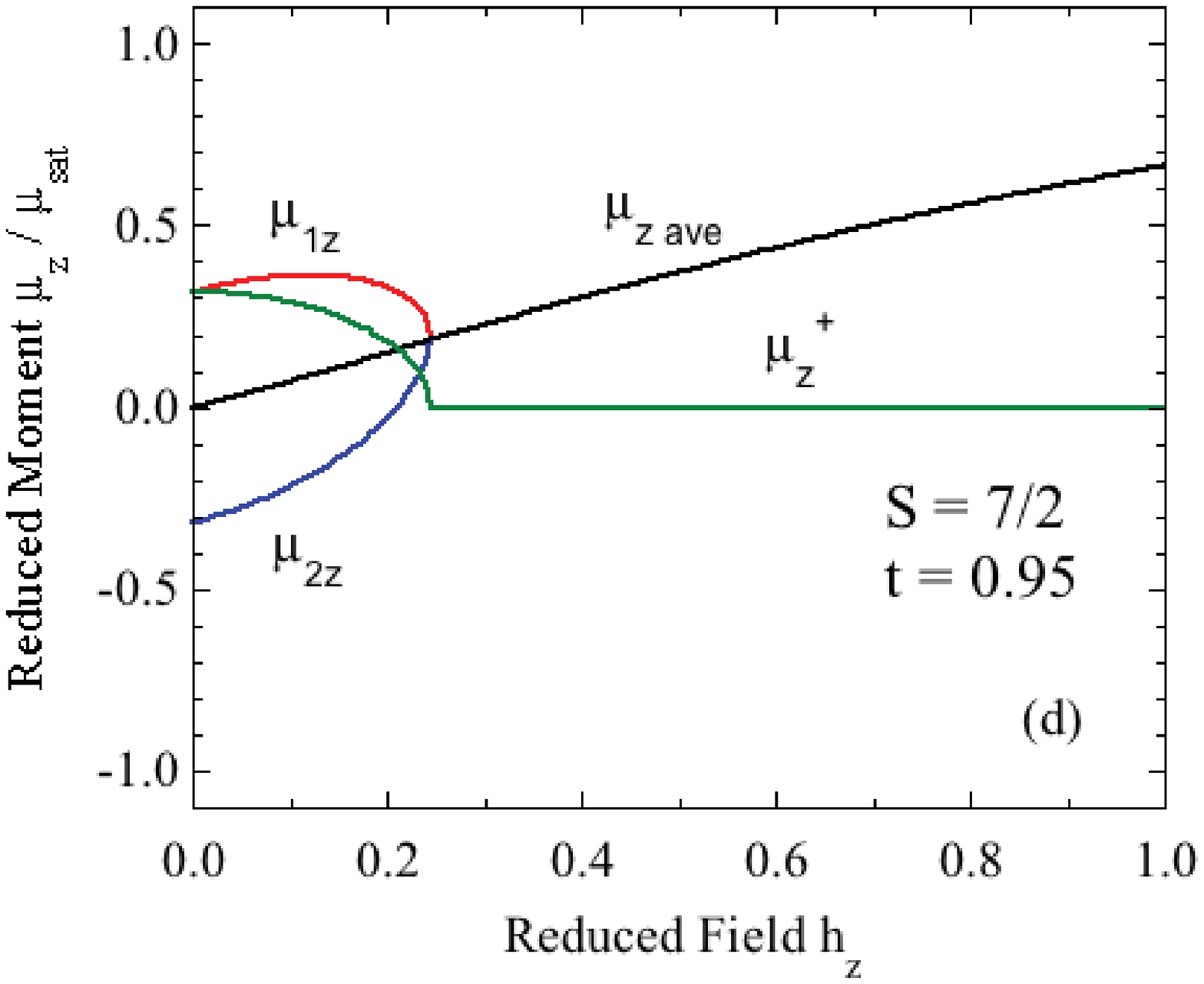}
\caption{(Color online) Same as Fig.~\ref{Fig:muzAFMhA10fJm1S12} except for spin $S=7/2$.  Note the factor of three difference in the abscissa scale between this figure and that one.}
\label{Fig:muzAFMhA10fJm1S72}
\end{figure*}

\begin{figure*}
\includegraphics [width=2.75in]{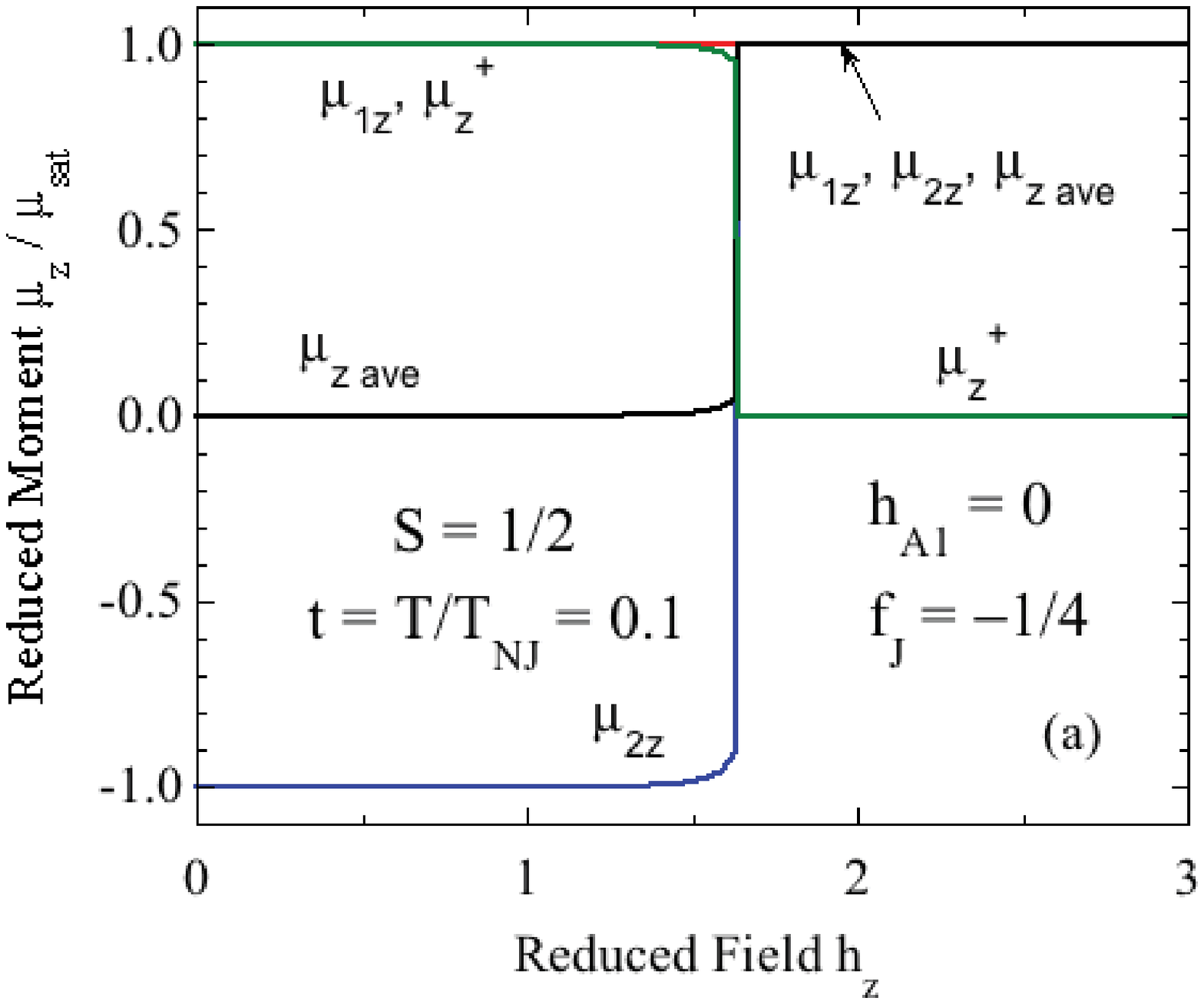}\includegraphics [width=2.75in]{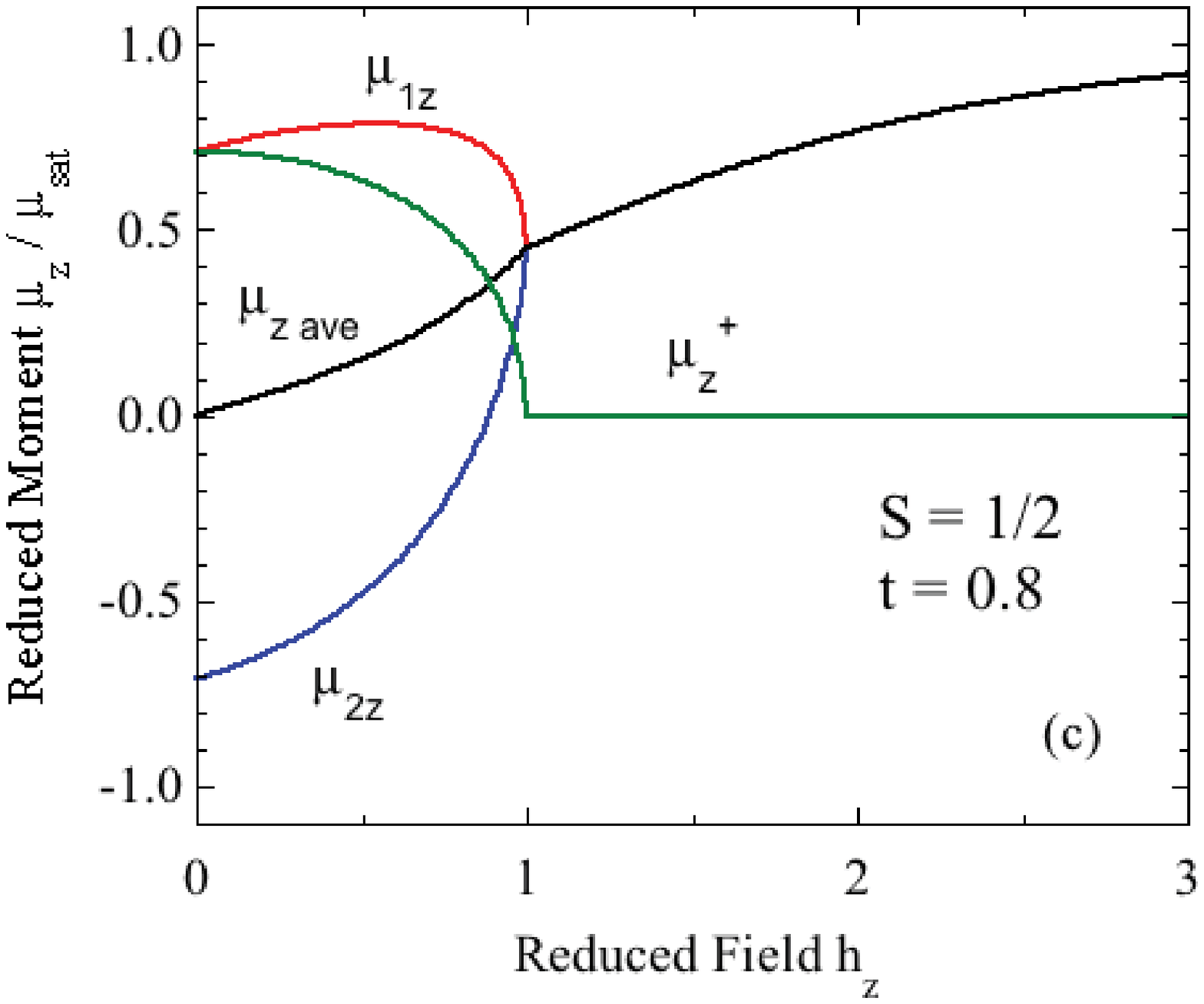}
\includegraphics [width=2.75in]{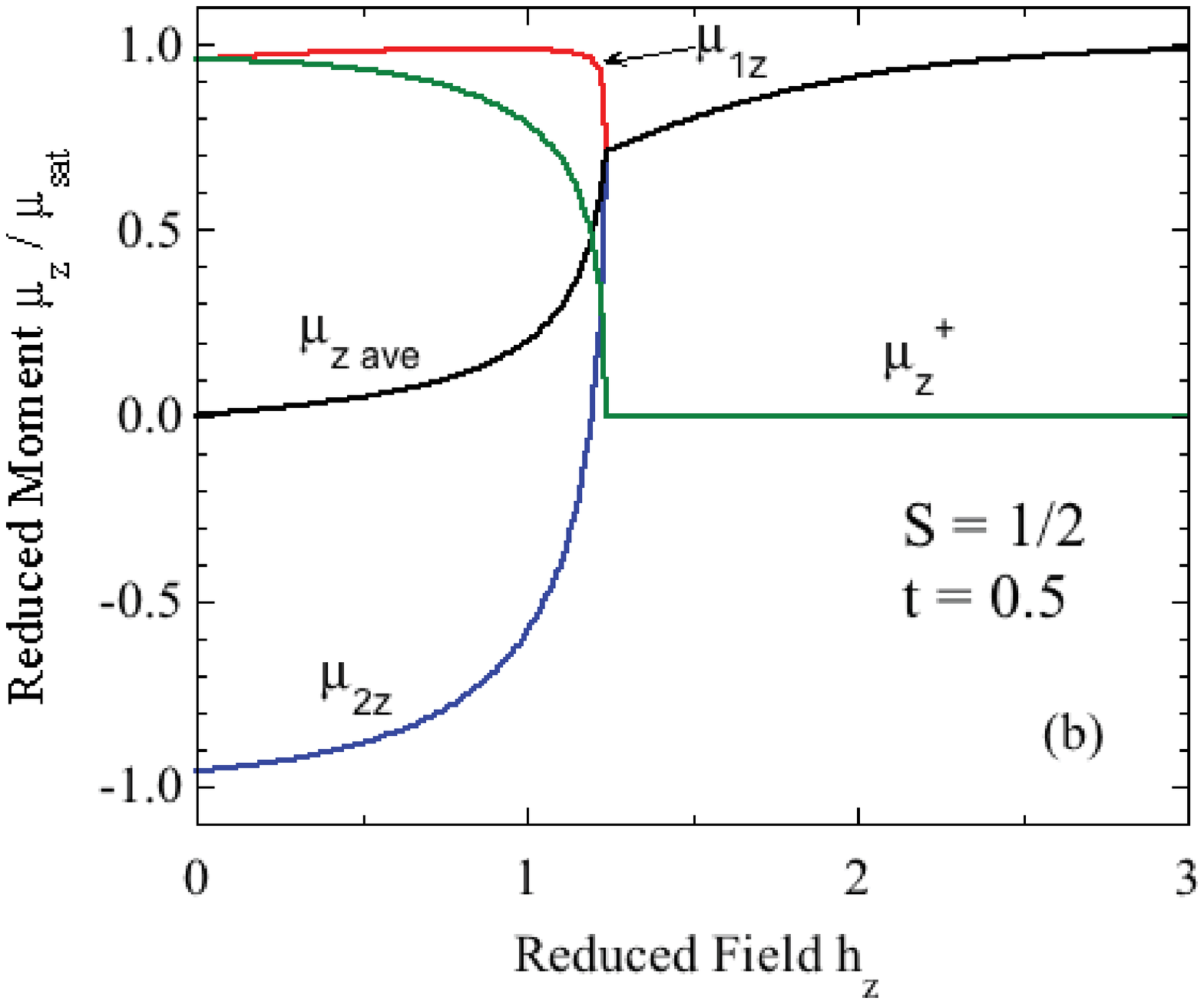}\includegraphics [width=2.75in]{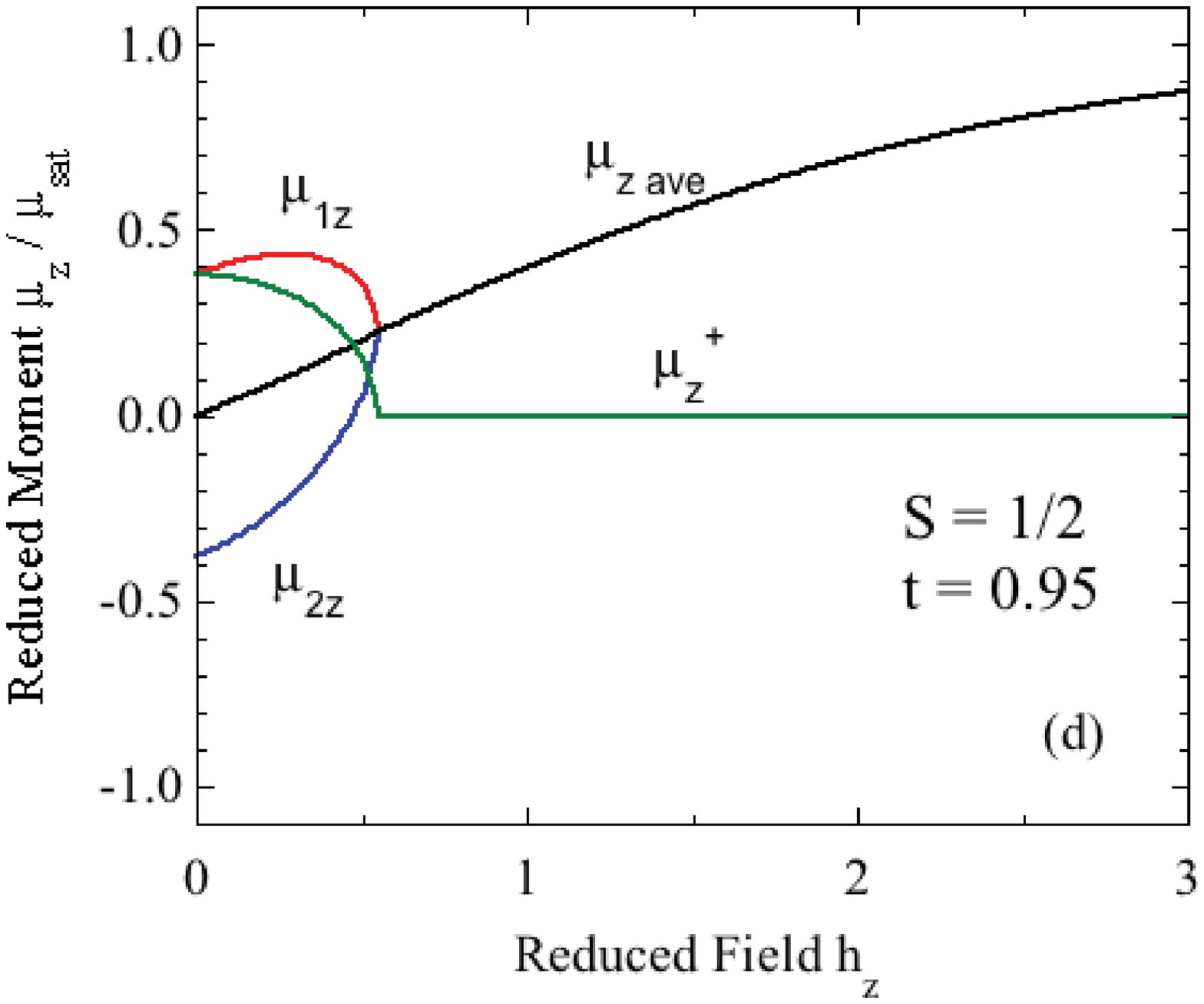}
\caption{(Color online) Same as Fig.~\ref{Fig:muzAFMhA10fJm1S12} except that here $f_J=-1/4$. The data for $t=0.1$ show strongly first-order transitions, for $t=0.5$ weakly first-order transitions, and second-order transitions for $t=0.8$ and~0.95.}
\label{Fig:muzAFMhA10fJm25S12}
\end{figure*}

Equations~(\ref{Eqs:mu12zvsH}) were solved for $\bar{\mu}_{1z}$ and~$\bar{\mu}_{2z}$ versus~$h_z$ for given values of $S$, $t$, $f_J$, and $h_{\rm A1}$ using an iterative procedure \cite{Johnston2017}.  Starting with $h_z=0$, the initial value of $\bar{\mu}_{1z}$ was set to~1 and $\bar{\mu}_{2z}$ solved for.  Then for that value of $\bar{\mu}_{2z}$, $\bar{\mu}_{1z}$ was solved for. These steps were iterated until the differences in $\bar{\mu}_{1z,2z}$ between subsequent iterations were each less than~$10^{-10}$.  Typically the number of iterations needed was less that 10, but occasionally up to $\sim 10^4$ iterations were needed when approaching a phase transition.  Once $\bar{\mu}_{1z}$  and $\bar{\mu}_{2z}$ were determined, $\bar{\mu}_{z{\rm ave}} = (\bar{\mu}_{1z} + \bar{\mu}_{2z})/2$ and $\bar{\mu}_z^\dagger = (\bar{\mu}_{1z} - \bar{\mu}_{2z})/2$ were determined.  This sequence was repeated for the next value of $h_z$, where the starting value of $\bar{\mu}_{1z}$ was the final value from the previous value of~$h_z$.

Shown in Figs.~\ref{Fig:muzAFMhA10fJm1S12} and~\ref{Fig:muzAFMhA10fJm1S72} are plots of $\bar{\mu}_{1z}$, $\bar{\mu}_{2z}$, $\bar{\mu}_{z{\rm ave}}$, and~$\bar{\mu}_z^\dagger$ versus~$h_z$ for $f_J=-1$, $h_{\rm A1}=0$, $t=0.1$, 0.5, 0.8, and 0.95 for spins $S=1/2$ and $S=7/2$, respectively.  The data versus~$h_z$ for $S=1/2$ and $S = 7/2$ have similar evolutions of the shapes on decreasing temperature, but the abscissa ranges for $S=7/2$ are a factor of three smaller than for $S=1/2$.  Qualitative plots of $\bar{\mu}_{iz}~(i=1$,~2) similar to those in Figs.~\ref{Fig:muzAFMhA10fJm1S12} and~\ref{Fig:muzAFMhA10fJm1S72} were shown in Fig.~11 of Ref.~\cite{Shapira1970}. The boundary between the AFM and PM states occurs with increasing field when $\bar{\mu}_z^\dagger\to0^+$.  We denote this reduced critical field by $h_{\rm c\, AFM}$.  Thus for $h_z \geq h_{\rm c\, AFM}$, one has $\bar{\mu}_{1z} = \bar{\mu}_{2z}$ and $\bar{\mu}_z^\dagger=0$.  Second-order transitions at $h_{\rm c\, AFM}$ are observed for full the temperature range $0 < t_{\rm A} \leq 1$ for $f_J=-1$ and  $h_{\rm A1}=0$.

First-order transitions between the AFM and PM phases can occur over a range of low temperatures ending at a tricritical point temperature above which the transitions are second-order.  For example, we changed $f_J$ from $-1$ to the value of $-1/4$ while leaving $h_{\rm A1}=0$ as in Fig.~\ref{Fig:muzAFMhA10fJm1S12}.  Numerical solutions for $\bar{\mu}_{iz}\ (i=1$, 2), $\bar{\mu}_z^\dagger$ and $\bar{\mu}_{z\,{\rm ave}}$ are plotted versus~$h_z$ in Fig.~\ref{Fig:muzAFMhA10fJm25S12} for reduced temperatures $t_{\rm A} = 0.1$, 0.5, 0.8 and~0.95.  At high $T$, the AFM to PM transitions are seen to be second order.  However, at $t=0.5$ and~0.1, the transitions are strongly and weakly first order, respectively, where a discontinuous change in the AFM order parameter $\bar{\mu}_z^\dagger$ occurs at the transition.

\begin{figure*}
\includegraphics [width=3.in]{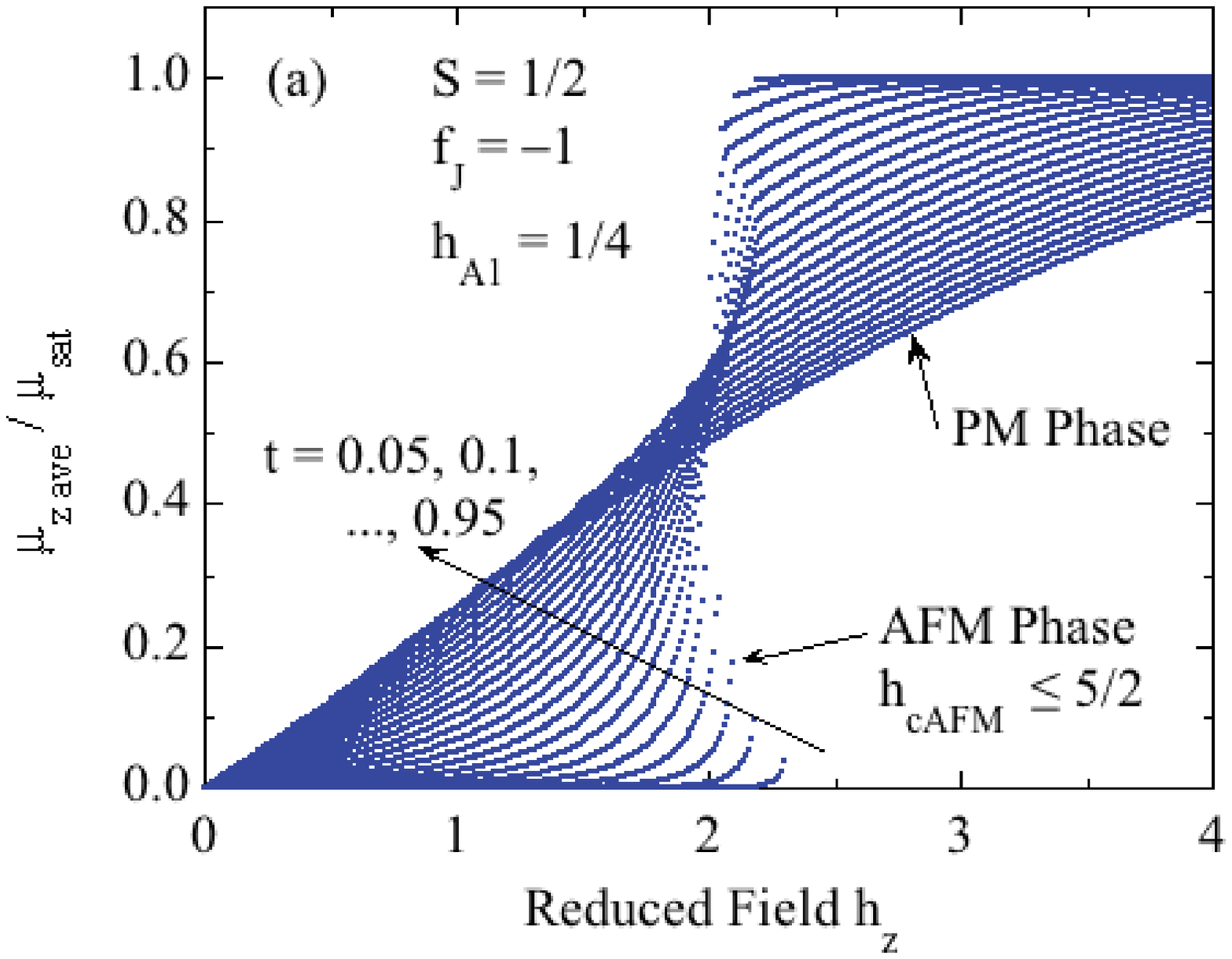}\includegraphics [width=3.in]{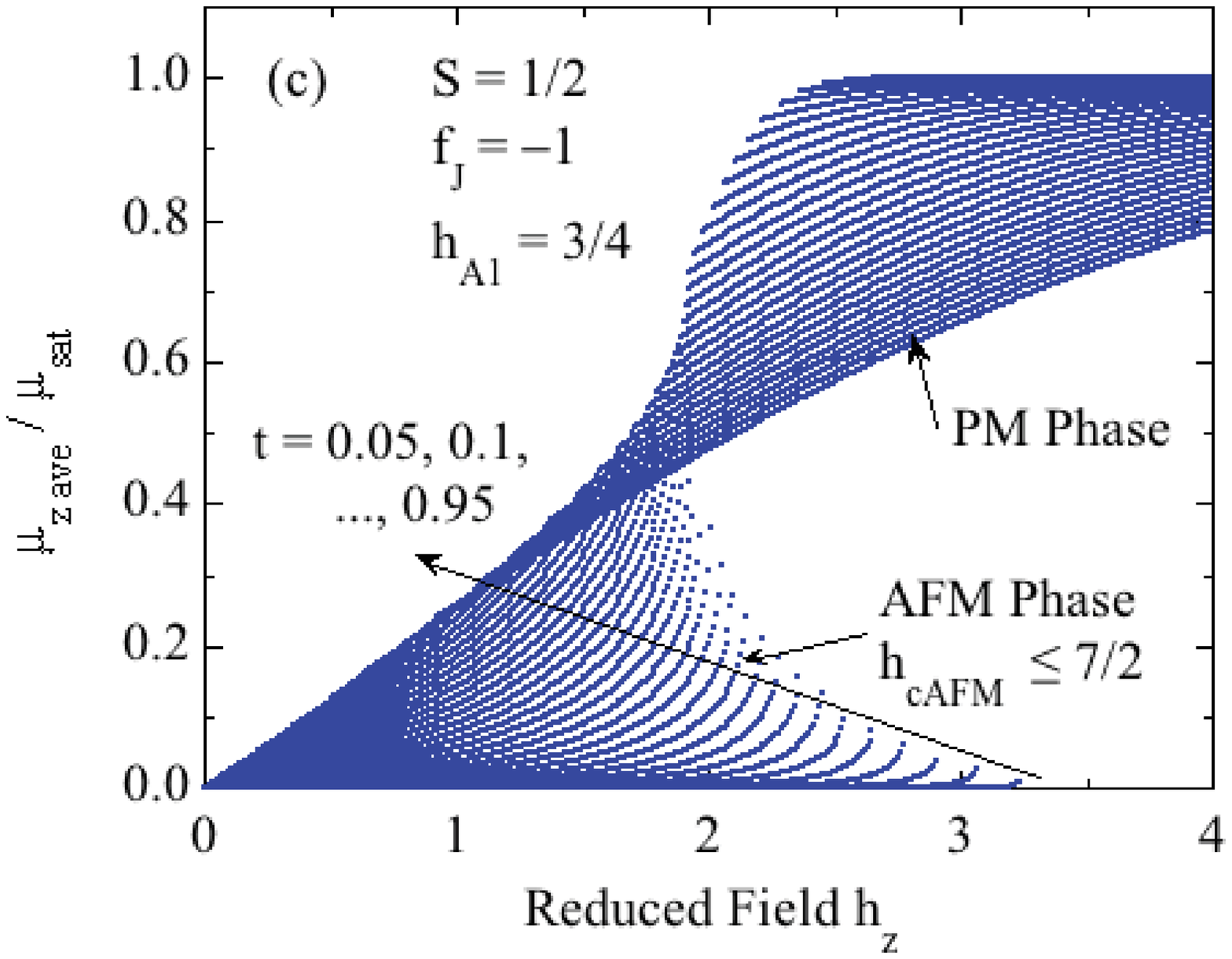}
\includegraphics [width=3.in]{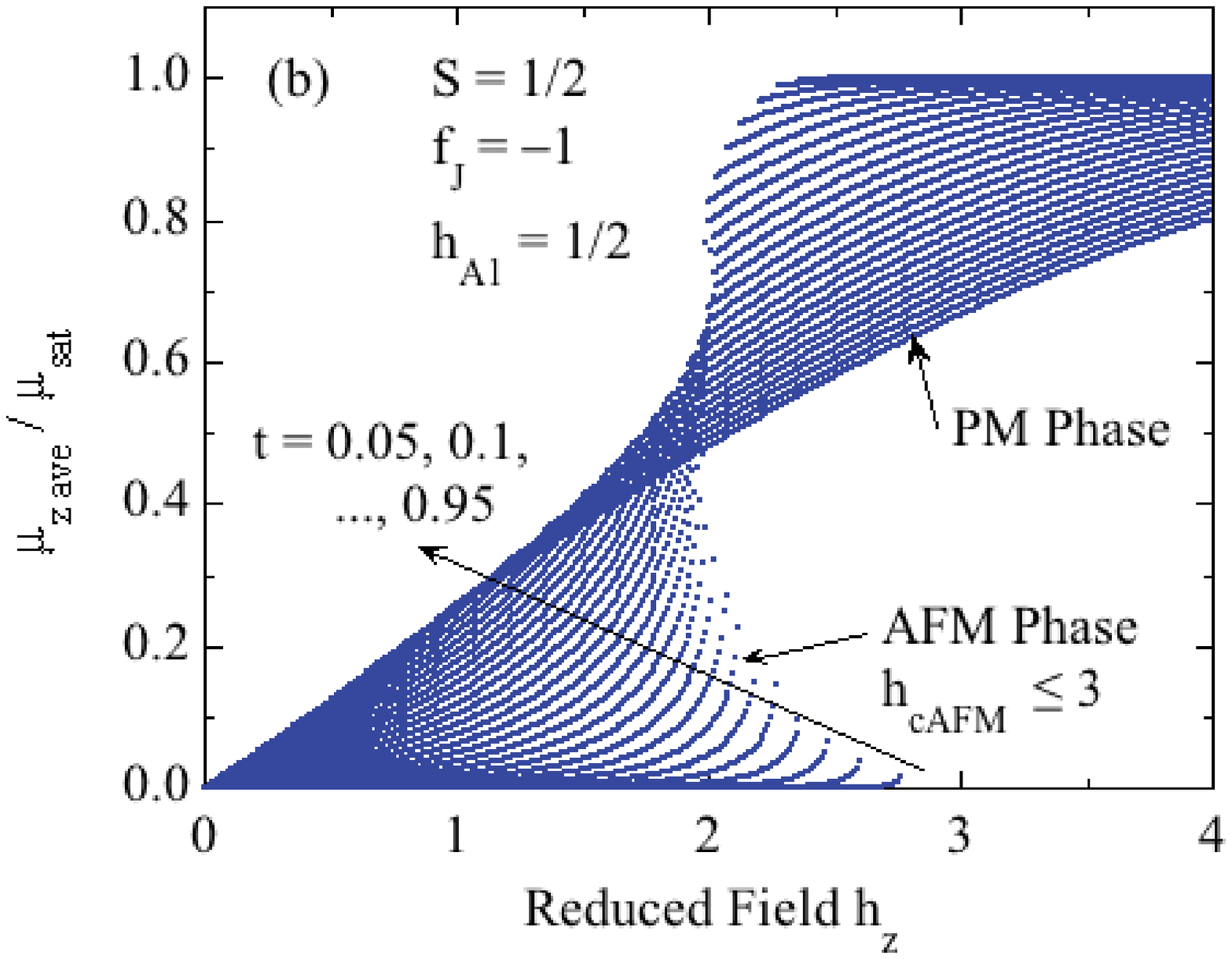}\includegraphics [width=3.in]{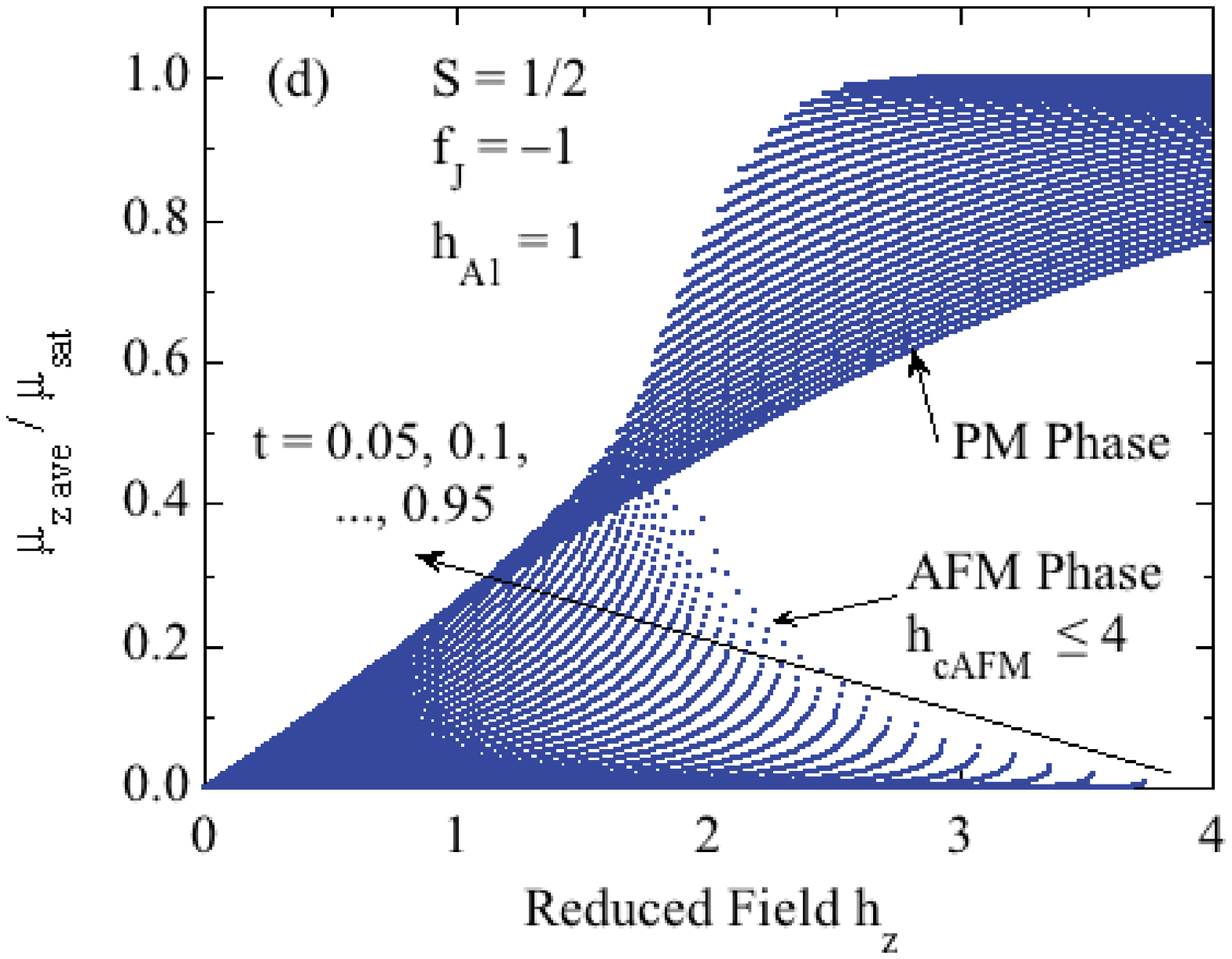}
\caption{(Color online) Reduced average $z$-axis moment per spin $\bar{\mu}_{z{\rm ave}} \equiv \mu_{z\rm ave}/\mu_{\rm sat}$ for the low-field AFM and high-field PM phases of a collinear antiferromagnet versus reduced field~$h_z$ for spin~$S=1/2$ and $f_J=-1$ at reduced temperatures $t = T/T_{{\rm N}J}$ as shown for reduced anisotropy fields (a)~$h_{\rm A1}=1/4$, (b)~$h_{\rm A1}=1/2$, (c)~$h_{\rm A1}=3/4$, and (d)~$h_{\rm A1}=1$ calculated using Eqs.~(\ref{Eqs:mu12zvsH}) and~(\ref{Eq:muzave}).  }
\label{Fig:muAveDagAFMhA125fJm1S12FixA}
\end{figure*}

\begin{figure}
\includegraphics [width=3.3in]{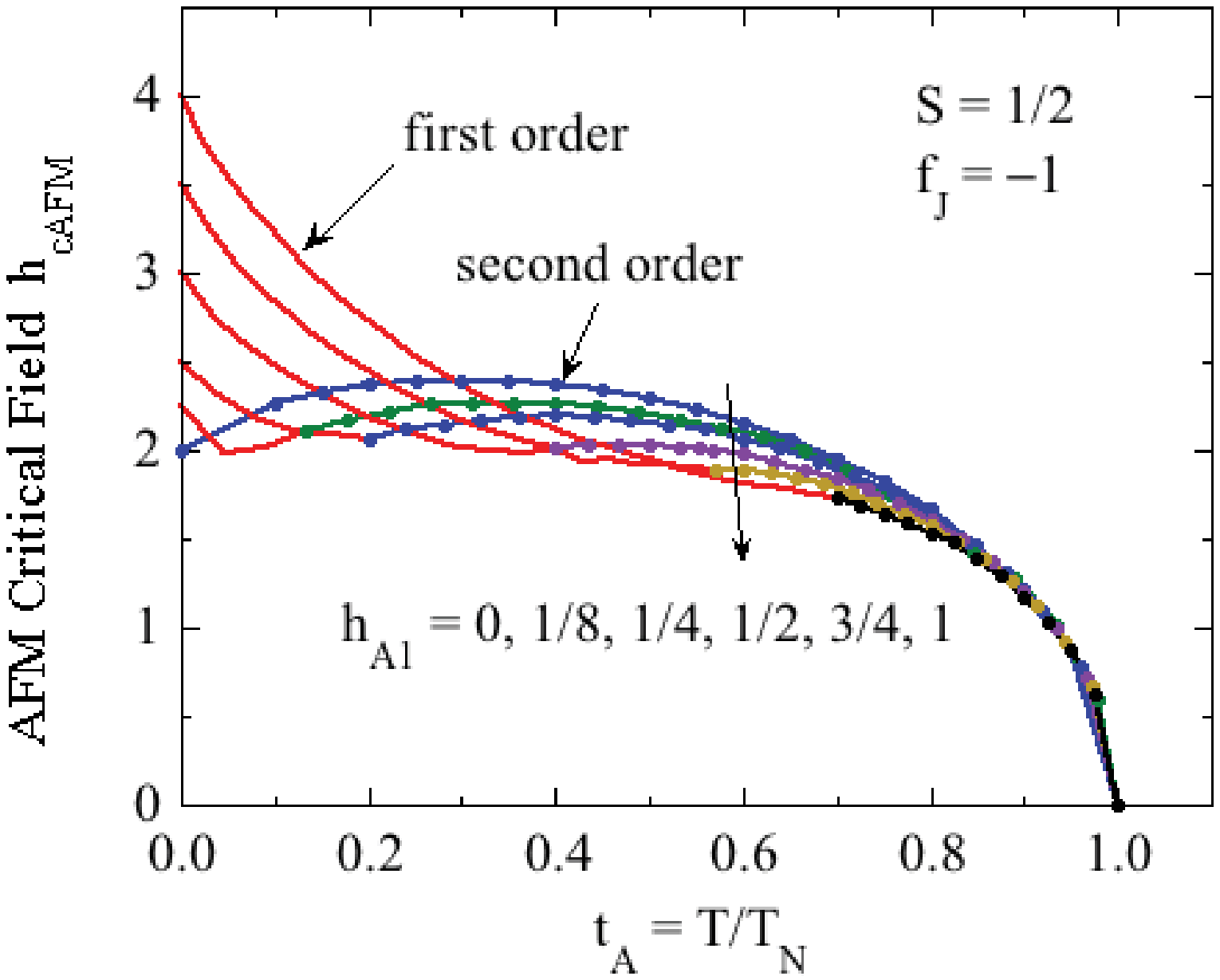}
\caption{(Color online) AFM critical field $h_{\rm cAFM}$ versus reduced temperature $t_{\rm A}$ for $S=1/2$ with \mbox{$f_J=-1$} and $h_{\rm A1}$ from~0 to~1 as indicated.  First-order transition lines are in red without data points and second-order transitions are in other colors with data points.  A tricritical point temperature separates the first- and second-order transitions on the transition line for each $h_{\rm A1}\geq 1/8$. }
\label{Fig:hcAFMvstAS12fJm1hA1xx}
\end{figure}

We carried out additional calculations of $\bar{\mu}_z^\dagger$ and~$\bar{\mu}_{z{\rm ave}}$ versus $h_z$ and reduced temperature~$t=T/T_{{\rm N}J}$.  Plots of $\bar{\mu}_{z{\rm ave}}$ versus~$h_z$ for spin~$S=1/2$ and $f_J=-1$ for $t = 0.05$ to~0.95 for reduced anisotropy fields $h_{\rm A1}=1/4$, 1/2, 3/4, and~1 calculated using Eqs.~(\ref{Eqs:mu12zvsH}) are shown in Fig.~\ref{Fig:muAveDagAFMhA125fJm1S12FixA}.  One sees a clear evolution from first-order to second-order transitions with increasing temperature.  The values of the AFM critical field $h_{\rm cAFM}$ were determined from Fig.~\ref{Fig:muAveDagAFMhA125fJm1S12FixA} as the value of $h_z$ at which $\bar{\mu}^\dagger\to0$ with increasing~$h_z$.  Second-order transitions are characterized by a continuous change for $\bar{\mu}^\dagger\to0$, whereas a first-order transition shows a discontinuous change as noted above.  After converting $t$ to $t_{\rm A}$ using Eq.~(\ref{Eq:tADef}), plots of the resulting $h_{\rm cAFM}$ versus~$t_{\rm A}$ are shown in Fig.~\ref{Fig:hcAFMvstAS12fJm1hA1xx} for $S=1/2$, $f_J=-1$, and $h_{\rm A1}$ values from~0 to~1.  The first-order transition data are represented by solid red curves, and the second-order data by solid curves connecting data points of different colors.  These plots are not phase diagrams, which are given in Fig.~\ref{Fig:PhaseDiagram_S12fJm1hA10} below for the same values of~$h_{\rm A1}$ as in Fig.~\ref{Fig:muAveDagAFMhA125fJm1S12FixA} and also for $h_{\rm A1} = 0$ and~1/8.

\subsection{Magnetic Free Energy}

\begin{figure*}
\includegraphics [width=3.in]{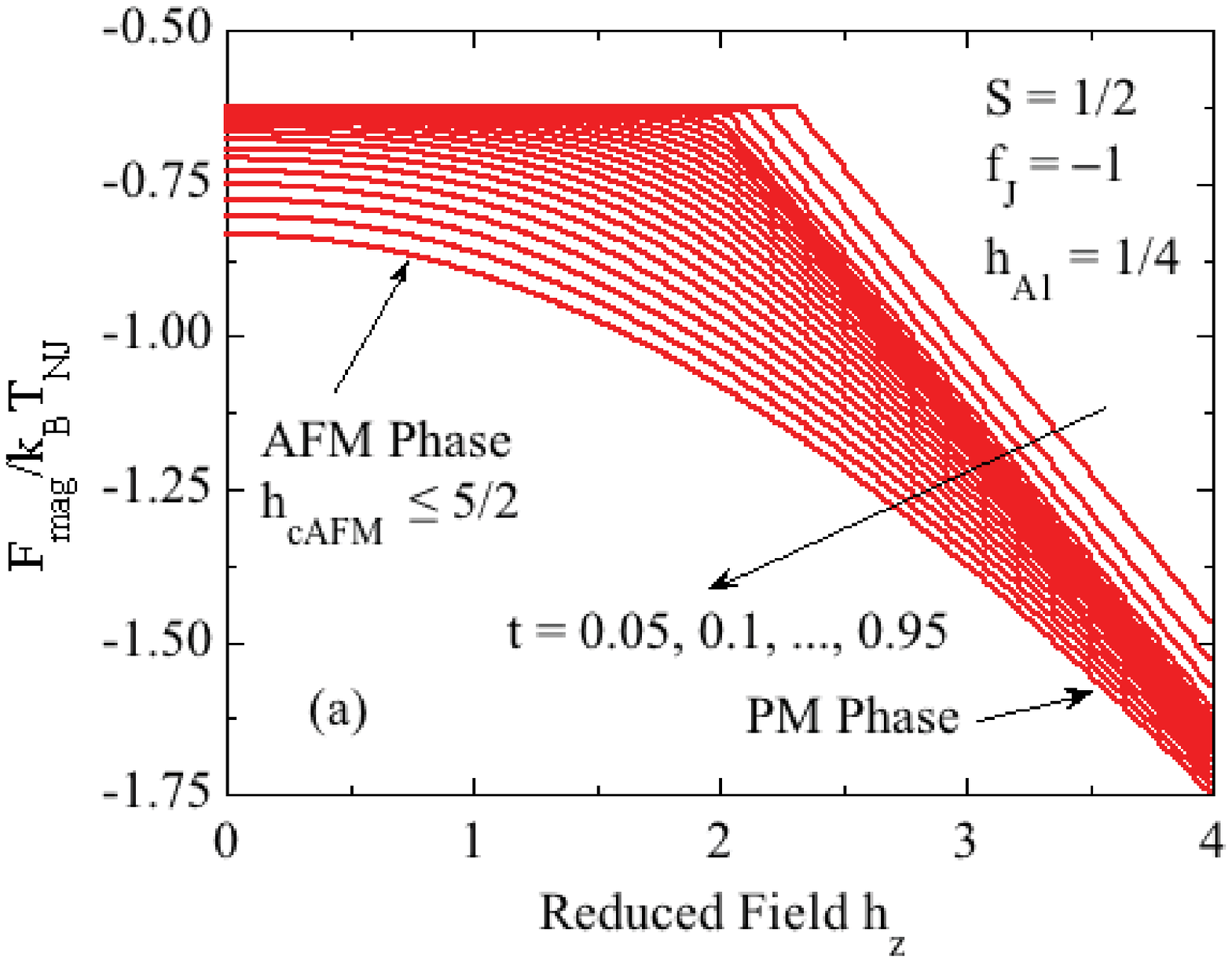}\includegraphics [width=3.in]{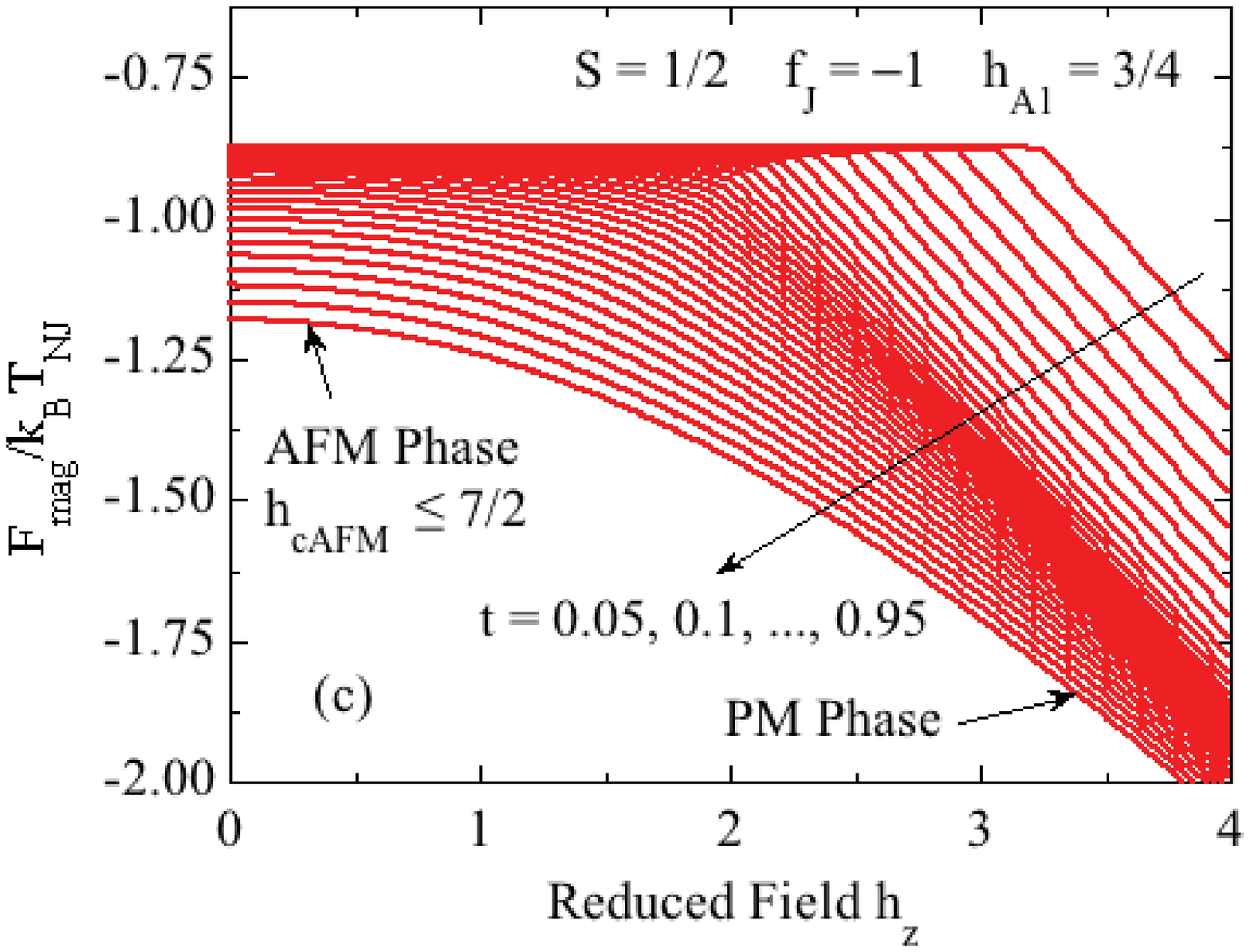}
\includegraphics [width=3.in]{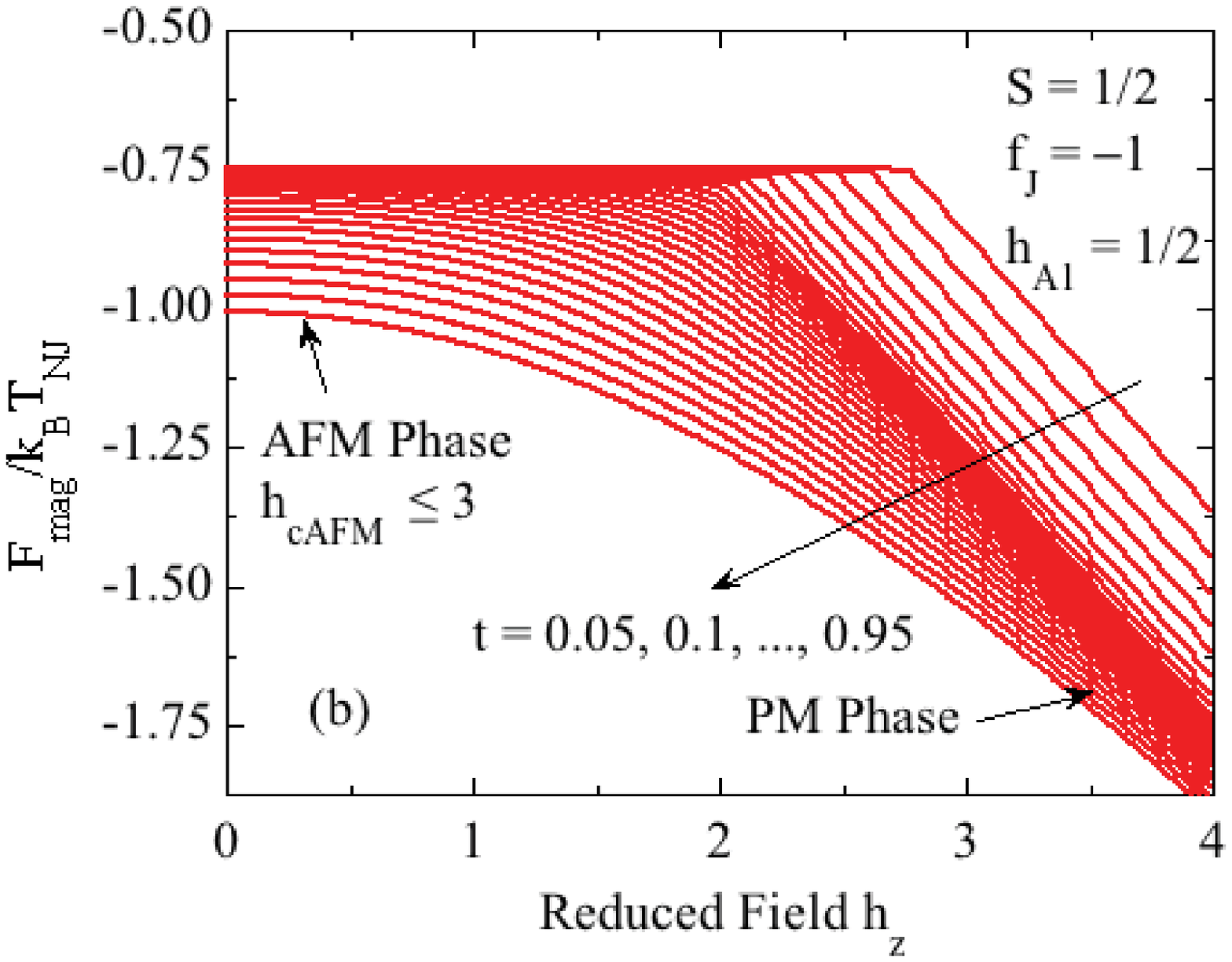}\includegraphics [width=3.in]{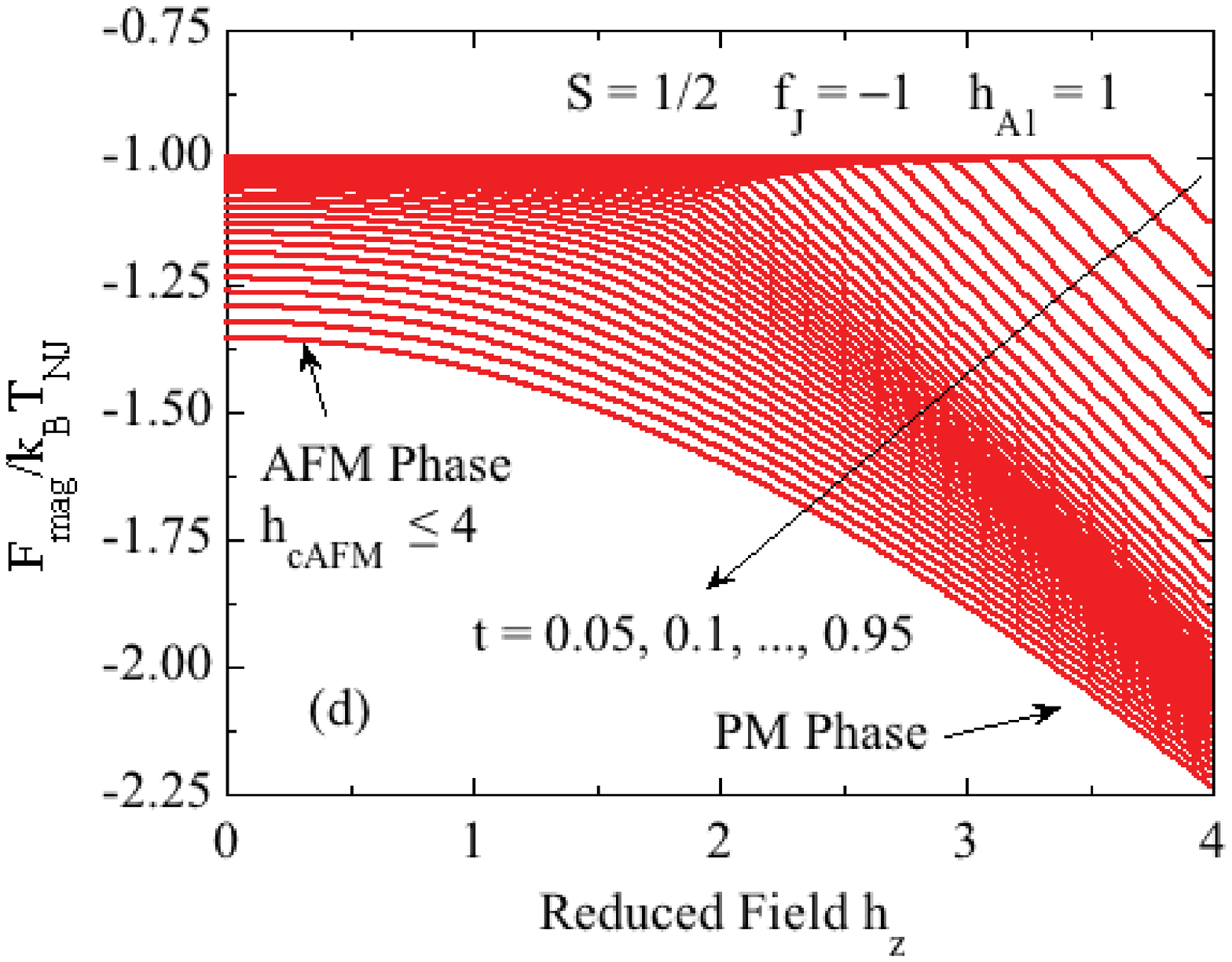}
\caption{(Color online) Reduced magnetic free energy per spin $F_{\rm mag}/k_{\rm B}T_{{\rm N}J}$ for the low-field AFM and high-field PM phases of a collinear antiferromagnet versus reduced field~$h_z$ for spin~$S=1/2$ and $f_J=-1$ at reduced temperatures $t = T/T_{{\rm N}J}$ as shown for reduced anisotropy fields (a)~$h_{\rm A1}=1/4$, (b)~$h_{\rm A1}=1/2$, (c)~$h_{\rm A1}=3/4$, and (d)~$h_{\rm A1}=1$ calculated using Eqs.~(\ref{Eqs:Thermodynamics}) and data such as in Fig.~\ref{Fig:muAveDagAFMhA125fJm1S12FixA}.  }
\label{Fig:muAveDagAFMhA125fJm1S12Fix}
\end{figure*}

Once $\bar{\mu}_{z{\rm ave}}$ is determined as described above, the reduced magnetic free energy of the AFM phase $F_{\rm magAFM}/k_{\rm B}T_{{\rm N}J}$ is calculated versus~$t,\ h_z,$ and~$h_{\rm A1}$ using Eqs.~(\ref{Eqs:Thermodynamics}).  Plots of $F_{\rm magAFM}/k_{\rm B}T_{{\rm N}J}$ versus~$h_z$ for $f_J=-1$, $S=1/2$ and reduced temperatures~$t$ from 0.05 to~0.95 are shown for reduced anisotropy fields $h_{\rm A1}=1/4$ to~1 in Fig.~\ref{Fig:muAveDagAFMhA125fJm1S12Fix}.  One sees that at low temperatures for each value of $h_{\rm A1}$, $F_{\rm magAFM}/k_{\rm B}T_{{\rm N}J}$ shows  a discontinuity in slope at the respective $h_{\rm cAFM}$ corresponding to the first-order discontinuity in $\bar{\mu}_z$ in Fig.~\ref{Fig:muAveDagAFMhA125fJm1S12FixA}, whereas at the higher temperatures $F_{\rm magAFM}/k_{\rm B}T_{{\rm N}J}$ varies smoothly through $h_{\rm cAFM}$, corresponding to a second-order transition in $\bar{\mu}_z$, as quantified in Fig.~\ref{Fig:hcAFMvstAS12fJm1hA1xx}.

\section{\label{Sec:PhaseDiags} Phase Diagrams}

The phase diagrams discussed here are those with the anisotropy field oriented along the $z$~axis as in Fig.~\ref{Fig:chiParallel}, for which the ground state in $h_z=0$ is a collinear AFM aligned along that axis, and with a reduced external field~$h_z$ in the $+z$~direction.   We first discuss the zero-temperature properties and phase diagrams of Heisenberg systems with classical anisotropy fields and then extend the discussion to finite-temperature phase diagrams.  Because phase diagrams for $S=1/2$ are not relevant when uniaxial quantum $DS_z^2$ anisotropy is present in Heisenberg spin systems~\cite{Johnston2017}, here we emphasize phase diagrams for this spin value.

\subsection{\label{Sec:T0PhaseDiags} Zero-Temperature Phase Diagrams and Magnetizations versus Field}

The zero-temperature properties and phase diagrams are determined from the relative free energies of SF and AFM phases and their dependences on the parameters $S$, $f_J$, $h_{\rm A1}$, and~$h_z$.  The PM phase appears at and above the critical field of the phase with the lower free energy.

\subsubsection{Spin-Flop Phase}

For $t\to0$, the entropy of the SF phase in $H_z=0$ is zero due to the nondegenerate ground state arising from the nonzero exchange field, so Eqs.~(\ref{Eqs:Thermodynamics}) yield
\bse
\label{Eqs:FmagSFT0}
\bea
\frac{F_{\rm magSF}(h_z=0,t\to0)}{k_{\rm B}T_{{\rm N}J}} &=& \frac{U_{\rm magSF}(h_z=0,t\to0)}{k_{\rm B}T_{{\rm N}J}}\\*
&& -\ S\int_0^{h_z}\bar{\mu}_{z\rm SF}(h_z,t\to0)dh_z.\nonumber
\eea
Equation~(\ref{UmagAFMHz0t0}) gives the first term as
\be
\frac{U_{\rm magSF}(h_z=0,t\to0)}{k_{\rm B}T_{{\rm N}J}} = -\frac{3S}{2(S+1)},
\label{UmagAFMHz0t02}
\ee
and Eqs.~(\ref{Eqs:muVsHAFPMFlop}) give
\be
\bar{\mu}_{z{\rm SF}}(h_z,t\to0) = 
\begin{cases}
h_z/h_{\rm cSF}\quad (h_z \leq h_{\rm cSF})\\
1, \hspace{0.5in} (h_z \geq h_{\rm cSF}),
\end{cases}
\label{Eq:muzvshzSF}
\ee
where Eq.~(\ref{Eq:HcFlopDef}) gives the SF critical field as
\be
h_{\rm cSF}(t\to0) = \frac{3(1-f_J-h_{\rm A1})}{S+1}
\label{Eq:HcFlopDef2}
\ee
\ese
using $\bar{\mu}_{\rm SF}=1$ for $t\to0$.  Thus Eq.~(\ref{Eq:FmagGen}) gives the normalized free energy of the SF phase versus~$h_z$  for $t\to0$ as
\begin{widetext}
\be
\frac{F_{\rm magSF}(h_z,t\to0)}{k_{\rm B}T_{{\rm N}J}} =
\begin{cases}
-\frac{3S}{2(S+1)} -S\frac{h_z^2}{2h_{\rm cSF}}\hspace{1.13in}(h_z \leq h_{\rm cSF})\\*
\\*
-\frac{3S}{2(S+1)} -S\left[\frac{h_{\rm cSF}}{2}+(h_z-h_{\rm cSF})\right]\quad(h_z \geq h_{\rm cSF}).
\end{cases}
\label{FmagSFto0}
\ee
\end{widetext}

\subsubsection{Antiferromagnetic Phase}

For the AFM phase at $t\to0$, the moments cannot respond to the field without a spin-flip transition to the PM phase.  Also, the entropy is zero at $t\to0$ because the ground state is nondegenerate on account of the presence of the exchange and anisotropy fields.  Thus using Eq.~(\ref{Eq:Ui}) with $\bar{\mu}_0=1$, the reduced free energy per spin is
\bea
\frac{F_{\rm magAFM}(h_z,t\to0)}{k_{\rm B}T_{{\rm N}J}} &=& \frac{U_{\rm magAFM}(h_z,t\to0)}{k_{\rm B}T_{{\rm N}J}}\nonumber\\*
&&\hspace{-1in} =-\frac{3S(1+h_{\rm A1})}{2(S+1)} \quad (h_z\leq h_{\rm cAFM}).\label{Eq:eps0AFM}
\eea
Thus if $h_{\rm A1}=h_z=0$, the free energies of the SF and AFM phases in Eqs.~(\ref{FmagSFto0}) and~(\ref{Eq:eps0AFM}), respectively, are the same, as required.  The AFM critical field $h_{\rm cAFM}$, at which $\bar{\mu}_{2z}= -1$ flips to the PM state with  $\bar{\mu}_{2z}=\bar{\mu}_{1z}=  +1$ with increasing~$h_z$, is determined next.  

The spin-flip field to the PM state (the $t=0$ AFM critical field $h_{\rm cAFM}$) is determined by the conditions under which $\bar{\mu}^\dagger$ in Eq.~(\ref{Eq:muzdagger}) goes to zero with increasing~$h_z$.  This was carried out by solving Eqs.~(\ref{Eq:barmu2z}) at $t=0.01$ for various values of $S$, $h_{\rm A1}>0$ and~$-1\leq f_J<1$.  In this way, we obtain
\be
h_{\rm cAFM} = \frac{3(1+h_{\rm A1})}{S+1} \quad (t\to0,\ -1\leq f_J<1),
\label{Eq:hcAFM}
\ee
which is independent of $f_J$ in the given $f_J$ range.  This expression is in agreement with our numerical data for the AFM to PM spin-flip transition field at $t\to0$ obtained from numerical calculations such as the extrapolations to $t=0$ in Fig.~\ref{Fig:hcAFMvstAS12fJm1hA1xx} above for $S=1/2$, $f_J=-1$, and various values of $h_{\rm A1}$, and in the phase diagram in Fig.~\ref{Fig:PhaseDiagram_S12fJm1hA10}(f) below for $S=1/2$, $f_J=-1$, and $h_{\rm A1}=1$.

Using Eqs.~(\ref{Eqs:Thermodynamics}) and~(\ref{Eq:eps0AFM}) we obtain the field dependence of the free energy per spin of the AFM phase (and high-field PM phase) as
\bea
\frac{F_{\rm magAFM}(t\to0)}{k_{\rm B}T_{{\rm N}J}} &=& -\frac{3S(1+h_{\rm A1})}{2(S+1)} \quad (h_z\leq h_{\rm cAFM}),\nonumber\\*
\label{Eq:FmagAFMT0}\\*
\frac{F_{\rm magAFM}(t\to0)}{k_{\rm B}T_{{\rm N}J}}  &=& -\frac{3S(1+h_{\rm A1})}{2(S+1)} -S(h_z-h_{\rm cAFM})\nonumber\\*
&&\hspace{1in}(h_z\geq h_{\rm cAFM}).\nonumber
\eea

\subsubsection{Comparison of the Free Energies of the Spin-Flop and Antiferromagnetic Phases}

\begin{figure*}
\includegraphics [width=3.in]{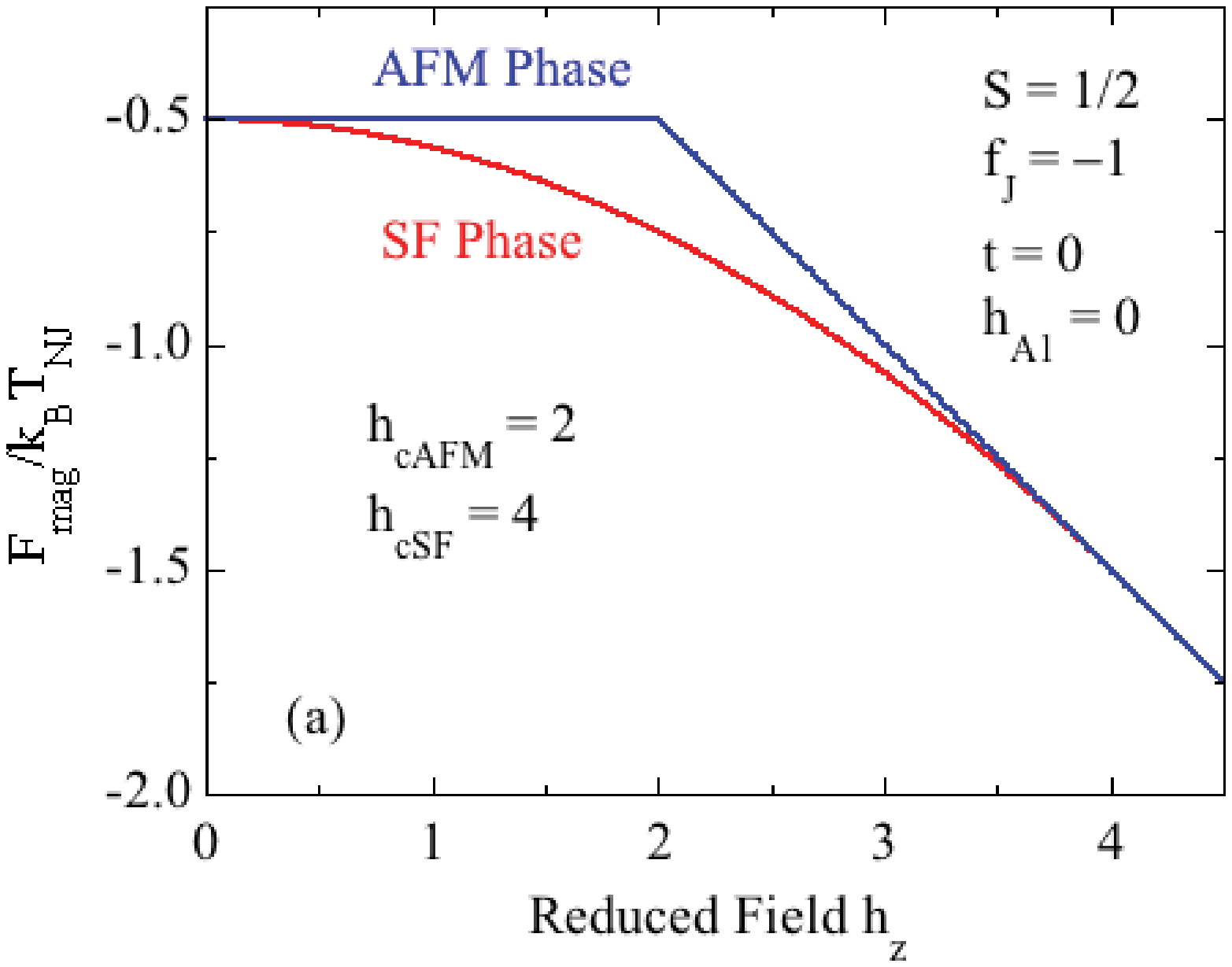}\includegraphics [width=3.in]{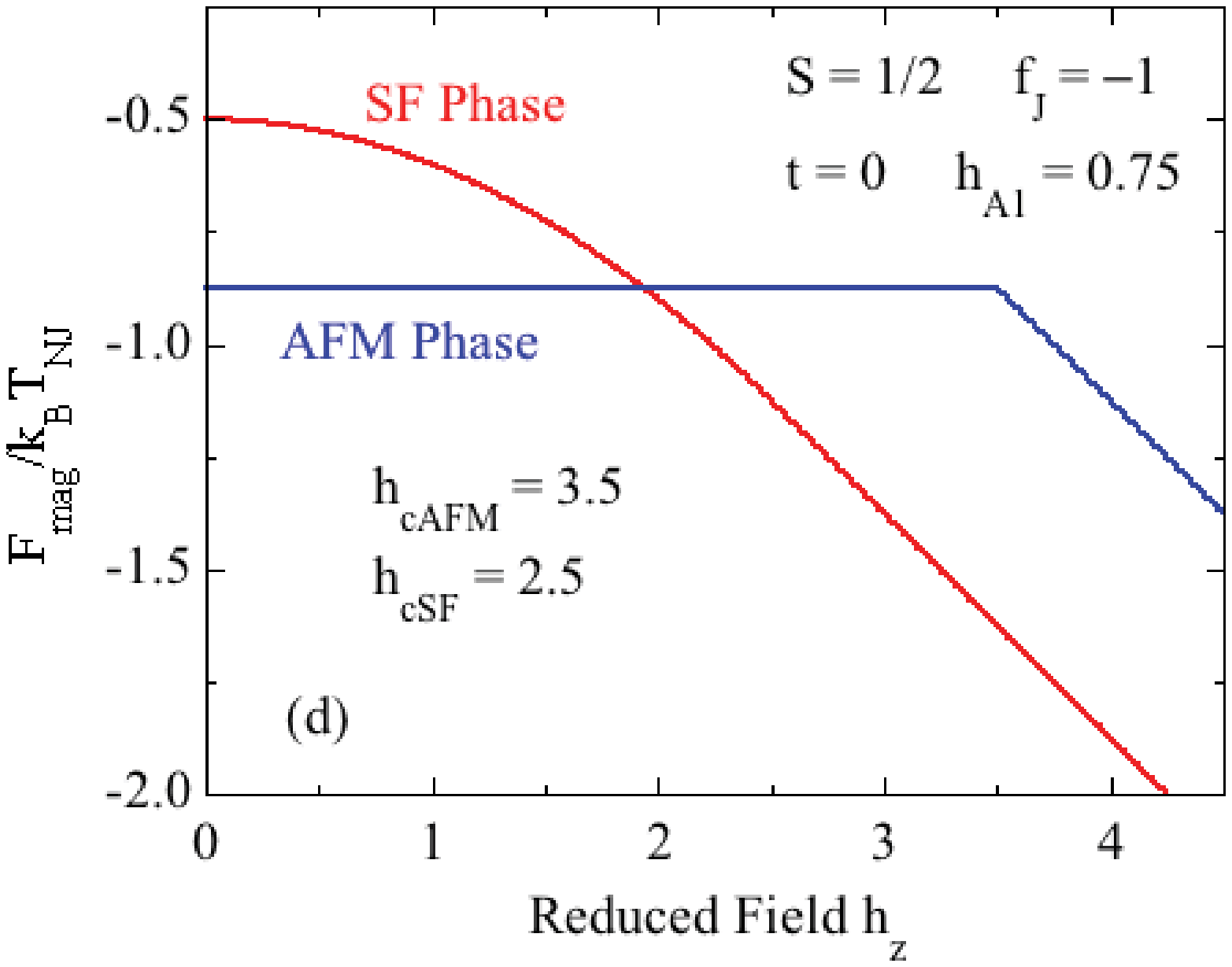}
\includegraphics [width=3.in]{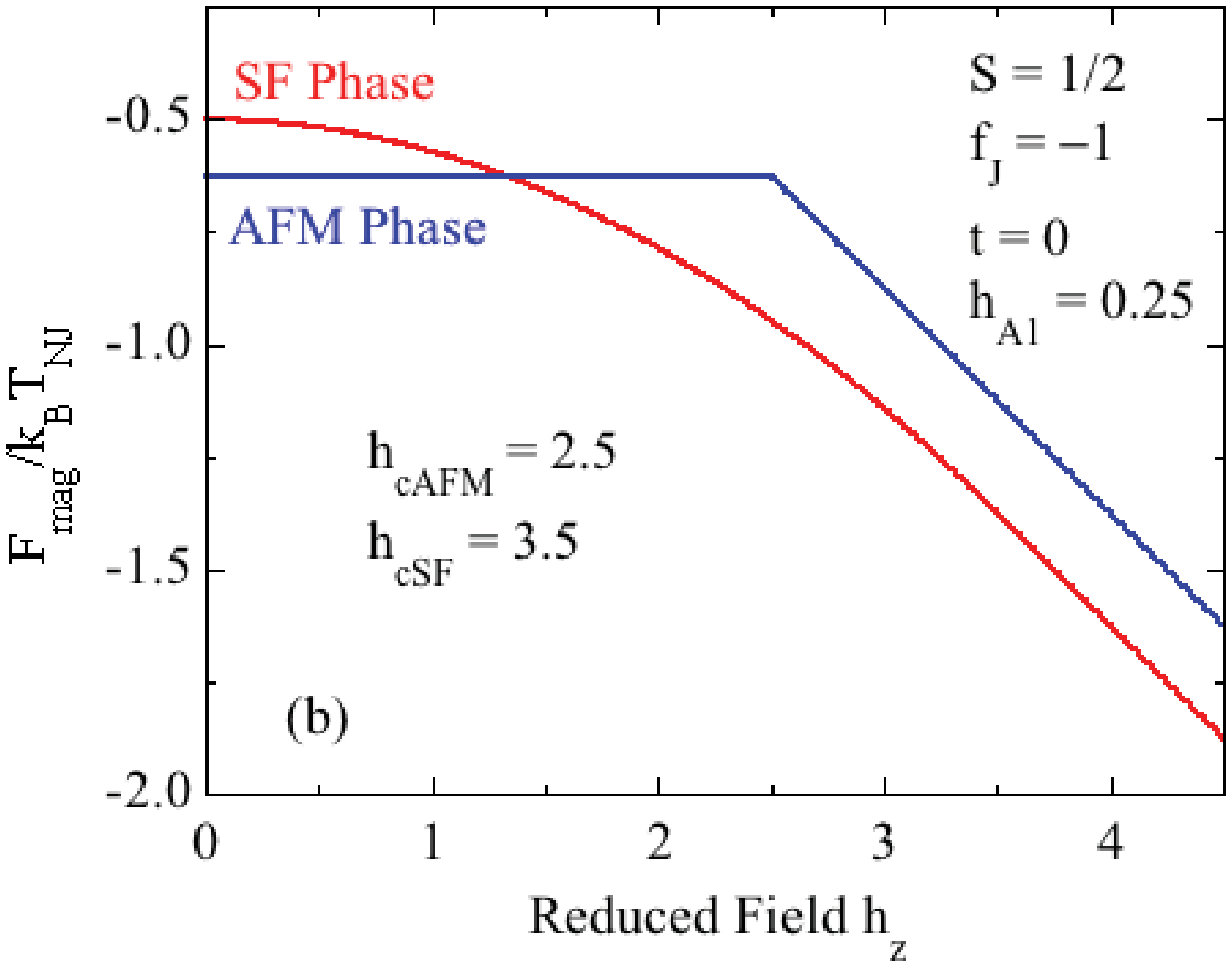}\includegraphics [width=3.in]{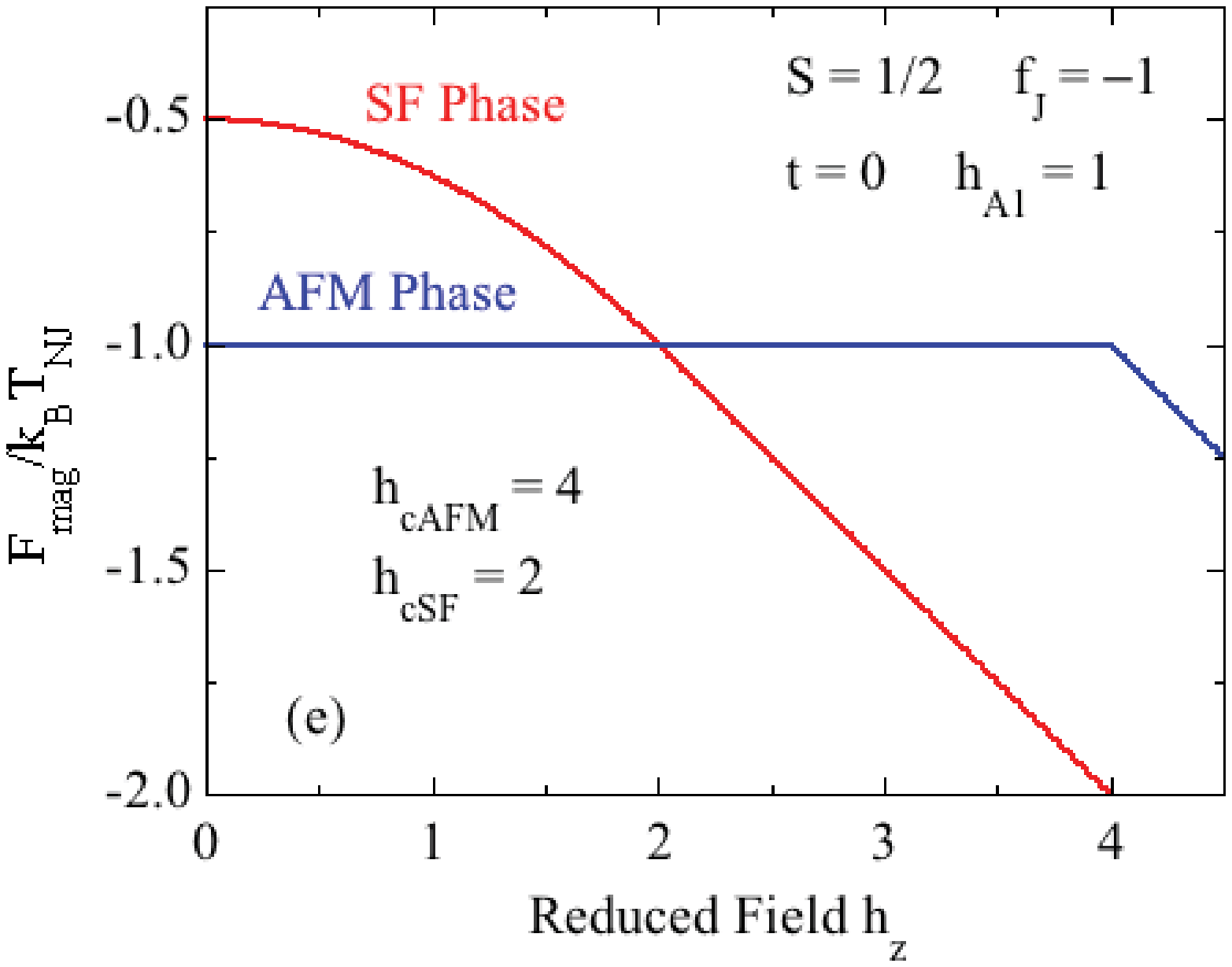}
\includegraphics [width=3.in]{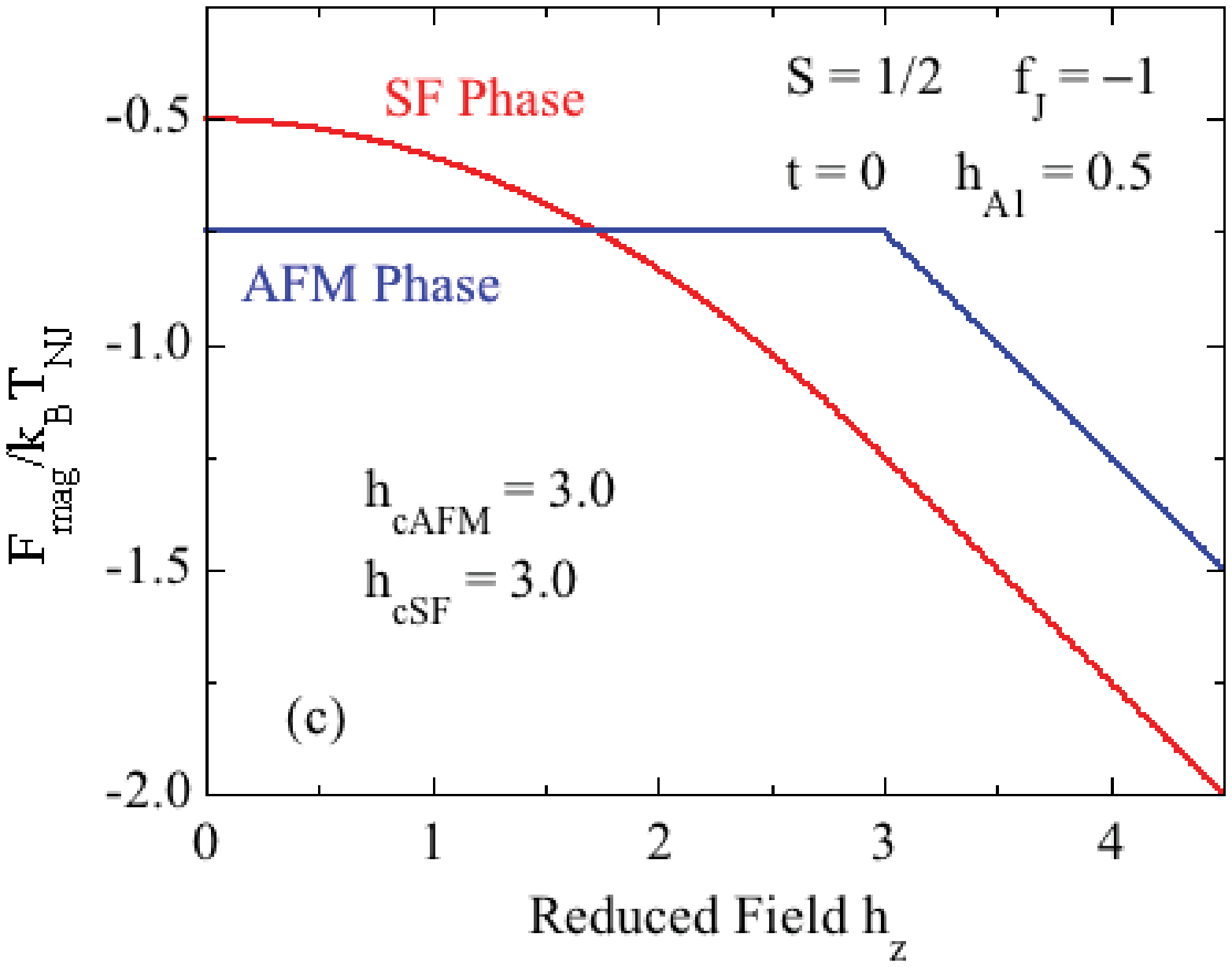}\includegraphics [width=3.in]{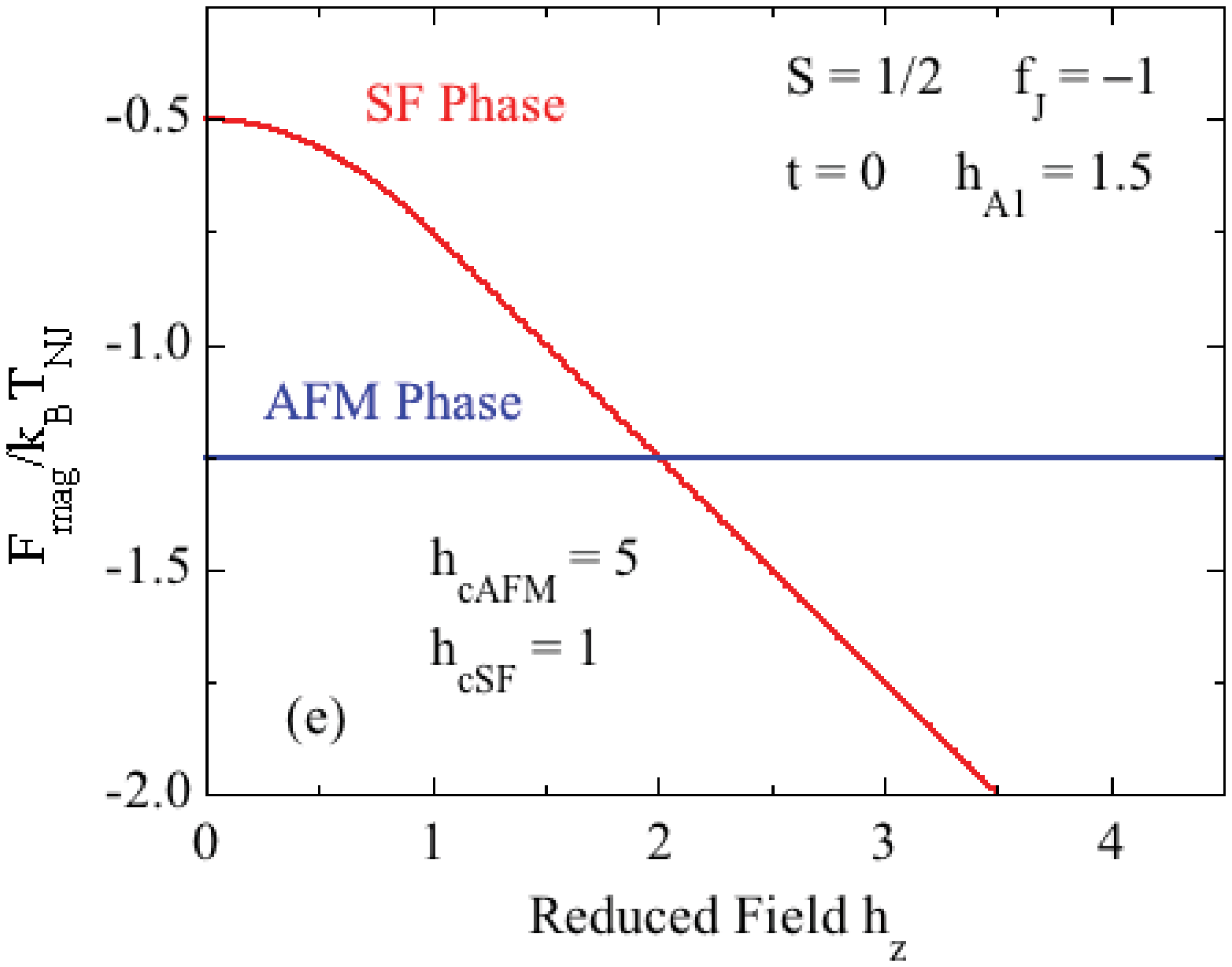}
\caption{(Color online) Reduced free energy $F_{\rm mag}/k_{\rm B}T_{{\rm N}J}$ at reduced temperature $t\to0$ versus reduced magnetic field $h_z$ for collinear $z$-axis AFMs with spin~$S=1/2$ and reduced anisotropy fields~$h_{\rm A1}$ of (a)~0, (b)~1/4, (c)~1/2, (d)~3/4, (e)~1, and~(f)~3/2.  The phases in competition are the collinear antiferromagnetic (AFM) and spin-flop (SF) phases. The PM phases occur above the respective critical fields of the AFM and SF phases as listed. The data were calculated from Eqs.~(\ref{FmagSFto0}) and~(\ref{Eq:FmagAFMT0}).}
\label{Fig:FmagT0SFAFMvsHzS12fJm1hA1xx}
\end{figure*}

Figure~\ref{Fig:FmagT0SFAFMvsHzS12fJm1hA1xx} illustrates the free energies $F_{\rm mag}$ per spin versus reduced field~$h_z$ of the SF and AFM phases (and their high-field PM phases) for $t\to0$, given in Eqs.~(\ref{FmagSFto0}) and~(\ref{Eq:FmagAFMT0}), respectively, for $f_J=-1$ and anisotropy parameters $h_{\rm A1} = 0$ to~1.5.  For $h_{\rm A1}=0$ the lowest-energy phase for $h_z>0$ is the SF phase.  Upon increasing~$h_{\rm A1}$, one sees an evolution where the AFM phase is more stable at low fields, but  transforms to the SF phase at increasing values of $h_z$, where the AFM to SF phase transition is first order due to the discontinuity in slope of $F_{\rm mag}$ versus $h_z$ at the transition point, which corresponds to a discontinuity in the magnetization there.

\begin{figure}
\includegraphics [width=3.in]{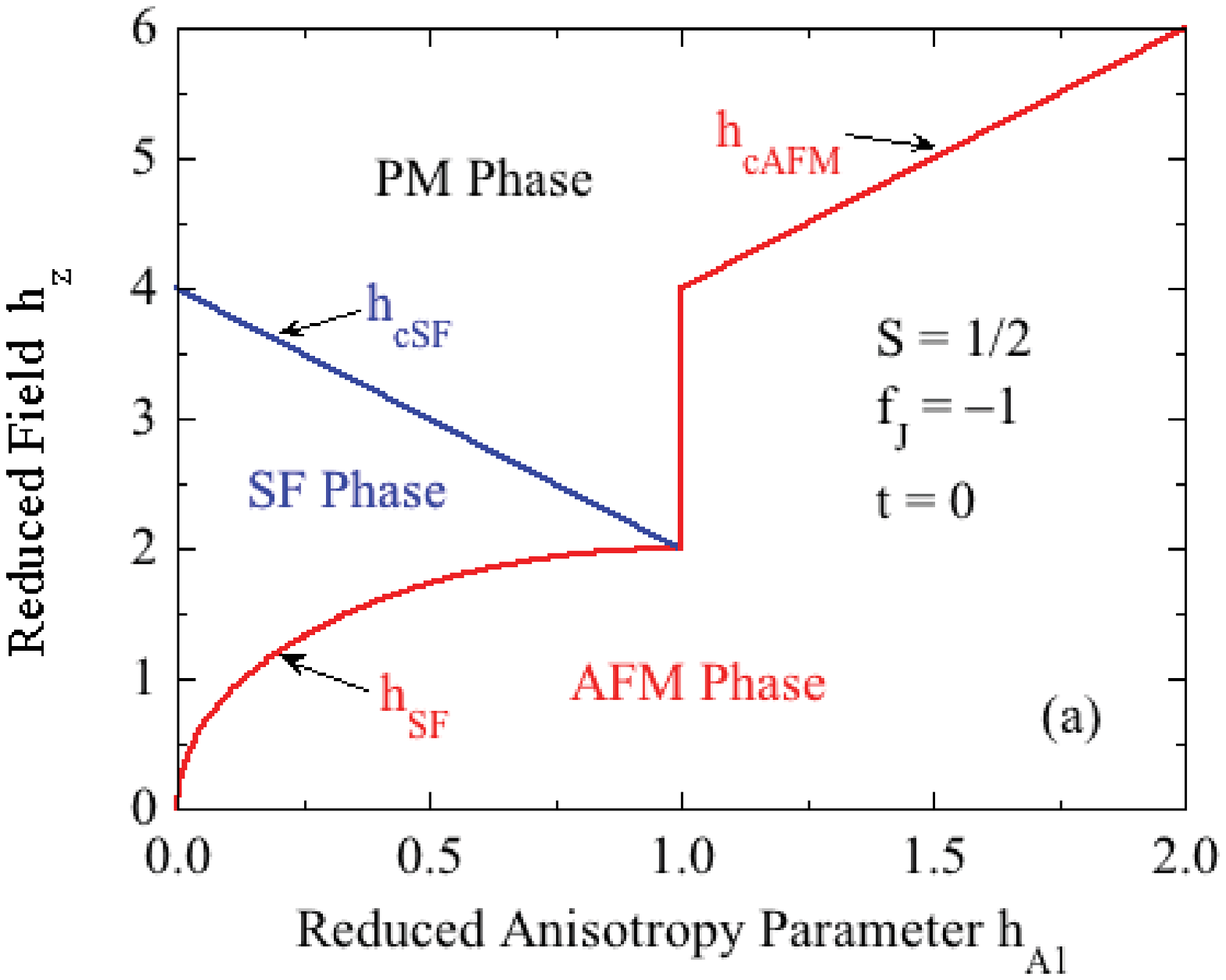}
\includegraphics [width=3.in]{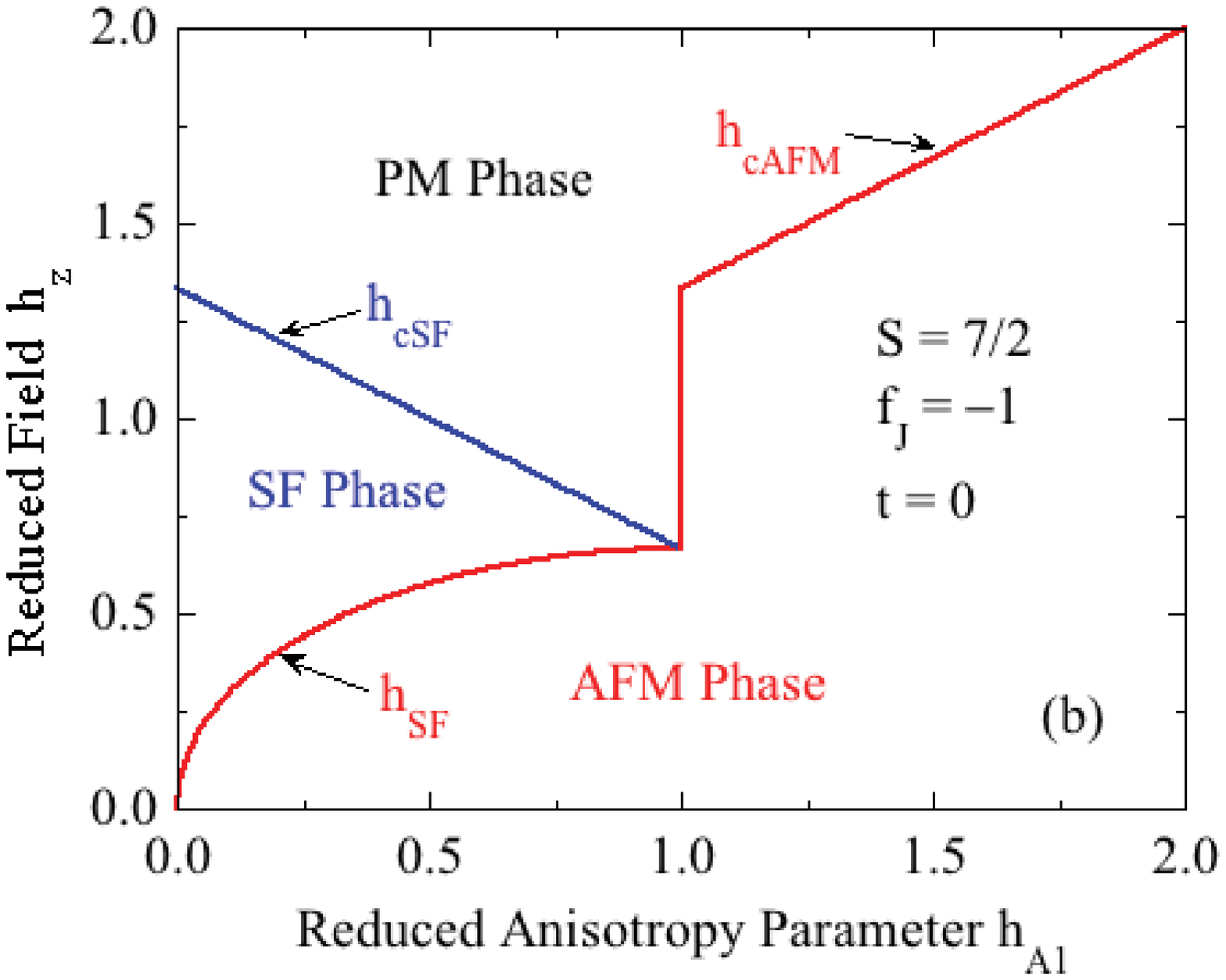}
\caption{(Color online) Zero-temperature phase diagrams in the $h_z$-$h_{\rm A1}$ plane for collinear $z$-axis AFMs with $f_J=-1$ and for spins~(a)~$S=1/2$ and (b)~$S=7/2$.  The phases in competition are the collinear $z$-axis antiferromagnetic (AFM) and spin-flop (SF) phases, with the paramagnetic~(PM) phase in each case above the respective critical field $h_{\rm cAFM}$ and~$h_{\rm cSF}$. Note that the ordinate axes are different for the two spin values.  The transitions from AFM to PM and AFM to SF are first order, and from SF to PM are second order.}
\label{Fig:T0PhaseDiagramS1272}
\end{figure}

Shown in Fig.~\ref{Fig:T0PhaseDiagramS1272} are zero-temperature phase diagrams in the $h_z$--$h_{\rm A1}$ plane for collinear $z$-axis AFMs with $f_J=-1$ and for spins $S=1/2$ and~$S=7/2$, obtained by determining which of the AFM and SF phases (and associated high-field PM phases) has the lower free energy using Eqs.~(\ref{FmagSFto0}) and~(\ref{Eq:FmagAFMT0}).  One sees that the phase diagrams are the same for $S=1/2$ and~$S=7/2$, apart from a reduction in ordinate scale by a factor of three for $S=7/2$ compared to that for $S=1/2$.  For $h_{\rm A1}>1$ the AFM phase undergoes a spin-flip transition directly to the PM phase with increasing~$h_z$, sidestepping the intermediate SF phase. 

The analytic behavior of the AFM--SF transition field $h_{\rm SF}$ for $f_J=-1$ such as in Fig.~\ref{Fig:T0PhaseDiagramS1272} in the region $0\leq h_{\rm A1} \leq 1$ is found to be
\be
h_{\rm SF} = \frac{3}{S+1}\sqrt{2h_{\rm A1}-h_{\rm A1}^2}\ .
\label{Eq:hSF}
\ee
However, this expression is only valid for $f_J=-1$, which corresponds to a bipartite AFM with only nearest-neighbor exchange interactions of equal value.  If \mbox{$f_J\neq-1$,} we find
\bea
h_{\rm SF} &=& \frac{3}{S+1}\sqrt{h_{\rm A1}(1-f_J)-h_{\rm A1}^2}\ ,\label{Eq:hSF2}\\*
0 &<& h_{\rm A1} < (1-f_J)/2,\qquad -3\lesssim f_J<1,\nonumber
\eea
where the upper $h_{\rm A1}$ limit is the maximum value for which $h_{\rm SF} < h_{\rm cSF}$, the lower limit on $f_J$ is obtained by requiring $h_{\rm SF} < h_{\rm cAFM}$ for the given $h_{\rm A1}$ range, and the upper limit on $f_J$ is required for any AFM, where the value $f_J=1$ corresponds to a FM rather than an AFM\@.

Thus the deviation of $f_J\equiv \theta_{{\rm p}J}/T_{{\rm N}J}$ from the value \mbox{of $-1$} usually assumed can have a very significant influence on the variation of $h_{\rm SF}$ with~$h_{\rm A1}$ according to Eq.~(\ref{Eq:hSF2}), a situation not investigated previously to our knowledge.  This is important in view of the fact that within MFT one can have $-\infty < f_J<1$ for AFMs.  Indeed, most real AFMs are not bipartite with more than nearest-neighbor interactions.

The reduced fundamental exchange parameter $h_{\rm A1}$ is expressed in terms of the reduced exchange field $h_{\rm A0}$ at $T=0$ using  Eq.~(\ref{Eq:HA0}), the $t=0$ value $\bar{\mu}_i=1$, and the definition in Eq.~(\ref{Eq:halphaDef}) as
\be
h_{\rm A1} = \frac{S+1}{3}h_{\rm A0}.
\ee
Inserting this into Eq.~(\ref{Eq:hSF}) gives
\be
h_{\rm SF} = \sqrt{2\left(\frac{S+1}{3}\right)h_{\rm A0}-h_{\rm A0}^2}\ .
\label{Eq:hSF33}
\ee
Now using Eq.~(\ref{Eq:Hexch02}) for the exchange field together with Eq.~(\ref{Eq:halphaDef}) gives the reduced exchange field at $T=0$ as
\be
h_{\rm exch0} = \frac{S+1}{3}.
\ee
Substituting this into Eq.~(\ref{Eq:hSF33}) gives
\be
h_{\rm SF} = \sqrt{2h_{\rm exch0}h_{\rm A0}-h_{\rm A0}^2}\ .
\label{Eq:hSF4}
\ee
In terms of the unreduced fields one has
\be
H_{\rm SF} = \sqrt{2H_{\rm exch0}H_{\rm A0}-H_{\rm A0}^2}\ .
\label{Eq:hSF5}
\ee
This expression is identical to the standard equation for $H_{\rm SF}$ obtained using spin-wave theory assuming $f_J=-1$ \cite{Keffer1966}.  A more accurate expression obtained from Eq.~(\ref{Eq:hSF2}) is
\be
H_{\rm SF} = \sqrt{H_{\rm exch0}H_{\rm A0}(1-f_J)-H_{\rm A0}^2}\ .
\label{Eq:hSF6}
\ee
As noted above, $f_J<1$ for an AFM.

\subsubsection{Magnetization versus Field}

\begin{figure}
\includegraphics [width=3.in]{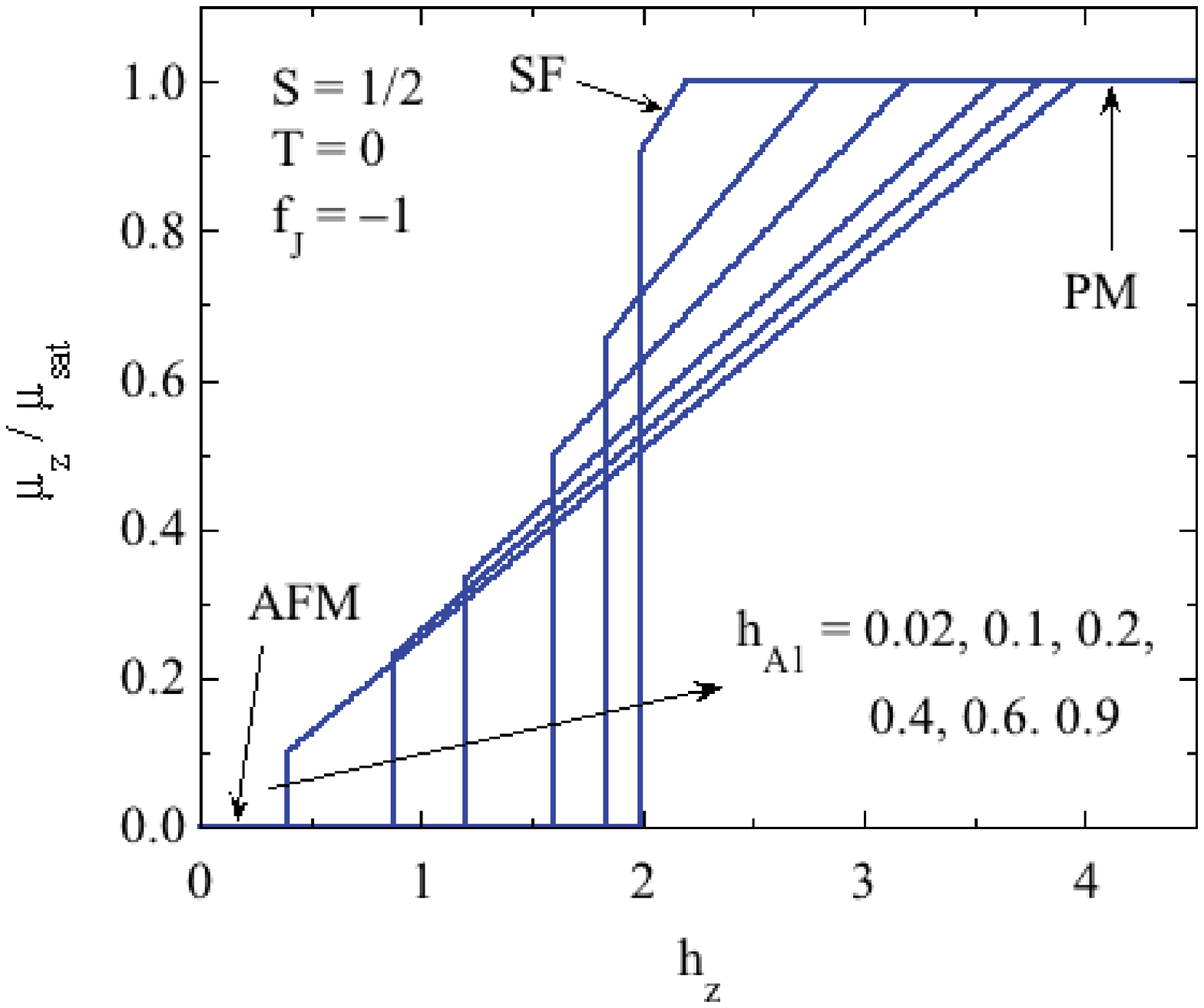}
\caption{(Color online) Reduced $z$-axis moment $\bar{\mu}_z\equiv \mu_z/\mu_{\rm sat}$ per spin versus reduced field~$h_z=g\mu_{\rm B}H_z/k_{\rm B}T_{{\rm N}J}$ for spins $S=1/2$ at zero temperature for anisotropy parameters $h_{\rm A1}$ as listed and $f_J\equiv \theta_{{\rm p}J}/T_{{\rm N}J} = -1$.}
\label{Fig:T0muzVShzAFMSFS12fJm1hA1X}
\end{figure}

The magnetization of the SF phase is proportional to field according to Eq.~(\ref{Eq:barmuzSF}), which at $T=0$ reads 
\be
\bar{\mu}_z = \frac{h_z}{h_{\rm cSF}} \qquad (h_z\leq h_{\rm cSF}),
\label{Eq:barmuzSFT0}
\ee
where the spin-flop critical field is given by Eq.~(\ref{Eq:HcFlopDef}) with $\bar{\mu}_{\rm SF}=1$ at $T=0$ as
\be
h_{\rm cSF} = \frac{3(1-f_J-h_{\rm A1})}{S+1}.
\label{Eq:HcFlopDefT0}  
\ee
According to Eqs .~(\ref{Eq:hSF2}), if $h_{\rm A1} > (1-f_J)/2$ the AFM phase undergoes a first-order transition with $\bar{\mu}_{z{\rm ave}}=0$ to the fully-saturated PM state with $\bar{\mu}_{z{\rm ave}}=1$ at the $T=0$ transition field $h_z = h_{\rm cAFM}$ in Eq.~(\ref{Eq:hcAFM}), whereas if $h_{\rm A1} < (1-f_J)/2$, the AFM state instead has a first-order transition to the SF phase at $h_{\rm SF}$ until the SF phase saturates at $h_z=h_{\rm cSF}$ to $\bar{\mu}_z=1$ after which it remains constant at $\bar{\mu}_z(h_z) = 1$.  With these criteria, the $\bar{\mu}_z(h_z)$ behaviors were determined as shown in Fig.~\ref{Fig:T0muzVShzAFMSFS12fJm1hA1X} for $S=1/2,\ f_J=-1$ and a range of $h_{\rm A1}$ values from~0.02 to~0.9 as shown.  Changing the value of $f_J$ results in no qualitative change in the $\bar{\mu}_z$ versus~$h_z$ plots, but where the corresponding ranges of $h_{\rm A1}$ values and ordinate scales giving similar-looking plots as in Fig.~\ref{Fig:T0muzVShzAFMSFS12fJm1hA1X} are changed appropriately.

\subsubsection{Perpendicular Magnetic Fields}

\begin{figure}
\includegraphics [width=3.in]{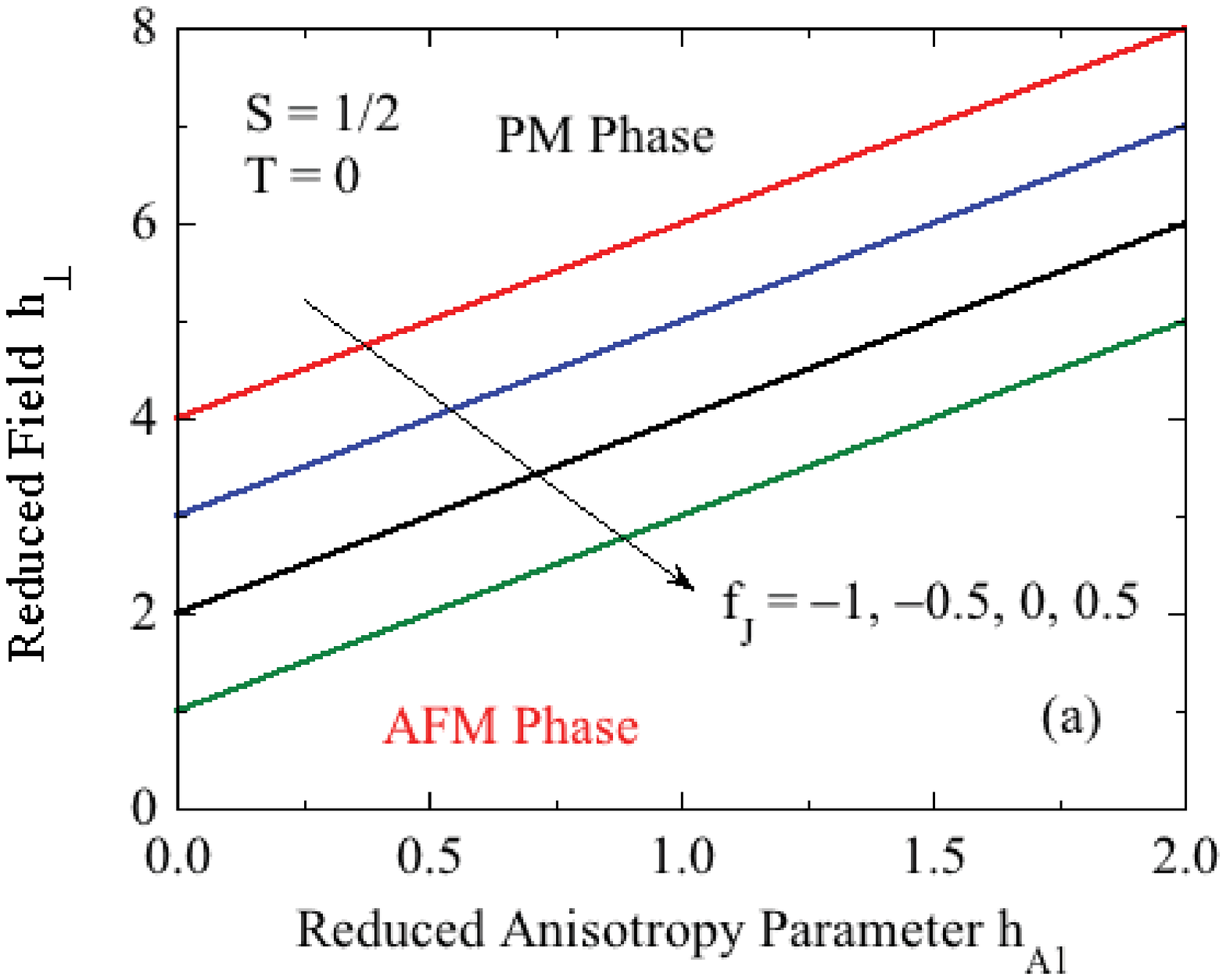}
\includegraphics [width=3.in]{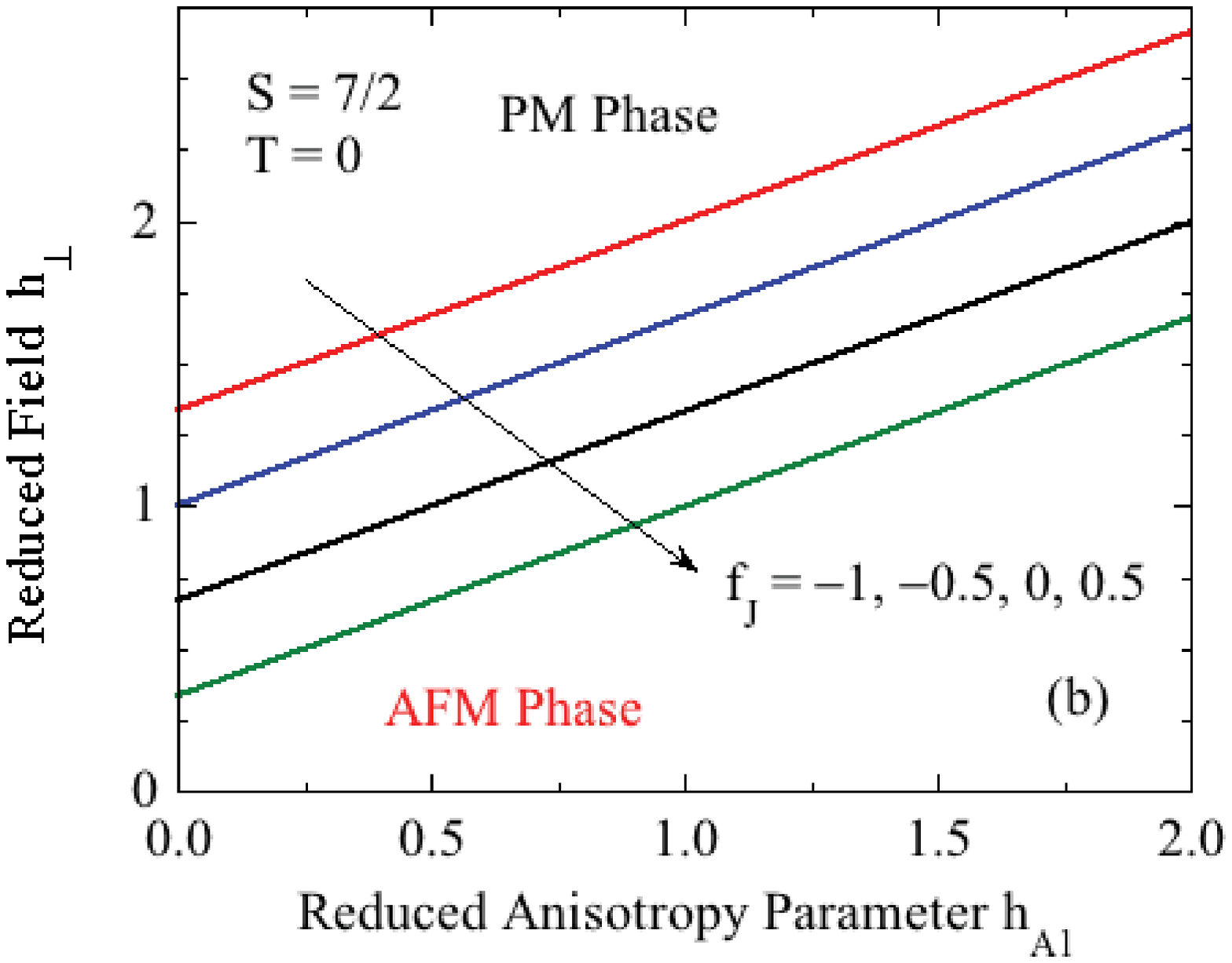}
\caption{(Color online) Zero-temperature phase diagrams in the $h_\perp$-$h_{\rm A1}$ plane for collinear $z$-axis AFMs with $f_J=-1$ to~0.5 and for spins~(a)~$S=1/2$ and (b)~$S=7/2$.  The phases in competition are the canted antiferromagnetic (AFM) and the paramagnetic~(PM) phase that occurs above the respective critical field $h_{\rm c\perp}$. The plots are drawn according to Eq.~(\ref{Eq:hcPerp0}).  The ordinate axes are different for the two spin values.  The transitions from canted AFM to PM are second order.}
\label{Fig:hcPerp0S12fJxxhA1xx}
\end{figure}

When the applied field is perpendicular to the easy axis or easy plane of a collinear or noncollinear AFM as shown in Fig.~\ref{Fig:High_Perp_Field_Structs}, only one transition versus field occurs which is a second-order transition from the canted AFM phase to the PM phase at the perpendicular critical field $h_{\rm c\perp AFM}$ given by Eq.~(\ref{Eq:hcPerpRed}) at $T=0$ as
\be
h_{\rm c\perp AFM0} = \left(\frac{3}{S+1}\right)(1+h_{\rm A1}-f_J).
\label{Eq:hcPerp0}
\ee
The phase diagrams in the $h_\perp$--$h_{\rm A1}$ plane for spins $S=1/2$ and $S=7/2$ are shown in Fig.~\ref{Fig:hcPerp0S12fJxxhA1xx}, where the AFM--PM transition lines vary linearly with $h_{\rm A1}$ for each value of $S$ and~$f_J$\@.

\subsection{\label{Sec:Hz-T phase diagrams} Field versus Temperature Phase Diagrams for Fields Along the Easy Axis of Collinear Antiferromagnets}

\begin{figure*}
\includegraphics [width=3.in]{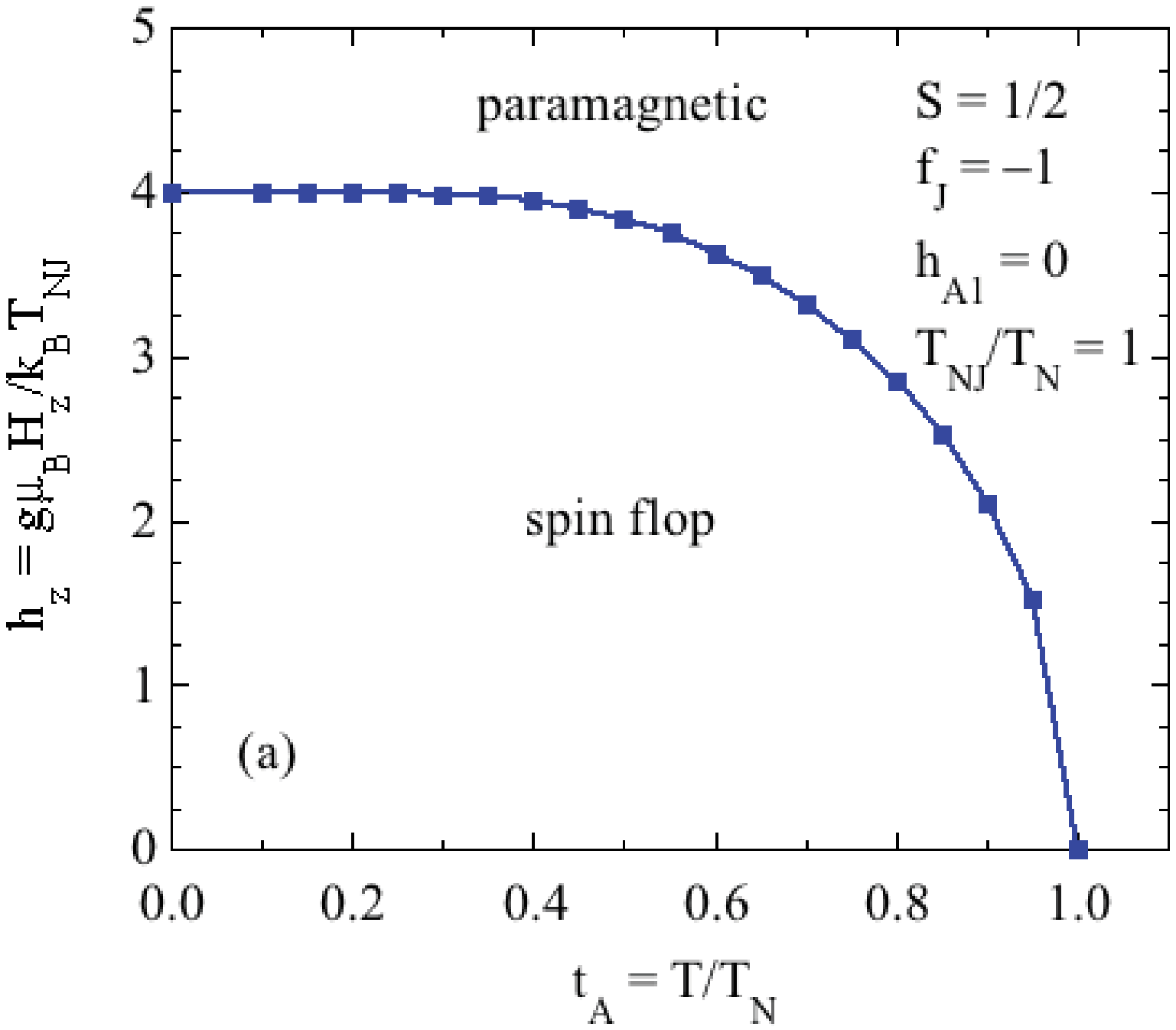}\includegraphics [width=3.in]{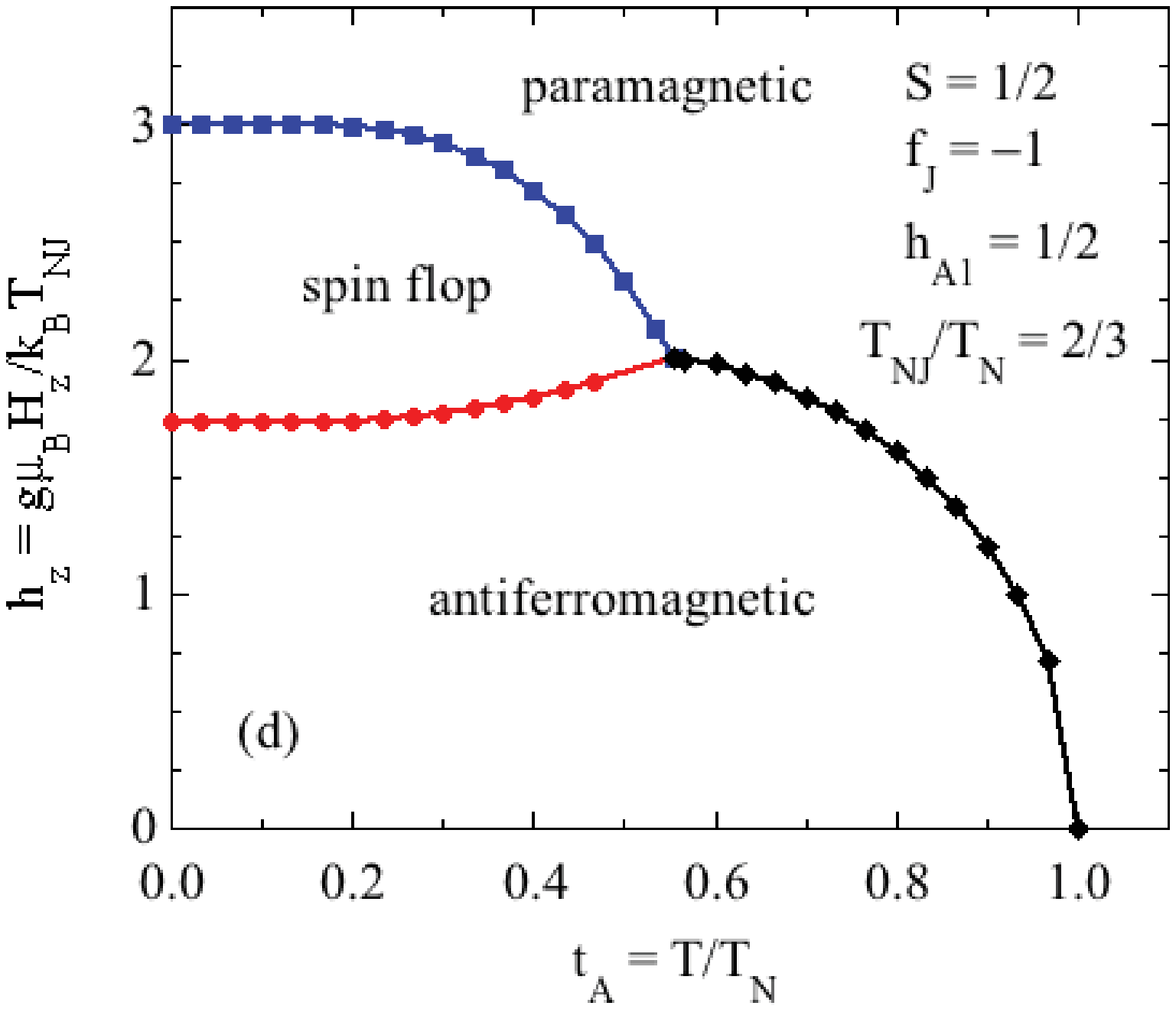}
\includegraphics [width=3.in]{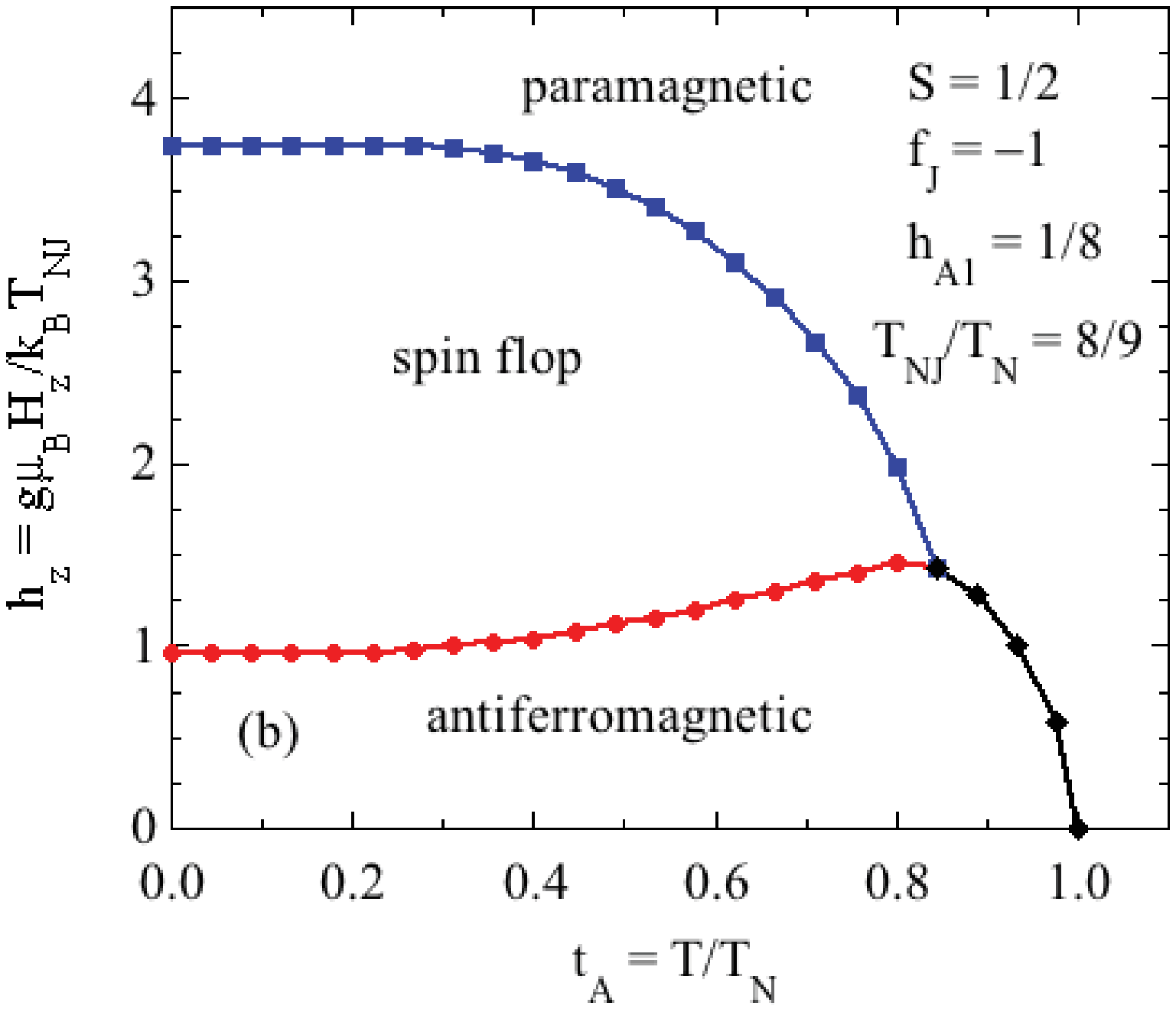}\includegraphics [width=3.in]{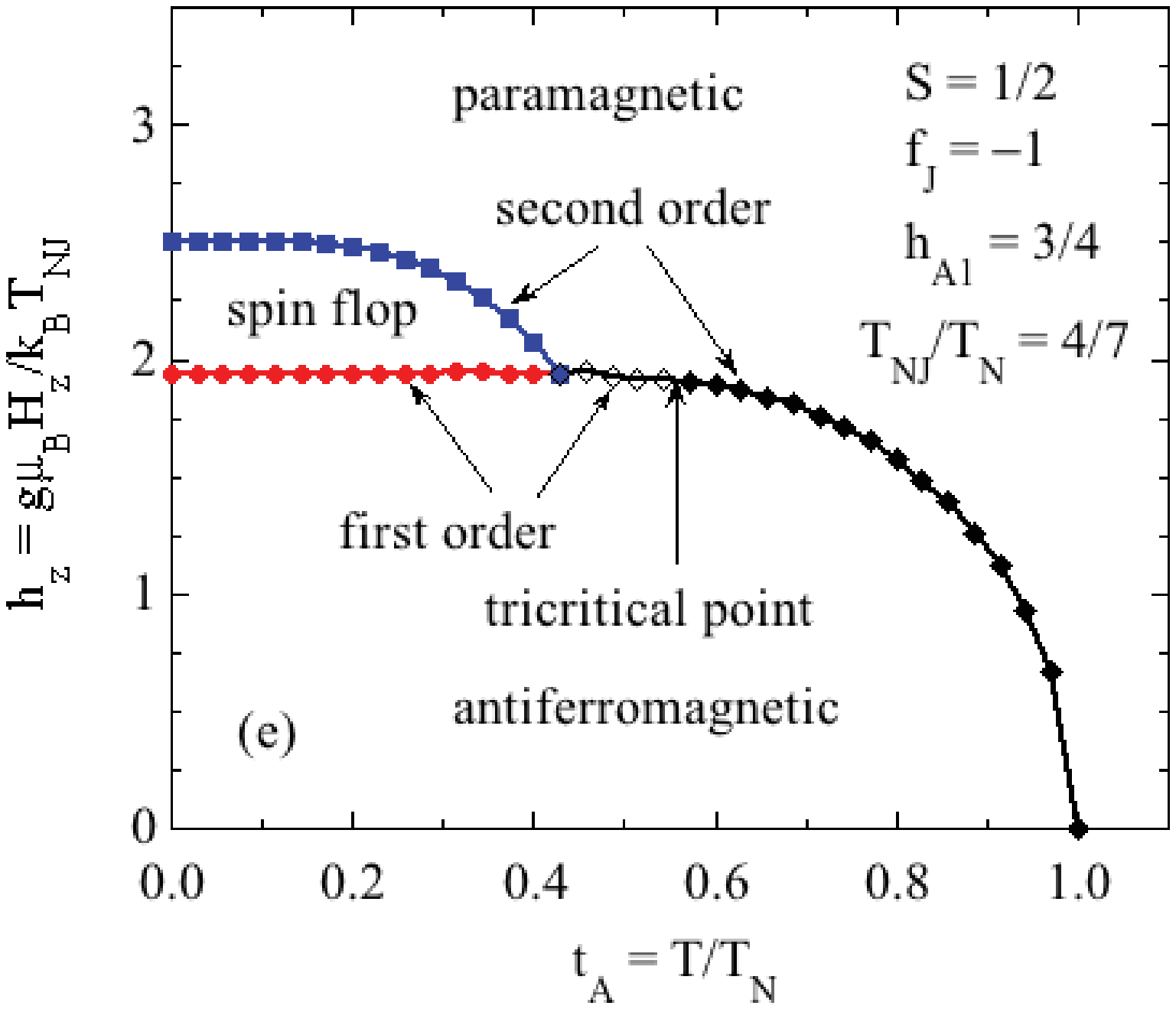}
\includegraphics [width=3.in]{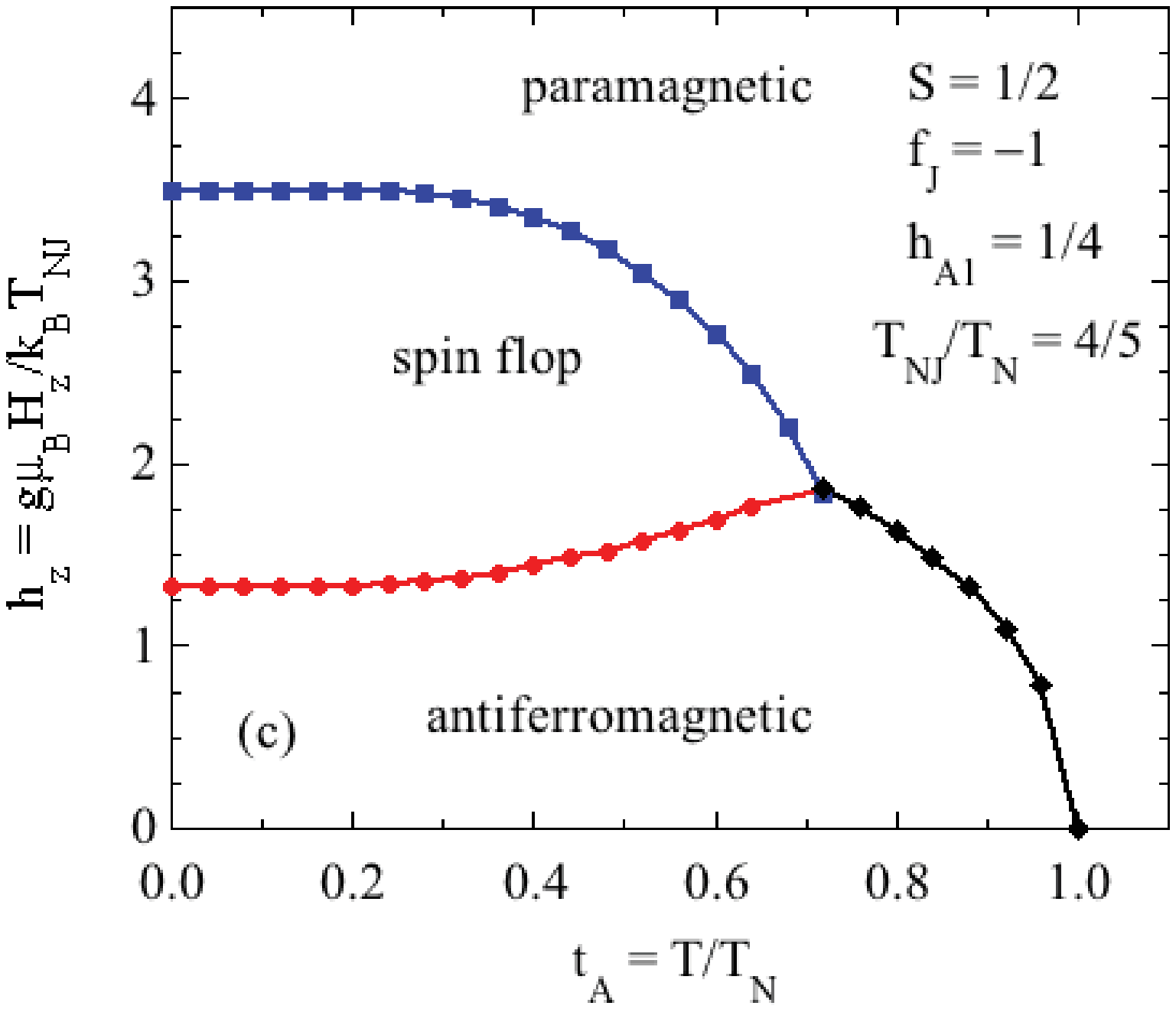}\includegraphics [width=3.in]{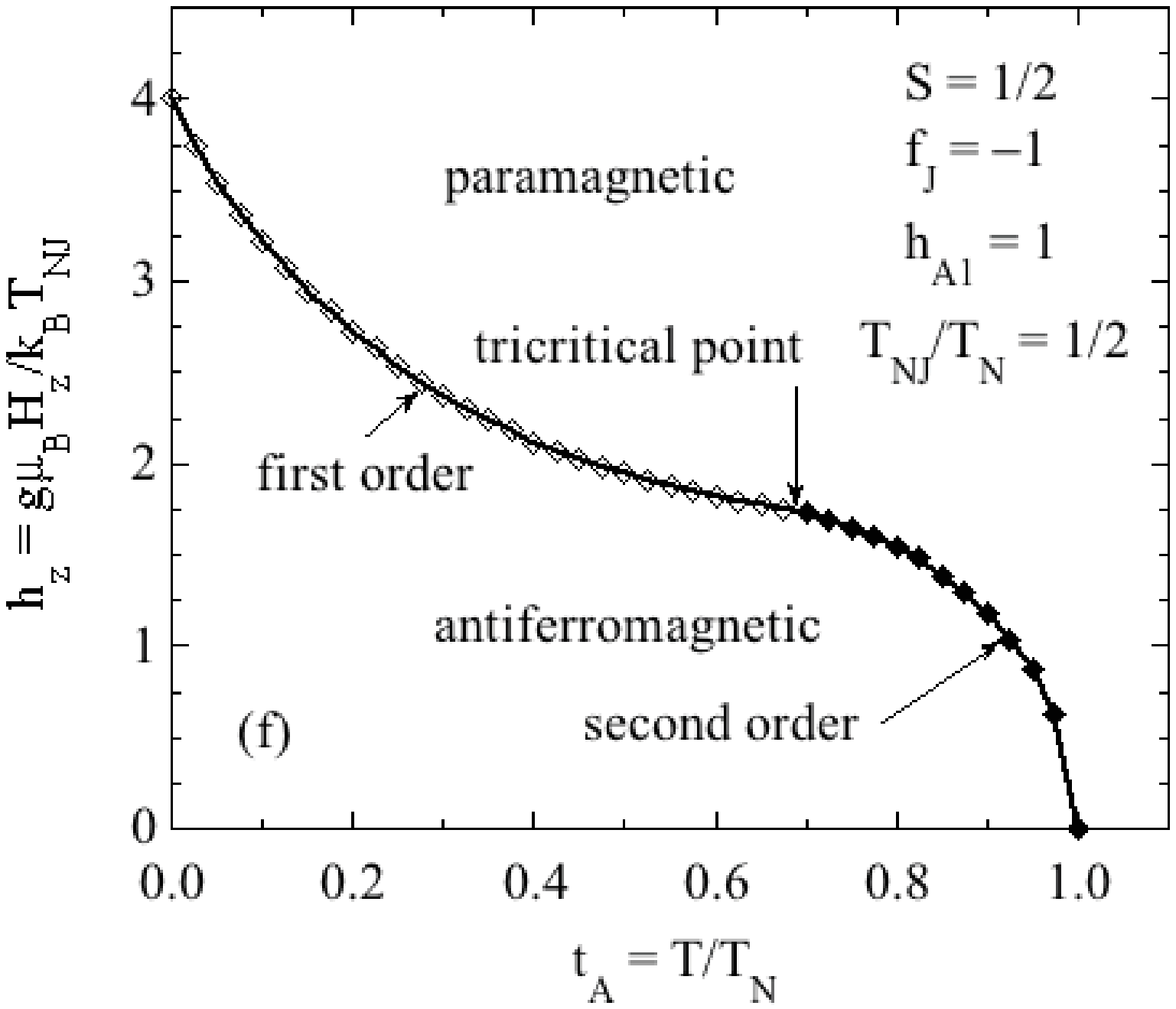}
\caption{(Color online) Reduced parallel magnetic field $h_{\rm z}$ versus reduced temperature~$t_{\rm A}$ phase diagrams for spin~$S=1/2$ and reduced anisotropy fields~$h_{\rm A1}$ equal to (a)~0, (b)~1/8, (c)~1/4, (d)~1/2, (e)~3/4, and (f)~1 obtained from numerical calculations.  The SF to PM transitions are second order and the AFM to SF transitions are first order.  The AFM to PM transitions can be second order [(a)--(d)], or both first and second order in different field ranges separated by a tricritical point~[(e),~(f)].  The lines are guides to the eye.}
\label{Fig:PhaseDiagram_S12fJm1hA10}
\end{figure*}

In order to determine the phase diagrams in the field versus temperature plane for given values of $S,\ f_J$, and~$h_{\rm A1}$, one must determine which of the AFM or SF phases and associated PM phases have the lowest free energy at each temperature and field for given values of $S$, $h_{\rm A1}$, and~$f_J$ using information such as illustrated above in Figs.~\ref{Fig:FmagVsHSFhA150fJm1S12} and~\ref{Fig:muAveDagAFMhA125fJm1S12Fix}.  The transitions from the AFM to the SF phase are always first order.  For transitions of the SF or AFM phase to the associated PM phase, the transition field is  determined as the field at which the angle $\theta\to0$ or $\mu_z^\dagger\to0$, respectively. First-order transitions have discontinuities in these quantities on crossing a transition line.

Shown in Fig.~\ref{Fig:PhaseDiagram_S12fJm1hA10} are the $h_z$ versus $t_{\rm A}$ phase diagrams for $S=1/2$, $f_J=-1$, and six values of the reduced anisotropy parameter $h_{\rm A1}$ from~0 to~1.  The phase diagrams were initially constructed versus~$t=T/T_{{\rm N}J}$ but the abscissa was then converted to $t_{\rm A} = T/T_{\rm N}$ using Eq.~(\ref{Eq:tADef}).  The $t=0$ transition fields obtained from Fig.~\ref{Fig:T0PhaseDiagramS1272} are included in Fig.~\ref{Fig:PhaseDiagram_S12fJm1hA10}.  For $h_{\rm A1}=0$ the phase diagram contains no $z$-axis-aligned AFM phase because for any finite field the ordered moments flop to form a canted AFM phase, the spin-flop phase.  Even a rather small value $h_{\rm A1}=1/8$ gives rise to a SF phase in a large area of the phase diagram in Fig.~\ref{Fig:PhaseDiagram_S12fJm1hA10}(b) and a bicritical point appears where the AFM, SF, and PM phase lines meet.  With further increase of $h_{\rm A1}$, the SF phase region shrinks, as shown for $h_{\rm A1}=1/4$,~1/2, and~3/4 in Figs.~\ref{Fig:PhaseDiagram_S12fJm1hA10}(c)--\ref{Fig:PhaseDiagram_S12fJm1hA10}(e).  In addition, for $h_{\rm A1}=3/4$ a tricritical point occurs at $t_{\rm A}\approx 0.56$ separating second- and first-order AFM to PM transitions, as shown.  Finally, for $h_{\rm A1}=1$ in Fig.~\ref{Fig:PhaseDiagram_S12fJm1hA10}(f), the spin-flop region disappears and the tricritical point moves to lower temperature with respect to $T_{\rm N}$ compared to that for $h_{\rm A1}=3/4$.  We note that in Fig.~\ref{Fig:PhaseDiagram_S12fJm1hA10}(e) for $h_{\rm A1}=1$, the $T=0$ value of the AFM to PM transition field is larger than for lower $h_{\rm A1}$ values at higher temperatures, and is the same as the $T=0$ value of the SF to PM transition field in Fig.~\ref{Fig:PhaseDiagram_S12fJm1hA10}(a).

\begin{figure}
\includegraphics [width=3.in]{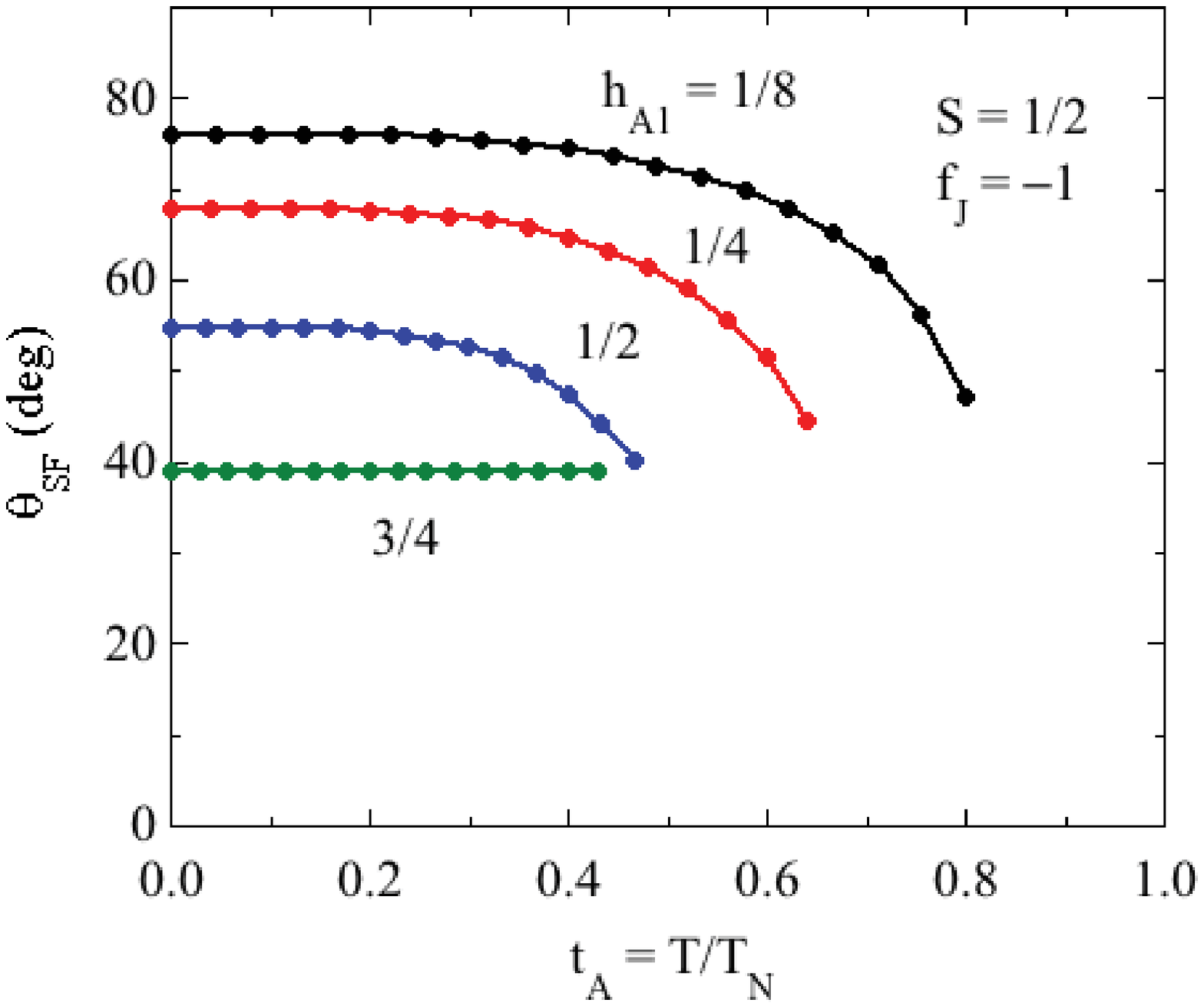}
\caption{(Color online) The angle $\theta_{\rm SF}$ that the ordered moments in the spin-flop phase make with the applied field along the $z$~axis on the first-order transition line between the AFM and SF phases in Fig.~\ref{Fig:PhaseDiagram_S12fJm1hA10} versus reduced temperature $t_{\rm A}$ for the same reduced anisotropy parameters~$h_{\rm A1}$ for which the phase diagrams in Figs.~\ref{Fig:PhaseDiagram_S12fJm1hA10}(b)--\ref{Fig:PhaseDiagram_S12fJm1hA10}(e) were constructed.}
\label{Fig:SpinFlopAngle_S12fJm1}
\end{figure}

In a spin-flop transition of an otherwise collinear antiferromagnet, the spins flop from alignment along the $z$~axis to what is generally thought to be an approximately perpendicular orientation.  An interesting question is how close to a $\theta = 90^\circ$ angle the moments in the SF phase make with the $z$~axis ($\theta_{\rm SF}$) on the (first-order) transition line between the AFM and SF phases.  Shown in Fig.~\ref{Fig:SpinFlopAngle_S12fJm1} are plots of $\theta_{\rm SF}$ versus reduced temperature~$t_{\rm A}$ for the parameters in the phase diagrams in Figs.~\ref{Fig:PhaseDiagram_S12fJm1hA10}(b)--\ref{Fig:PhaseDiagram_S12fJm1hA10}(e).  These data were obtained as part of the calculations required to construct the phase diagrams in Fig.~\ref{Fig:PhaseDiagram_S12fJm1hA10}.  One sees rather strong dependences of $\theta_{\rm SF}$ on both $t_{\rm A}$ and the anisotropy parameter~$h_{\rm A1}$.  Futhermore, the maximum angle of the moments from the $z$~axis on the transition line versus temperature depends strongly on $h_{\rm A1}$, varying from only about 40$^\circ$ for $h_{\rm A1}=3/4$ to about 77$^\circ$ for $h_{\rm A1}=1/8$.  Thus when a spin-flop transition occurs, the angle that the moments make with the $z$~axis is generally not close to 90$^\circ$.  According to Fig.~\ref{Fig:SpinFlopAngle_S12fJm1}, this discrepancy increases with increasing~$h_{\rm A1}$.

\subsection{Magnetization versus Field Isotherms for Fields Along the Easy Axis of Collinear Antiferromagnets}

\begin{figure*}
\includegraphics [width=3in]{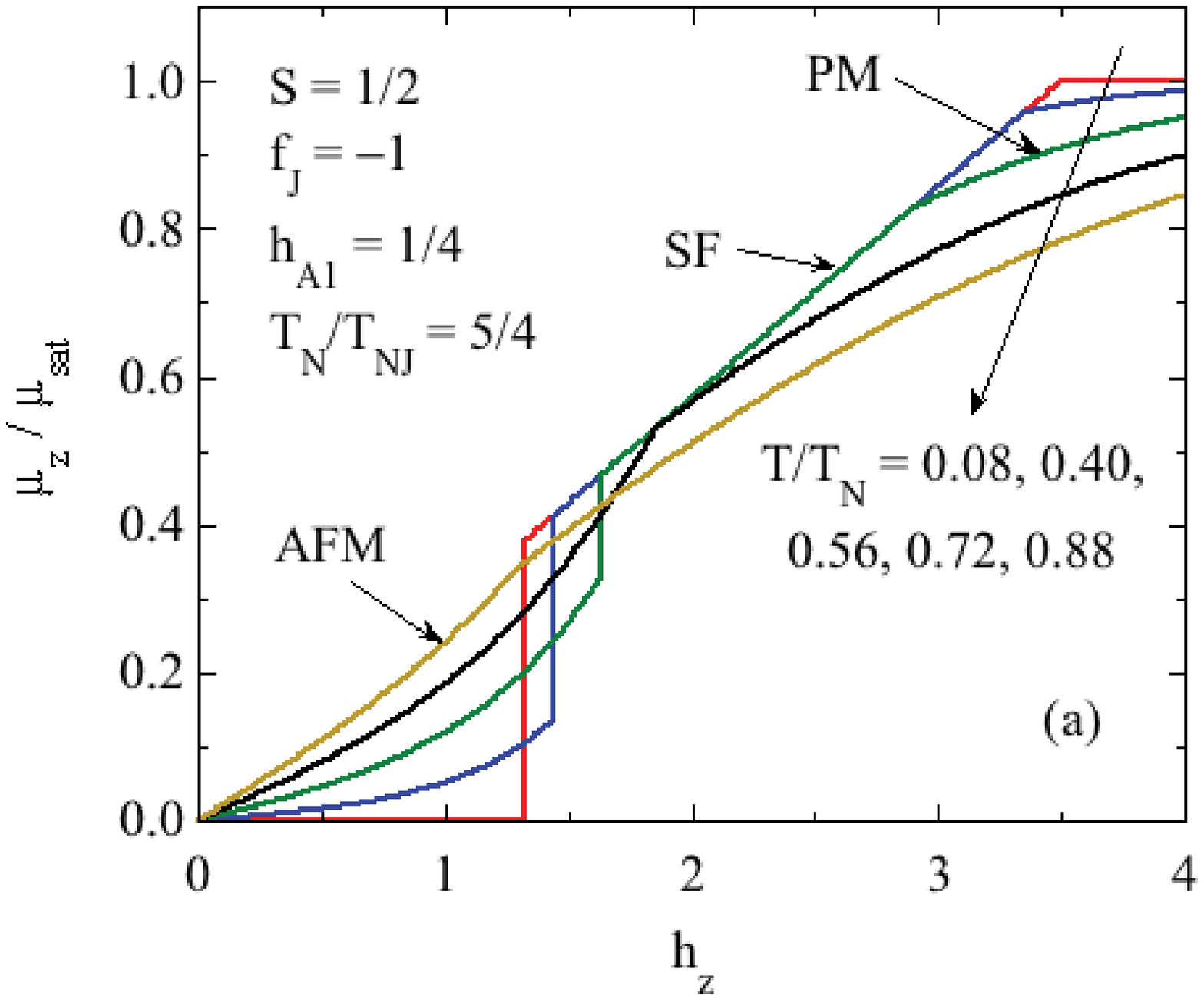}\includegraphics [width=3in]{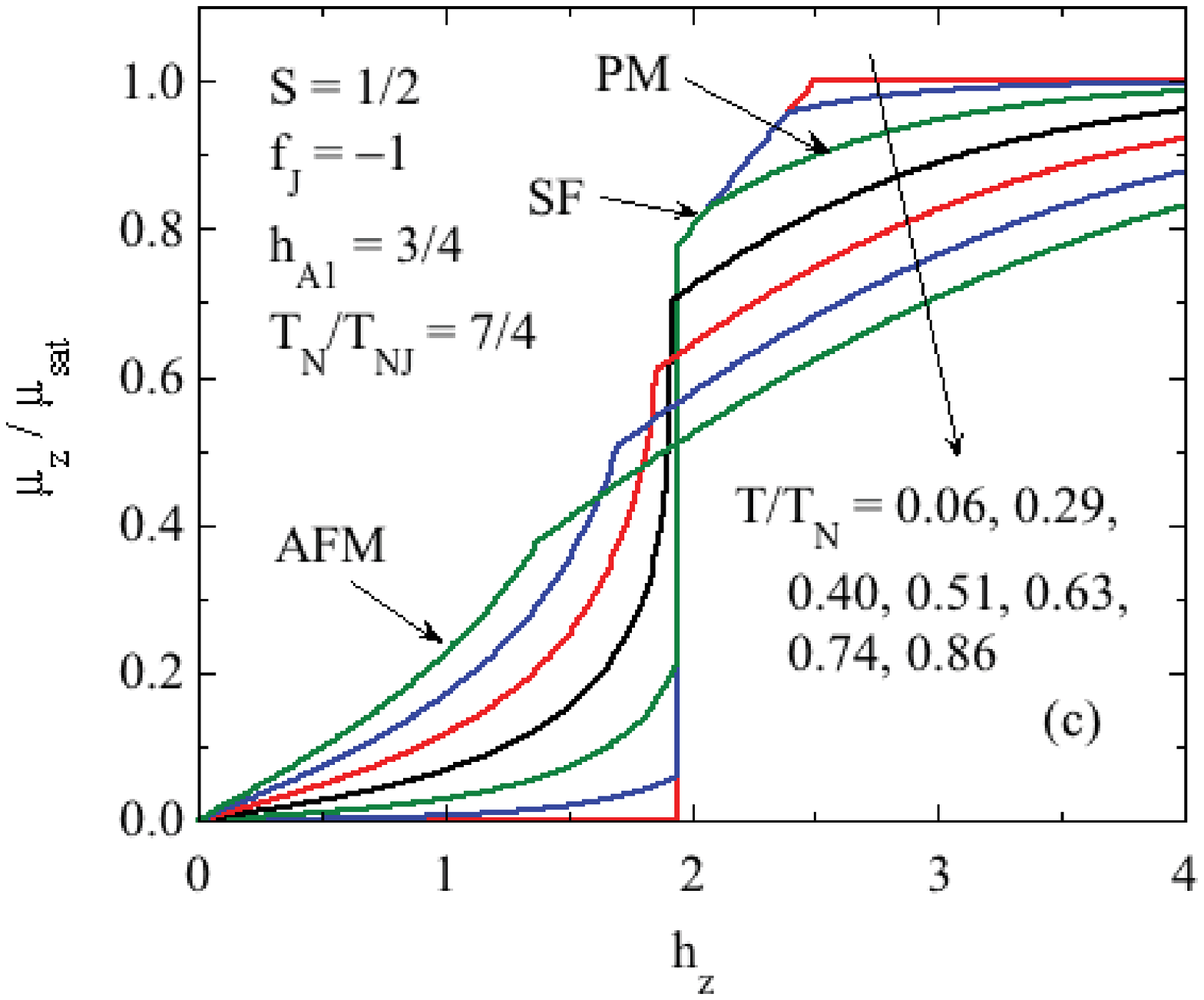}
\includegraphics [width=3in]{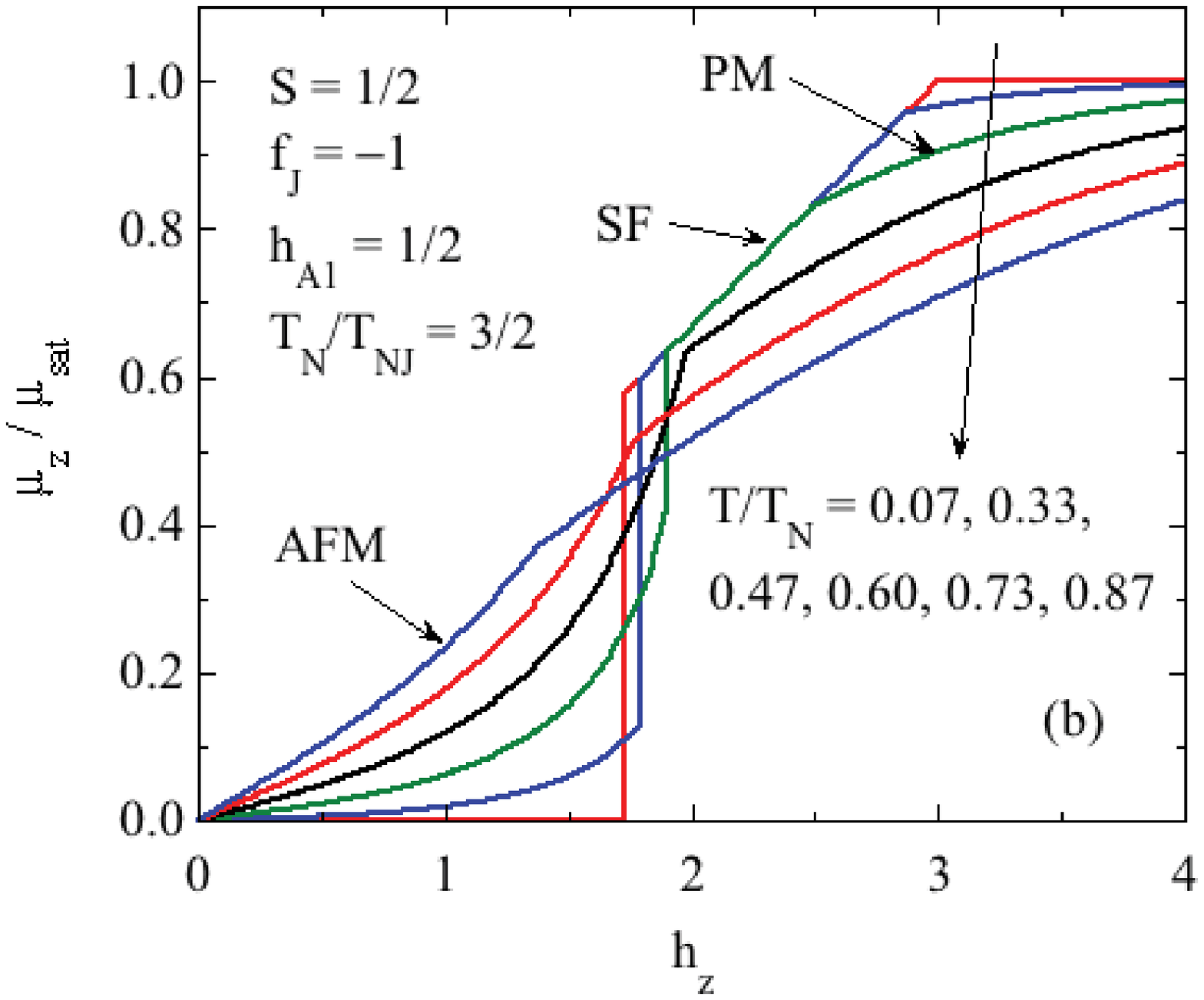}\includegraphics [width=3in]{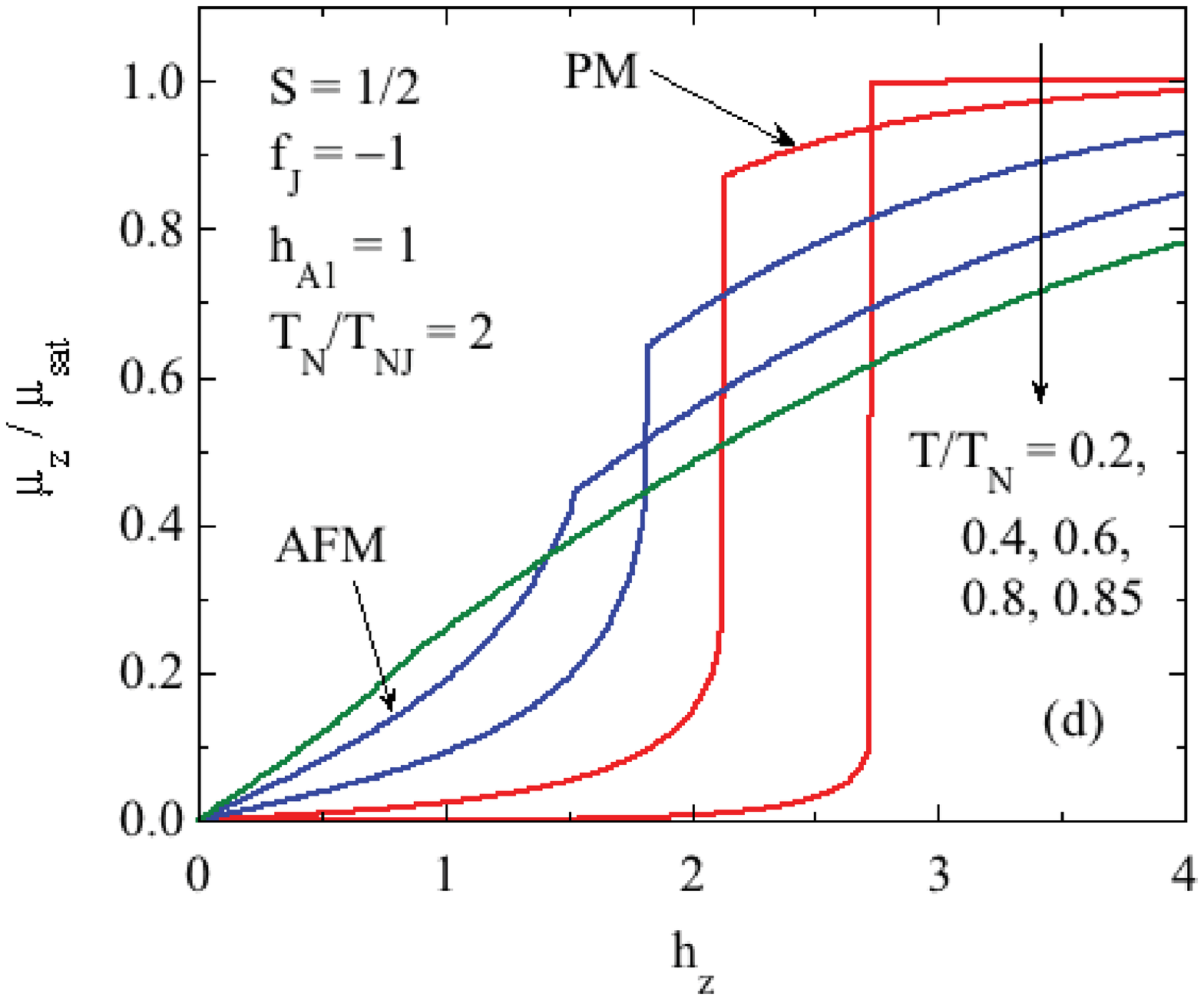}
\caption{(Color online) Reduced $z$-axis magnetic moment~$\bar{\mu}_z\equiv\mu_z/\mu_{\rm sat}$ versus reduced magnetic field $h_z = g\mu_{\rm B}H_z/k_{\rm B}T_{{\rm N}J}$ at the listed reduced temperatures~$t_{\rm A} = T/T_{\rm N}$ for spins~$S=1/2$, $f_J=-1$ and with reduced anisotropy parameters~$h_{\rm A1}$ equal to (a)~1/4, (b)~1/2, (c)~3/4, and (d)~1.  The SF to PM transitions are second order and the AFM to SF transitions are first order.  The AFM to PM transitions can be second order [(a), (b)], or either first or second order in different field ranges separated by a tricritical point~(c, d) (see the phase diagram in Fig.~\ref{Fig:PhaseDiagram_S12fJm1hA10}). }
\label{Fig:muzVShzAFMSFPMhA125fJm1S12s}
\end{figure*}

High-field magnetization versus field $M(H)$ isotherm measurements are basic to characterizing the magnetic properties of AFMs.  Here we utilize the above information specifying the conditions for phase transitions between the AFM, SF, and PM phases with fields along the easy $z$~axis to calculate magnetization versus field data at particular temperatures below the respective~$T_{\rm N}$.  These calculations allow direct comparisons to experimental $M_z(H)$ data on single crystals.

For anisotropy parameter $h_{\rm A1}=0$, for the spin-flop phase plots of $\bar{\mu}_{z{\rm SF}}$ versus~$h_z$ for a fixed temperature~$t_{\rm A}\equiv T/T_{\rm N}=1/2$ and a selection of anisotropy parameters $h_{\rm A1} = 0$ to~1 were presented in Fig.~\ref{Fig:MFT_Flop_muzVShS12fm1} for spins $S=1/2$ and~$S=7/2$, which included both the SF and PM regimes.  Plots of $\bar{\mu}_{z{\rm SF}}$ versus~$h_z$ for fixed $h_{\rm A1}=1/2$ with different values of $t=T/T_{{\rm N}J}$ were presented in Fig.~\ref{Fig:muVsHtSFhA150fJm1S12} for $S=1/2$, 2, and~7/2.

The behaviors of $\bar{\mu}_z$ versus~$h_z$ for $S=1/2$ and $f_J=-1$ were calculated for a values of $t_{\rm A}$ from $\sim0.1$ to~0.9 and $h_{\rm A1}$ values in the range $1/4 \leq h_{\rm A1} \leq 1$, including the influence of phase transitions as applicable.  The calculations are shown in Fig.~\ref{Fig:muzVShzAFMSFPMhA125fJm1S12s}, where the first or second-order nature of the phase transitions are reflected in the field dependence of the magnetization.

\subsection{Phase Diagrams for Fields Perpendicular to the Easy Axis or Plane of Collinear or Planary Noncollinear Antiferromagnets}

\begin{figure}
\includegraphics [width=3.in]{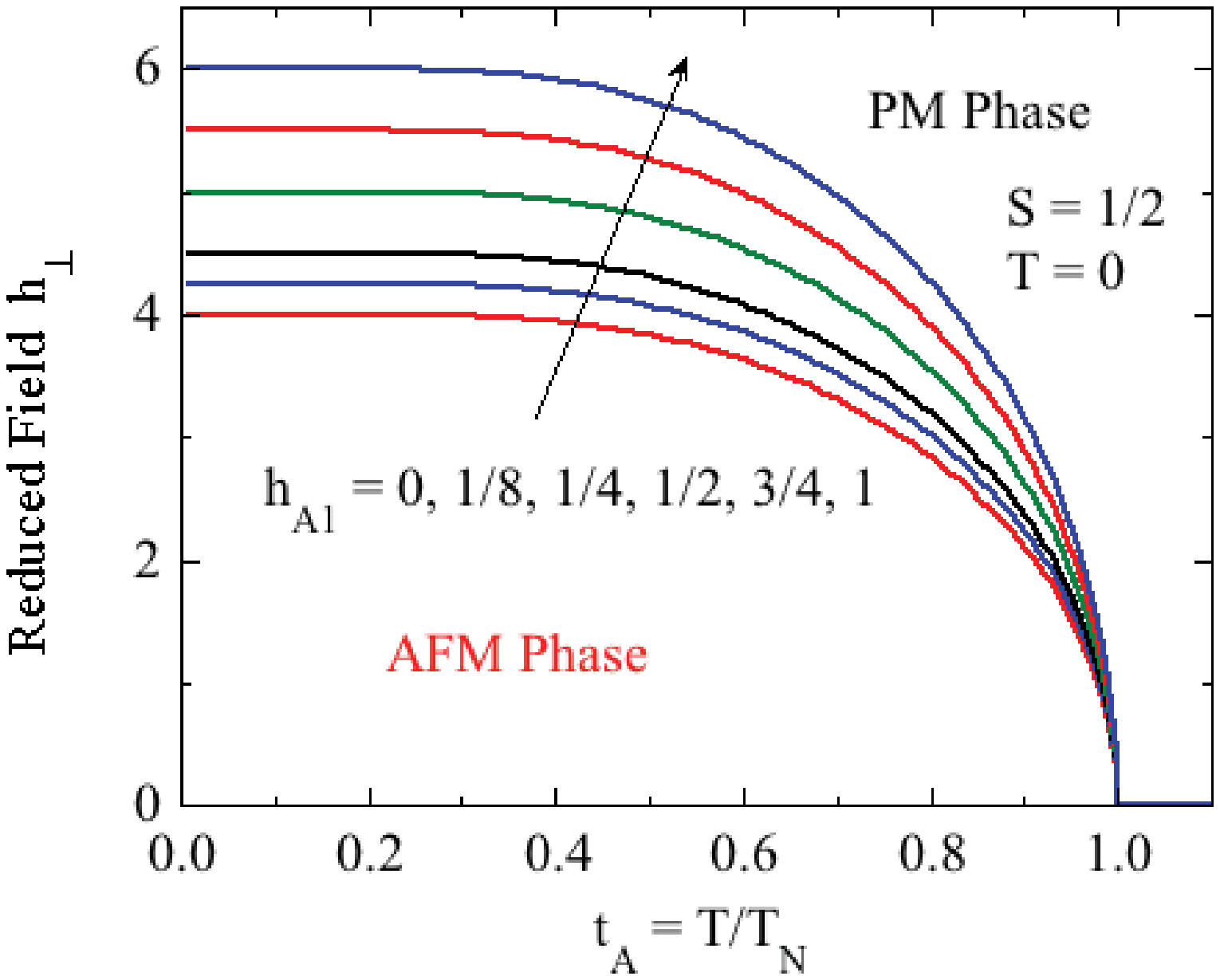}
\caption{(Color online) Phase diagram in the reduced perpendicular field $h_\perp$ versus reduced temperature~$t_{\rm A}$ plane for easy-axis or easy-plane collinear or planar noncollinear AFMs.  Plots of data obtained using Eq.~(\ref{Eq:hcPerpRed}) are shown for the same values of reduced anisotropy parameter $h_{\rm A1}$ for which the phase diagrams in Fig.~\ref{Fig:PhaseDiagram_S12fJm1hA10} were constructed.}
\label{Fig:hcPerpfJm1S12hA1xx}
\end{figure}

The critical field $h_{\rm c\perp AFM}$ dividing the canted AFM from the PM state of collinear or planar noncollinear AFMs versus reduced anisotropy~$h_{\rm A1}$ and $f_J$ parameters for fields perpendicular to the easy axis or plane of collinear or planar noncollinear AFMs is given in Eq.~(\ref{Eq:hcPerpRed}).  Plots of  $h_{\rm c\perp AFM}$ versus~$t_{\rm A}$ are shown in Fig.~\ref{Fig:hcPerpfJm1S12hA1xx} for the same values of $h_{\rm A1}$ for which the phase diagrams in Fig.~\ref{Fig:PhaseDiagram_S12fJm1hA10} were constructed.  From a comparison of the two figures, one sees that for each value of $h_{\rm A1}>0$, the $h_{\rm c\perp AFM}(t_{\rm A})$ value in Fig.~\ref{Fig:hcPerpfJm1S12hA1xx} lies at a higher field than the maximum transition field in Fig.~\ref{Fig:PhaseDiagram_S12fJm1hA10} at the same temperature.

\section{\label{Sec:Summary} Summary}

The main purpose of this work is to enable an estimate of the amount of uniaxial or planar anisotropy that exists in an otherwise isotropic Heisenberg spin system to be made from experimental magnetic susceptibility and/or high-field magnetization data.  The systems described contain identical crystallographically-equivalent spins.  Another important goal was to provide a classical description of magnetic anisotropy of quantum $S=1/2$ systems for which quantum uniaxial $DS_z^2$ single-ion anisotropy is not applicable.  In this paper the anisotropy is quantified by the fundamental reduced anisotropy parameter~$h_{\rm A1}$ in Eq.~(\ref{Eq:hA1Def}) which depends on~$S$ and the unreduced anisotropy field $H_{\rm A1}$, normalized by the N\'eel temperature in the absence of anisotropy~$T_{{\rm N}J}$, but not on the temperature~$T$\@. The $T$ dependence is included via the $T$ dependence of the reduced ordered and/or field-induced moment~$\bar{\mu}$ in Eq.~(\ref{Eq:HA0}).  The present treatment is strictly valid for local-moment antiferromagnets but not for itinerant ones.

There are several ways to extract $h_{\rm A1}$ from experimental data for single crystals of local-moment collinear antiferromagnets with uniaxial or planar anisotropy.  Indeed, if one has single-crystal low-field magnetic susceptibility versus temperature data as well as high-field magnetization isotherm data, this parameter is overdetermined and one can compare the values obtained from analyses of the respective data sets.  Since $g$~anisotropy is not included in the present treatment, the single-spin Curie constant~$C_1$ in the Curie-Weiss law~(\ref{Eqs:CWLaw}) is the same for fields parallel and perpendicular to the easy axis or easy plane for the known value of~$S$.  However, $g$~anisotropy for the AFM and PM phases is easily accomplished by substituting the appropreate values of $g_\alpha$ for~$g$ in the expression for the Curie constant if the values of $g_\alpha$ are known from independent measurements such as electron spin resonance.

\subsection{Analysis of Single-Crystal Magnetic Susceptibility Data}

An easy way to determine~$h_{\rm A1}$ is to measure the anisotropy of the Weiss temperature $\theta_{\rm p}$ in the Curie-Weiss law~(\ref{Eqs:CWLaw}) for the paramagnetic susceptibility at $T\geq T_{\rm N}$ of single crystals.  Here we only consider uniaxial $z$-axis anisotropy, since $xy$-plane anisotropy gives the same expression for $h_{{\rm A}1}$.  From Eqs.~(\ref{Eqs:ChiabAxAnis}) and~(\ref{Eqs:chicA}), respectively, the Weiss temperatures in the Curie-Weiss law for the $xy$~plane and $z$-axis field directions at temperatures $T\geq T_{\rm N}$ are
\bse
\bea
\theta_{{\rm p}xy} &=& \theta_{{\rm p}J},\label{Eq:thetapxy}\\*
\theta_{{\rm p}z} &=& \theta_{{\rm p}J} + h_{\rm A1}T_{{\rm N}J},
\eea
\ese
so
\be
\theta_{{\rm p}z} - \theta_{{\rm p}xy} =  h_{\rm A1}T_{{\rm N}J}.
\label{Eq:ThetaDiff}
\ee
Then using Eq.~(\ref{Eq:TNATNRatio}), one obtains
\be
\theta_{{\rm p}z} - \theta_{{\rm p}xy} = \frac{h_{\rm A1}T_{\rm N}}{1+h_{\rm A1}},
\ee
which allows one to easily solve for $h_{\rm A1}$ from the two measured Weiss temperatures and the measured N\'eel temperature~$T_{\rm N}$.

Another parameter of the theory is $f_J\equiv \theta_{{\rm p}J}/T_{{\rm N}J}$, the ratio of the Weiss and N\'eel temperatures due to exchange interactions alone.  This is not measurable directly but can be derived as follows.  Using Eqs.~(\ref{Eq:TNATNRatio}) and~(\ref{Eq:thetapxy}), one obtains
\be
\frac{\theta_{{\rm p}xy}}{T_{\rm N}} = \frac{\theta_{{\rm p}J}}{T_{{\rm N}J}(1+h_{\rm A1})} = \frac{f_J}{1+h_{\rm A1}},
\label{Eq:thetaTNratio}
\ee
from which $f_J$ can be obtained using $h_{{\rm A}1}$ from above.

Another expression useful for determining the values of $h_{\rm A1}$ and $f_J$ for collinear $z$-axis AFMs is Eq.~(\ref{Eq:ChiParPerpTNARatio}), which gives
\be
\frac{\chi_z(T_{\rm N})}{\chi_{xy}(T_{\rm N})} = 1 + \frac{h_{\rm A1}}{1-f_J},
\label{Eq:ChiParPerpTNARatio2}
\ee

Thus any of the combinations of two of Eqs.~(\ref{Eq:ThetaDiff}), (\ref{Eq:thetaTNratio}), and~(\ref{Eq:ChiParPerpTNARatio2}) can be used to solve for $h_{\rm A1}$ and~$f_J$.  Self-consistency can be checked by comparing the derived sets with each other, and/or with values derived from high-field  magnetization data for collinear AFMs as described in the following section.

\subsection{Analysis of High-Field $z$-Axis Magnetization Data}

According to Figs.~\ref{Fig:mubarVStA} and~\ref{Fig:OrderedMomentMFT_HA} for AFM and SF phases, respectively, for $T\lesssim 0.2\,T_{\rm N}$ the zero-field reduced ordered moment is nearly saturated at the value of unity, irrespective of the spin value.   It is this low-temperature range of collinear antiferromagnets aligned along the $z$~axis for which the high-field behavior is examined in this section.

For~$h_{\rm A1}>0$, according to Eq.~(\ref{Eq:hSF2}) and Figs.~\ref{Fig:T0muzVShzAFMSFS12fJm1hA1X} and~\ref{Fig:muzVShzAFMSFPMhA125fJm1S12s}(a)--\ref{Fig:muzVShzAFMSFPMhA125fJm1S12s}(c), a spin-flop (SF) transition from the AFM phase to the SF phase occurs at the reduced SF field
\bea
h_{\rm SF} = \frac{3}{S+1}\sqrt{h_{\rm A1}(1-f_J - h_{\rm A1})}.
\label{Eq:HcFlopDef4}
\eea
This transition is easy to see in $M_z(H)$ isotherm measurements because it is first order. In the SF phase, the magnetization is proportional to field according to Eq.~(\ref{Eq:muzvshzSF}), which we reproduce here
\bse
\be
\bar{\mu}_z(h_z,t\to0) = 
\begin{cases}
h_z/h_{\rm cSF}\quad (h_z \leq h_{\rm cSF})\\
1 \hspace{0.55in} (h_z \geq h_{\rm cSF}),
\end{cases}
\ee
where the SF critical field at which the SF phase undergoes a second-order transition to its PM phase is
\be
h_{\rm cSF} = \frac{3(1-f_J-h_{\rm A1})}{S+1}.
\label{Eq:HcFlopDef5}
\ee
\ese

From Eqs.~(\ref{Eq:HcFlopDef4}) and~(\ref{Eq:HcFlopDef5}), one has the ratio
\be
\frac{h_{\rm cSF}}{h_{\rm SF}} = \sqrt{\frac{1-f_J-h_{\rm A1}}{h_{\rm A1}}}.
\ee
Thus if both $h_{\rm cSF}$ and~$h_{\rm SF}$ can be measured at low temperatures, an additional equation that does not involve the spin~$S$ is available to solve for $f_J$ and $h_{\rm A1}$.

For $h_z<h_{\rm cSF}$, the reduced single-spin susceptibility $\bar{\chi}_{z{\rm SF}}$ for the spin-flop phase is given by Eq.~(\ref{Eq:chibarzSF}) as
\be
\bar{\chi}_{z{\rm SF}} \equiv \frac{\chi_{z{\rm SF}} T_{{\rm N}J}}{C_1} =\frac{1}{1-f_J-h_{\rm A1}},
\label{Eq:chiRedDef3}
\ee
where the single-spin Curie constant given in Eq.~(\ref{Eq:CurieConst2}) is assumed to be known from the fit of the high-temperature susceptibility by the Curie-Weiss law, and $\chi_{z{\rm SF}}$ is often measurable at fields above $h_{\rm SF}$ if the SF transition is observed.

\subsection{Analysis of High-Field Perpendicular Magnetization Data}

The present section discusses the magnetic response to high fields applied perpendicular to the easy axis or plane of a collinear or planar noncollinear antiferromagnet.  The reduced perpendicular susceptibility per spin $\bar{\chi}_{\rm\perp AFM}$ is given by Eq.~(\ref{Eq:barchiperp}) as
\be
\frac{\chi_{\rm \perp AFM} T_{{\rm N}J}}{C_1}  = \frac{1}{1-f_J+h_{\rm A1}}.
\ee
Comparing this equation with Eq.~(\ref{Eq:chiRedDef3}) shows that \mbox{$\chi_{\rm \perp AFM}<\chi_{z{\rm SF}}$,} with
\be
\frac{\chi_{\rm \perp AFM}}{\chi_{z{\rm SF}}} = \frac{1-f_J-h_{\rm A1}}{1-f_J+h_{\rm A1}}.
\label{ChiperpAFMSFRatio}
\ee

Finally, the critical field for the AFM to PM transition, if it occurs instead of a transition to a SF phase, is given by Eq.~(\ref{Eq:hcperT0}) as
\be
h_{\rm c\perp AFM}  = \frac{3(1-f_J+h_{\rm A1})}{S+1}.
\ee
This field is somewhat larger than $h_{\rm cSF}$ in Eq.~(\ref{Eq:HcFlopDef5}), the difference being
\be
h_{\rm c\perp AFM}-h_{\rm cSF} = \frac{3h_{\rm A1}}{S+1}.
\ee
This expression is very useful because it does not contain~$f_J$.  The drawback is that these two critical fields are often too large to measure except for materials with low~$T_{\rm N}$.  Alternatively, the ratio of the two critical fields is
\be
\frac{h_{\rm cSF}}{h_{\rm c\perp AFM}} = \frac{H_{\rm cSF}}{H_{\rm c\perp AFM}} = \frac{1-f_J-h_{\rm A1}}{1-f_J+h_{\rm A1}}.
\ee
The right side is the inverse of the respective ratio of the susceptibilities obtained from Eq.~(\ref{ChiperpAFMSFRatio}).

\subsection{Comparison of Classical Anisotropy with Quantum $DS_z^2$ Anisotropy Predictions}

Finally we compare the predictions of the present work for $T_{\rm N}$ and~$\theta_{\rm p}$ with those for quantum $DS_z^2$ anisotropy~\cite{Johnston2017}.  In the present case, the N\'eel temperature is simply described by Eq.~(\ref{Eq:TNATNRatio}) as
\be
T_{\rm N} = T_{{\rm N}J}(1+h_{\rm A1}),
\label{Eq:TNClass22}
\ee
which is a linear function of $h_{\rm A1}$ irrespective of its value.  However, for $-DS_z^2$ anisotropy, where a positive sign of $D$ is defined such that $z$-axis collinear AFM ordering is favored over $xy$-plane ordering, and with $d\equiv D/k_{\rm B}T_{{\rm N}J}$, one obtains a nonlinear dependence of $T_{\rm N}$ on~$d$.  On the other hand, for small~$d$ one obtains \cite{Johnston2017}
\be
T_{\rm N} = T_{{\rm N}J}\left[1+\frac{d(2S-1)(2S+3)}{15}\right].
\label{Eq:TNDSz2}
\ee
In contrast to Eq.~(\ref{Eq:TNClass22}), this linear dependence on~$d$ also depends explicitly on~$S$ for $S\geq1$. Comparison of Eqs.~(\ref{Eq:TNClass22}) and~(\ref{Eq:TNDSz2}) indicates that for weak anisotropy one can relate the anisotropy parameters in the present classical anisotropy model to that in the quantum $-DS_z^2$ model for $S\geq1$ by
\be
h_{\rm A1} = \frac{d(2S-1)(2S+3)}{15}.
\label{Eq:hA1tod}
\ee

Similarly, the Weiss temperature in the Curie-Weiss law with the field applied along the easy axis of a uniaxial antiferromagnet is given by Eq.~(\ref{Eq:thetapJ+thetapA}) as
\be
\theta_{\rm p} = \theta_{{\rm p}J} + T_{{\rm N}J}h_{\rm A1}.
\ee
In the case of uniaxial $DS_z^2$ anisotropy one also obtains a linear dependence on $d$ given by \cite{Johnston2017}
\be
\theta_{\rm p} = \theta_{{\rm p}J} + T_{{\rm N}J}\frac{d(2S-1)(2S+3)}{15},
\ee
where here again the second term depends on $S$, is zero for $S=1/2$, and gives the same correspondence as in Eq.~(\ref{Eq:hA1tod}).

\acknowledgments

I thank N. S. Sangeetha for helpful discussions.  This research was supported by the U.S. Department of Energy, Office of Basic Energy Sciences, Division of Materials Sciences and Engineering.  Ames Laboratory is operated for the U.S. Department of Energy by Iowa State University under Contract No.~DE-AC02-07CH11358.

\appendix

\section{\label{Sec:UnifiedMFT} Unified Molecular-Field Theory in the Absence of Anisotropy}

Here we review the properties of Heisenberg AFMs within the context of the unified MFT \cite{Johnston2012, Johnston2015, Johnston2017} in the absence of any type of anisotropy that are needed for the theoretical development in the presence of classical anisotropy fields.  All spins are assumed to be identical and crystallographically equivalent.

\subsection{Curie-Weiss Law}

The  Curie-Weiss law for the magnetic susceptibility~$\chi_\alpha$ in the paramagnetic (PM) state in the $\alpha$ principal-axis direction at temperatures $T \geq T_{\rm N}$, where $T_{\rm N}$ is N\'eel temperature resulting from the combined influences of the anisotropy and Heisenberg exhange interactions, is written for a representative spin by
\bse
\label{Eqs:CWLaw}
\be
\chi_\alpha = \frac{C_1}{T-\theta_{\rm p\alpha}},
\ee
where the Weiss temperature~$\theta_{\rm p\alpha}$ depends in general on~$\alpha$,
\be
C_1 = \frac{g^2S(S+1)\mu_{\rm B}^2}{3k_{\rm B}}
\label{Eq:CurieConst2}
\ee
\ese
is the single-spin Curie constant, $g$ is the spectroscopic splitting factor ($g$~factor), $\mu_{\rm B}$ is the Bohr magneton and $k_{\rm B}$ is Boltzmann's constant.   For simplicity it is assumed in this paper that the $g$~factor is isotropic.  For moments that are aligned along a principal axis~$\alpha$, $g$~can be replaced by a variable $g_\alpha$ in the respective equations.  Here we consider isotropic Weiss temperatures arising from exchange interactions only, denoted as $\theta_{{\rm p}J}$.

\subsection{Exchange Field}

In MFT, one replaces the sum of the Heisenberg exchange interactions acting on a representative central spin~$i$ by an effective magnetic field called the Weiss molecular field or ``exchange field'' ${\bf H}_{{\rm exch}i}$ and treats it as an applied field where the exchange energy $E_{{\rm exch}\,i}$ for spin~$i$ is 
\be
E_{{\rm exch}\,i} = -\vec{\mu}_i\cdot{\bf H}_{{\rm exch}i}.
\label{Eq:EexchtoHexch}
\ee
Taking into account the exchange interactions of $\vec{\mu}_i$ with all neighbors $\vec{\mu}_j$ with which it interacts, the exchange field is given in general by 
\be
{\bf H}_{{\rm exch}\,i} = -\frac{1}{g^2\mu_{\rm B}^2}\sum_j J_{ij} \vec{\mu}_{j},\\*
\label{Eq:HexchiDef}
\ee
where $J_{ij}$ is the Heisenberg exchange interaction between spins~$i$ and~$j$ and a positive (negative) value corresponds to an AFM (ferromagnetic FM) interaction.  Since all magnetic moments are assumed to be identical and in crystallographically equivalent positions in the lattice, each spin has the same local exchange field in {\bf H = 0}, irrespective of the orientation of the spin with respect to those of the other spins in the system.  The component of ${\bf H}_{{\rm exch}\,i}$ in the direction of $\vec{\mu}_{i}$ is
\be
H_{{\rm exch}\,i} = \hat{\mu}_i\cdot {\bf H}_{{\rm exch}\,i} = -\frac{1}{g^2\mu_{\rm B}^2}\sum_j J_{ij}\mu_j\cos\alpha_{ji},
\label{Eq:HexchDef3}
\ee
where $\alpha_{ji}$ is the angle between $\vec{\mu}_{j}$ and $\vec{\mu}_{i}$ when $H\neq0$. If $H=0$ we denote these angles instead by $\phi_{ji}$. 

In the ordered magnetic state in {\bf H} = 0, the component of the local ${\bf H}_{{\rm exch}\,i0}$ in the direction of $\vec{\mu}_i$, and also its magnitude, is
\be
H_{{\rm exch}0} = -\frac{\mu_0}{g^2\mu_{\rm B}^2}\sum_j J_{ij}\cos\phi_{ji},
\label{Eq:Hexch0Def3}
\ee
where we dropped the subscript~$i$ because of the equivalence of each moment in $H=0$ and $\mu_0$ is the magnitude of the $T$-dependent ordered moment in {\bf H} = 0 which is the same for all spins because of their crystallographic equivalence.

\subsection{Antiferromagnetic Ordering}

For $H\to0$, the AFM ordering temperature $T_{{\rm N}J}$ and the Weiss temperature $\theta_{{\rm p}J}$ in the Curie-Weiss~(\ref{Eqs:CWLaw}) law due to exchange interactions alone are respectively given by
\bse
\label{Eqs:TNqpJ}
\bea
T_{{\rm N}J} &=& -\frac{S(S+1)}{3k_{\rm B}}\sum_j J_{ij}\cos\phi_{ji},\label{Eq:TmGeneral}\\*
\theta_{{\rm p}J} &=& -\frac{S(S+1)}{3k_{\rm B}}\sum_j J_{ij}\label{Eq:WeissTemp},
\eea
\ese
where  the sums are over all neighbors~$j$ of a given central spin~$i$, the subscript $J$ on the left sides signifies that these quantities arise from exchange interactions only, and $\phi_{ji}$ is the angle between moments $j$ and~$i$ in the AFM structure at $T<T_{{\rm N}J}$ with $H=0$.  The ratio~$f_J$ is defined as
\be
f_J\equiv \frac{\theta_{{\rm p}J}}{T_{{\rm N}J}} = \frac{\sum_j J_{ij} }{\sum_j J_{ij}\cos\phi_{ji}},
\label{Eq:fRatioDef}
\ee
where to obtain the second equality Eqs.~(\ref{Eqs:TNqpJ}) were used.  For a FM, $\phi_{ji}=0$ for all~$j$, and hence $f_J=1$.  For AFMs, at least one of the $J_{ij}$ must be positive (AFM interaction) and at least one of the $\phi_{ji}\neq0$, leading to $f_J<1$.  Thus within MFT, for AFM ordering one has
\be
-\infty<f_J<1.
\label{Eq:fRange}
\ee

By comparing Eqs.~(\ref{Eq:Hexch0Def3}) and~(\ref{Eq:TmGeneral}), one can write the zero-field exchange field ${\bf H}_{{\rm exch}0}$ seen by each magnetic moment $\vec{\mu}_{i0}$ as
\bea
{\bf H}_{{\rm exch}\,i0} &=& \frac{3k_{\rm B}T_{{\rm N}J}\vec{\mu}_{i0}}{g^2\mu_{\rm B}^2S(S+1)} = \frac{T_{{\rm N}J}}{C_1}\vec{\mu}_{i0},\label{Eqs:Hexchi050}\\*
H_{{\rm exch}0} &=& \frac{3k_{\rm B}T_{{\rm N}J}\mu_0}{g^2\mu_{\rm B}^2S(S+1)} =\frac{T_{{\rm N}J}}{C_1}{\mu}_0,
\nonumber
\eea
where the single-spin Curie constant $C_1$ is defined in Eq.~(\ref{Eq:CurieConst2}).

Within MFT the thermal-average ordered and/or field-induced magnetic moment $\vec{\mu}_i$ is in the direction of its local magnetic induction ${\bf B}_i = {\bf H}_{{\rm exch}i}+{\bf H}$.  When a classical anisotropy field is present, one adds ${\bf H}_{{\rm A}i}$ to this. The magnitude $\mu_i$ of $\vec{\mu}_i$ in that direction is determined using the Brillouin function $B_S(y)$ according to the self-consistency requirement
\begin{subequations}
\label{Eq:BS(y)}
\be
\mu_i = \mu_{\rm sat}B_S(y_i)
\label{Eq:muvsBrill}
\ee
where
\be
y_i = \frac{g\mu_{\rm B}B_i}{k_{\rm B}T}
\label{Eq:yDef}
\ee
\end{subequations}
and $B_i$ is the component of ${\bf B}_i$ in the direciton of $\vec{\mu}_i$.  Our unconventional definition of the Brillouin function is~\cite{Reif1965}
\bse 
\label{Eqs:BS}
\be
B_S(y) = \frac{1}{2S} \left\{(2S+1)\coth\left[(2S+1)\frac{y}{2}\right]-\coth\left(\frac{y}{2}\right)\right\}
\label{Eq:BrillouinFunction},
\ee
for which the lowest-order Taylor-series expansion about $y=0$ is
\be
B_S(y) = \frac{(S+1)y}{3} + {\cal O}(y^3).
\label{Eq:BSyTaylor}
\ee
The derivative of $B_S(y)$ is
\bea
&&B_S^\prime(y) \equiv \frac{dB_S(y)}{dy}\label{Eq:dBSy0}\\*
 &&=  \frac{1}{4S}\bigg\{{\rm csch}^2\left(\frac{y}{2}\right) -\ (2S+1)^2{\rm csch}^2\left[(2S+1)\frac{y}{2}\right]\bigg\}.\nonumber
\eea
From Eq.~(\ref{Eq:BSyTaylor}), the lowest-order term of a Taylor-series expansion of $B_S^\prime(y)$ about $y=0$ is
\be
B_S^\prime(y) = \frac{S+1}{3} + {\cal O}(y^2).
\label{Eq:BSprimeExpand}
\ee
\ese

We define the reduced temperature~$t$ and reduced zero-field ordered moment~$\bar{\mu}_0(t)$ in $H=0$ as
\bse
\label{Eqs:tmu0musat}
\bea
t &=& \frac{T}{T_{{\rm N}J}},\label{Eq:tDef}\\*
\bar{\mu}_0 &=& \frac{\mu_0}{gS\mu_{\rm B}}\label{Eq:barmu0Def},
\eea
where the saturation moment~$\mu_{\rm sat}$ of each spin is
\be
\mu_{\rm sat} = gS\mu_{\rm B}.
\label{Eq:musat}
\ee
\ese
Using Eq.~(\ref{Eq:barmu0Def}), one can write the magnitude of the zero-field exchange field in Eq.~(\ref{Eqs:Hexchi050}) as
\be
H_{{\rm exch}0} = \frac{3k_{\rm B}T_{{\rm N}J}\bar{\mu}_0}{g\mu_{\rm B}(S+1)}.
\label{Eq:Hexch02}
\ee

For $H=0$, with $B_i = H_{{\rm exch}0}$ in Eq.~(\ref{Eq:Hexch02}), Eq.~(\ref{Eq:muvsBrill}) for calculating the ordered moment versus~$T$ in $H = 0$ becomes
\be
\bar{\mu}_0 = B_S(y_0), \quad{\rm with}\quad  y_0 =\frac{3\bar{\mu}_0}{(S+1)t}.
\label{Eq:mubar0}
\ee
This zero-field expression is valid within MFT for a FM and any type of AFM containing identical crystallographically-equivalent spins.  The total derivative $d\bar{\mu}_0/dt$ is obtained from Eq.~(\ref{Eq:mubar0}) as
\be
\frac{d\bar{\mu}_0}{dt} = -\frac{\bar{\mu}_0}{t\left[\frac{(S+1)t}{3B_S^\prime(y_0)} - 1\right]},
\label{Eq:dbarmu0dt}
\ee
where $\bar{\mu}_0(t)$ is obtained by numerically solving Eq.~(\ref{Eq:mubar0}) and the $B_S(y)$ and $B_S^\prime(y)$ functions are given in Eqs.~(\ref{Eqs:BS}).

\subsection{Internal Energy and Heat Capacity for AFM Ordering in Zero Field}

The internal energy per spin $U_{\rm mag}$ in zero field is given for any AFM containing identical crystallographically-equivalent spins by
\be
U_{\rm exch0} = -\frac{1}{2}\mu_0H_{\rm exch0},
\label{Eq:Umag0hA10}
\ee
where the factor of 1/2 compensates for the fact that $H_{\rm exch0}$ arises from exchange interactions between a central spin and each of its interacting neighbors, and hence arises from pairs of spins, whereas $U_{\rm mag}$ is per spin.  Writing $U_{\rm mag}$ in reduced parameters using Eqs.~(\ref{Eqs:tmu0musat}) and using Eq.~(\ref{Eq:Hexch02}) gives
\be
\frac{U_{\rm exch0}}{k_{\rm B}T_{{\rm N}J}} = -\frac{3S\bar{\mu}_0^2}{2(S+1)}.
\label{Eq:UmagAFMH0}
\ee
The magnetic heat capacity per spin is given in reduced units by
\be
\frac{C_{\rm mag}}{k_{\rm B}} = \frac{d(U_{\rm exch0}/k_{\rm B}T_{{\rm N}J})}{dt} = -\left[\frac{3S\bar{\mu}_0}{(S+1)}\right]\frac{d\bar{\mu}_0}{dt},
\label{Eq:Cmag0AFM}
\ee
where $\bar{\mu}_0(t)$ is obtained by solving Eq.~(\ref{Eq:mubar0}) and $d\bar{\mu}_0/dt$ is given by Eq.~(\ref{Eq:dbarmu0dt}).

\subsection{Magnetization in the Paramagnetic State}

Let the applied field be in the $\alpha$ principle-axis direction.  In the paramagnetic state above $T_{{\rm N}J}$, the thermal average of each magnetic moment is in the direction of the applied field.  Hence $\alpha_{ji}=0$ in Eq.~(\ref{Eq:HexchDef3}) and one obtains 
\be
H_{{\rm exch}\,\alpha} = -\frac{\mu_\alpha}{g^2\mu_{\rm B}^2}\sum_j J_{ij}  = -\frac{\bar{\mu}_\alpha S}{g\mu_{\rm B}}\sum_j J_{ij},
\label{Eq:Hexchipara}
\ee
where we dropped the subscript~$i$ because all induced moments are equivalent in the PM state.  As in Eq.~(\ref{Eq:barmu0Def}), we define the reduced moment in the $\alpha$ direction as
\be
\bar{\mu}_\alpha \equiv \frac{\mu_\alpha}{gS\mu_{\rm B}}.
\label{Eq:barmualphaDef}
\ee
Then using Eq.~(\ref{Eq:WeissTemp}), Eq.~(\ref{Eq:Hexchipara}) becomes
\bse
\be
H_{{\rm exch}\,\alpha} = \frac{3\bar{\mu}_\alpha k_{\rm B}\theta_{{\rm p}J}}{g\mu_{\rm B}(S+1)},
\label{Eq:Hexch:z0}
\ee
so
\be
\frac{g\mu_{\rm B}H_{{\rm exch}\,\alpha}}{k_{\rm B}T} = \frac{3\bar{\mu}_\alpha\theta_{{\rm p}J}}{(S+1)T}.
\label{Eq:Hexch:z}
\ee
\ese
Including the applied field~$H_\alpha$ in $B_i$, Eqs.~(\ref{Eq:BS(y)}) give
\be
\bar{\mu}_\alpha = B_S\left[\frac{3\bar{\mu}_\alpha\theta_{{\rm p}J}}{(S+1)T} + \frac{g\mu_{\rm B}H_\alpha}{k_{\rm B}T}\right].
\label{Eq:muvsBrill2}
\ee

For $H_\alpha\to0$, using Eq.~(\ref{Eq:WeissTemp}) and the first-order Taylor series expansion in Eq.~(\ref{Eq:BSyTaylor}), Eq.~(\ref{Eq:muvsBrill2}) becomes
\bse
\be
\mu_\alpha = \frac{C_1 H_\alpha}{T-\theta_{{\rm p}J}},
\ee
where $C_1$ is the single-spin Curie constant in Eq.~(\ref{Eq:CurieConst2}), which yields an isotropic Curie-Weiss law (\ref{Eqs:CWLaw}) given by
\be
\chi_{\rm PM\alpha}(T) = \frac{\mu_\alpha}{H_\alpha} = \frac{C_1}{T-\theta_{{\rm p}J}},
\label{Eq:ChialphaPM}
\ee
yielding
\be
\chi_{\rm PM\alpha}(T_{{\rm N}J}) = \frac{C_1}{T_{{\rm N}J}-\theta_{{\rm p}J}}.
\label{Eq:chiTNJHeis}
\ee
\ese
We define the reduced magnetic field $h_\alpha$ in the $\alpha$ principal-axis direction as
\be
h_\alpha \equiv \frac{g\mu_{\rm B}H_\alpha}{k_{\rm B}T_{{\rm N}J}}.
\label{Eq:barhDef}
\ee
Then in reduced variables Eq.~(\ref{Eq:muvsBrill2}) becomes
\be
\bar{\mu}_\alpha = B_S\left[\frac{3\bar{\mu}_\alpha f_J}{(S+1)t} + \frac{h_\alpha}{t}\right],\qquad (t \geq 1)
\label{Eq:muvsBrill3}
\ee
where the ratio $f_J = \theta_{\rm p}/T_{\rm N}$ is given in terms of the exchange constants and the magnetic structure in Eq.~(\ref{Eq:fRatioDef}).  Equation~(\ref{Eq:muvsBrill3}) must be solved numerically for $\bar{\mu}_\alpha$ for given values of $S$, $f_J$, $h_\alpha$ and $t$.

\subsection{\label{Sec:MperpNoHA} Magnetization of a Planar AFM in a Perpendicular Field}

To determine the perpendicular component~$\mu_\perp$ of a magnetic moment in a collinear or planar noncollinear AFM oriented in the $xy$~plane, the net torque $\vec{\tau}$ on a representative mmoment $\vec{\mu}_i$ is set to zero according to
\be
\vec{\tau} = \vec{\mu}_i \times {\bf H}_{{\rm exch}\,i} + \vec{\mu}_i \times {\bf H} = 0.
\label{Eq:tauDef}
\ee
The magnetic moment vectors are written in spherical coordinates as
\bse
\label{Eqs:muimujExpand}
\bea
\vec{\mu}_i &=& \mu\big[\sin\theta(\cos\phi_i\,\hat{\bf i} + \sin\phi_i\,\hat{\bf j}) + \cos\theta\,\hat{\bf k}\big]\label{Eq:HiHMuTEqs}\\*
\vec{\mu}_j &=& \mu\big[\sin\theta(\cos\phi_j\,\hat{\bf i} + \sin\phi_j\,\hat{\bf j}) + \cos\theta\,\hat{\bf k}\big]\nonumber\\*
&=& \mu\Big\{\sin\theta\Big[(\cos\phi_i\cos\phi_{ji} - \sin\phi_i\sin\phi_{ji})\,\hat{\bf i}\nonumber \\*
&&\hspace{0.2in}+ (\sin\phi_i\cos\phi_{ji} + \cos\phi_i\sin\phi_{ji})\,\hat{\bf j}\Big] + \cos\theta\,\hat{\bf k}\Big\},\nonumber
\eea
where in the last equality we used trig identities with
\be
\phi_{ji} = \phi_j - \phi_i.
\ee
\ese
Using the definition of the exchange field in Eq.~(\ref{Eq:HexchiDef}) and the requirement that $\sum_jJ_{ij}\sin\phi_{ji}=0$ for stability of an AFM structure~\cite{Johnston2015}, the first term in Eq.~(\ref{Eq:tauDef}) is found to be 
\bea
\vec{\mu}_i \times {\bf H}_{{\rm exch}\,i} &=& -\frac{3\bar{\mu}^2Sk_{\rm B}}{S+1}\sin\theta\cos\theta(T_{{\rm N}J}-\theta_{{\rm p}J}) \nonumber\\*
&& \hspace{0.5in} \times\ (\sin\phi_i\,\hat{\bf i} - \cos\phi_i\,\hat{\bf j}).\label{Eq:Term1}
\eea
Taking ${\bf H} = H_\perp\hat{\bf k}$, the second term in Eq.~(\ref{Eq:tauDef}) is
\be
\vec{\mu}_i \times {\bf H} = \bar{\mu}g\mu_{\rm B}S H_\perp\sin\theta(\sin\phi_i\,\hat{\bf i} - \cos\phi_i\,\hat{\bf j}).
\label{Eq:Term2}
\ee

Substituting Eqs.~(\ref{Eq:Term1}) and~(\ref{Eq:Term2}) into~(\ref{Eq:tauDef}) gives
\be
\frac{3\bar{\mu}k_{\rm B}}{S+1}\cos\theta(T_{{\rm N}J}-\theta_{{\rm p}J}) = g\mu_{\rm B}H_\perp.
\label{Eq:tauequalszero}
\ee
Using $\bar{\mu}\equiv \mu/(gS\mu_{\rm B})$ one obtains
\be
\frac{3k_{\rm B}}{g\mu_{\rm B}S(S+1)}\mu\cos\theta(T_{{\rm N}J}-\theta_{{\rm p}J}) = g\mu_{\rm B}H_\perp.
\label{Eq:tauequalszeroB}
\ee
Referring to Fig.~\ref{Fig:chiPerp2}, the perpendicular component $\mu_\perp$ of the induced magnetic moment of each spin is
\be
\mu_\perp = \mu\cos\theta,
\label{Eq:muzVsCosTheta}
\ee
where $\mu(T)$ is the magnitude of the ordered moment.  Then Eq.~(\ref{Eq:tauequalszeroB}) gives
\bse
\bea
\mu_\perp &=& \frac{C_1H_\perp}{T_{{\rm N}J}-\theta_{{\rm p}J}}\label{Eq:M(H)perp}\\*
&\equiv& \chi_{\perp J} H_\perp,\nonumber\\*
\chi_{\perp J} &\equiv& \frac{\mu_\perp}{H_\perp} = \frac{C_1}{T_{{\rm N}J}-\theta_{{\rm p}J}}.\label{Eq:ChiPerpJ}
\eea
\ese
The applies for fields~$H_\perp$ less than the critical field $H_{{\rm c\perp}J}(T)$ at which the moments become parallel and the system exhibits a second-order transition into the PM state.  From Eq.~(\ref{Eq:M(H)perp}), the critical field is given by
\be
H_{{\rm c\perp}J}(T) = \frac{\mu(T)}{\chi_{\perp J}},
\ee
where $\mu(T)$ is the ordered moment in the AFM state versus~$T$\@.

Comparing Eqs.~(\ref{Eq:ChiPerpJ}) and~(\ref{Eq:chiTNJHeis}) one sees that
\be
\chi_{\perp J}(T\leq T_{{\rm N}J}) = \chi_{{\rm PM}J}(T_{{\rm N}J}).
\ee
Thus $\chi_{\perp J}$ in the AFM state at $T\leq T_{{\rm N}J}$ is independent of $T$ with the value $\chi_{{\rm PM}J}$ of the PM state at $T=T_{{\rm N}J}$.

Dividing each side of Eq.~(\ref{Eq:tauequalszero}) by $k_{\rm B}T_{{\rm N}J}$ gives
\be
\bs
\frac{3\bar{\mu}\cos\theta}{S+1}\left(1-f_J\right) &= h_\perp,\\*
\frac{3\bar{\mu}\cos^2\theta}{(S+1)t}\left(1-f_J\right) &= \frac{h_\perp\cos\theta}{t}.
\end{split}
\label{Eq:mubarbarh}
\ee
The magnitude of the induced moment is
\bea
\bar{\mu} &=& B_S\left[\left(\frac{g\mu_{\rm B}}{k_{\rm B}T}\right)(H_{{\rm exch}\,i} + H_\perp\cos\theta)\right]\label{Eq:barmu}\\*
&=& B_S\left\{\frac{3\bar{\mu}}{(S+1)t}\big[1-(1-f_J)\cos^2\theta\big] + \frac{h_\perp\cos\theta}{t}\right\},\nonumber
\eea
where $H_\perp\cos\theta$ is the component of {\bf H} in the direction of each of the magnetic moments, the reduced field is $h_\perp\equiv g\mu_{\rm B}H_\perp/k_{\rm B}T_{{\rm N}J}$ from Eq.~(\ref{Eq:barhDef}) and the reduced temperature is $t\equiv T/T_{{\rm N}J}$ according to Eq.~(\ref{Eq:fRatioDef}).

Substituting the left-hand side of Eq.~(\ref{Eq:mubarbarh}) for $h_\perp\cos(\theta)/t$ into Eq.~(\ref{Eq:barmu}) and simplifying yields
\be
\bar{\mu}=B_S\left[\frac{3\bar{\mu}}{(S+1)t}\right].
\label{Eq:ordMoment}
\ee
This is identical to Eq.~(\ref{Eq:mubar0}) for determining $\bar{\mu}_0(t)$ with $H=0$.  Hence the ordered moment magnitude  is independent of field for $h_\perp$ less than the reduced perpendicular critical field $h_{\rm c\perp}$, which is given by the first of Eqs.~(\ref{Eq:mubarbarh}) with $\theta=0$ as 
\be
h_{\rm c\perp} = \frac{3\bar{\mu}(1-f_J)}{S+1},
\label{Eq:hcPerp}
\ee 
where the ordered reduced moment $\bar{\mu}$ is temperature dependent and hence so is $h_{\rm c\perp}$.

\end{document}